\documentclass[mnsc,nonblindrev]{informs3_no_remark} 
\OneAndAHalfSpacedXI % current default line spacing
\usepackage{natbib}
 \bibpunct[, ]{(}{)}{,}{a}{}{,}%
 \def\bibfont{\small}%

\usepackage{subfigure}
\usepackage{mwe}
\usepackage{lscape,booktabs,longtable}
\usepackage{amsmath}
\usepackage[pdftex,colorlinks=true,urlcolor=blue,citecolor=blue,pdfstartview=FitH]{hyperref}
\usepackage{comment}
\usepackage{enumerate}
\usepackage[ruled,vlined]{algorithm2e}
\usepackage{tikz}

\usepackage{enotez}
\let\footnote=\endnote

\newcommand{\E}[1]{{\mathrm{E}\left[#1\right]}}

\newcommand{\PP}[1]{\mathrm{P}\left( #1 \right)}
\newcommand{\Var}[1]{{\mathrm{Var}\left(#1\right)}}

\newcommand{\edit}{}

\TheoremsNumberedThrough     % Preferred (Theorem 1, Lemma 1, Theorem 2)
\ECRepeatTheorems

\EquationsNumberedThrough    % Default: (1), (2), ...
\MANUSCRIPTNO{MS-SMS-21-03979.R1}

\begin{document}
%%%%%%%%%%%%%%%%

%\setcounter{page}{1}
%\DOI{10.1287/xxxx.0000.0000}%


\RUNTITLE{Staffing for Batches}

% \TITLE{Bundling Information Goods of Decreasing Value}
\TITLE{
How to Staff When Customers Arrive in Batches}
\ARTICLEAUTHORS{%
\AUTHOR{Andrew Daw$\,^{*,1}$, Robert C. Hampshire$\,^{2,3}$, Jamol J. Pender$\,^{4}$}
%, \URL{}}
\AFF{$\,^{1}$ Data Sciences and Operations,  University of Southern California Marshall School of Business;  $\,^{2}$ Gerald R. Ford School of Public Policy, University of Michigan; $\,^{3}$ University of Michigan Transportation Research Institute; $\,^{4}$ School of Operations Research and Information Engineering,  Cornell University; \EMAIL{*andrew.daw@usc.edu}} 
%\AUTHOR{Authors blinded for review}
} % end of the block

%\tableofcontents
%\newpage

\ABSTRACT{%(no more than 300 words)
In \edit{many different settings}, requests for service can arrive in near or true simultaneity with one another. This creates batches of arrivals to the underlying queueing system. In this paper, we study the staffing problem for the batch arrival queue. We show that batches place a \edit{dangerous and deceptive} stress on services, \edit{requiring} a high amount of resources \edit{and exhibiting a fundamentally larger tail in those demands}. \edit{This uncovers a service regime in which a system with large batch arrivals may have low utilization but will still have non-trivial waiting.}
%In fact, we find that there is no economy of scale as the number of customers in each batch increases, creating a stark contrast with the square root safety staffing rules enjoyed by systems with individual arrivals of customers. Furthermore, when customers arrive both quickly and in batches, an economy of scale can exist, but it is weaker than what is typically expected. 
Methodologically, these staffing results follow from novel large batch and large batch-and-rate limits of the multi-server queueing model. In the large batch limit, we establish the first formal connection between \edit{general} multi-server queues and storage processes, another family of stochastic \edit{models}. By consequence, we show that the batch scaled queue length process is not asymptotically normal, and that, in fact, the fluid and diffusion-type limits coincide. \edit{Hence,} the (safety) staffing of this system must be directly proportional to the batch size just to achieve a non-degenerate probability of wait. \edit{In exhibition of the existence and insights of this large batch regime, we apply our results to data on Covid-19 contact tracing in New York City. In doing so, we identify significant benefits produced by the tracing agency's decision to staff above national recommendations, and we also demonstrate that there may have been an opportunity to further improve the operation by optimizing the arrival pattern in the public health data pipeline.}

}
%\KEYWORDS{} 

%\input{response.tex}
%\thispagestyle{empty} 
\clearpage
\maketitle

%%%%%%%%%%%%%%%%%%%%%%%%%%%%%%%%%%%%%%%%%%%%%%%%%%%%%%%%%%%%%%%%%%%%%%

% Samples of sectioning (and labeling) in MNSC
% NOTE: (1) \section and \subsection do NOT end with a period
%       (2) \subsubsection and lower need end punctuation
%       (3) capitalization is as shown (title style).
%
%\section{Introduction.}\label{intro} %%1.
%\subsection{Duality and the Classical EOQ Problem.}\label{class-EOQ} %% 1.1.
%\subsection{Outline.}\label{outline1} %% 1.2.
%\subsubsection{Cyclic Schedules for the General Deterministic SMDP.}
%  \label{cyclic-schedules} %% 1.2.1
%\section{Problem Description.}\label{problemdescription} %% 2.

% Text of your paper here

\section{Introduction}\label{intro} %%1.

\textit{Even the best laid plans can go astray.} Perhaps the most universal pillar of queueing theory's contributions is the recognition that ``congestion [is] a stochastic phenomenon'' \citep{kingman2009first}. Uncertainty and variability are inherent to service systems, and thus probability models have become natural tools to understand the structure of these operations and identify managerial insights within them. Some of the clearest and most useful examples of this preparation for randomness lie in staffing decisions. Often referred to as safety staffing rules, these regimes prescribe a certain level of additional staffing on top of the expected number of customers in the system if the staffing was unlimited. Much like an inventory safety stock may balance the tradeoff of holding costs and ordering costs, the safety staffing level weighs some quality-of-service metric, such as the fraction of customers that wait, against some measure of operational efficiency, such as the total staffing cost or simply the total number of servers employed. In this paper, we show that batch arrivals place a \edit{dangerously and deceptively} strong stress on service systems. \edit{That is, the stress brought by large batches (or bursts) of arrivals is of fundamentally higher order and  significantly different nature compared to  what occurs under similarly large rates of single-file arrivals.}
%requiring a high level of resources to achieve what is typically only a tradeoff-comprimising quality of service. 

Classically,  three of the most commonly contrasted staffing paradigms are the quality driven (QD), quality-and-efficiency driven (QED), and efficiency driven (ED) regimes. All three of these are designed for heavily trafficked services that receive a large rate of \edit{individually arriving} customers. As the names suggest, these each land at different points on the \edit{tradeoff between} quality and efficiency. The QD regime places a premium on the customer experience, maintaining a high number of servers to ensure that essentially no customer will wait. On the other hand, the ED regime prioritizes system efficiency, with a relatively low number of servers yielding that virtually all customers will wait. Striking a balance between these two, the QED regime uses a medium level of safety staffing to create an environment in which some, but not all, customers wait. We will formally review each of these regimes and their surrounding literature in Section~\ref{regimeReview}; an excellent and extensive review is available in \citet{van2019economies}. In the language of these regimes, our results show that batch arrivals of customers  require QD-like staffing just to achieve QED-like probability of delay. \edit{Additionally, not only is the safety staffing of higher order, it also demands a larger coefficient, because the tail of the service demand (or stochastic offered load) is fundamentally heavier than the tail of the analogous distribution under large arrival rates.}

\edit{In this paper, we establish the large batch limiting regime for the multi-server queue, and we use this to construct staffing methodology for queues with batch arrivals. We characterize properties of this regime that appear almost paradoxical operationally: even if systems with large batch arrivals are well staffed, there will be non-trivial waiting yet low system utilization. Surprising as that may seem, we need not look far to find examples that show that this regime exists and that, moreover, it is highly consequential in practice. As an illustrative example for the occurrence and consequences of large batches, we will apply our theoretical results to a contact tracing case study using New York City (NYC) Test \& Trace data \citep{blaney2022covid}. Indeed, we show that Covid-19 contact tracing presents an excellent recent example of the impacts of large batch arrivals. In this service system, batches are inescapable: because tests are conducted in batches, positive cases are found in batches, and thus contact tracers receive jobs by the batch. Hence, this case study shows how the public health question, ``how many contact tracers does my community need?'' is equivalent this paper's focal question, ``how do you staff a batch arrival queue?'' Furthermore, the data combines with our analytical contributions to show that the answer to this question is meaningfully different from the familiar answers to the seemingly similar question: ``how do you staff a queue?''\footnote{\edit{A brief numerical demonstration of why staffing for batches must present an inherently different challenge can be found in Appendix~\ref{batchDemo}.}}}

%\tr{characterize the regime: batch arrivals = non-trivial waiting, with relatively low utilization}

%\tr{this regime exists, and it matters!}

%\tr{as an illustrative example for the occurrence and consequences of large batches, we will apply these results to a contact tracing case study using New York City (NYC) Test \& Trace data \citep{blaney2022covid}.}

%\tr{maybe reusue: Because tests are conducted in batches, positive cases are found in batches}

Of course, there are \edit{many} additional applications of batch arrival queues across a myriad of other contexts. \edit{For another likely familiar and pandemic-related} example, in the \edit{return to in-person instruction in the} Covid-19 era, many universities allocated quarantine space for infected students to isolate in campus hotels \citep{fox2021response,giufurta2021full,gluckman2021some}. \edit{Here,} the batch staffing question is equivalent to asking if enough hotel rooms have been set aside. Many  classic examples of batch arrival queues lie in transportation, especially in public or mass transit. For a particular example that these authors have (unfortunately) experienced, customers from the same flight arrive to an airport car rental desk all at once, which places a large stress on the system and thus can create lengthy wait times. Beyond truly simultaneous batches, it is increasingly common that heavily trafficked services may actually receive an arrival stream that is gradual then sudden, varying significantly across time. This has been observed as a naturally occurring phenomenon of customer arrivals across many different application domains \citep[see, e.g.,][]{aksin2007modern,kim2014call,ibrahim2016modeling}. In such settings, arrival rates have been observed to be over-dispersed over short intervals of time, leading to many arrivals in a brief period. Intuitively, sufficiently rapid bursts of arrivals should function like a truly simultaneous batch arrival; we make formal arguments to support this in \edit{Appendix~\ref{burst}}. \edit{There are also many related challenges in modern computing settings, and, while those models may differ in certain ways, we argue in Section~\ref{batchReview} that their corresponding staffing levels must be \emph{at least} as high as the levels needed to support the batch arrival queue.}

%\tr{maybe cut:}

%\tr{refer to brief numerical demonstration of why staffing for batches is a different challenge is in Appendix~\ref{batchDemo}}

%straight to contributions?

%applications:
%\begin{itemize}
%\item autonomous vehicle remote assistance / teleoperations - mention that it was our original motivating application, maybe including frank/bruce line, talk about the variety of flavors, emerging technology, namedrop, level 4 v level 5, tie to human(s)-in-the-loop ML in general (fact checkers? data labelling? others?) and use that to segue to the next application, cite some papers / articles
%\item modern computing - Qiaomin's paper is on Monte Carlo Tree Search (MCTS), which is well studied in the RL literature and uses simulations to approximate the value function of a given state (or more precisely, at each given state? - that would be the kicker here), that alg is behind the wild success of AlphaGo, etc. A particularly relevant example here, because the simulation asks for a certain number of replications to be completed, but it does not require which servers should complete these jobs (ie it is not fork-join). note that this also supports driverless vehicles
%\item quarantine beds - could talk about Cornell using the Statler to house students that had tested positive, would it be enough? probably need to find other related examples
%\item 
%\end{itemize}

\subsection{Contributions and Organization}

%\tr{the lessons are about the nuances of the arrival pattern}

%\tr{batch arrivals are dangerous, largely because they can be deceptive}

%\tr{avoid focusing on QD/QED/ED -- save that for lit review section. focus on utilization, waiting, etc.}

%\tr{underlying components (aka infinite server) are meaningfully different!}

%\tr{make sure to mention contact tracing case study -- lessons learned from covid}

%\tr{linear in the batch size, square root in the arrival rate}

%\tr{top-level lesson is that safety staffing should be proportional to $n\sqrt{\lambda}$, rather than $\sqrt{m}$ where $m = \lambda n$ is the effective arrival rate. this alone implies that the staffing under batch arrivals must be higher than that under solitary arrivals }

%\tr{the more nuanced lesson supporting this is that the fundamental service demands under batch arrivals are heavier in their tail than the demands from single-file arrivals. that is, there is no asymptotic normality under large batches}

%\tr{emphasize queueing analysis methodology consequences of large batch regime, because the infinite server scaffolding is so important}

%\tr{contact tracing example is truly meant to illustrate -- serves both as an entry to the literature on operational lessons learned from the Covid-19 pandemic, and as a demonstration of how batch arrivals impact real-world queueing systems}

%\tr{put this somewhere, maybe contributions -- We also believe our results have high applicability for \textit{burst} arrival queues, and we make brief arguments in favor of this in Appendix~\ref{burst}.}

\edit{The foremost lessons of this paper are in the nuances of the customer arrival pattern and the consequences that result for  service operations. The managerial insights of our theoretical contributions summarize as the two-pronged message -- batch arrivals can be \emph{dangerous} and \emph{deceptive}: 
\begin{enumerate}[i)]
\item \emph{Dangerous}: The top-level takeaway is that the \textbf{safety staffing must be square root in the arrival rate but linear in the batch size} (Theorems~\ref{fDelayConv} and~\ref{bothLimitMulti}). That is, the safety staffing should be proportional to $n\sqrt{\lambda}$ rather than $\sqrt{m}$, where $m = \lambda n$ is the \emph{effective} arrival rate for $\lambda$ as the batch arrival rate and $n$ as the batch size. This alone implies that the staffing under batch arrivals must be higher than that under single-file arrivals.
\item \emph{Deceptive}: The more subtle companion to this is that the fundamental \textbf{service demands under batch arrivals are heavier in their tail than the demands from single-file arrivals} of the same overall volume (Theorems~\ref{fBatchScale} and~\ref{bothLimitInf}). Hence, if the batch size $n$ dominates the arrival rate $\lambda$, not only should the safety staffing be of order $n$, it must have a coefficient that is meaningfully larger than that for the same effective arrival rate  with $\lambda$ large.
\end{enumerate}}
\noindent
Methodologically, the large batch staffing is theoretically justified by a novel batch scaling limit of the multi-server queue, the first of its kind for a finite server model and significantly more general than prior infinite server results. This scaling yields a \textbf{connection from batch arrival queues to storage processes}, another family of continuous time stochastic processes. The limit is similar to a fluid scaling in that we scale inversely by the index of the limit, but the normalization is through the batch size, rather than the arrival rate of batches. Hence, the arrival epochs are untouched, which preserves the randomness in the limit. \edit{The intuition for this lies in the scaffolding of the result from the analogous infinite server queue. Through this simpler case, we can recognize that \textbf{there is no asymptotic normality} and that, moreover, \textbf{the fluid- and diffusion-type limits coincide} (up to centering) in the large batch regime. To the best of our knowledge, this regime presents the first such observation of these scalings being identical, and thus contributes to the stochastic models literature. %The contrast of these underlying components is sufficient to explain the difference between the large batch regime and the classic scalings from which popular staffing rules have been derived.
}

\edit{To ground these insights in reality, we demonstrate our theoretical results in a \textbf{case study on Covid-19 contact tracing} operations. This example is meant to illustrate. It serves as both a demonstration of how batch arrivals impact real-world queueing systems and as an entry to the literature on operational lessons learned from the Covid-19 pandemic. For the former, we see that this data does indeed exhibit the hallmark properties for this regime: \textbf{large batches, low utilization, yet non-trivial wait}.  For the latter, our simulation experiments show both that (1) NYC Test \& Trace made a wise decision to staff well above national guidelines and (2) there may have been an opportunity to further improve the contact tracing operations simply through careful management of the arrival pattern of cases within the public health data pipeline.}

\edit{The remainder of the paper is organized as follows. In Section~\ref{litReview}, we survey the literature. Then, we precisely define the batch arrival queueing model in Section~\ref{model}. In Section~\ref{batchSec}, we prove the novel {large batch limit} of the multi-server queueing model in essentially full generality (Section~\ref{batchMultiSec}), and this is built from establishing the infinite server limit under the same generality (Section~\ref{batchInfSec}). To contextualize this regime as an endpoint a spectrum from classic heavy traffic results in a straightforward way, in Section~\ref{hybridSec}, we also prove novel {large batch-and-rate limits} for the infinite (Section~\ref{hybridInfSec}) and multi-server (Section~\ref{hybridMultiSec}) Markovian queueing systems. We contrast the large batch and batch-and-rate regimes in Section~\ref{contrastSec} and identify managerial insights in their differences. Building from this analysis, we conduct the {Covid-19 contact tracing case study} in Section~\ref{contactSec}. Finally, we conclude in Section~\ref{concludeSec}.} All proofs are contained in the appendix.

\section{Service Operations Context and Literature Review}\label{litReview}

%Introduce the idea of the queueing models broadly, review the background: QD/QED/ED and batch queues

\edit{We begin by surveying this paper's context. In this section, let us review the literature on asymptotic staffing regimes (Section~\ref{regimeReview}), batch arrival queues and storage processes (Section~\ref{batchReview}), and contact tracing and Covid-19 operations (Section~\ref{contactReview}).}

%There are two primary streams of literature from which this paper draws inspiration, and to which it hopes to contribute: operational staffing regimes and the analysis of batch arrival queues. We survey the former of these in Section~\ref{regimeReview} and the latter in Section~\ref{batchReview}. \edit{Then, because we also aim to contribute to the literature on Covid-19 era operations through our case study, we survey this literature in Section~\ref{contactReview}.}

\subsection{Operational Regimes and Arrival-Rate-Many-Server Asymptotics}\label{regimeReview}

%\tr{mention somewhere that we will often compare to QD and QED}

To set the stage for this paper, let us briefly review the operational context of the ED, QED, and QD regimes, \edit{as we contrast with the latter two throughout}. For more depth than the space here allows, readers should see \citet{van2019economies} for a recent survey and nice tutorial, containing much of the state of the art of this methodology and its use in surrounding problems. 
%In this subsection, we will first sketch the three regimes and then highlight prior works of particular interest and relevance.

\edit{Each} of these regimes are based upon the theory of heavy traffic limits, meaning asymptotics of the multi-server queue length \edit{$Q(\lambda)$} as the arrival rate \edit{$\lambda$} grows large. 
%Hence, we will frame these three in terms of this arrival rate, say $\lambda$. 
Walking through the arrival rate, staffing level, and  system performance, these three regimes can be summarized as:
\begin{itemize}
\item \textbf{Quality Driven (QD):}
\begin{itemize}
\item individual arrivals at rate proportional to $\lambda$
\item number of servers $c(\lambda)$ proportional to $\lambda + \delta \lambda$ for some $\delta > 0$ 
\item in this regime, no customers wait and $\PP{Q(\lambda) \geq c(\lambda)} \to 0$ \edit{as $\lambda \to \infty$}
\end{itemize}
\item \textbf{Efficiency Driven (ED):}
\begin{itemize}
\item individual arrivals at rate proportional to $\lambda$
\item number of servers $c(\lambda)$ proportional to  $\lambda + \delta$ for some $\delta > 0$ 
\item in this regime, all customers wait and $\PP{Q(\lambda) \geq c(\lambda)} \to 1$ \edit{as $\lambda \to \infty$}
\end{itemize}
\item \textbf{Quality and Efficiency Driven (QED):}
\begin{itemize}
\item individual arrivals at rate proportional to $\lambda$
\item number of servers $c(\lambda)$ proportional to $\lambda + \delta \sqrt{\lambda}$ for some $\delta > 0$
\item in this regime, some customers wait and $\PP{Q(\lambda) \geq c(\lambda)} \to h(\delta) \in (0,1)$ \edit{as $\lambda \to \infty$}
\end{itemize}
\end{itemize}
The second bullet under each of the regimes provides context for the staffing problem we will consider throughout the paper, and together the second and third bullets  explain the nomenclature. That is, the QD regime bears the name ``quality'' because of the absence of waiting in the limit. By contrast, the ED regime achieves ``efficiency'' because the staffing is just slightly above the incoming rate of customer traffic, protecting system stability without committing to much else.  Finally, QED strikes a balance between the two with its so called ``square root (safety) staffing rule.'' Some customers wait, but not all do, and the staffing level grows with arrival rate, but it's not directly proportional to it. In other words, doubling the arrival rate would call for staffing less than double the number of servers.
See Figure 12 in \citet{gans2003telephone} for a comprehensive summary of the operational performance of these different regimes, as well as an excellent survey of relevant call center research, where these regimes have been of particular practicality.

There has been a quite rich history of work analyzing these operational regimes; we provide here a brief sampling. Starting with the foundation, \citet{halfin1981heavy} is universally recognized as the analytical cornerstone of QED staffing. Indeed, the QED regime is also often referred to as the Halfin-Whitt regime, and it is the heavy traffic limit established by \citet{halfin1981heavy} that first formally established the square root staffing rule, solidifying what had been folklore for many years prior. Furthermore, the analysis therein continues to serve as a blueprint for research in these settings, as will be the case here. Also quite relevant here are \citet{borst2004dimensioning} and \citet{garnett2002designing}, which both classify and contrast the three operational regimes (QD, ED, and QED).\footnote{In some early works that classify the regimes, such as \citet{borst2004dimensioning,garnett2002designing}, QED staffing is called the ``rationalized'' regime, referring to its balance between quality and efficiency.} \citet{borst2004dimensioning} considers the fully Markovian Erlang-C ($M/M/c$ in Kendall notation), while \citet{garnett2002designing} studies the effect of abandonment through the Erlang-A model ($M/M/c+M$). \citet{zeltyn2005call} extends this to general patience distributions. \citet{jennings1996server,green2007coping,feldman2008staffing} find similar success of square root staffing in the case of time-varying arrival rates; see \citet{whitt2007you} for a survey specifically devoted to the staffing problem in time-varying settings. There has also been a great deal of work extending these concepts from their Markovian model origins to greater generality of arrival and service time distributions. To this end, the analysis translating the QED regime and Halfin-Whitt limits to the $G/GI/c$ queue in \citet{reed2009g} will be of particular importance to our analysis here, since the $G_t^B/GI/c$ queue is the setting for our large batch limits.

Our work here also joins a stream of literature that considers alternatives outside of the QD, ED, and QED regimes, as well as ones that live between them. For example, \citet{bassamboo2010capacity} considers parameter uncertainty in the arrival rate and finds that for low uncertainty the square root staffing rule still performs well, but in the case of high levels of uncertainty it can be outperformed by newsvendor-style methods. \citet{gurvich2010staffing} approaches demand rate uncertainty through a chance-constrained optimization problem, where quality of service requirements are addressed through high probability guarantees, rather than average performance guarantees. \citet{jongbloed2001managing,mathijsen2018robust} are similarly interested in addressing uncertain and over-dispersed demand. The non-degenerate slowdown regime in \citet{atar2012diffusion,atar2011asymptotically,atar2014scheduling} is of particular relevance for our hybrid batch-and-arrival-rate limits, as here we also construct the limit through an extra parameter housed in the exponent of two separate components of the queue. By comparison to \citet{atar2012diffusion}, these two are the arrival rate and the batch size, rather than the number of servers and the service rate. Another important connection for both our large batch and hybrid scalings are peakedness approximations \citep[see, e.g.,][and references therein]{whitt1992understanding,massey1996stationary,pang2012impact}.\footnote{\citet{eckberg1983generalized} appears to be a foundational reference for peakedness approximations, but we have had trouble locating this manuscript.} The staffing levels we identify in this work align with the order of magnitude of peakedness approximation staffing for both the large batch and the hybrid scalings, but we make two important emphases here. First, in both scalings, the terms which are analogs of the peakedness parameter are also increasing in the limit as the mean arrival rate increases. Second, and more importantly, in the large batch scaling the peakedness approximation does identify the correct order of magnitude, but there is no asymptotic normality. In fact, as we remark upon in Section~\ref{batchSec} \edit{and analyze in depth in Section~\ref{contrastSec}}, the \edit{tail of the} limiting distribution in the large batch scaling is strictly greater than sub-Gaussian. 

%\tr{likely want to add peakedness details here, if not later}

\subsection{Batch Arrival Queues and Storage Processes}\label{batchReview}

The large batch setting in this paper separates our work from previous queueing theoretic studies on batch arrival multi-server queues, which either assume a less general arrival process than we consider here or, perhaps more critically, assume bounded batch sizes. For example, \citet{neuts1978algorithmic} and \citet{baily1981algorithmic} each use matrix-geometric approaches to study the $GI^B/M/c$ queue under the assumption that the batch size distribution $B$ is bounded, with \citet{neuts1978algorithmic} considering the stationary setting and \citet{baily1981algorithmic} the transient. In each setting, the bounded-ness of the batch size is essential, as this bound dictates the size of the underlying matrices. This bounded batch size assumption is also used to study the $GI^B/M/c$ model in \citet{zhao1994analysis} and \citet{chaudhry2016analytically}, with the former giving explicit expressions for the generating function and an equation satisfied by the steady-state probabilities and the latter providing efficient  computational methods while also simplifying the approach of the former. Again the bound is essential, as these approaches are built upon root-finding methods where the number of roots is equal to the batch size. By comparison, the general unbounded batch size setting has often called for approximate approaches, such as the bounds on the $GI^B/G/c$ system that were constructed in \citet{yao1984bounds} through comparison to single arrival queues. \citet{yao1985some} then gives tighter bounds for the $M^B/M/c$ queue using the $M^B/G/1$ system and demonstrates that these bounds can be used to approximate the $GI^B/G/c$. Computational methods have also been provided in \citet{cromie1979further} for the fully Markovian setting, the $M^B/M/c$ queue, although these were only done for three specific batch distributions: constant size, geometric, and Poisson.

In studying this large batch setting we will prove batch scaling limits of the queue, in which the batch size and the number of servers grow large and the queue length is scaled inversely. Through the batch scaling limits, we connect the general batch arrival queueing models to storage processes, another class of stochastic processes. Similar albeit less general scalings have been explored recently in \citet{de2017shot,daw2019distributions}. Specifically, the limits we prove in this work for the $G_t^{B}/GI/c$ queue generalize the batch scaling results of $M^{B}/M/\infty$ queueing systems shown in \citet{de2017shot,daw2019distributions}, which converge to shot noise processes with exponential decay. Let us emphasize that the limits in \citet{de2017shot,daw2019distributions} are inherently tied to the Markovian assumption. Both prior works draw upon tools specifically for Markov processes, like the infinitesimal generator and the associated partial and ordinary differential equations. Here, in addition to extending to the multi-server setting, we also adopt greater generality of distributions and thus do not have such methodology available to us. To prove this generalization beyond the Markovian setting, we develop an approach that is entirely agnostic to the arrival epoch process, which is what enables our results to be immediately applicable to queues with time-varying and/or correlated inter-arrival times. This approach of leveraging the infinite server queue to understand the multi-server system is similar to the techniques used by \citet{reed2009g} to extend the Halfin-Whitt heavy traffic limits to non-Markovian service durations. Indeed, \citet{reed2009g} serves as a key predecessor  and inspiration for our proof methodology in the large batch scaling. The limiting relationship between infinite server queues and shot noise processes was also discussed as motivation in \citet{kella1999linear}, although this relationship was presented without proof. This connection allows us to make use of a broad literature on storage processes, which can be seen as a generalization of  shot noise processes.

Storage processes, which can also be referred to as dams, content processes, or even fluid queues, are positive valued, continuous time stochastic processes in which the process level will jump upwards by some amount at epochs given by a point process. Between jumps the process will decrease according to some function of its state. In generality, the release dynamics may also be a function of the history of the process rather than just the current state; such a setting will be necessary to study the multi-server queue's limiting form in the case of general service.  Many of the results that will be most relevant here are focused on the stationary distributions of storage processes. Even on its own\edit{,} the study of stationary distributions of storage processes has a rich history, with early work including expressions of stationary distributions for shot noise processes given in \citet{gilbert1960amplitude}. Later work found similar results for more general settings, including \citet{cinlar1972dams,yeo1974finite,yeo1976dam,rubinovitch1980level,kaspi1984storage}. A line of study that will be particularly useful for us can be found in \citet{brockwell1977stationary,brockwell1982storage}, as these works find integral equations for the stationary distributions of storage processes in generality. These forms will be of great use to us in our staffing analysis. For precursors to this work in a different but no less interesting setting, see \citet{harrison1976stationary, harrison1978recurrence}. Connections between queues and storage processes are not new in general, as the single server queue has been known to have a workload process that is a storage process. However, to the best of our knowledge, our work is the first connection between multi-server queues and storage processes. For an overview of the \edit{pre-existing} connections and  related ideas, see \citet{prabhu2012stochastic}.

Another interesting and closely related process to the batch arrival queue is the fork-join queueing model, which may hold relevance for many similar applications while also being inherently distinct from the models we study here. At its most elemental, the fork-join queue functions as follows. Upon each arrival a job is split into $k$ parallel tasks, each one routed to one of $k$ separate servers. Each task waits for service, is served, and then waits to be re-joined with the rest of the tasks in the job. Once all $k$ tasks have been completed by their respective servers and re-joined, the job is considered complete. Aside from simply being an intriguing stochastic model to analyze \citep[e.g.,][]{baccelli1989fork,lu2017heavy}, there are many interesting variations and relaxations of this problem, such as scheduling control for multiple job and server types \citep[e.g.,][]{atar2012control,ozkan2019control} and correlation and redundancy among the tasks and sub-queues \citep[e.g.,][]{gardner2017better,gardner2017redundancy,wang2019delay,hong2021sharp}.  A survey on the fork-join queue is available in \citet{thomasian2014analysis}. In comparing fork-join and batch arrival queues, the key differences lie in the structuring of the waiting and in the post-processing. For the former, it is an issue of centralization versus decentralization: the batch arrival model has one pooled queue that feeds jobs or tasks to servers as they become available, whereas the fork-join model has one queue per server or station. Then, for the latter, the batch arrival queue by default does not include a synchronization step re-joining the jobs at departure. For the purposes of the staffing problem, we consider the centralization to be the more important difference. While the synchronization may prompt a different focus in performance metrics, if the step can be automated it may not actually require its own server. By comparison, the pooling principle posits that there are meaningful differences between centralized and decentralized queueing structures, and, following that intuition, the staffing requirements we find here should be a lower bound on what is needed in matching fork-join systems. \edit{We also believe our results have high applicability for \textit{burst} arrival queues, and we make brief arguments in favor of this in Appendix~\ref{burst}.}
%\tr{work in multi-server jobs?}

\subsection{\edit{Covid-19 Contact Tracing and Operations of Pandemic Case Investigation}}\label{contactReview}

\edit{Operations researchers have made tremendous contributions to the global effort against the Covid-19 pandemic.\footnote{\edit{For a small but representative samples of the many different ways have the operations community has risen to the many challenges brought by Covid-19, see this M{\&}SOM virtual special issue: \url{https://pubsonline.informs.org/page/msom/the-impact-of-operations-in-the-covid-19-pandemic}.}} Through the case study we present in Section~\ref{contactSec}, we claim that although staffing for contact tracing may appear to be a classic operations-type question, this answer has unexpectedly novel characteristics. What is immediately clear, however, is that the public health context makes the answer of critical importance. For example, at the pandemic's onset, there were many broad and public calls to drastically grow the national contact tracing workforce in the United States \citep[see, e.g.,][]{watson2020national,naccho2020building}. In fact, the lower-end of these national guidelines was roughly 50 times the staff on hand at the start of the outbreak \citep{ruebush2021covid}. As we will demonstrate through operational data on Covid-19 contact tracing in NYC \citep{blaney2022covid,nyc2023nyc,nyc2022covid}, even those projected levels may have been too conservative, and our analysis of the batch arrival of cases exposes a fundamentally higher staffing need.}

\edit{It is also now clear that \emph{successful} Covid-19 contact tracing was highly impactful. For example, through a natural experiment that occurred due to a brief IT error in England, conservative causal estimates in \citet{fetzer2021measuring} show that contact tracing would have provided a 63\% reduction in new infections and a 66\% reduction in Covid-19 related deaths in the weeks following this data error. Similar estimates from \citet{wang2022effectiveness} suggest that contact tracing led to a 40-50\% reduction in new cases in Austin, TX. Likewise, the modeling-based nationwide calculations in \citet{rainisch2022estimated} estimate that contact tracing averted over one million cases and approximately thirty thousand hospitalizations over just a sixty day period.
}

\edit{It is thus intuitive that staffing decisions  can be highly consequential in the context of this global emergency. In fact, staffing is a common thread of many of the proposed future research directions in the public health operations literature before Covid-19 \citep{gupta2022om}. In many ways, this is also well-documented in the course of the Covid-19 pandemic. For example, \citet{lash2021covid} documents many struggles with contact tracing efficacy across the United States, with an overload of work as a common cause. The data from \citet{blaney2022covid} shows that NYC is somewhat of an exception to this, as the Test \& Trace operation achieved higher rates of success relative to peer agencies nationwide. We examine staffing as a possible explanation for this in Section~\ref{contactSec}. Our analysis will also emphasize the importance of properly managing the public health data pipeline, and this aligns with findings from \citet{chen2011managing}.}

\edit{Of course, at this point, many Covid-19 contact tracing operations have ceased or shifted focus. This includes the subject of our study, NYC Test \& Trace, which has now re-aligned to testing and treating rather than testing and tracing \citep{vasan2022ensuring}. Hence, the insights we identify in this space are intended to be \emph{from} this pandemic rather than \emph{for} it. }

\section{Model and Preliminaries}\label{model}

%something here?

To distinguish our setting from the literature on batch arrival queues and their close relatives, let us clearly define the batch arrival queueing model we analyze here in full generality. 
%We do so in Section~\ref{modelGen}, in which we establish the notation we will use throughout the paper. 

%\subsection{General Batch Arrival Multi-Server Queueing Model}\label{modelGen}

In Kendall notation, the general model we study in this paper is the $G_t^B/GI/c$ queue\edit{, or, to reflect the scaling, the $G_t^{B(n)}/GI/cn$}. That is, arrivals to the queueing system occur in batches drawn from a sequence of independent and identically distributed positive discrete random variables, denoted $\{B_i(n) \mid i \in \mathbb{Z}^+\}$ where $n \in \mathbb{Z}^+$ is such that $\E{B_1(n)} \in O(n)$. We will refer to $n$ as the relative batch size; the limits in Section~\ref{batchSec} will be indexed by $n$. These batch arrivals occur at epochs given by some general and possibly time-varying point process, hence we will let $N_t$ be the number of epochs that have occurred by time $t$ for all $t \geq 0$. Similarly, let $\{A_i \mid i \in \mathbb{Z}^+\}$ denote the arrival epochs. Then, we will let $c > 1$ be such that $cn \in \mathbb{Z}^+$ is the number of servers, meaning that the staffing level grows with the relative batch size. The servers will serve customers in a first-come-first-serve discipline. There is unlimited waiting space. We will assume that service durations are drawn from a sequence of independent and identically distributed positive random variables, $\{S_{i,j} \mid i \in \mathbb{Z}^+, 1 \leq j \leq B_i(n)\}$, indexed first by the batch in which the specific customer arrived ($i$) and then by the \edit{order} in which customers from this batch entered service ($j$). We will let $G(x) = \PP{S_{1,1} \leq x}$ for all $x \geq 0$, which does not depend on $n$. Additionally, let $\bar{G}(x) = 1 - G(x)$ for all $x$. Following the same indexing, we will let $W_{i,j}$ be the time that the $j^\text{th}$ customer within the $i^\text{th}$ batch waits before beginning service. 

Using these components, the specific stochastic process we will study will be $Q_t^C(n)$, which is the queue length process (meaning the total number in system, including the customers in service and those waiting) for the general batch arrival multi-server queue at time $t \geq 0$ for relative batch size $n$. The $C$ superscript refers to this model as the general batch arrival analog of the classical Erlang-C model. This superscript is of particular relevance in distinguishing the model from a close cousin that we will use as a stepping stone in our analysis: the  $G_t^B/GI/\infty$ queue. Let us define $Q_t^\infty(n)$ as the analogous infinite server queueing model as a counterpart to $Q_t^C(n)$. That is, $Q_t^\infty(n)$ tracks the queue length or total number in system under the same assumptions on batch sizes, arrival epoch process, and service distributions, with the single (but important) difference being that there is an unlimited number of servers. Hence, no customers will wait for service, rendering the $Q_t^\infty(n)$ an idealized form of $Q_t^C(n)$ that is more tractable for analysis. We will assume that the initial conditions, $Q_0^C(n)$ and $Q_0^\infty(n)$, are known for all $n$.

With this notation in hand, we can also now define the staffing problem that is at the heart of this paper. For some $\epsilon > 0$, we seek to find a $c > 1$ such that the probability that the queue length exceeds or equals the number of servers is at most $\epsilon$, i.e.
\begin{align}
\PP{Q_t^C(n) \geq cn } \leq \epsilon
\label{exProbDef}
.
\end{align}
We will refer to $\PP{Q_t^C(n) \geq cn}$ as the \emph{exceedance probability}. This can also be thought of as the probability that all servers are busy at time $t$. The individual arrival analog of the exceedance probability, say $\PP{Q_t^C(1) \geq c}$, is often referred to as the ``delay probability'' in the case of stationary Poisson process arrivals, since the famous PASTA theorem implies that this is also the probability that an arriving customer will have to wait for service \citep{wolff1982poisson}. Even setting the lack of a Poisson assumption aside, we can observe that the event $\{Q_t^C(n) \geq cn \}$ does not offer such guarantees in the case of batch arrivals. In the batch setting, this would correspond to the event that \textit{all} customers within an arriving batch would have to wait. Hence, one could instead consider other performance metrics that require a higher standard of service, such as the probability that \textit{some} customers wait, $\PP{Q_t^C(n) + B(n) \geq cn}$, or more generally, $\PP{Q_t^C(n) + pB(n) \geq cn}$ for some $p \in (0,1)$. We will \edit{primarily} stick to the exceedance probability as defined in \eqref{exProbDef}, as even the weakest of these requirements will be enough to create the strong staffing requirements to which we have alluded, \edit{and} our analysis can be carried through directly for similar events \edit{as a continuous mapping}. \edit{Indeed, in Section~\ref{contrastSec}, we will also discuss the $\PP{Q_t^C(n) + B(n) \geq cn}$ exceedance probability.} %\tr{focus on \eqref{exProbDef}, but will also contrast with $\PP{Q_t^C(n) + B(n) \geq cn}$ in Section BLAH}

%\tr{many other models for bursts, these are just to make a point, should mention we can combine a la Dynamic Contagion}

%\tb{each $\alpha$ and $\beta$ are rate parameters, limit is like thinking of bursts on the scale of seconds whereas staffing decisions are typically on the scale of several hours}

%\tb{hence, we stick to batches}

\section{Staffing for Large Batches: From Queues to Storage Processes}\label{batchSec}

%\tb{
%\begin{itemize}
%\item QD-type staffing yields QED-type performance
%\item holds in very general settings: $G_t^B/GI/c$ queue
%\item built by connecting queues to storage processes in a batch scaling limit
%\item starts with establishing a result for the infinite server variant, the $G_t^B/GI/\infty$ queue
%\item index the queue by the mean batch size $n$: introduce the asymptotic regime and the Kendall notation for the pre-limit object
%\item make sure to connect limit to staffing problem and/or QDvsQED
%\item need to introduce the time 0 notation
%\item make assumption about limit of $B_i(n)\slash n$ here
%\item need to motivate storage and shot noise processes here
%\item say Theorem~\ref{fDelayConv} is the main result of this section
%\end{itemize}
%}

To rigorously determine how to staff queues with large batch arrivals, we must first simply understand the behavior of the queue in this setting. From the general batch arrival queueing models defined in~\ref{model}, we have a natural sequence of systems indexed by the relative batch size $n$. In the multi-server model, $Q_t^C(n)$, both the batch sizes, $\{B_i(n) \mid i \in \mathbb{Z}^+\}$, and the number of servers, $cn$, depend on $n$, whereas in the infinite server model, $Q_t^\infty(n)$, only the batch sizes do. Because we want to reason about the system as the batch size grows large, let us suppose that there exists a sequence of positive independent and identically distributed random variables $\{M_i \mid i \in \mathbb{Z}^+\}$ such that $B_1(n) \slash n \stackrel{\mathsf{D}}{\Longrightarrow} M_1$ as $n \to \infty$. 
%Applying the limit indexing to the Kendall notation, we are considering a sequence of $G_t^{B(n)}/GI/cn$ systems. As this notation implies, this staffing level is inherently QD in style, as the number of servers, $cn$, is exactly proportional effective arrival rate, $\E{N_t \slash t} \E{B_1(n)} \in O(n)$. 
The staffing results we find for this large batch setting will follow as an immediate consequence from a connection we prove between batch arrival queues and storage processes, the continuous time stochastic processes we have surveyed in Section~\ref{litReview}.

\subsection{Scaffolding from the Infinite Server Queue}\label{batchInfSec}

To build intuition for how the batch arrival queue behaves as the relative batch size increases, let us start by decomposing the infinite server queue length. Because the sheer abundance of servers implies that no customer will wait for service, any customer in the system at time $t$ is in service. In other words, the queue length at time $t$ is the collection of customers that arrived before $t$ but complete service after $t$. Summing over the arrival epochs and batches so far, this equivalence yields the equation
\begin{align}
Q_t^\infty(n)
&=
\sum_{j=1}^{Q_0^\infty(n)} 
\mathbf{1}\{t < S_{0,j}\}
+
\sum_{i=1}^{N_t}
\sum_{j=1}^{B_i(n)}
\mathbf{1}\{t < A_i+S_{i,j}\}
.
\label{qInfDef}
\end{align}
Here, the queueing dynamics are plain: $Q_t^\infty(n)$ will jump up by the amount of the $i^\text{th}$ batch size $B_i(n)$ upon the $i^\text{th}$ arrival epoch, and it will then jump down a unit size upon each service completion. As the relative batch size increases, the up-jumps will dwarf the down-jumps \edit{in size}, but the down-jumps will also become increasingly frequent. Following that intuition, let us define an alternate stochastic process, $\psi_t^\infty$, \edit{as} 
\begin{align}
\psi_t^\infty
&=
\psi_0^\infty \bar{G}_0(t)
+
\sum_{i=1}^{N_t}
M_i \bar{G}(t- A_i)
,
\label{fPsiDef}
\end{align}
where $\psi_0^\infty$ is a known initial condition. The idea here is similar to \eqref{qInfDef}, with jumps upward at epochs given by a distributionally equivalent point process $N_t$ (hence the duplicated notation), but this is the only source of stochasticity in the model. That is, otherwise, the behavior is deterministic. Thus, $\psi^\infty_t$ is a shot noise process, with the behavior between arrival epochs determined by the tail CDF of the queue's service distribution, $\bar{G}(\cdot)$. This suggests the connection between the two models. \edit{Given $A_i$, the} expected value of an indicator function in the queue decomposition is $\E{\mathbf{1}\{t < A_{\edit{i}} +S_{i,j}\}} = \PP{t - A_i < S_{i,j}} = \bar{G}(t-A_i)$. Hence, the law of large numbers links the infinite server queue and the shot noise process: if normalized by the batch size, the sum over the distributionally identical indicator functions should converge to the tail CDF as $n \to \infty$. To formalize this reasoning, let us assume that the known initial conditions converge, i.e.~$Q_0^\infty(n)\slash n \to \psi_0^\infty$ as $n \to \infty$. Then, in Theorem~\ref{fBatchScale} we prove that under this large batch scaling regime, the general batch arrival infinite server queue converges to a general shot noise process.

\begin{theorem}\label{fBatchScale}
As $n \to \infty$, the batch scaling of the $G_t^{B(n)}/GI/\infty$ queue $Q_t^\infty(n)$ yields
\begin{align}
\frac{Q_t^\infty(n)}{n}
\stackrel{D}{\Longrightarrow}
\psi_t^\infty
,
\end{align}
pointwise in $t \geq 0$, where $\psi_t^\infty$ is a shot noise process with the $i^\text{th}$ jump having size $M_i$ for each $i \in \mathbb{Z}^+$, as defined in Equation~\eqref{fPsiDef}.
\end{theorem}

Conceptually, this large batch scaling bears similarity to a fluid scaling, as we are ``shrinking'' customers while also increasing the inflow of customers. However, by comparison to a traditional fluid limit, this increase in inflow is through a greater amount of simultaneous arrivals rather than a faster rate of individual arrivals. Hence, the randomness of the arrival epoch point process is preserved, and this is what yields the random limit. If $Q_t^\infty(n)$ answers the question ``how many customers are in the system at time $t$?'' then after normalizing by $n$, $Q_t^\infty(n)\slash n$ must answer ``how many \edit{(relative)} \emph{batches of} customers are in the system at time $t$?'' As the relative batch size grows large, Theorem~\ref{fBatchScale} implies that the latter of these questions is effectively answered by $\psi_t^\infty$ as well. Because $\psi_t^\infty$ is deterministic between arrival epochs, it offers a more amenable platform for analysis of the queue with large batch arrivals. In fact, if $\bar{G}(\cdot)$ is continuous then $\psi_t^\infty$ will be continuous as well. We will leverage these concepts in the next subsection for our batch arrival queue staffing methodology. 

Before doing so, let us first briefly provide a little more intuition about the shot noise process $\psi_t^\infty$. If the arrival epochs follow a time  homogeneous Poisson process, then we can leverage conditional uniformity to provide a closed form expression for the limiting generating function.

\begin{corollary}\label{mgfCor}
If $N_t$ is a stationary Poisson process with rate $\lambda > 0$, the moment generating function of $\frac{Q_t^\infty(n)}{n}$ converges to
\begin{align}
\E{e^{\frac{\theta}{n}Q_t^\infty(n)}}
\longrightarrow
e^{\theta\psi_0^\infty \bar{G}_0(t) + \lambda \int_0^t \left(\E{e^{\theta M_1 \bar{G}(x)}}-1\right)\mathrm{d}x}
,
\end{align}
for each $t \geq 0$ as $n \to \infty$.
\end{corollary}

Our assumption here that the arrival process is Poisson is only temporary, but it gives us valuable insight into the properties of the limiting distribution. In particular, the nested exponential form of the moment generating function implies that the distribution of the shot noise process is \emph{not} sub-Gaussian\edit{.}
%, meaning that the tail is greater than the tail of the normal. 
This fact will be important throughout our staffing analysis \edit{because} the exceedance probability is inherently a tail probability event, and it will further gain relevance in contrast with the asymptotic normality that we discuss in Section\edit{s}~\ref{hybridSec} \edit{and~\ref{contrastSec}}.

\subsection{Batch Scaling Limit of the General Multi-Server Queue}\label{batchMultiSec}

Having gained intuition from the connection of the infinite server batch arrival queue and the general shot noise process, let us now turn to the multi-server model that at the heart of our staffing problem. Setting aside the initial customers in the system for the moment for the sake of space on the page, in Equation~\eqref{qCdef} we can see that we can also decompose the multi-server queue length into a sum over which customers are still in the system. However, by comparison to the infinite server model, the customers present in the multi-server system at time $t$ are not only the customers actively being served, but also the ones who are waiting to receive service. Hence, we must correct the sum over prior arrivals to also include the effect of waiting. This yields
\begin{align}
Q_t^C(n)
&=
\underbrace{\sum_{i=1}^{N_t}\sum_{j=1}^{B_i(n)} \mathbf{1}\{t < A_i + S_{i,j}\}}_{\text{Queue length ignoring wait}}
+
\underbrace{\sum_{i=1}^{N_t} \sum_{j=1}^{B_i(n)} \mathbf{1}\{ A_i + S_{i,j} \leq t < A_i + S_{i,j} + W_{i,j}\}}_{\text{Waiting time correction}}
,
\label{qCdef}
\end{align}
and if we were to include the customers present at time 0, we would simply mimic both summations for that population of customers. Inspecting~\eqref{qCdef}, we can recognize a familiar form. The first summation is identical to what we have seen in the infinite server queue length decomposition, and thus by Theorem~\ref{fBatchScale} we know that as the relative batch size increases this should resemble the shot noise process if properly normalized. Hence, we turn our attention to the waiting time correction. 

To start, let us reason about when waiting should occur, since $W_{i,j} > 0$ is necessary for these indicator functions to ever be equal to 1 for some value of $t$. By definition, a customer will wait in the multi-server queueing model when the number of  present customers is greater than the number of servers, and the delay of their start of service will be longer when the number of excess customers is higher. If the number of customers is no more than the number of servers, then the multi-server queue will behave just like the infinite server queue and there will be no waiting. However, whenever the queue length exceeds the staffing level, the service will be bottlenecked by the number of servers. Following this intuition, let us introduce a storage process model, $\psi_t^C$, that modifies the shot noise process in an analogous fashion. If the storage process is below a capacity level $c$, it should behave just like the shot noise process, but when the storage process exceeds the capacity its behavior should be limited by that level. This leads us to
\begin{align}
\psi_t^C
&=
\underbrace{
\left(\psi_0^C \wedge c\right) \bar{G}_0 (t)
+
\sum_{i=1}^{N_t} M_i \bar{G}(t-A_i)
}_{\text{Shot noise process without capacity}}
+
\underbrace{
\left(\psi_0^C - c\right)^+ \bar{G}(t)
+
\int_0^t \left( \psi_{t-s}^C - c \right)^+ \mathrm{d} G(s)
}_{\text{Capacity correction}}
\label{fPsiCGen}
.
\end{align}
Like the relationship between $Q_t^\infty(n)$ and $\psi_t^\infty$, we can see that $\psi_t^C$ mimics the behavior of $Q_t^C(n)$: up-jumps at the arrival epochs and  decreases between, with the rate of decrease being limited by the staffing or capacity level. Furthermore, just like the shot noise process $\psi_t^\infty$, here we can see that $\psi_t^C$ has deterministic behavior between arrival epochs, capturing the relative lack of variability seen at a large scale. With an analogous initial condition, $Q_0^C(n) \slash n \to \psi_0^C$ as $n \to \infty$, this leads us to our first main result, the general batch scaling limit of the multi-server queue in Theorem~\ref{fDelayConv}.

\begin{theorem}\label{fDelayConv}
As $n \to \infty$, the batch scaling of the $G_t^{B(n)}/GI/cn$ queue $Q_t^C(n)$ yields
\begin{align}
\frac{Q_t^C(n)}{n} \stackrel{D}{\Longrightarrow} \psi_t^C
,
\end{align}
pointwise in $t \geq 0$, where $\psi_t^C$ is a generalized storage process as defined in Equation~\eqref{fPsiCGen}.
\end{theorem}

Just like we remarked for the comparison of $Q_t^\infty(n)$ and $\psi_t^\infty$, $Q_t^C(n)$ answers the question ``how many customers are in the system?'' while $\psi_t^C$ answers ``how many batches are in the system at time $t$?'' We can also reframe the staffing problem in the same manner. Rather than searching for a staffing level $cn$ that delivers a sufficiently low probability that the number of customers will exceed the number of servers, Theorem~\ref{fDelayConv} allows us to instead seek a level $c$ such that the probability that the number of \emph{batches} will not exceed it is sufficiently low. This shows us that, in the presence of large batches, the staffing level must be directly proportional to the batch size. That is, by direct consequence of the convergence of the batch arrival queue to the storage process, itself a stochastic model, the queue's exceedance probability at staffing level $cn$ converges to a non-degenerate probability.

\begin{corollary}\label{staffCor}
In the $G_t^{B(n)}/GI/cn$ queue as $n \to \infty$ with $c > 1$,
\begin{align}
\lim_{n\to\infty} \PP{Q_t^C(n) \geq cn} 
=
\PP{\psi^C_t \geq c}
\in (0,1)
,
\end{align}
for each $t \geq 0$ such that the arrival epoch process $N_t$ satisfies
\begin{align}
\PP{
\sum_{i=1}^{N_t} M_i \bar{G}(t-A_i)
+
\int_0^t \left( \psi_{t-s}^C - c \right)^+ \mathrm{d} G(s)
> 
c
-
\left(\psi_0^C \wedge c\right) \bar{G}_0 (t)
-
\left(\psi_0^C - c\right)^+ \bar{G}(t)
}
\in (0,1)
,
\label{arrCond}
\end{align}
meaning that arrival process does not render the storage exceeedance probability trivially degenerate.
\end{corollary}

Plainly, if the relative batch size doubles, Theorem~\ref{fDelayConv} and Corollary~\ref{staffCor} show that the queue's staffing level should precisely double as well. This shows the lack of an economy of scale: unlike when an arrival rate grows large, there is not a labor savings benefit as the batch size grows large. 
%To this same end, this shows the strong demands of simultaneous arrivals to service systems. 
The staffing in this system is directly proportional to the effective arrival rate, like in the QD regime, but Theorem~\ref{fDelayConv} shows that the limit is still random and Corollary~\ref{staffCor} emphasizes that this high level of staffing still yields a non-degenerate exceedance probability, like in the QED regime. %We will further explore the comparison of large batches and large arrival rates in Section~\ref{hybridSec}.

\begin{figure}[htb]
\centering
\includegraphics[width=\columnwidth]{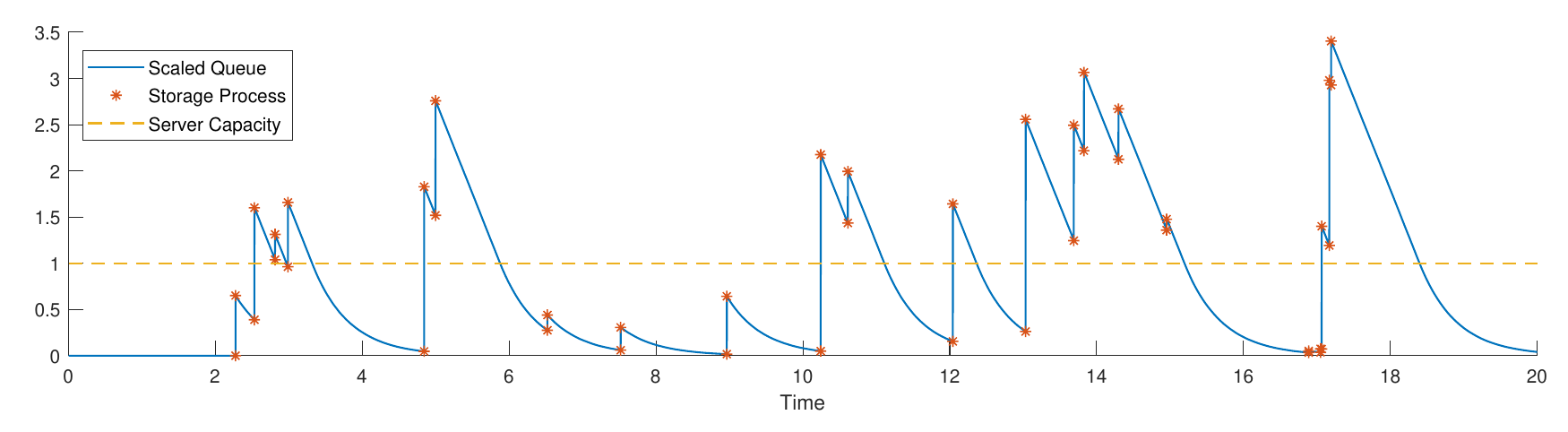}
%\vspace{-1.3pc}
\caption{A comparison of the simulated scaled queue length process and the calculated storage process sample paths defined on the same arrival epochs and jumps.}~\label{fig:samplepaths}
\end{figure}

Let us note that the arrival epoch process condition in \eqref{arrCond} is hardly restrictive. For example, it is immediately satisfied for $N_t$ as a (possibly non-stationary) Poisson process with nonzero rate. Still, the storage process may be somewhat opaque as defined in \eqref{fPsiCGen}. To provide some intuition about this stochastic process, let us consider the case of exponentially distributed service. For this example, suppose that $\bar{G}(x) = e^{-\mu x}$ for some $\mu > 0$. Then, Equation~\eqref{fPsiCGen} yields that
$$
\psi_t^C
=
\psi_0^C e^{-\mu t}
+
\sum_{i=1}^{N_t} M_i e^{-\mu(t-A_i)}
+
\int_0^t \left( \psi_{t-s}^C - c \right)^+ \mu e^{-\mu s} \mathrm{d} s
.
$$
Multiplying and dividing by $e^{-\mu t}$ inside the integral, we can re-express this as
$$
\psi_t^C
=
\psi_0^C e^{-\mu t}
+
\sum_{i=1}^{N_t} M_i e^{-\mu(t-A_i)}
+
e^{-\mu t}\int_0^t \left( \psi_{t-s}^C - c \right)^+ \mu e^{\mu (t-s)} \mathrm{d} s
,
$$
and by changing the variable of integration to be $s$ instead of $t-s$, we furthermore have
$$
\psi_t^C
=
\psi_0^C e^{-\mu t}
+
\sum_{i=1}^{N_t} M_i e^{-\mu(t-A_i)}
+
e^{-\mu t}\int_0^t \left( \psi_{s}^C - c \right)^+ \mu e^{\mu s} \mathrm{d} s
.
$$
Since we know that the process jumps by $M_i$ at the $i^\text{th}$ arrival, let us take $t \in (A_i, A_{i+1})$ and focus on the behavior between jumps. Because storage processes are deterministic on  inter-jump intervals, we can take the derivative with respect to time and observe that for $t \in (A_i, A_{i+1})$,
\begin{align*}
\frac{\mathrm{d}\psi_t^C}{\mathrm{d}t}
&=
-
\mu \psi_0^C e^{-\mu t}
-
\mu \sum_{i=1}^{N_t} M_i e^{-\mu(t-A_i)}
-
\mu e^{-\mu t}\int_0^t \left( \psi_{s}^C - c \right)^+ \mu e^{\mu s} \mathrm{d} s
+
\mu \left( \psi_{t}^C - c \right)^+
\\
&=
- \mu \psi_t^C
+
\mu \left( \psi_{t}^C - c \right)^+
\\
&=
- \mu \left( \psi_t^C \wedge c \right)
.
\end{align*}
Hence, in the case of exponential service the inter-jump dynamics of this process can be easily summarized. If $\psi_t^C$ is above the threshold $c$, it drains linearly, if it is below $c$, it decays exponentially. This precisely matches what we would expected from a $G_t/M/c$ queue: departures at a rate proportional to the minimum of the number in system and the number of servers. As an example of the limiting threshold dynamics, in Figure~\ref{fig:samplepaths} we plot a simulated scaled queue length sample path along with the calculated storage process values when given the same arrival epochs. 

Let us note that our focus in this section has been to establish the key managerial takeaway showing the strong demand that batch arrivals place on service systems, as captured in Theorem~\ref{fDelayConv}'s insight that the QD-style staffing will yield QED-style performance. This shows the absence of an economy of scale. Furthermore, Corollary~\ref{staffCor} implies that the staffing problem for the batch arrival queue can be directly translated to a staffing problem for the storage process limit. While we have not explicitly said \textit{how} to staff a storage process here, this is the focus of Appendix~\ref{staffingSec}, in which we build upon results available in the storage process literature. In what follows of the main body of the paper, we are interested in finding a second key managerial insight, specifically \edit{through contrast with} the case in which customers arrive both en masse \textit{and} quite frequently.

\section{Large Batches and Large Arrival Rates: Revealing a Spectrum}\label{hybridSec}

%\tr{maybe reduce number of QD/QED comparisons}

Following Section~\ref{batchSec}'s consequences for queues with large batch arrivals, it is natural to wonder what interplay exists between large batches and large arrival rates. This section's results will reveal a spectrum between the large batch limit in Theorem~\ref{fDelayConv} and the classical QED heavy traffic limit, originating in \citet{halfin1981heavy}. To identify the interior between these extremes in sufficient clarity, we will now focus our attention on  the $M^n/M/c$ queue in steady-state. %That is, we will suppose the batches are all of a fixed size and that they occur at arrival epochs according to a homogeneous Poisson process. 
The limiting regimes we now consider here will be what we will refer to as a \emph{hybrid} scaling \edit{or \emph{large batch-and-rate} regime}, in which the effective arrival rate grows large through both the batch size and batch arrival rate. 

Since we are now considering a more specific setting, let us define updated notation. First, let us introduce $m$ as the relative effective arrival rate; this will be the index for our hybrid limits. Following the steady-state assumption, we will drop the $t$ subscripts  and let $Q^\infty(m)$ and $Q^C(m)$ be the infinite and multi-server queue lengths, respectively. Perhaps the most important parameter of this scaling will be $\nu \in [0,1]$, which dictates the relative weight of the batch size and the batch arrival rate within the effective arrival rate. That is, we will suppose that all batches are of size $\edit{n(m) = n_0 m^\nu}$ for some constant $\edit{n_0 > 0}$, and likewise we will let the batch arrival rate be $\edit{\lambda(m) = \lambda_0 m^{1-\nu}}$ for some $\edit{\lambda_0 > 0}$. Hence, the product of the batch size and the batch arrival rate, $\edit{\lambda_0 n_0 m}$, is the effective arrival rate of customers to the service system, \edit{and we will refer to $m$ as the relative effective arrival rate}. By convention, we will assume that $\edit{n_0 m^\nu}$ is an integer to avoid cumbersome expressions, but this analysis can be carried through with a rounded quantity as the batch size instead, such as $\edit{\lceil n_0 m^\nu \rceil}$. If $\nu = 1$, this scaling reduces to the large batch \edit{regime from} Section~\ref{batchSec}, and if $\nu = 0$, we will recover the QED regime.
%\tb{
%\begin{itemize}
%\item index by mean effective arrival rate $m$, proportional to the product of the arrival rate and the batch size
%\item also establish results for the infinite server here
%\item at $\nu = 1$ the fluid and diffusion limits coincide, which points back the the prior section's observation that QD-type staffing yields QED-type performance
%\item steady-state analysis, dropping the $t$ subscript
%\item use a convention that batch sizes and staffing levels given by non-integer quantities are actually evaluated at the integer ceiling of these values.
%\item say Theorem~\ref{bothLimitMulti} is the main result of this section
%\end{itemize}
%}

%$m$ is proportional the product of the mean arrival rate and the batch size. That is,  $\lambda n m$ is the mean rate of customers' entry into the system

%We will assume 

\subsection{Hybrid \edit{Batch-and-Rate} Limits for the Infinite Server Queue}\label{hybridInfSec}

To begin building intuition on the interplay of the arrival rate and the batch size in both queueing models, let us first review some properties of the infinite server queue.  In particular, since the classical QD and QED regimes are closely related to law of large numbers and central limit theorem type results, the mean and variance of the queue length hold particular relevance. Hence, in Proposition~\ref{infMeanVar} we provide the mean and variance \edit{in} this hybrid scaling parameter setting. 

\begin{proposition}\label{infMeanVar}
In an $M^n/M/\infty$ queue with arrival rate $\edit{\lambda(m) = \lambda_0 m^{1-\nu}}$ and batch size $\edit{n(m) = n_0 m^{\nu}}$, the mean steady-state queue length is given by
\begin{align}
\E{Q^\infty(m)}
&=
\frac{\edit{\lambda_0 n_0} m}{\mu}
,
\label{isMean}
\end{align}
and the steady-state variance is equal to
\begin{align}
\Var{Q^\infty(m)}
&=
\frac{\edit{\lambda_0 n_0} m (\edit{n_0} m^{\nu} + 1)}{2\mu}
,
\label{isVar}
\end{align}
where $\nu \in [0, 1]$.
\end{proposition}

Look at how these quantities relate for different values of $\nu$, or specifically, how the mean and standard deviation compare as $\nu$ changes. Starting with the classical, at $\nu = 0$ the standard deviation is of order $\sqrt{m}$, whereas the mean is of order $m$. On the other hand, at $\nu = 1$ for the pure large batch scaling setting, the mean and standard deviation are both of order $m$. On the interior, for $\nu \in (0,1)$, the standard deviation is then of order $m^{(1+\nu)/2}$, making it not quite on the same level as the mean but greater than the square root of it. Hence, as long as $\nu < 1$,  the order of the mean dominates the order of the standard deviation, and through this we can find different limits for the two different orders of normalization.

\begin{theorem}\label{bothLimitInf}
Let $\nu \in [0,1)$. As $m\to\infty$ in the $M^n/M/\infty$ queue with arrival rate $\edit{\lambda(m) = \lambda_0 m^{1-\nu}}$ and batch size $\edit{n(m) = n_0 m^{\nu}}$, the steady-state queue length converges to a constant when normalized by $m$:
\begin{align}
\frac{Q^\infty(m)}{m}
\stackrel{\mathsf{a.s.}}{\longrightarrow}
\frac{\edit{\lambda_0 n_0}}{\mu}
\edit{.}
\end{align}
However, when centered by its mean and normalized by $m^{\frac{1+\nu}{2}}\sqrt{1 + \frac{1}{\edit{n_0} m^{\nu}}}$, the steady-state queue length converges to
\begin{align}
\frac{Q^\infty(m) - \frac{\edit{\lambda_0 n_0} m}{\mu}}{m^{\frac{1+\nu}{2}}\sqrt{1 + \frac{1}{\edit{n_0} m^{\nu}} } }
\stackrel{D}{\Longrightarrow} 
X
\label{diffusionEq}
,
\end{align}
as $m \to \infty$, where $X \sim \mathsf{Norm}(0, \edit{\lambda_0 n_0^2} \slash 2\mu)$.
\end{theorem}

\begin{figure}[htb]
\centering
\subfigure[$\nu = 0$]{
\includegraphics[width=0.31\textwidth]{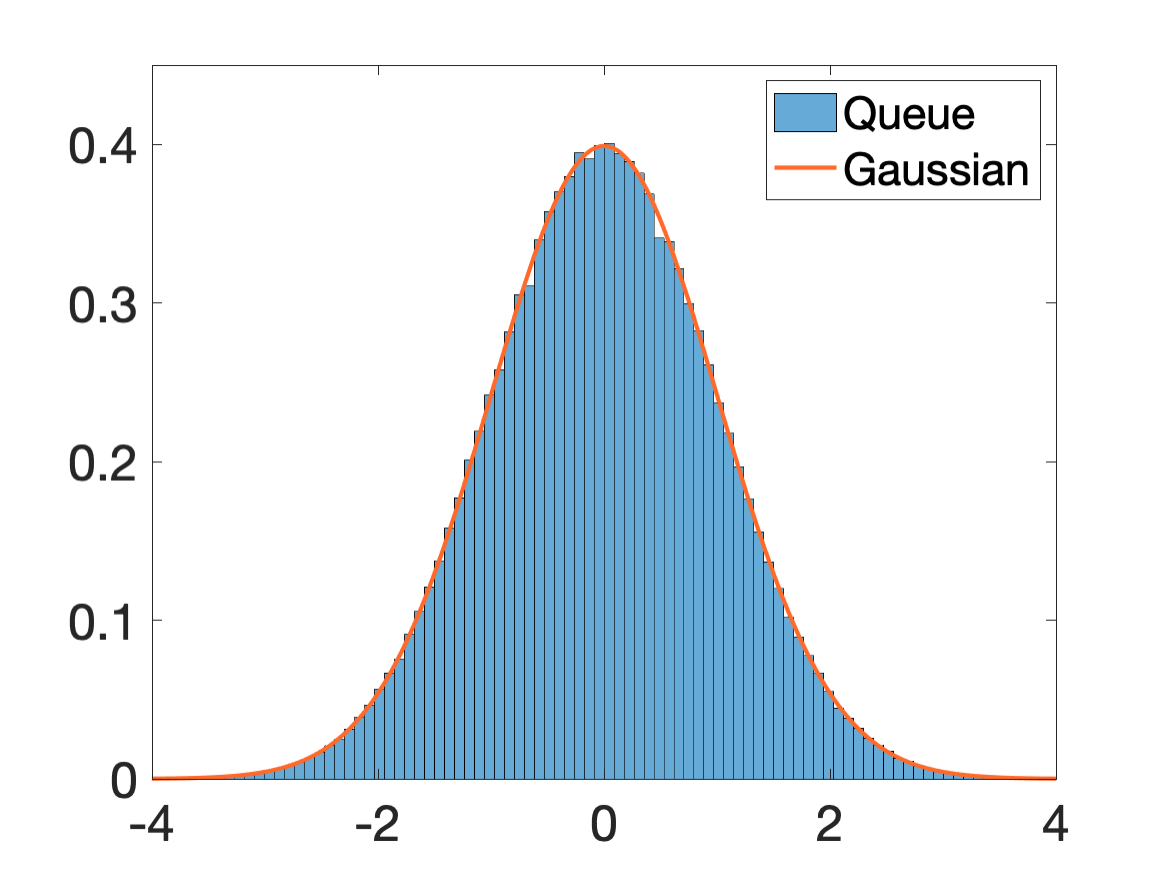}
}
\subfigure[$\nu = 1\slash 5$]{
\includegraphics[width=0.31\textwidth]{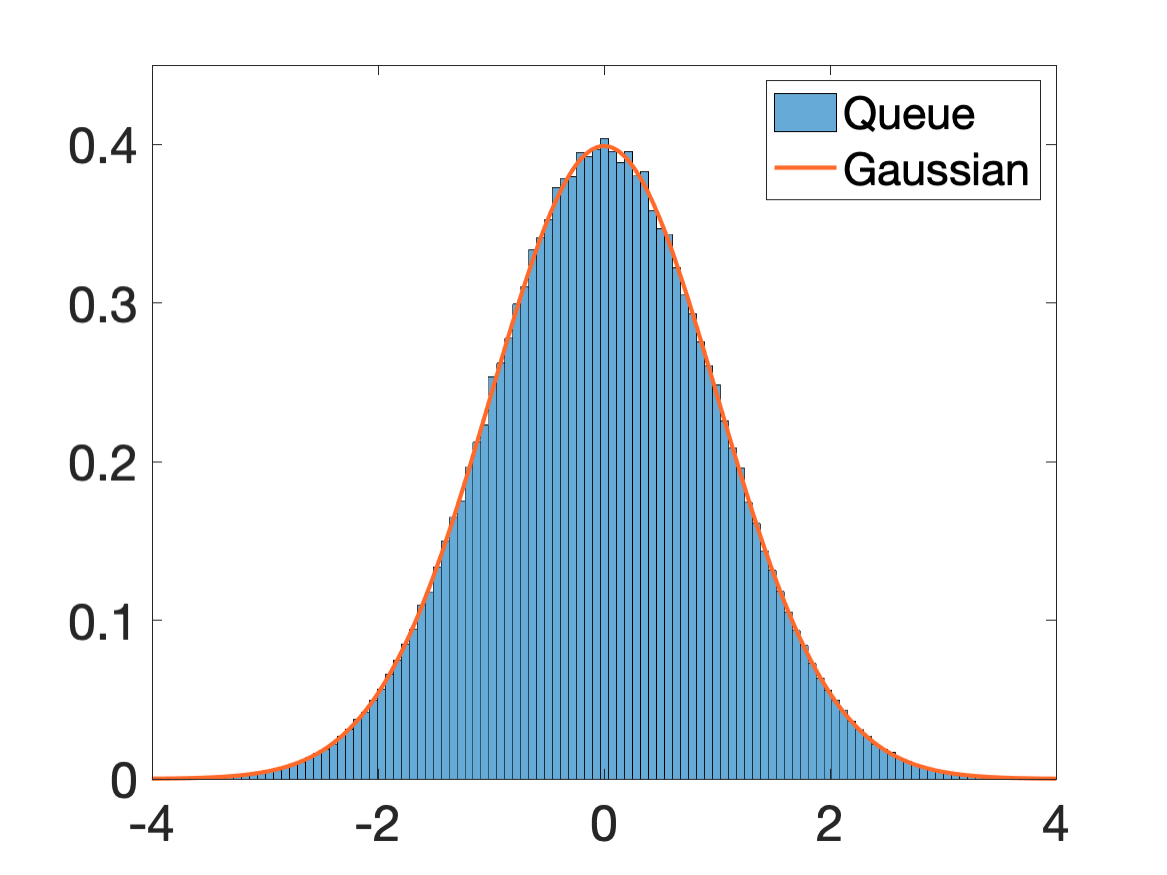}
}
\subfigure[$\nu = 2\slash 5$]{
\includegraphics[width=0.31\textwidth]{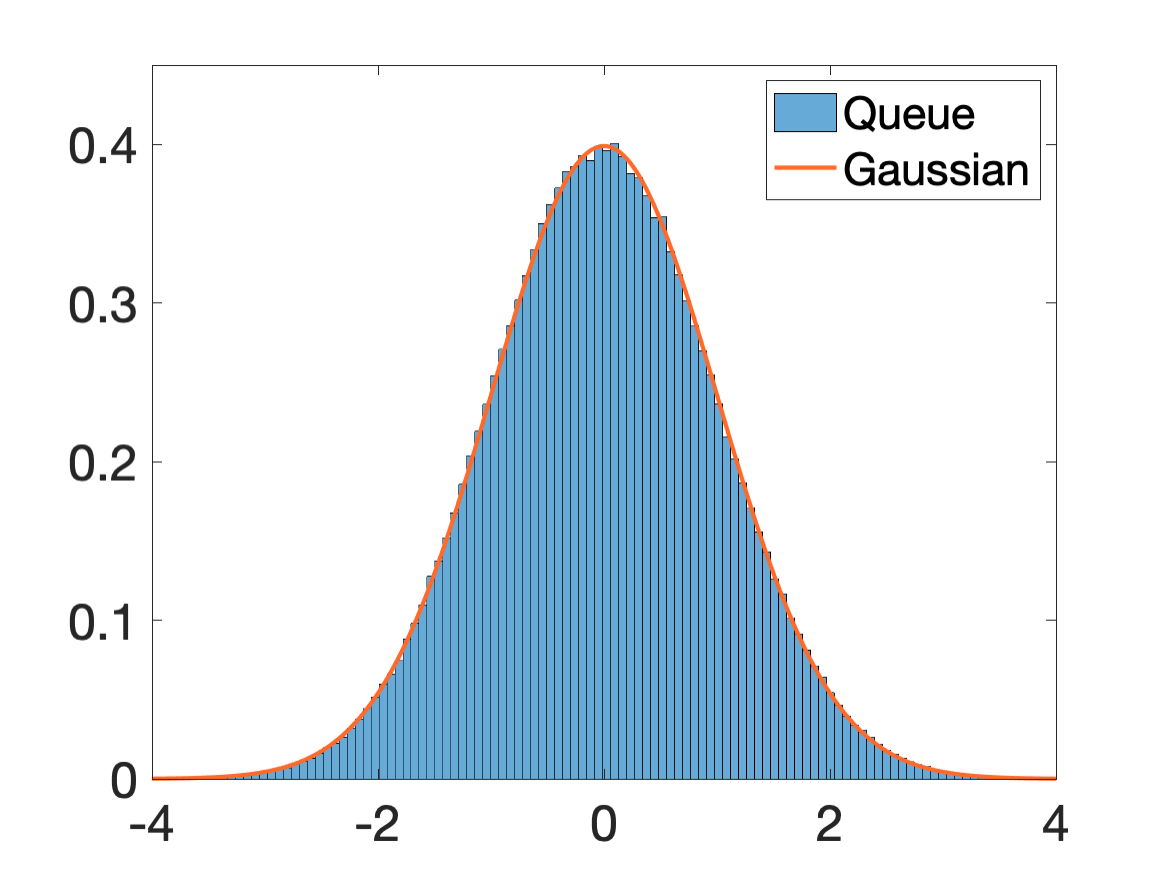}
}
\subfigure[$\nu = 3\slash 5$]{
\includegraphics[width=0.31\textwidth]{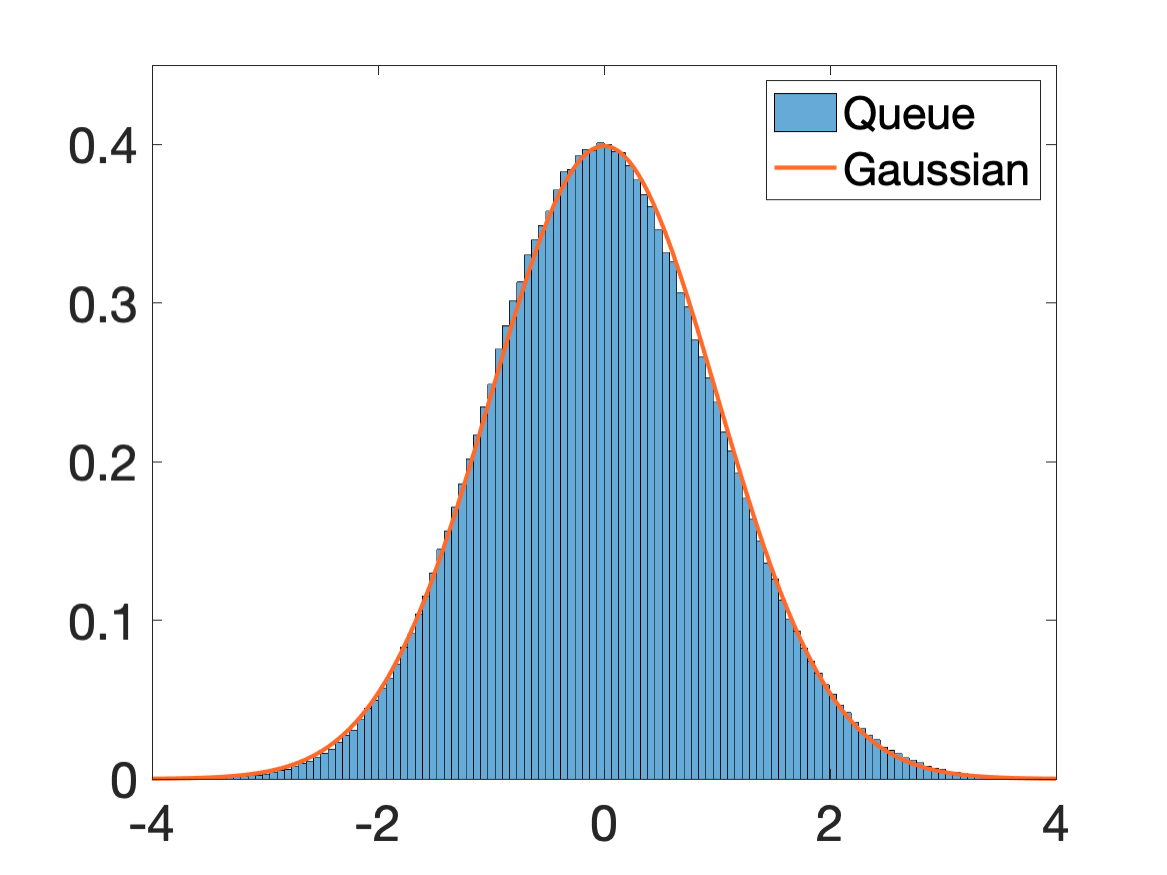}
}
\subfigure[$\nu = 4\slash 5$]{
\includegraphics[width=0.31\textwidth]{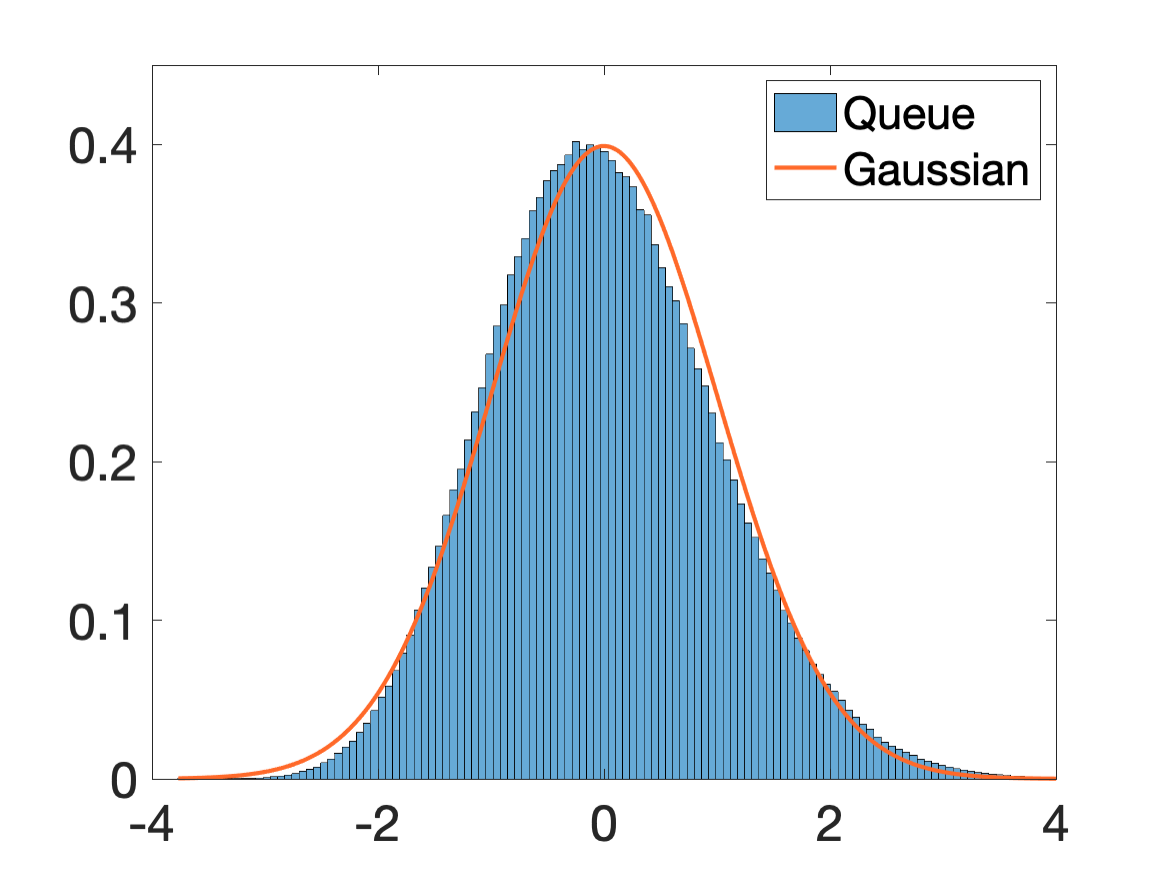}
}
\subfigure[$\nu = 1$]{
\includegraphics[width=0.31\textwidth]{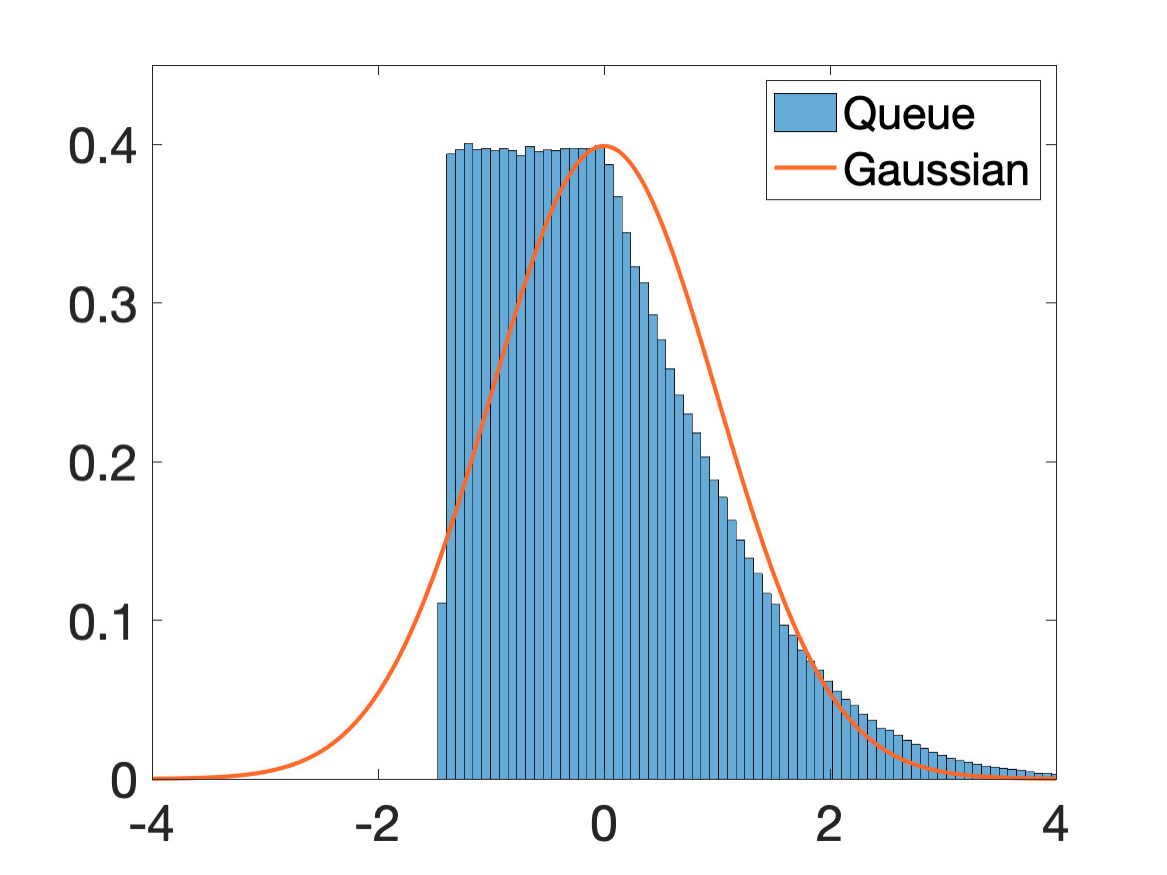}
}
\caption{Comparison of standard normal and simulated hybrid scaled infinite server queue lengths centered and normalized \edit{according to Equation~\eqref{diffusionEq}} for varying values of $\boldsymbol{\nu}$, where $\boldsymbol{m = 10^5}$.}\label{normFig}
\end{figure}

Let us make the contrast clear: if $\nu = 1$, there is only one limit, and this is given by Theorem~\ref{fBatchScale}. Furthermore, as made plain by Corollary~\ref{mgfCor} and the surrounding remarks, this limit is random but not normal. Rather, the limiting distribution dominates Gaussian or sub-Gaussian distributions. As a demonstration of this, we compare the density function of a standard normal to the simulated histograms of the centered and normalized infinite server queue under a range of hybrid scaling settings in Figure~\ref{normFig}. As we can see, the asymptotic normality of the hybrid scaling is overwhelming when the arrival rate is near the same or higher order than the batch size;  it is close even when the batch size is 10,000 and the arrival rate is 10. But, when the limit is fully on the batch size, the distribution of the scaled and centered queue is clearly not normal.

%for $\nu \in (0,1)$, the order of the normalization is strictly between $1\slash 2$ and 1

%\begin{corollary}
%When centered by its mean and normalized by $m^{\frac{1+\nu}{2}}$ for $\nu \in (0,1)$, the steady-state queue length converges to
%\begin{align}
%\frac{Q^\infty(m) - \frac{\lambda n m}{\mu}}{m^{\frac{1+\nu}{2}}}
%\stackrel{D}{\Longrightarrow} 
%X
%,
%\end{align}
%as $m \to \infty$, where $X \sim \mathsf{Norm}(0, \lambda n^2 \slash 2\mu)$.
%\end{corollary}

Because the leading order of the standard deviation is $m^{(1+\nu)\slash 2}$, one can think of the normalization as being of this order (without the multiplied square root term) for simplicity, if preferred. We will make use of this condensed presentation in our subsequent staffing analysis in the next subsection. Following Theorem~\ref{bothLimitInf}, we see how this hybrid large batch and large arrival rate limit recovers a Goldilocks takeaway like what QED offers, albeit still at a higher order. If $\nu < 1$ in the hybrid regime, scaling by $m$ is too much, scaling by $\sqrt{m}$ is too little, but scaling by $m^{(1+\nu)/2}$ is just right. However, if $\nu = 1$, there is no choice to be made; there is only one scaling and it is of order $m$\edit{, and this is a consequence of the fact that if $\nu = 1$, then Theorem~\ref{fBatchScale} shows that the fluid- and diffusion-type limits coincide at the shot noise process}.

\subsection{Hybrid \edit{Batch-and-Rate} Limits for the Multi-Server Queue}\label{hybridMultiSec}

Now, to properly connect these results to staffing decisions, we must move beyond the infinite server queue to the multi-server. So, for the second of our two main results, we will apply this hybrid scaling limit to the case of finitely many servers. We will \edit{again} set the arrival rate and batch size as $\edit{\lambda(m) = \lambda_0 m^{1-\nu}}$ and $\edit{n(m) = n_0 m^\nu}$ for the limit indexed by $m$ with spectrum parameter $\nu$, and because of Theorem~\ref{bothLimitInf} we will set the staffing level as $\edit{c(m) = \lambda(m)n(m)} \slash \mu + \delta m^{(1+\nu)/2}$ for some $\delta > 0$.\footnote{We could sharpen the convergence (to the same result) by having the staffing level instead as $\edit{\lambda_0 n_0} m \slash \mu + \delta m^{(1+\nu)/2} \sqrt{1 + 1 \slash (\edit{n_0}m^\nu)}$, but we stick with a safety staffing of $\delta m^{(1+\nu)\slash 2}$ for a simplified and clear presentation.} In this case, we prove a steady-state exceedance probability limit fashioned in the style of Proposition 1 of \citet{halfin1981heavy} with higher order safety staffing.

%\tr{batches need more}

\begin{theorem}\label{bothLimitMulti}
Let $\nu \in (0,1)$. As $m \to \infty$ in the $M^n/M/c$ queue with arrival rate $\edit{\lambda(m) = \lambda_0 m^{1-\nu}}$, batch size $\edit{n(m) = n_0 m^\nu}$, and staffing level $\edit{c(m) = \lambda_0 n_0} m \slash \mu + \delta m^{(1+\nu)/2}$ for some $\delta > 0$, the steady-state exceedance probability converges to
\begin{align}
\lim_{m \to \infty}
\PP{Q^C(m) \geq \edit{c(m)}}
%\frac{\edit{\lambda_0 n_0} m }{ \mu} + \delta m^{\frac{1+\nu}{2}}}
=
\frac{
\frac{\edit{n_0}}{2\delta}  \sqrt{\frac{\edit{\lambda_0}}{\mu \pi}} e^{-\frac{\delta^2}{{\edit{\lambda_0 n_0^2}}\slash{\mu}}}
}
{
\Phi\left(\frac{\delta}{\edit{n_0} } \sqrt{\frac{2\mu}{\edit{\lambda_0}}}  \right)
+
\frac{\edit{n_0}}{2\delta}  \sqrt{\frac{\edit{\lambda_0}}{\mu \pi}}  e^{-\frac{\delta^2}{{\edit{\lambda_0 n_0^2}}\slash{\mu}}}
}
,
\label{hybLimEq}
\end{align}
where $\Phi(\cdot)$ is the cumulative distribution function of a standard normal random variable.
\end{theorem}

The asymptotic normality on the interior of the spectrum makes the limit more closely related to QED staffing than may have been obvious. In fact, the expression for the limiting value in the right hand side of Equation~\eqref{hybLimEq} exactly matches the right hand side of Proposition 1 of \citet{halfin1981heavy} for a Halfin-Whitt parameter $\beta = {\delta}\slash{\edit{n_0}}\sqrt{{2\mu}\slash{\edit{\lambda_0}}}$. \edit{This yields the corresponding result for case of $\nu = 0$.} Hence, arrivals in large batches and at large rates recover precise QED performance as these batch sizes and arrival rates grow large simultaneously, but to properly account for the batches the order of the safety staffing must be higher than what QED prescribes. 

For these reasons, we say that an economy of scale does exist in the presence of large batches and large arrival rates, but that it is weaker than what is typically expected when only the arrival rate is large. That is, if the effective arrival rate $\edit{\lambda_0 n_0} m$ doubles through \edit{the relative effective arrival rate} $m$ doubling, the resulting staffing will be less than double the prior level so long as $\nu < 1$. In the case that $\nu = 1$, we revert to the pure large batch setting in Section~\ref{batchSec}, in which there is no economy of scale and the new staffing is precisely double the level before.

%- neat! above QED staffing (greater than square root of mean arrival rate $\lambda n m$) but yields QED performance, exactly! Note $\nu \in [0,1)$

%\tr{universal interpretation of Theorems~\ref{fDelayConv} and~\ref{bothLimitMulti}: add a commentary about safety staffing being proportional to $n(m) \sqrt{\lambda(m)}$ -- linear in the batch size, square root in the arrival rate -- note that staffing at $\lambda(m)n(m)/\mu + \delta n(m)\sqrt{\lambda(m)}$ should deliver the approximately the same exceedance probability for any $\lambda(m) n(m) = m$ with $m$ sufficiently large, so long as $\nu < 1$ -- large batch and batch-and-rate agree in the order of the safety staffing, but let's look at the nuanced contrast between them}

\edit{With both Theorems~\ref{fDelayConv} and~\ref{bothLimitMulti} now in hand, let us observe universal batch staffing guidelines. In interpreting the two results, we see that both limits agree that the safety staffing should be proportional to $n \sqrt{\lambda}$. That is, the safety staffing is linear in the batch size and square root in the arrival rate. Theorem~\ref{bothLimitMulti} tells us that staffing at $\lambda(m)n(m)/\mu + \delta n(m)\sqrt{\lambda(m)}$ should deliver the approximately the same exceedance probability for any $\lambda(m) n(m) = m$ with effective arrival rate $m$ sufficiently large, so long as $\nu < 1$. If $\nu = 1$, however, Theorem~\ref{fDelayConv} agrees that $n \sqrt{\lambda}$ is the correct order, but the nature of the exceedance probability need not be the same as the $\nu < 1$ case. In fact, these can differ substantially. Next, we inspect the nuanced contrast between these regimes.}

\section{\edit{Contrasting the Large Batch and Batch-and-Rate Regimes}}\label{contrastSec}

\edit{To fully emphasize the challenges of large batch arrivals, let us compare the large batch regime of Section~\ref{batchSec} with Section~\ref{hybridSec}'s batch-and-rate regime as $\nu \to 1$. Like in Section~\ref{hybridSec}, let us continue to use the Markovian system with constant batch sizes for straightforward comparison.}

%As we alluded to in Section~\ref{litReview}, this both matches and justifies the peakedness approximations such as in \citet{whitt1992understanding}, while also providing a cautionary tale that service managers should remain wary that peakedness parameters may not be fixed and may increase in tandem with the effective arrival rate. 

\edit{As we alluded to in Section~\ref{litReview}, the  safety staffing of order $n \sqrt{\lambda}$ both matches and justifies the peakedness approximations such as in \citet{whitt1992understanding}, but this comes with important caveats. Let us first review the peakedness approximation. Using the present notation, equations (12) and (13) from \citet{whitt1992understanding} provide a staffing guideline of $c(m) = m/\mu + \delta \sqrt{z m}$ for $m = \lambda n$, where $z$ is
\begin{align}
z 
&=
\frac{\mathsf{c}_a^2 + 1}{2}
,
\end{align}
with $\mathsf{c}_a^2$ as the squared coefficient-of-variation for the arrival process. For the $M^n/M/c$ or $M^n/M/\infty$ systems (suppressing any dependence of $\lambda$ or $n$ on $m$), $\mathsf{c}_a^2$ can be calculated through
\begin{align}
\mathsf{c}_a^2
&=
\lim_{t\to\infty}
\frac{
\Var{ n N_t }
}{
\E{ n N_t }
}
=
n
.
\end{align}
Hence, the peakedness approximation suggests a staffing of $c(m) = \lambda n /\mu + \delta \sqrt{\lambda n(n+1)/2}$, where now this shows that the peakedness parameter $z$ will change with $n$. Moreover, this is precisely the order of staffing we have now rigorously justified both in the batch-and-rate setting (Theorem~\ref{bothLimitMulti}) and the large batch regime (Theorem~\ref{fDelayConv}). In fact, we can even recognize the peakedness parameter sitting within the infinite server variance given in Proposition~\ref{infMeanVar}, which applies to both settings.
}

\edit{However, while this perspective shows that these regimes agree in their safety staffing order when phrased in terms of the arrival rate and batch size, there remains an important difference between the regimes. That is, this peakedness approximation is predicated on the observation that the stochastic offered load, or number in system for the infinite server model, is ``typically asymptotically normally distributed'' as the arrival volume increases \citep{whitt1992understanding}. This is indeed what Theorem~\ref{bothLimitInf} confirms for the batch-and-rate regime, but contrast this with Theorem~\ref{fBatchScale}. Instead of a Gaussian limit, Theorem~\ref{fBatchScale} yields a shot-noise process, and, as we've remarked, the tail of this distribution is fundamentally heavier than a normal. We have seen this stark contrast in Figure~\ref{normFig}.
} 

\edit{Hence, the critical flaw of the peakedness approximation for the large batch setting is the associated assumption of asymptotic normality. Because the underlying infinite server systems have this vital difference in limiting objects, we find that the scaffolded multi-server limits differ as well. This manifests itself in the coefficient $\delta$. In either case, setting $\delta$ according to some target exceedance probability, such as in Equation~\eqref{exProbDef}, essentially reduces to solving a tail inversion problem. Hence, $\delta$ can vary to a significant degree between the two regimes. Let us demonstrate.}

%\tr{expand on peakedness -- matches and justifies rigorously, but look what happens at the boundary -- the difference is in the nature of the infinite server limits (both are scaffolded from this)}

%\tr{the contrast is in the infinite server scaffolding! Theorem~\ref{fBatchScale} versus Theorem~\ref{bothLimitInf}: shot-noise versus gaussian. these have different tails, and that leads to different staffing}

%\tr{maybe move peakedness approx discussion here -- flaw of using peakedness for large batches is in the underlying normal approximation of the infinite server}

\begin{figure}[htb]
\centering
\includegraphics[width=\textwidth]{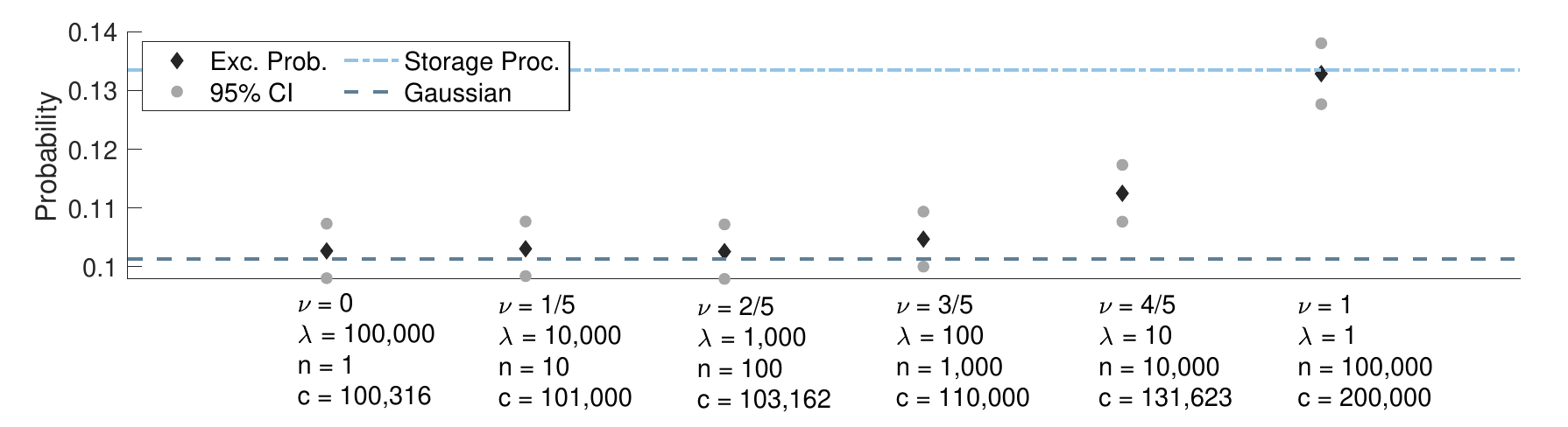}
\caption{Comparison of simulated performance \edit{when staffed at $\lambda n + n \sqrt{\lambda}$} and theoretical exceedance probabilities given by the Gaussian and storage process limits for varying values of $\boldsymbol{\nu}$, where $\boldsymbol{m = 10^5}$.}\label{hybCompFig}
\end{figure}

\edit{To see the effects of these divergent tails, let us inspect the staffing performance as the batch-and-rate regime approaches the large batch scaling, meaning as $\nu \to 1$. In Figure~\ref{hybCompFig}, we have the finite-server sequel to Figure~\ref{normFig}. Here we again simulate six cases, each with an effective arrival rate of $m = 100,000$, and now we take safety staffing of amount $n \sqrt{\lambda}$ above the offered load. In the first four cases, $\nu = 0, 1/5, 2/5,$ and $3/5$, we can see that the hybrid limit is clearly manifesting and that the simulated exceedance probabilities seem to be converging to the Gaussian-based value from above. However, at $\nu = 4/5$, we see that the simulated value strays from both the normal and storage limits, and then at $\nu = 1$, the performance aligns with the storage process calculation, which itself is considerably higher than the target derived from asymptotic normality.
}

\edit{To build intuition for what we see in Figure~\ref{hybCompFig} (and, similarly, on a first-order level in Figure~\ref{normFig}), let us inspect the utilization in each limiting regime. In the batch-and-rate setting given in Theorem~\ref{bothLimitMulti}, we can see that $\lim_{m\to \infty} \lambda(m)n(m) \slash \left(c(m) \mu\right) = 1$, but on the other hand under large batches according to Theorem~\ref{fDelayConv}, we find that $\lambda n \slash \left(c n \mu\right) = \lambda \slash (c\mu) < 1$ is fixed for all $n$. Hence, we have a simple and practical way of distinguishing the two regimes: If the utilization is near 1 with a large arrival volume through both fast rates and batches (or bursts), we find ourselves in the hybrid setting, and asymptotic normality should apply, just as we expect from classic QED or peakedness approaches. However, if we have batch (or burst) arrivals yet the utilization is lower, we are instead aligned with the large batch regime and its storage process limits. Indeed, this is what Figure~\ref{hybCompFig} shows. In the six scenarios of $\nu$ in Figure~\ref{hybCompFig}, the utilization is 99.7\%, 99.0\%, 96.9\%, 90.9\%, 76.0\%, and 50\%, respectively. The four cases that align well with Theorem~\ref{bothLimitMulti} are the four that have utilization above $90\%$; the middle ground case at $\nu = 4/5$ has moderate utilization at 76\%, and the storage-process-aligned case is idle as often as it is utilized.  Let us emphasize that the mean number of customers arriving per unit of time is constant at $m = 100,000$ across the six settings, yet it is the composition of the arrival pattern that changes not only the necessary staffing levels, but also the nature of the staffing performance at the corresponding levels.
}

\edit{In addition to the difference in performance (or, analogously, in the prescribed safety staffing coefficient), let us also observe that some important operational details may be lost in the batch-and-rate setting (and, equivalently, in the peakedness approximations). For example, let us recall that this paper's focal exceedance probability, $\PP{Q^C \geq c}$, could actually be replaced with the more-demanding $\PP{Q^C + n \geq c}$. In the $M^n/M/c$ setting,  $\PP{Q^C \geq c}$ is the probability that \emph{all} customers in an arriving batch must wait, whereas $\PP{Q^C + n \geq c}$ is the probability that \emph{some} customers in the batch wait. Of course, the complement $\PP{Q^C + n < c}$ gives the probability that no customers wait, and so, although we have focused on $\PP{Q^C \geq c}$ in our discussion, it may be quite natural for managers to use $\PP{Q^C + n \geq c}$ as the target instead. Theorem~\ref{fDelayConv} immediately applies the large batch regime to this case through continuous mapping, but, in the batch-and-rate setting, Proposition~\ref{coincideProp} shows that the two targets are actually the same in the limit.
}

%for $n=1$ and $\nu = 0$ this would coincide with Proposition 1 of HW, but the $\delta$ parameter needs to be multiplied by $\sqrt{2}$, which can be seen from Theorem~\ref{bothLimitInf}

\begin{proposition}\label{coincideProp}
\edit{Let $\nu \in (0,1)$. As $m \to \infty$ in the $M^n/M/c$ queue with arrival rate $\lambda(m) = \lambda_0 m^{1-\nu}$, batch size ${n(m) = n_0 m^\nu}$, and staffing level ${c(m) = \lambda_0 n_0} m \slash \mu + \delta m^{(1+\nu)/2}$, the some-wait and all-wait exceedance probabilities coincide asymptotically, i.e.,
\begin{align}
\PP{Q^C(m) + n(m) > c(m)}
%n_0 m^\nu \geq \frac{{\lambda_0 n_0} m }{ \mu} + \delta m^{\frac{1+\nu}{2}}}
-
\PP{Q^C(m) \geq c(m)}
%\frac{{\lambda_0 n_0} m }{ \mu} + \delta m^{\frac{1+\nu}{2}}}
\longrightarrow
0
\end{align}
as $m \to \infty$.
}
\end{proposition}

\begin{figure}[htb]
\centering
\includegraphics[width=.325\textwidth]{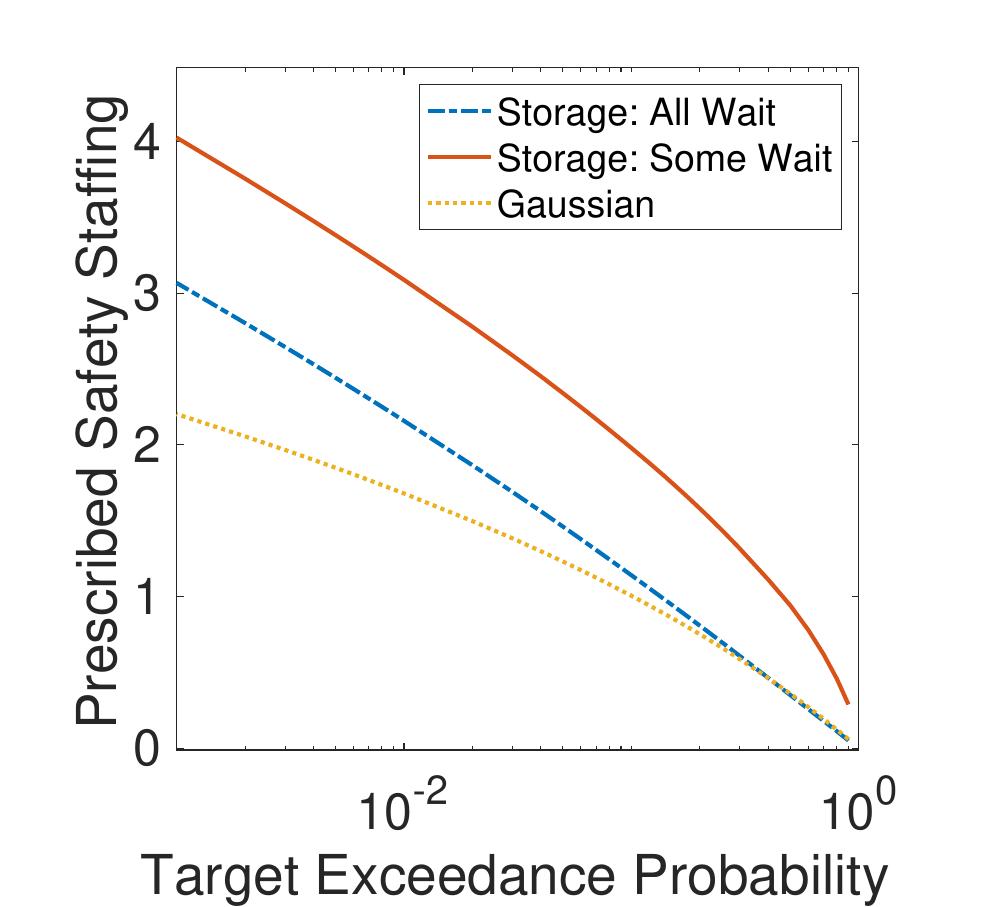}
\includegraphics[width=.325\textwidth]{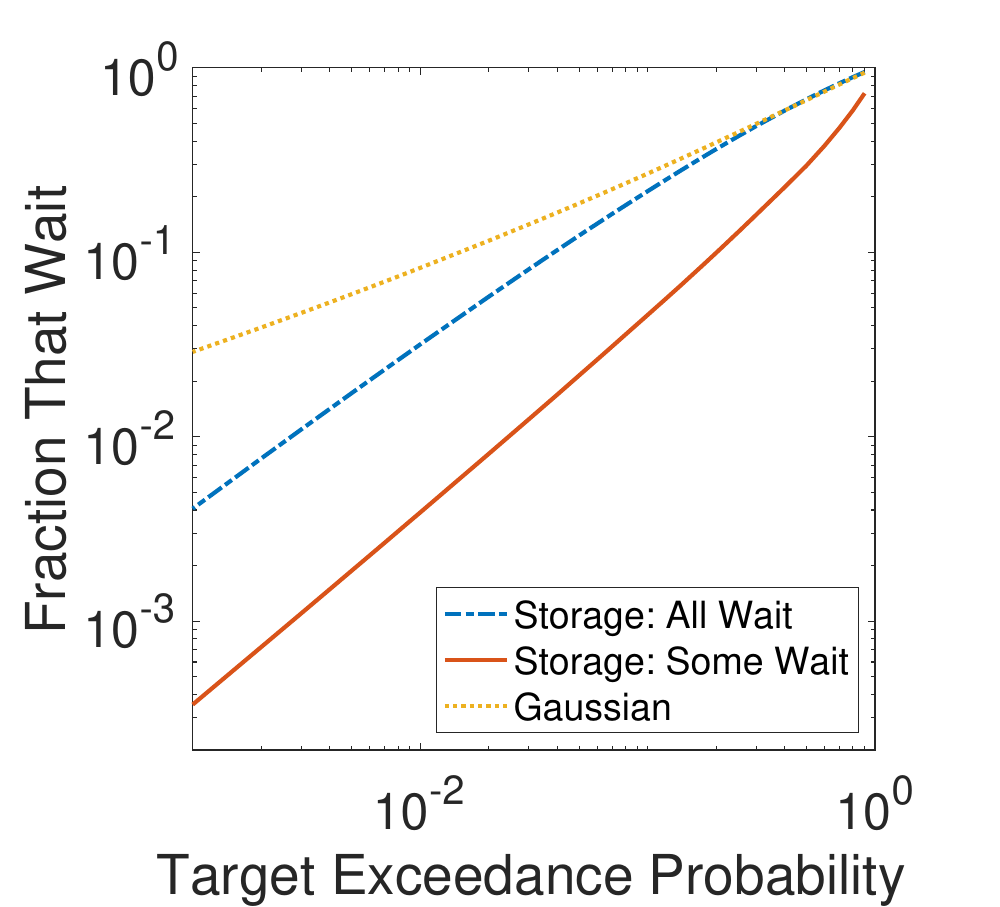}
\includegraphics[width=.325\textwidth]{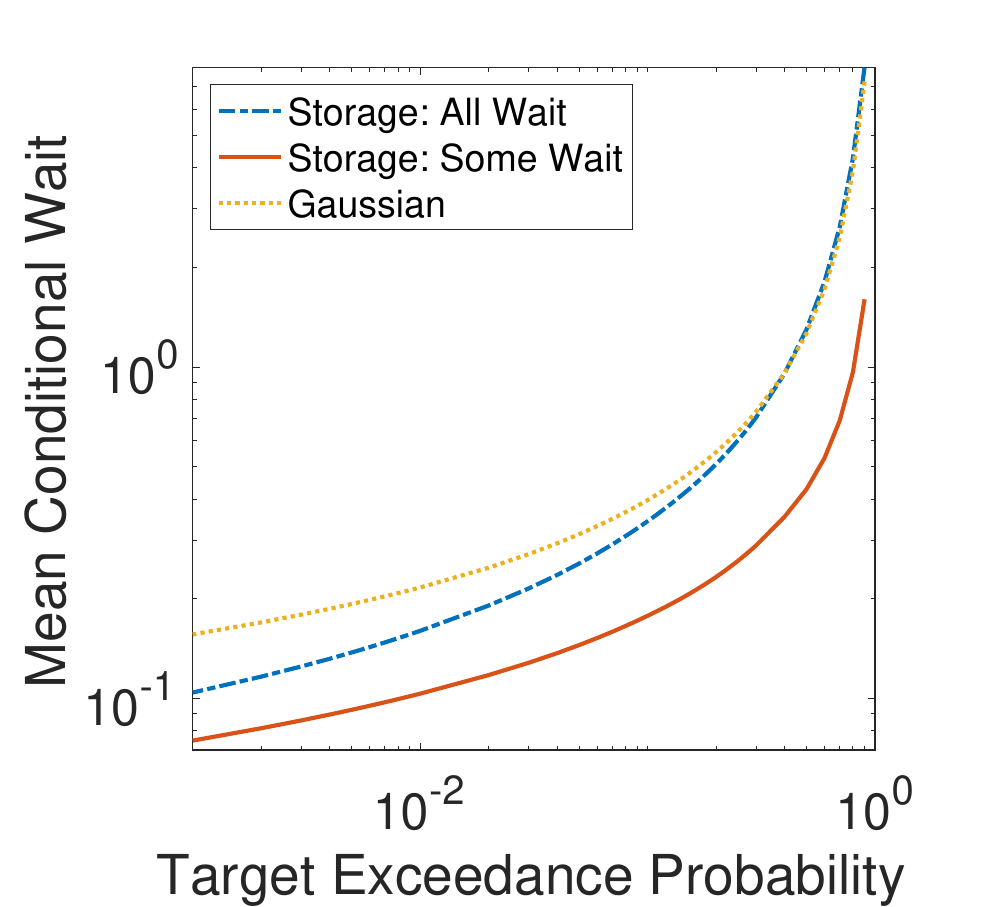}
\caption{\edit{Contrast of the different safety staffing levels prescribed by the storage process (Theorem~\ref{fDelayConv}) and hybrid Gaussian-based (Theorem~\ref{bothLimitMulti}) limiting objects, and the fractions of customers that wait and their mean waiting times that result under these staffing levels if the true setting is the large batch regime.}}\label{storageVsGaussianFig}
\end{figure}

\edit{Hence, in some sense, Proposition~\ref{coincideProp} shows that some of the nuances (and challenges) of the batch arrival context are washed out in the large batch-and-rate regime. On one hand, this is good news for managers, because if the service operation is truly in the batch-and-rate regime, then staffing at the all-wait target also delivers some-wait guarantees. However, on the other hand, Proposition~\ref{coincideProp} reveals what may be a tempting case of fool's gold. If the staffing is set according to batch-and-rate or peakedness approximation guidance, but the service setting actually more closely aligns with large batch regime, then managers may find the operation to be critically understaffed and doomed to  under-perform, particularly if the some-wait target is the true performance metric.
}

\edit{Figure~\ref{storageVsGaussianFig} shows how much these metrics can differ when calculating $M^n/M/c$ staffing. For the same target probability, the left-most plot shows the different safety staffing coefficients that would be given by the storage process limit under the two exceedance events, and by the Guassian-based limit under either event. The middle and right-most plots then compute the actual waiting fractions and conditional waiting times, respectively, in the storage process limit when using the staffing levels given from the three different approaches. In other words, these latter two figures show what would happen if the service environment is truly in the large batch regime (hence the storage process computations) but possibly not staffed accordingly. As we can see, not only is the waiting higher under Gaussian-based prescription, it does not improve at the same rate as the storage-based staffing when the target exceedance probability decreases.}

\section{\edit{Staffing for Contact Tracing: An Illustrative Case Study on NYC Test \& Trace Data}}\label{contactSec}

\edit{As we have seen in the preceding contrast in Section~\ref{contrastSec}, the large batch regime appears almost paradoxical operationally: the system may have low utilization, but customers still experience non-trivial wait. It would be fair to wonder if systems truly exhibit this modus operandi in reality; yet, we need not search far to find evidence. As an illustration of batch arrival queues in a prominent and recent setting, in this section we explore data from contact tracing for the Covid-19 pandemic, apply our models and methodology to it, and examine the managerial insights that result. }

%\tr{see contact tracing}

%\edit{Case Study on NYC Test and Trace Data}

%\tr{example contact tracing shows that the large batch regime does exist in reality, and, moreover, it shows how important and delicate these staffing decisions can be}

%\tr{high level motivate, then describe variables/parameters, then describe particular operation and associate particular parameter value}

\edit{We base this case study upon \citet{blaney2022covid}, which provides description and data from the first 17 months of the NYC Test \& Trace Corps, the Covid-19 contact tracing operation of NYC Health + Hospitals and the NYC Department of Health and Mental Hygiene (DOHMH). Serving the entire population of the NYC area, Test \& Trace conducted Covid-19 case investigations for over 940,000 people during these 74 weeks in the data, from June 1, 2020 to October 31, 2021. By law, all laboratory-based or point-of-care Covid-19 test results were reported to the DOHMH, and, from this registry of all probable (positive antigen) or confirmed (positive molecular) tests, Test \& Trace was responsible for case investigation and contact tracing for the city. Immediately, we can notice that batch arrivals of cases is an unavoidable characteristic of the contact tracing system. At the stage of the pandemic recorded in this study, tests themselves were processed in batches, meaning the positive cases among them would inevitably be passed to Test \& Trace in batches. Hence, it is not possible to guarantee single-file arrivals (without significant sacrifices of operational efficiency). Furthermore, IT management policies led to further conglomeration before the positive cases were truly received for tracing, compounding the batches into sizes that are inarguably large. We will closely inspect the impacts of this aggregation.}

\edit{\citet{blaney2022covid} describes Test \& Trace as what is essentially a tandem service system. The full team of contact tracers were divided into two separate roles, case investigators and monitors. For each positive case and the associated patient, case investigators provided education and resources, identified contacts and locations of exposure, and evaluated symptoms and need for support.
%``provided education, assessed symptoms, elicited close contacts (i.e., persons within 6 feet of the person undergoing case investigation while potentially infectious for 10 or more cumulative minutes), identified locations the person visited while infectious, evaluated need for supportive services, and advised on duration of isolation'' \citep[][pg. 2]{blaney2022covid}. 
The contacts provided during case investigation were then assigned to monitors. While both roles are certainly critical to the public health mission, we will focus on the case investigators in this study, as this first phase of service receives the batch arrivals directly. Here, the precise batch staffing question is ``{how many case investigators should NYC Test \& Trace have?}''}

\edit{In this setting, $Q_t^C$ becomes the number of cases either in investigation or awaiting it at time $t \geq 0$. (Note we are not indexing by $n$ or $m$ in this section because the case study is not concerned with a limit.) Similarly, we will let $c$ be the number of case investigators; $B_i$ becomes the number of cases to trace in the $i$th arrival, and $\lambda$ will be the rate of arrivals of new batches of cases. Then, $S_{i,j}$ is the duration of case investigation for the $j$th patient within the $i$ batch of cases. Because ``timeliness is key to the success of any contact tracing operation'' \citep[][pg. 7]{blaney2022covid}, the primary performance metric in our  study will be waiting times. Specifically, to match the data, we will use mean wait per case and the number (or fraction) of cases that wait more than one day.}

\edit{The goal of this case study is  two-fold. First, in Section~\ref{caseStudy1}, we will demonstrate what NYC Test \& Trace did well. Namely, \citet{blaney2022covid} describes a staffing level that exceeded national guidelines, and we can quantify the impacts of this through the batch arrival queueing model. Second, in the spirit of identifying lessons learned from Covid-19, Section~\ref{caseStudy2} explores where there may have been opportunities to improve the case investigation operation. In particular, in a system with end-to-end control like this public health administration had, we show that there is an opportunity to \emph{optimize the arrival pattern}. Across these two pursuits, this case study will also demonstrate how this example showcases the large batch regime, and we will detail how this service system exhibits the hallmark characteristics of a queue with large batches.}

\subsection{\edit{Investigating Performance Under Other Staffing Guidelines}}\label{caseStudy1}

%\tr{part 1) what NYC did well -- went above the national recommended staffing level for contact tracers (a la storage vs Gaussian, but don't formally claim that: it's two points and we're talking about two functions) -- Figure~\ref{samplePathSvG}}

\edit{Expanding the national staffing level of contact tracers was a first-order priority at the start of the Covid-19 pandemic, as there were only 2,200 specialists (trained for other diseases like HIV and tuberculosis) employed in these roles in public health agencies at the start of 2020 \citep{ruebush2021covid}. 
%For example, on April 10, 2020, experts at the Johns Hopkins Center for Health Security and the Association of State and Territorial Health Officials issued an estimation of needing 100,000 contact tracers in the United States \citep{watson2020national}, and, similarly, 
April 16, 2020 guidance from the \citet{naccho2020building} projected 30 contact tracers needed per 100,000 people. By comparison, NYC Test \& Trace built up a workforce of 4,147 contact tracers, or approximately 47 per 100,000 people in the city.}

\edit{In some sense, both 30 per 100k and 47 per 100k are recommendations in the style of the large batch regime (or, likewise, the QD regime), in that they are directly proportional to the expected arrival volume, rather than proportional to the offered load plus a lower order safety staffing. However, like how the storage process staffing dominates the Gaussian-based staffing at $\nu = 1$ as contrasted in Section~\ref{contrastSec}, here we see that NYC Test \& Trace staffed at a level over 50\% more than what \citet{watson2020national} or \citet{naccho2020building} advised. (Of course, we are not claiming that this is \emph{exactly} a storage versus Gaussian comparison, as that would be akin to fitting two separate functions to two separate points.) 
}

\edit{To contrast Test \& Trace's level with the national guidelines, we will simulate the case investigation phase as a $D^B/M/c$ queueing system at two values of $c$: 937 investigators, which is the actual amount per \citet{blaney2022covid}, and 599 investigators, which is the proportionally equivalent number of case investigators if the overall contact tracing workforce followed the 30 per 100k guide. We assume that each of the $c$ contact tracers works an 8 hour shift each day. Because \citet{blaney2022covid} describes that DOHMH exported records of positive cases to DOHMH once daily, we model the batches of cases as arriving at deterministic $1/\lambda = 1$ day intervals. The daily case counts are obtained from \citet{nyc2023nyc}, and they range from 131 to 8,077 with a mean of 1,775.0.\footnote{\edit{We will scrutinize the design decision of once daily case aggregation and reporting in the following subsection, but for now let us just emphasize that this places this case study squarely in the territory of large batches.}} Because this case study is meant to be illustrative, we will assume exponentially distributed service; through this assumption the waiting time performance metrics can be computed using Propositions~\ref{waitLimit} and~\ref{waitID} in Appendix~\ref{waitApp}.
}

\edit{The mean case investigation service duration is not entirely clear from the \citet{blaney2022covid} data. It is reported that there are on average 0.1 days between first case investigation call attempt and true end of the case investigation phase, but any given contact attempt may not be successful. On the other hand, material used in training of  Test \& Trace contact tracers includes prompts to the patient that the expected call length is 20 minutes. Rather than explicitly modeling the possible pre-emption and re-entry, we will instead conduct a sensitivity analysis over possible mean case investigation duration values, iterating over possibilities for $\E{S}$ at 30 minute multiples between 20 minutes and 140 minutes $\approx 0.1$ days while comparing to the known actual performance metrics.}

%\edit{$D^B/M/c$ queue}

%\edit{deterministic arrivals because \citet{blaney2022covid} says once daily aggregation of cases -- will explore this more in the next section}

%\edit{exponential service for straightforward calculations, since this is meant to be illustrative}

%\tr{waiting times can be computed using Propositions~\ref{waitLimit} and~\ref{waitID} in Appendix~\ref{waitApp}.}

%\edit{sensitivity analysis across five mean service durations since it is not exactly specified -- 80 mins gets pretty close to reported performance (justify this in terms of mean wait and fraction that wait more than a day)}

%\edit{comparing 937 (actual) and 599 (hypothetical based on national guidelines) servers, which are the case investigators}

%\edit{could also model the monitors down stream, but the case investigators receive the batch arrivals and thus this is the relevant phase of the service for this paper's scope}

\begin{figure}[htb]
\centering
\includegraphics[width=.49\textwidth]{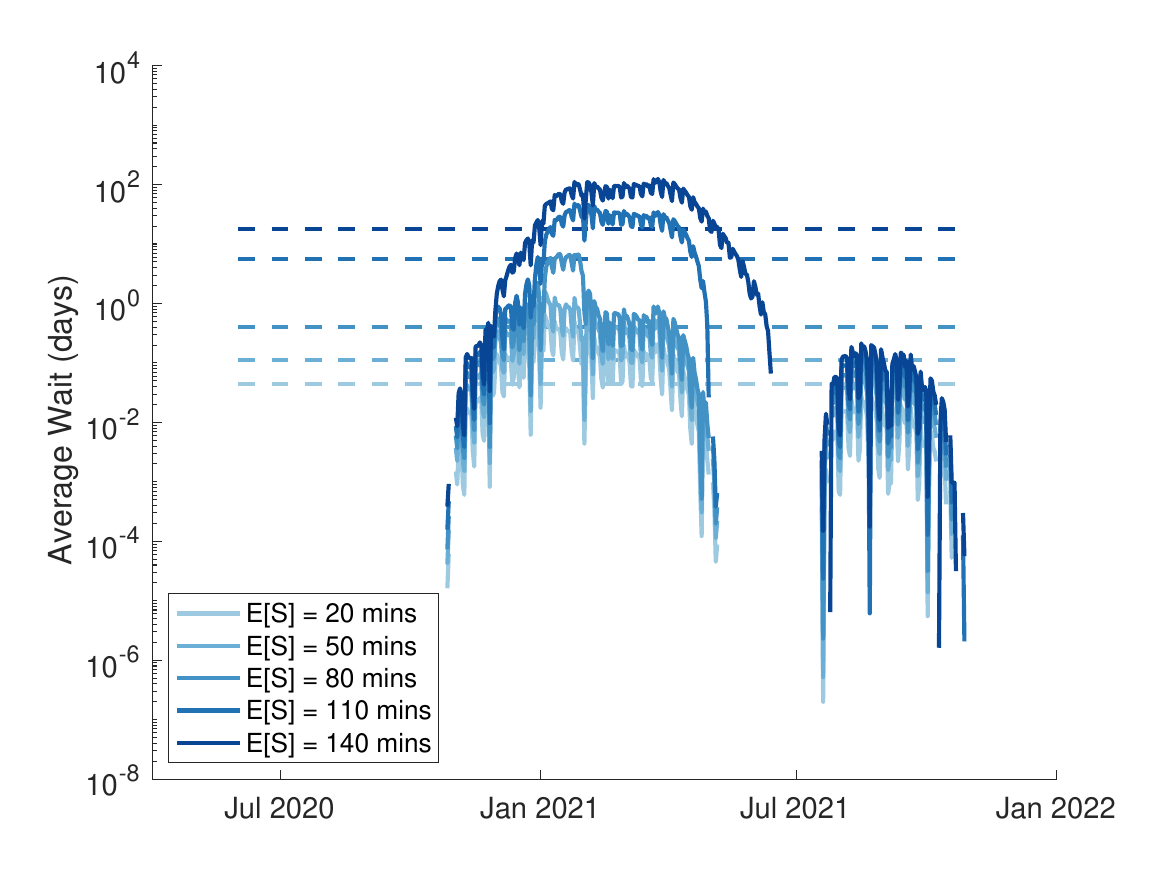}
\includegraphics[width=.49\textwidth]{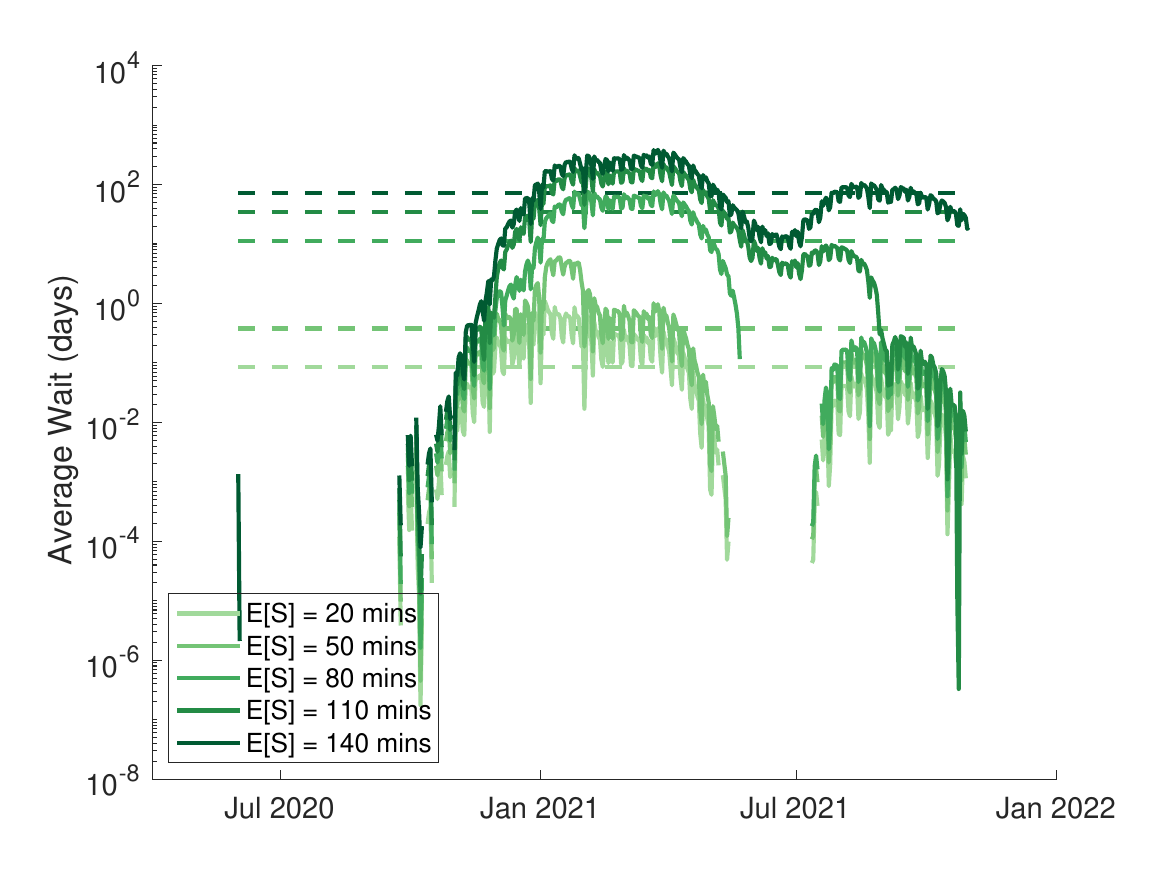}
\\
\includegraphics[width=.49\textwidth]{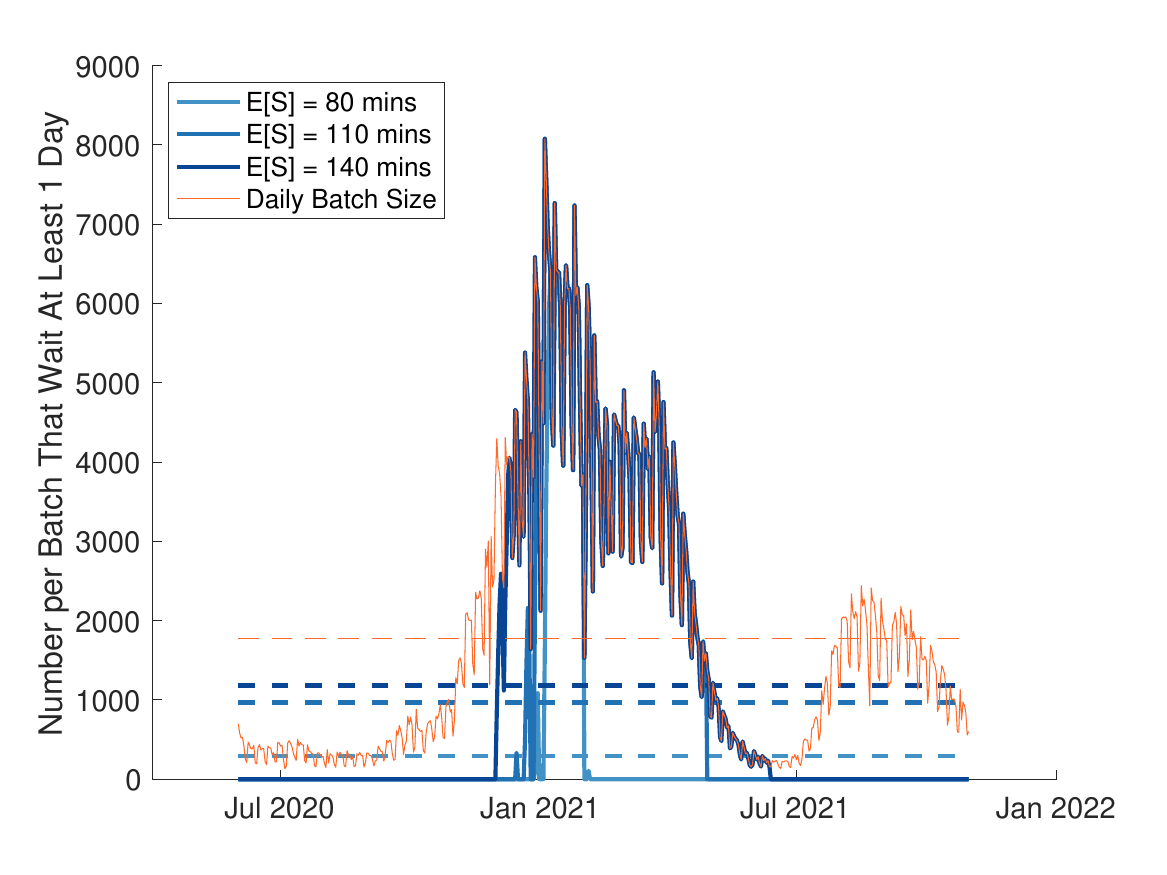}
\includegraphics[width=.49\textwidth]{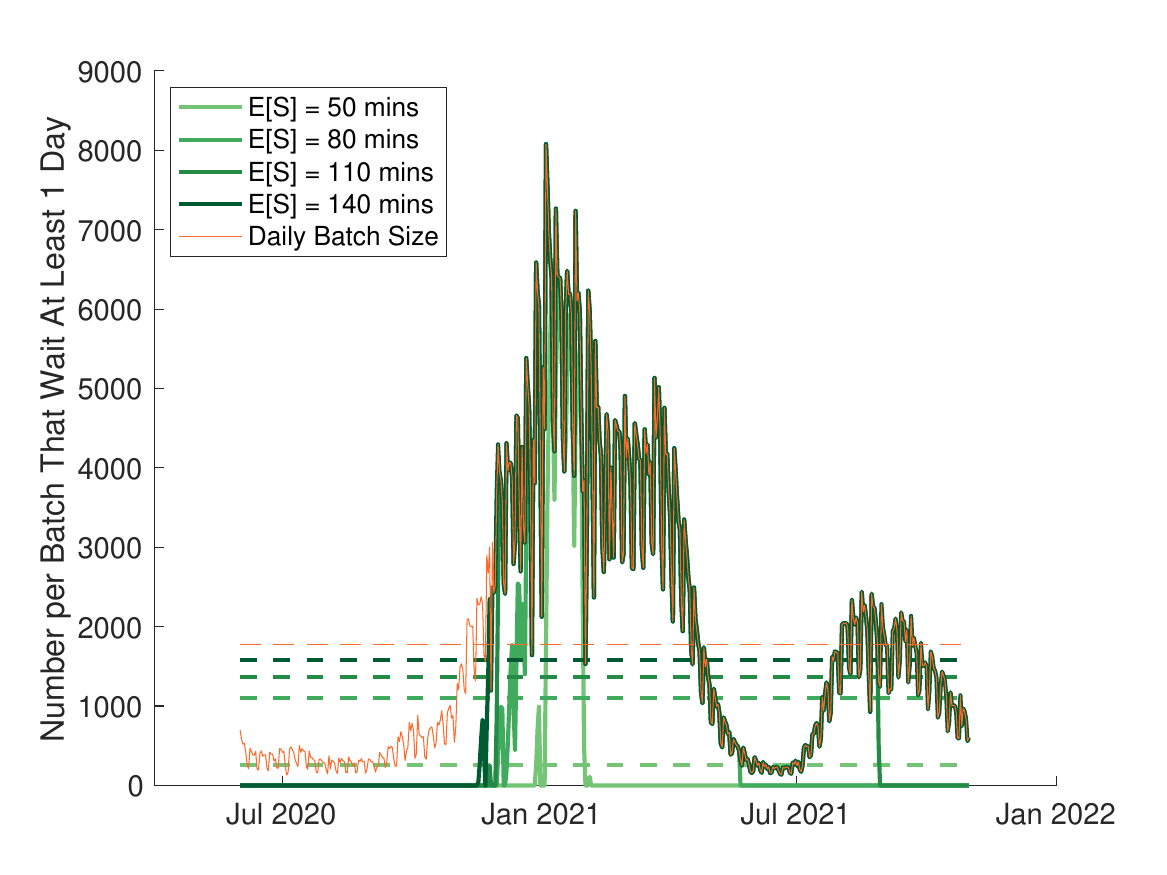}
\caption{\edit{Average waiting time (first row) and number that wait a day or more (second row) within each day's batch of cases modeled across five hypothetical mean investigation durations using the true NYC Test and Trace daily caseload as batch sizes. The left figures have 937 case investigators (actual), while the right has 599 (hypothetical based on national guidelines). Dashed lines show the average across days in each scenario.}}\label{samplePathSvG}
\end{figure}

\edit{In Figure~\ref{samplePathSvG}, we plot the expected wait per case in each day's batch (top row) and the number of cases each day that wait at least 1 day before case investigation begins. The left-hand curves (in blue) show the performance under the various mean durations with the actual case investigation staffing level, $c=937$, and the right-hand side plots (in green) show the national guidelines alternative, $c=500$. Darker curves constitute longer mean case investigation durations, and dashed lines are averages across days. First, let us focus on the $c=937$ case and compare to the actual performance. From \citet{blaney2022covid}, we know that cases waited 0.6 days on average from positive test upload in the Test \& Trace system to first call attempt. Furthermore, 6.4\% of cases waited more than one day. Among the five considered values, $\E{S} = 80$ minutes most closely replicates this: The middle dashed horizontal line in the top left figure shows an average wait of correct order, and, compared to the thin dashed orange line, we can see the same in the lowest dashed horizontal line in the bottom left figure. Of course, the model is not a perfect reproduction,
% (the mean wait at $\E{S} = 80$ minutes is approximately 0.4 days, while the fraction that wait a day or more is 16.3\%)
but it is clear that this value is the closest qualitative match to the data. That is, the other options are off by at least one order of magnitude in at least one metric. Hence, we will consider 80 minutes as the most realistic mean duration.}

\edit{Focusing now on $\E{S} = 80$ minutes, let us compare the model performance under the actual staffing to the recommended hypothetical. In the top row figures, we can see that the average wait rises close to two orders of magnitude from $c=937$ to $c=599$ (0.41 days compared to 11.3). Moreover, the actual staffing level eliminates wait earlier and more often than the hypothetical. In the bottom row, we can similarly see that drastically more patients wait at least one day at $c=599$ (61.8\%, versus 16.3\% at $c=937$). In fact, one can make the case that at the mean wait of over  $11$ days and more than $50\%$ of patients waiting longer than one day, Covid-19 contact tracing would effectively be moot. The performance at $\E{S} = 50$ minutes for $c = 599$ is much closer to the actual and best $c=937$ model approximation, which suggests that staffing at the national guidelines could have necessitated that case investigators offer fewer services or solicit fewer potential exposures. This may offer a partial explanation for the success of Test \& Trace relative to other agencies in the United States. In comparison to a cross-sectional study of contact tracing in the U.S. in 2020~\citep{lash2021covid}, the NYC contact tracing operation had an approximately 25\% higher case investigation completion rate (75\% to 59\%) and nearly doubled contact identification rate (60\% of NYC case investigations named contacts, compared to 33\%).}

%\tr{Figure~\ref{samplePathSvG} -- credit to NYC: at what appears to be the most realistic service duration (80 mins, middle line in the waiting time figures), close to two orders of magnitude difference in mean waiting and close to 5 times as many cases that wait at least one day between the actual staffing NYC used and the level recommended by the national guidelines -- note that at the mean wait of $\sim11$ days and over $50\%$ waiting more than one day, Covid-19 contact tracing would effectively be moot}

\edit{Now, in the context of this paper's broader goals, let us step back and observe what is happening here. This data has shown exactly the same properties that our theory identified as hallmarks of the large batch regime: large batches, low utilization, yet non-trivial wait. For the actual staffing level $c = 937$, the utilizations (i.e., $\lambda \E{B} \E{S} / c$) under the five hypothetical service durations are 7.9\%, 19.7\%, 31.6\%, 43.4\%, and 55.3\%, respectively. However, not only is the mean wait non-trivial; in the latter four durations, the wait dominates the service duration. Comparing the left figures to those on the right, we see the tail-sensitivity of this staffing decision on the system performance. In particular, let us draw attention to the middle three durations. Here, the utilization is still moderate (30.9\%, 49.4\%, and 67.9\%, respectively), but the wait far surpasses that at $c=937$.
}

\begin{figure}[htb]
\centering
\includegraphics[width=.49\textwidth]{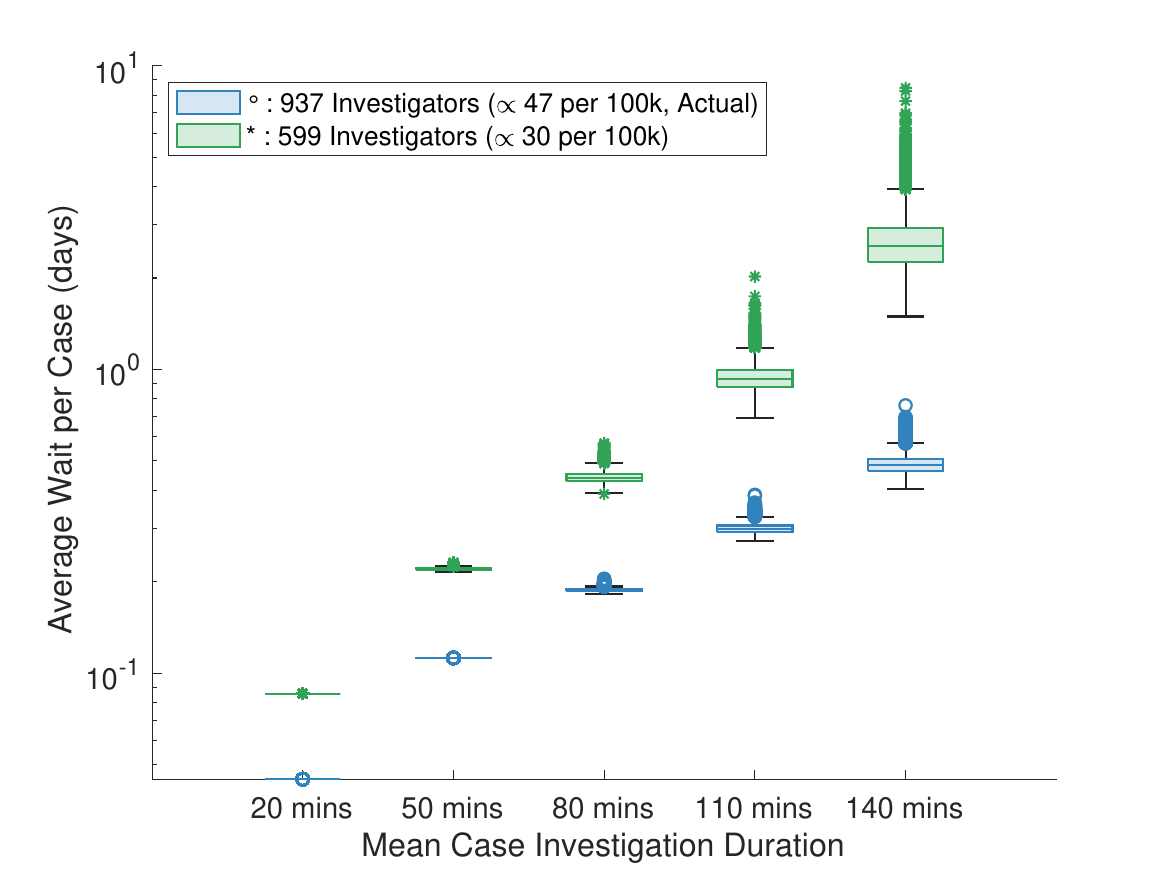}
\includegraphics[width=.49\textwidth]{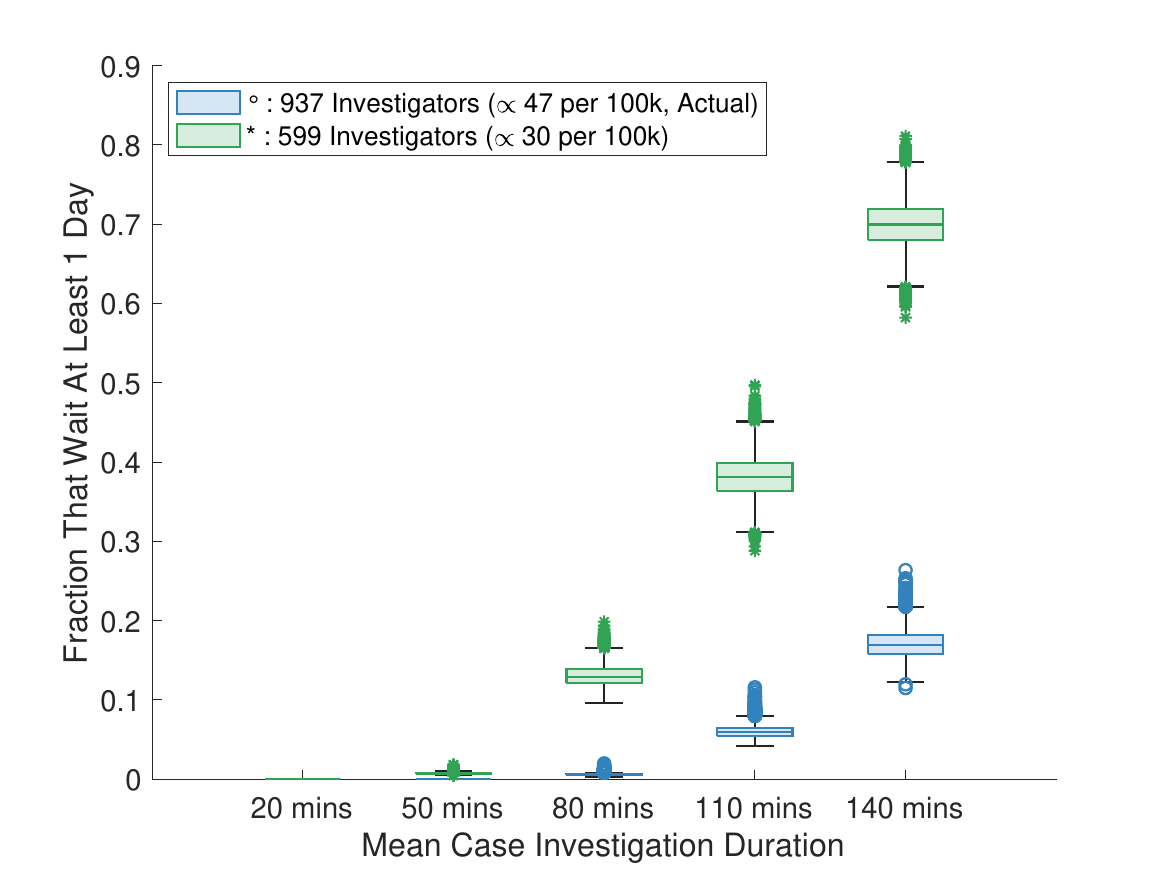}
\caption{\edit{Under uniformly random shufflings of NYC Test and Trace batch size sequences, average wait per case and fraction of cases that wait more than one day for $\boldsymbol{2^{14}}$ replications. }}\label{shuffleFig}
\end{figure}

\edit{Undoubtedly, some of the wait in this study is due to the obvious non-stationarity of the batch size distribution. However, let us briefly depart from reality to show that waiting persists even when accounting for the time variation. In Figure~\ref{shuffleFig}, we create synthetic stationarity while preserving the overall batch size distribution by simulating the $D^B/M/c$ system under a uniformly random shuffling of the batch sequence in each replication. Again, we compare $c=937$ and $c=599$ in terms of the mean wait and the fraction that wait more than one day. Even under these permuted batches, we see that the system experiences non-trivial waiting. Moreover, the mean waiting time is statistically significantly higher in the guidelines-based staffing across all duration scenarios, as is the case for the single-day waiting fraction for all durations above 20 minutes. }

%\tr{look what's happening here -- large batch regime!}

%\tr{utilization (mean batch size times mean duration divided by max servers) of the 5 hypothetical service durations: 7.9\%, 19.7\%, 31.6\%, 43.4\%, and 55.3\% respectively. }

%\tr{utilization under 599: 12.4\%, 30.9\%, 49.4\%, 67.9\%, and 86.4\% respectively. }

%\edit{use true ordering to motivate parameters, shuffle to show that waiting is not just caused by non-stationarity -- Figure~\ref{shuffleFig}}

%\edit{waiting is reduced when non-stationarity is removed through the shuffling, but still we see statistically significant differences in waiting between the two staffing levels in both performance metrics}

\subsection{\edit{Optimizing the Arrival Pattern for Integrated Operational Design}} \label{caseStudy2}

\edit{Reflecting on the waiting seen in Figures~\ref{samplePathSvG} and~\ref{shuffleFig} and on the structure of that model, we can recognize that some amount of wait is unavoidable, because the mean batch size is larger than either of the considered staffing levels. As we alluded to at the start of this section, this is an immediate consequence of the daily aggregation of cases. However, we can also now recognize that this was not a definition of the system; rather, at least to some degree, it was a choice.}

\edit{While in many cases the composition of the customer traffic may be beyond the scope of management, this contact tracing example presents an opportunity for end-to-end control of the arrival pattern. That is, because each municipality's Covid-19 testing process was closely managed and regulated by some combination of the corresponding city, county, and state public health agencies, the system could have been designed to produce more frequent batches of smaller size. 
This creates a tradeoff between the batch size and the arrival rate. For a given fixed arrival volume, faster and smaller batches present less acute staffing stress, but thus process batches more frequently, which may be costly. On the other hand, larger and less frequent batches would incur fewer processing costs, but would need more labor. 
}

%\tr{part 2) what NYC could have done better: optimizing the arrival pattern -- not daily arrivals, likely could have been even more responsive (note that, in fairness, practically this may be limitations of IT infrastructure)}

%\tr{tradeoff of batch size and arrival rate}

%\edit{while in many cases the composition of the customer traffic may be beyond the scope of management, this contact tracing example presents an opportunity for end-to-end control of the arrival pattern}

%\edit{that is, because the Covid-19 testing process is closely managed and regulated by the city, county, and state public health agencies, the system could have been designed to produce more frequent batches of smaller size}

%\edit{this naturally presents a tradeoff: for a given arrival volume, faster, smaller batches present less acute staffing stress but may incur more frequent processing costs per each batch}

\edit{To model this tradeoff, let us introduce what we will call the \emph{controlled arrival pattern problem} for the  $M^n/M/c$ system. Following the idea of centralized public health decision making in the contact tracing pipeline, we will assume that a central controller can decide both the batch size $n$ and the batch arrival rate $\lambda$, so long as the total effective arrival volume  $\lambda n = m$ is preserved. Following the results of this paper, we will set that the staffing level to be $c(m) = m\slash \mu + \delta n \sqrt{\lambda}$ for some constant $\delta > 0$ and some $\E{S} = 1/\mu$. Then, letting $\mathcal{C}_0$ be staffing cost per day per service agent, and $\mathcal{C}_1$ be a fixed processing cost per batched arrival, the objective of the controlled arrival problem will be to minimize service cost defined
\begin{align}
\mathcal{C}_0 c(m) + \mathcal{C}_1 \lambda
&=
\mathcal{C}_0 \left( \frac{m}{\mu} + \delta n \sqrt{\lambda} \right) + \mathcal{C}_1 \lambda
.
\label{serveCostObj}
\end{align}
For contact tracing, $\mathcal{C}_0$ can be thought of as the daily wages for each case investigator, and $\mathcal{C}_1$ would be the fixed cost to process each batch of cases, regardless of the batch size. These fixed costs may arise due to the costs of batch processing in groups of cases, like the standard of 96-well plates in reverse transcription–polymerase chain reaction (RT-PCR) \citep[e.g.,][]{emery2004real}, or simply from the time and effort needed to manage the aggregation in the contact tracing investigation system. Naturally, other costs may arise, such as a cost of materials per case. However, because  $m$ is held fixed, a per case cost will not change with $\lambda$ or $n$.\footnote{\edit{This controlled arrival pattern problem should apply to other integrated services, such as end-to-end transportation systems like airport shuttles and trains or Disneyland trams and monorails. In those cases, there are labor cost to staff security agents or ticket takers who serve customers that arrive by mass transit, and those mass transit systems incur first-order costs by how often they move, rather than how many customers they carry.}}}

%\edit{cost to run tests and aggregate into the system}

%\edit{cost to staff}

%\tr{make sure to refer to other arrival pattern control problem in appendix}

%\tr{make sure to explain problem in public health setting to tie back into numerics}

%\edit{applies to contact tracing, but also to end-to-end transportation systems like airport trains, Disneyland trams and monorails, etc.}

%\edit{Consider a $M^n/M/c$ queueing system in which the Suppose that the effective arrival rate $m = \lambda n$ is held fixed, but the arrival rate $\lambda$ and batch size $n$ are centrally controlled.}

\edit{In Proposition~\ref{serveCostProp}, we give the \emph{optimal arrival pattern} for this $M^n/M/c$ control problem.
\begin{proposition}\label{serveCostProp}
In the controlled $M^{n}/M/c(m)$  arrival pattern problem with fixed effective arrival rate $m \in \mathbb{R}_+$ such that $m = \lambda n$ and with staffing $c(m) = m/\mu + \delta n \sqrt{\lambda}$ , the expected service costs are minimized if and only if $\lambda = \mathcal{C}_* m^{{2}/{3}}$ and $n = \frac{1}{\mathcal{C}_*} m^{{1}/{3}}$, where $\mathcal{C}_* = \left(\delta \mathcal{C}_0 \slash 2\mathcal{C}_1\right)^{{2}/{3}}$.
\end{proposition}
}

\edit{Through Proposition~\ref{serveCostProp} we can see that staffing becomes a consequence of the arrival pattern control decision. More precisely, in the optimal arrival pattern, the safety staffing will be of order $m^{2/3}$. As an added benefit of controlling the arrival-rate-batch-size tradeoff, this choice of arrival pattern should ensure that the system is operating in the batch-and-rate regime, rather than the large batch regime, because $\lambda$ is of squared order relative to $n$ as functions of $m$. Hence, the staffing coefficient $\delta$ can safely be obtained through the lighter tailed Gaussian-based calculations in Theorem~\ref{bothLimitMulti}. Let us return to the case study data with the observations from this optimal arrival pattern in hand.}

\edit{To demonstrate how this arrival pattern control could have benefited the case investigation operation, we will now simulate the system as a $M^B/M/c$ queue where the arrival rate $\lambda$ can be changed each week to account for the upcoming expected arrival volume. By controlling the arrival rate, the public health agency also sets the staffing and the mean batch size. The policy for the rate-staffing decision in each week will be denoted as the pair $(\lambda, c)$.
%, where both $\lambda$ and $c$ depend on the effective daily arrival rate $m$, and $m$ may change from week-to-week.  
We again use the true \citet{nyc2023nyc} case count data as batch sizes, but we now use a stick-breaking-type procedure to split the batches by first sampling the number of batches per week and then dividing accordingly (see Algorithm~\ref{batchBreak} in Appendix~\ref{simApp}). This approach preserves the true caseload for each week while modeling the impact of $\lambda$. Because $1/\mu = 80$ minutes provided the most realistic performance values in the experiment in Section~\ref{caseStudy1}, let us adhere to this duration as the mean case investigation service time.
}

%\tr{non-stationary experiment here}

%\tr{simulation is $M^B/M/c$ (maintain batch size randomness)}

%\tr{Because 80 mins provided the most realistic performance values in the previous experiment, let us adhere to this duration as the mean case investigation service time}

%\edit{arrival volume non-stationarity}

\begin{figure}[htb]
\centering
\includegraphics[width=.325\textwidth]{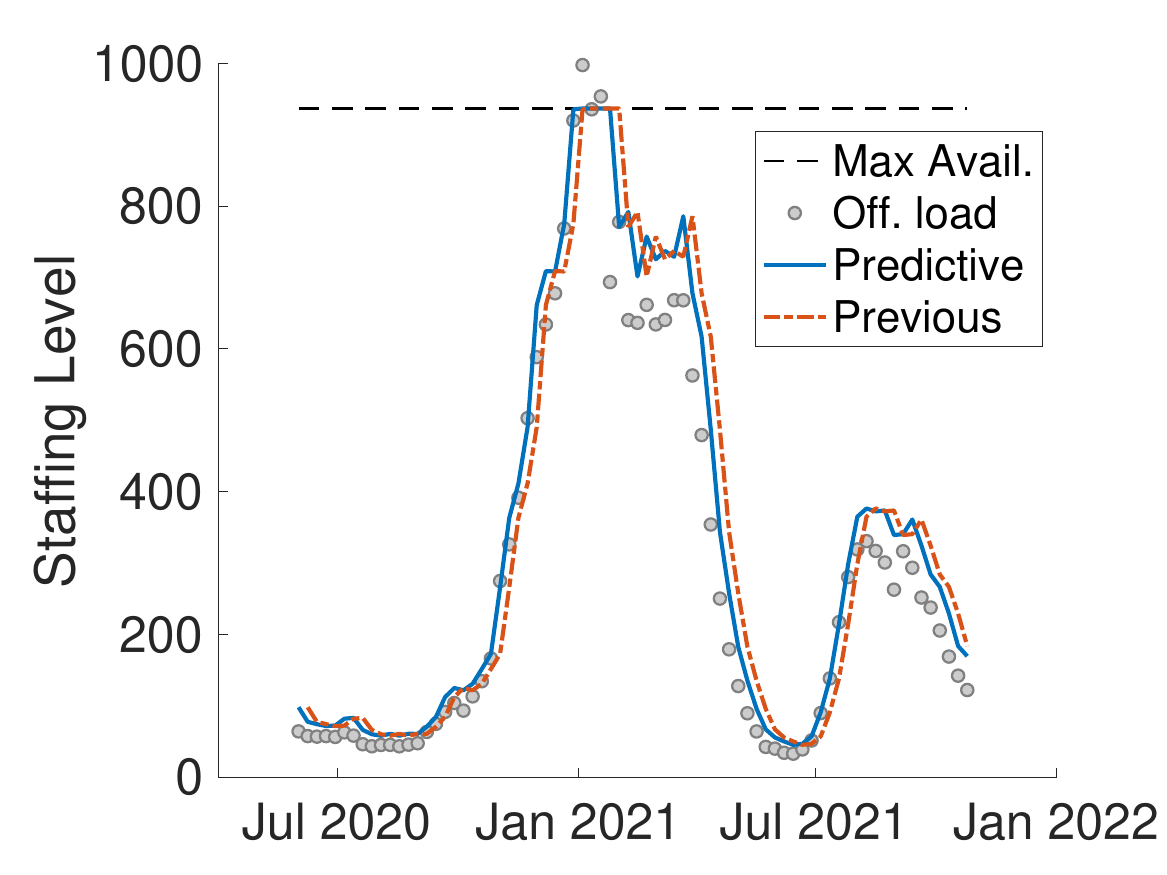}
\includegraphics[width=.325\textwidth]{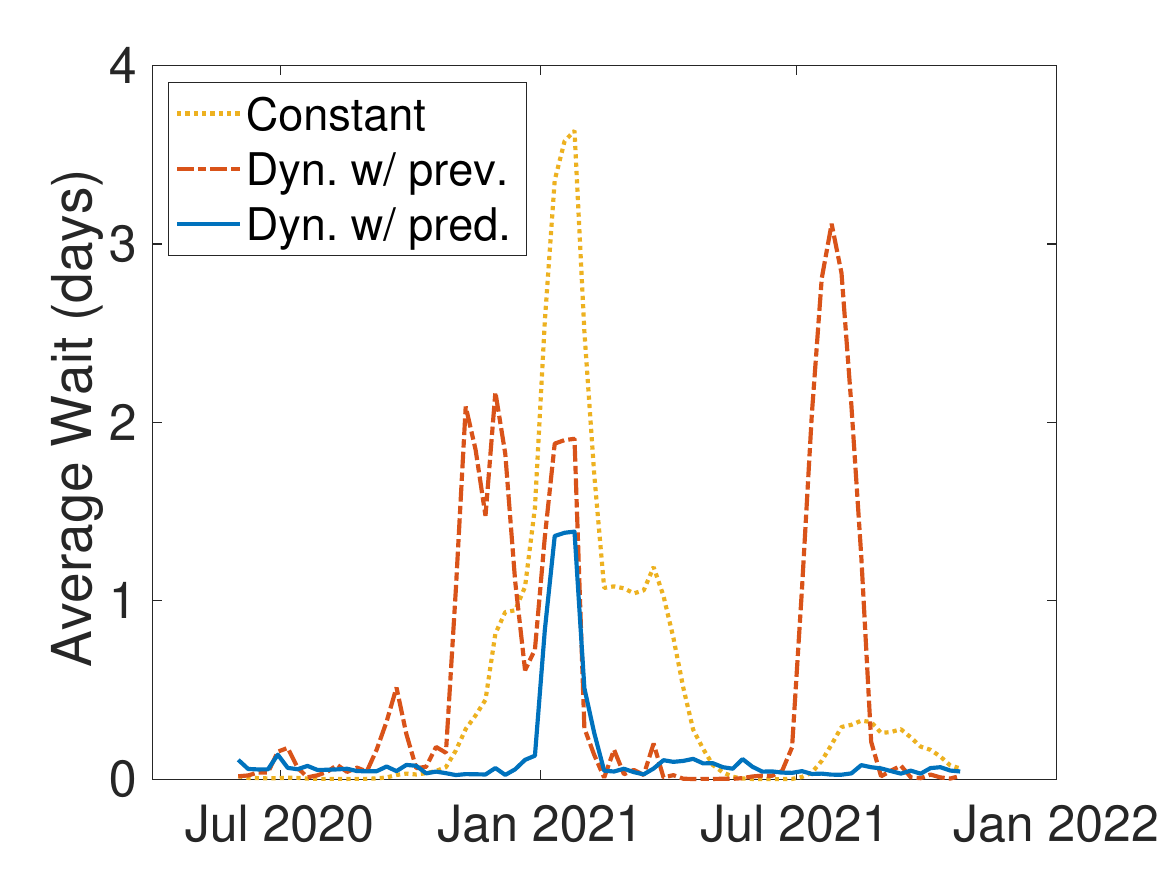}
\includegraphics[width=.325\textwidth]{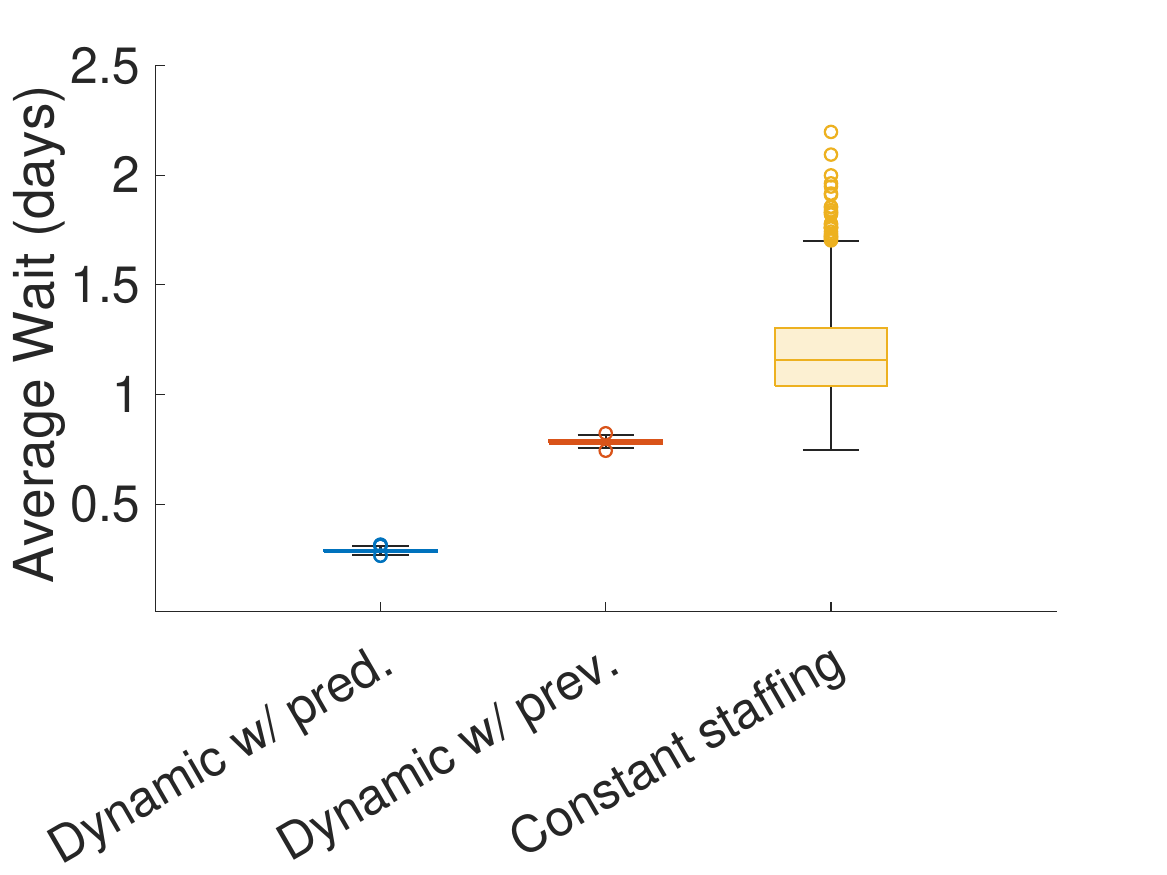}
\caption{\edit{Dynamic staffing level (compared to offered load), average wait per case in each day, and overall average wait per case across days in the three different rate-staffing policies. }}\label{CTdynFig}
\end{figure}

\edit{In Figure~\ref{CTdynFig}, we evaluate three different rate-staffing policies. First, as the analog to the true staffing we discussed in Section~\ref{caseStudy1}, we consider a constant staffing policy with daily arrivals, i.e. $(\lambda, c) = (1, 937)$ every week.\footnote{\edit{One can see that the average waiting is higher in the constant staffing in Figure~\ref{CTdynFig} than it was in the Section~\ref{caseStudy1} experiments; this is due to the change from deterministic intervals to Poisson epochs.}} Then, as a dynamic alternative inspired by the optimal arrival pattern from Proposition~\ref{serveCostProp} without using more labor than was truly available, we also set the ``dynamic with perfect predictions'' policy as $(\lambda, c) = (m^{2/3}, (m/\mu + \delta m^{2/3} \wedge 937))$, with $m$ as the true total caseload for the present week. Finally, to disentangle the benefit of the predictions and of the dynamic rate-staffing policy, we also consider a ``dynamic with previous predictions'' policy also of the form $(\lambda, c) = (m^{2/3}, (m/\mu + \delta m^{2/3} \wedge 937))$, but instead the present week $m$ is assumed to be unknown and the prior week's value is used as the estimate. Given the many powerful prediction methods employed during the Covid-19 pandemic \citep[see, e.g.,][and references therein]{cramer2022evaluation}, we consider use of the previous week's value as essentially the worst realistic case, and then, of course, the perfect predictions constitute the best.}

%\tr{ arrival pattern control with perfect prediction of next week's volume, arrival pattern control with last week's volume as prediction of next week's volume, and constant staffing at the max available with daily arrivals}

\edit{Though Figure~\ref{CTdynFig}, we can see the substantial benefit that comes with increasing the arrival rate and decreasing the batch size. Even though the total work remains the same, both the dynamic policies are able to deliver significantly less overall waiting per case while actually requiring only a fraction of the staffing for much of the 17 month data period. In a global emergency like the Covid-19 pandemic presented, resources may become particularly scarce or precious. This experiment suggests that it may have been possible to devote those people or dollars to some of the many other pressing public health needs, all while achieving the same or better performance. Given the context of this paper, this is precisely an observation about staffing for batch arrivals. By having more frequent and smaller arrivals, the dynamic policies in  Figure~\ref{CTdynFig} pull the contact tracing operation away from the purely large batch regime of Theorem~\ref{fDelayConv} and closer to the batch-and-rate regime of Theorem~\ref{bothLimitMulti}, where staffing performance is more robust (as established in Proposition~\ref{coincideProp}) and has a more favorable tail (as contrasted Figure~\ref{storageVsGaussianFig}).
}

%\section{Numerical Experiments and Demonstrations}

\begin{figure}[htb]
\centering
\includegraphics[width=.9\textwidth]{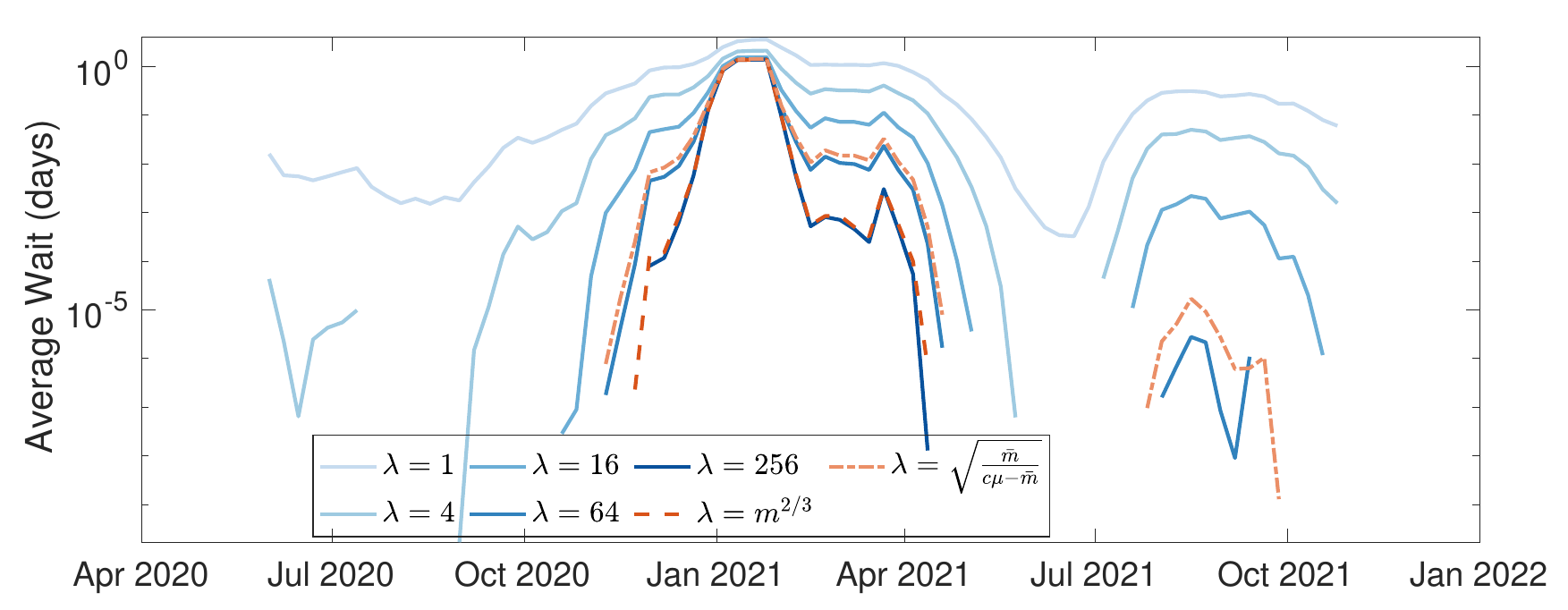}
\caption{\edit{For a fixed staffing at the true level (937), average wait within each days batch is plotted for increasing fixed arrival rates at powers of 4, and also plotted for a fixed arrival rate based on the overall case volume and for a dynamic arrival rate in the style of Proposition~\ref{serveCostProp}.}}\label{lamIncFig}
\end{figure}

\edit{As a closing thought for this case study, let us also remark that a pandemic may be rightfully deemed an ``all-hands-on-deck'' scenario, meaning that the staffing should not be dynamic, but instead static at the max available at all times. Even granting this reasonable position, though, we still believe there may have been an opportunity to improve performance through dynamic arrival rate control alone. In Figure~\ref{lamIncFig}, we show the average wait time across the case study time period under fixed staffing at $c = 937$ and different choices of $\lambda$. This includes a dynamic policy with $\lambda = m^{2/3}$, and also a fixed policy with $\lambda = \sqrt{\bar{m} / (c\mu - \bar{m})}$, where $\bar{m}$ is the average daily volume across the full horizon. We include this as an example of a different style of arrival pattern control problem, where the objective is based on wait and the staffing is held fixed. The full derivations of this alternate problem are available in Appendix~\ref{waitOptSec}, where we also extend the case study to a broader time horizon that includes the added challenges of the Omicron wave.}

%$\lambda = \sqrt{{m} / (c\mu - {m})}$

%\tr{on Figure~\ref{lamIncFig} -- even if a pandemic is rightfully deemed an ``all-hands-on-deck'' scenario and the staffing should be static at the max available at all times, there may have be a missed opportunity to control the rate-versus-batch tradeoff in the arrival pattern -- refer to more exploration of this in the appendix}

\section{Discussion and Conclusion}\label{concludeSec}

%\tr{what we've seen in this paper's analytical results: batch arrivals place significant stress on service systems, and when the batch size dominates the arrival rate, this stress has a heavier shape than if the same arrival volume occurred through fast rates \emph{and} batches}

%\tr{discuss lots of other applications (but not AV's), mention the intro ones again and ones from control problem, could also mention ports as a similar lessons learned from covid, and here arrivals are also in batches (many containers per ship)}

%\tr{\citet{tirmazi2020borg} -- low utilization also seemed in the related multi-server job setting}

%\tr{may be similar staffing questions for the testing and tracing challenges in other application areas, such as agricultural supply chains \citep[e.g.,][]{levi2020economically,dong2023impact}}

In this paper, we have found that service systems with large batches \edit{face dangerous and deceptive} operational challenges. %Our results show that such arrival patterns demand a high level of staffing, so much so that there may not even be an economy of scale as the effective arrival rate of customers grows. 
\edit{In particular, our analytical results show that batch arrivals place significant stress on service systems, and when the batch size dominates the arrival rate, this stress has a heavier shape than if the same arrival volume occurred through fast rates and smaller batches.}
In our two main results, we have seen that there is truly no economy of scale in the pure large batch limit (Theorem~\ref{fDelayConv}), and in the case of large batches and fast arrival rates, an economy of scale may occur but it will be weaker than what is typically expected (Theorem~\ref{bothLimitMulti}). 
%Comparing to the classic literature on staffing service operations, this means that for large batches it takes the high commitment of QD-style resources just to achieve what is typically a balanced compromise of performance, as classically offered by the QED regime. In the \edit{batch-and-rate} case, it remains true that QED-type performance will take more than typical QED-style staffing, but not quite as much as the QD regime. 
Under the hood, these  large batch insights are powered by a connection between batch arrival queues and storage processes. Our \edit{large batch-and-rate} limits reveal further that, at the extreme of this spectrum, the typically distinct law of large numbers (or, fluid) style and central limit theorem (diffusion) limits are actually the same in the case of the large batch regime. Here, the batch arrival queue's mean and standard deviation are of the same order, and thus there is no asymptotic normality, only the storage process limit. \edit{Hence, the coefficient of staffing must be higher than what is anticipated under typical Gaussian-based prescriptions.}

\edit{To provide an example of this large batch regime, we have applied our models and results to data on Covid-19 contact tracing in NYC. Here, we immediately found the hallmark properties of the regime: large batches, low utilization, yet non-trivial wait. We saw that we could reduce waiting by controlling the arrival pattern, as this could pull the operation towards the batch-and-rate regime, where performance is more robust. Nevertheless, NYC Test \& Trace admirably staffed at a high level relative to national guidance, and our simulations suggest this alone created considerable reductions in wait, which may have been critical to the agency's public health mission.}

\edit{We believe there are many other large batch lessons to be observed in applications. In fact, there are perhaps even many more to be learned from pandemic-era operations. For example, on a surface level, we can see a similar pattern in the well-documented backlogs that happened at ports in cities like Los Angeles, CA. Undoubtedly, bullwhip-type effects created the rise in arrival rates of {ships}, but ships can be thought of as simply \emph{batches of containers}. Hence, the bottlenecks around the port may have an insight not unlike that we have seen here, where perhaps capacity guidelines were set with lighter tails in mind. Similarly, there also may be staffing questions for testing and tracing challenges in other application areas, such as agricultural supply chains \citep[e.g.,][]{levi2020economically,dong2023impact}. Depending on how those tests are processed, batch staffing may also be relevant in those settings. Furthermore, in the introduction, we also discussed possible relevance for staffing or capacity planning in modern computing systems. \citet{tirmazi2020borg} has observed that low utilization also seems to be a consistent property of these settings. The fork-join, multi-server job, and redundancy models are certainly meaningfully different than this one, but perhaps the large batch regime can offer intuition towards these scenarios nonetheless. Then, in another health-related application, we can also recognize near-batch or burst structures in the arrival patterns for mass casualty events, in which emergency wards receive many new patients over a short time period \citep[e.g.,][]{mills2013resource,cohen2014minimizing,yomtov2014erlang}. In this setting, the batch-staffing principles we have developed may apply to both health worker labor and to reusable resources like hospital beds or medical devices.}

%\tr{refer to the many things in the appendix}

\edit{To close, let point out that many interesting questions remain. As one example, Section~\ref{caseStudy2} introduced the arrival pattern control problem, in which a central decision maker can set the arrival rate and batch size to minimize service costs, so long as the effective arrival rate remains fixed. We believe there may be many other relevant variants of this problem; indeed, in Appendix~\ref{waitOptSec}, we propose another based around waiting costs and apply that optimal arrival pattern to an extension of the Covid-19 case study. The interested reader may enjoy several other auxiliary and supporting results in the appendix.} %As another example, we believe there may be many interesting limiting regimes that relate to and connect with those we have explored here, like what Section~\ref{hybridSec} showed for the large batch regime and the classic QED. We are excited to explore this direction further.}

\section*{Acknowledgements}
\edit{We are grateful to Dr. Ted Long of New York City Health $+$ Hospitals and NYC Test \& Treat, and to Drs. Barbara Ferrer and Rita Singhal of the Los Angeles County Department of Public Health for their graciousness in answering questions, sharing information, and providing references.}
%Publicly available, preliminary drafts of this paper focused exclusively on staffing batch arrival queues in support of remote teleoperation for autonomous vehicles; we appreciate the anonymous referees who have helped us recognize the generality and applicability of our results.

\printendnotes

\bibliographystyle{informs2014} % outcomment this and next line in Case 1
\bibliography{remote.bib}

\begin{thebibliography}{87}
\providecommand{\natexlab}[1]{#1}
\providecommand{\url}[1]{\texttt{#1}}
\providecommand{\urlprefix}{URL }

\bibitem[{Aksin et~al.(2007)Aksin, Armony, \protect\BIBand{}
  Mehrotra}]{aksin2007modern}
Aksin Z, Armony M, Mehrotra V (2007) The modern call center: A
  multi-disciplinary perspective on operations management research.
  \emph{Production and operations management} 16(6):665--688.

\bibitem[{Andrews(1988)}]{andrews1988laws}
Andrews DW (1988) Laws of large numbers for dependent non-identically
  distributed random variables. \emph{Econometric theory} 4(3):458--467.

\bibitem[{Atar(2012)}]{atar2012diffusion}
Atar R (2012) A diffusion regime with nondegenerate slowdown. \emph{Operations
  Research} 60(2):490--500.

\bibitem[{Atar \protect\BIBand{} Gurvich(2014)}]{atar2014scheduling}
Atar R, Gurvich I (2014) Scheduling parallel servers in the nondegenerate
  slowdown diffusion regime: Asymptotic optimality results. \emph{The Annals of
  Applied Probability} 24(2):760--810.

\bibitem[{Atar et~al.(2012)Atar, Mandelbaum, \protect\BIBand{}
  Zviran}]{atar2012control}
Atar R, Mandelbaum A, Zviran A (2012) Control of fork-join networks in heavy
  traffic. \emph{2012 50th Annual Allerton Conference on Communication,
  Control, and Computing (Allerton)}, 823--830 (IEEE).

\bibitem[{Atar \protect\BIBand{} Solomon(2011)}]{atar2011asymptotically}
Atar R, Solomon N (2011) Asymptotically optimal interruptible service policies
  for scheduling jobs in a diffusion regime with nondegenerate slowdown.
  \emph{Queueing Systems} 69(3):217--235.

\bibitem[{Baccelli et~al.(1989)Baccelli, Makowski, \protect\BIBand{}
  Shwartz}]{baccelli1989fork}
Baccelli F, Makowski AM, Shwartz A (1989) The fork-join queue and related
  systems with synchronization constraints: Stochastic ordering and computable
  bounds. \emph{Advances in Applied Probability} 21(3):629--660.

\bibitem[{Baily \protect\BIBand{} Neuts(1981)}]{baily1981algorithmic}
Baily DE, Neuts MF (1981) Algorithmic methods for multi-server queues with
  group arrivals and exponential services. \emph{European Journal of
  Operational Research} 8(2):184--196.

\bibitem[{Bassamboo et~al.(2010)Bassamboo, Randhawa, \protect\BIBand{}
  Zeevi}]{bassamboo2010capacity}
Bassamboo A, Randhawa RS, Zeevi A (2010) Capacity sizing under parameter
  uncertainty: Safety staffing principles revisited. \emph{Management Science}
  56(10):1668--1686.

\bibitem[{Blaney et~al.(2022)Blaney, Foerster, Baumgartner, Benckert, Blake,
  Bray, Chamany, Devinney, Fine, Gindler et~al.}]{blaney2022covid}
Blaney K, Foerster S, Baumgartner J, Benckert M, Blake J, Bray J, Chamany S,
  Devinney K, Fine A, Gindler M, et~al. (2022) Covid-19 case investigation and
  contact tracing in new york city, june 1, 2020, to october 31, 2021.
  \emph{JAMA Network Open} 5(11):e2239661--e2239661.

\bibitem[{Borst et~al.(2004)Borst, Mandelbaum, \protect\BIBand{}
  Reiman}]{borst2004dimensioning}
Borst S, Mandelbaum A, Reiman MI (2004) Dimensioning large call centers.
  \emph{Operations research} 52(1):17--34.

\bibitem[{Brockwell(1977)}]{brockwell1977stationary}
Brockwell P (1977) Stationary distributions for dams with additive input and
  content-dependent release rate. \emph{Advances in Applied Probability}
  9(3):645--663.

\bibitem[{Brockwell et~al.(1982)Brockwell, Resnick, \protect\BIBand{}
  Tweedie}]{brockwell1982storage}
Brockwell PJ, Resnick SI, Tweedie RL (1982) Storage processes with general
  release rule and additive inputs. \emph{Advances in Applied Probability}
  14(2):392--433.

\bibitem[{Chaudhry \protect\BIBand{} Kim(2016)}]{chaudhry2016analytically}
Chaudhry ML, Kim JJ (2016) Analytically elegant and computationally efficient
  results in terms of roots for the {$GI^X/M/c$} queueing system.
  \emph{Queueing Systems} 82(1-2):237--257.

\bibitem[{Chen et~al.(2011)Chen, Brown, Hu, King, \protect\BIBand{}
  Chen}]{chen2011managing}
Chen YD, Brown SA, Hu PJH, King CC, Chen H (2011) Managing emerging infectious
  diseases with information systems: Reconceptualizing outbreak management
  through the lens of loose coupling. \emph{Information Systems Research}
  22(3):447--468.

\bibitem[{Cinlar \protect\BIBand{} Pinsky(1972)}]{cinlar1972dams}
Cinlar E, Pinsky M (1972) On dams with additive inputs and a general release
  rule. \emph{Journal of Applied Probability} 9(2):422--429.

\bibitem[{Cohen et~al.(2014)Cohen, Mandelbaum, \protect\BIBand{}
  Zychlinski}]{cohen2014minimizing}
Cohen I, Mandelbaum A, Zychlinski N (2014) Minimizing mortality in a mass
  casualty event: fluid networks in support of modeling and staffing. \emph{IIE
  Transactions} 46(7):728--741.

\bibitem[{Cramer et~al.(2022)Cramer, Ray, Lopez, Bracher, Brennen,
  Castro~Rivadeneira, Gerding, Gneiting, House, Huang
  et~al.}]{cramer2022evaluation}
Cramer EY, Ray EL, Lopez VK, Bracher J, Brennen A, Castro~Rivadeneira AJ,
  Gerding A, Gneiting T, House KH, Huang Y, et~al. (2022) Evaluation of
  individual and ensemble probabilistic forecasts of covid-19 mortality in the
  united states. \emph{Proceedings of the National Academy of Sciences}
  119(15):e2113561119.

\bibitem[{Cromie et~al.(1979)Cromie, Chaudhry, \protect\BIBand{}
  Grassmann}]{cromie1979further}
Cromie M, Chaudhry M, Grassmann W (1979) Further results for the queueing
  system {$M^X/M/c$}. \emph{Journal of the Operational Research Society}
  30(8):755--763.

\bibitem[{Davis(1984)}]{davis1984piecewise}
Davis MH (1984) Piecewise-deterministic markov processes: A general class of
  non-diffusion stochastic models. \emph{Journal of the Royal Statistical
  Society: Series B (Methodological)} 46(3):353--376.

\bibitem[{Daw \protect\BIBand{} Pender(2019)}]{daw2019distributions}
Daw A, Pender J (2019) On the distributions of infinite server queues with
  batch arrivals. \emph{Queueing Systems} 91(3-4):367--401.

\bibitem[{Daw \protect\BIBand{} Pender(2022)}]{daw2021ephemerally}
Daw A, Pender J (2022) An ephemerally self-exciting point process.
  \emph{Advances in Applied Probability} 54(2).

\bibitem[{de~Graaf et~al.(2017)de~Graaf, Scheinhardt, \protect\BIBand{}
  Boucherie}]{de2017shot}
de~Graaf W, Scheinhardt WR, Boucherie RJ (2017) Shot-noise fluid queues and
  infinite-server systems with batch arrivals. \emph{Performance evaluation}
  116:143--155.

\bibitem[{Dong et~al.(2023)Dong, Jiang, \protect\BIBand{} Xu}]{dong2023impact}
Dong L, Jiang P, Xu F (2023) Impact of traceability technology adoption in food
  supply chain networks. \emph{Management Science} 69(3):1518--1535.

\bibitem[{Eckberg(1983)}]{eckberg1983generalized}
Eckberg A (1983) Generalized peakedness of teletraffic processes. \emph{Proc.
  10th Intl. Teletraff. Congress}.

\bibitem[{Emery et~al.(2004)Emery, Erdman, Bowen, Newton, Winchell, Meyer,
  Tong, Cook, Holloway, McCaustland et~al.}]{emery2004real}
Emery SL, Erdman DD, Bowen MD, Newton BR, Winchell JM, Meyer RF, Tong S, Cook
  BT, Holloway BP, McCaustland KA, et~al. (2004) Real-time reverse
  transcription--polymerase chain reaction assay for sars-associated
  coronavirus. \emph{Emerging infectious diseases} 10(2):311.

\bibitem[{Esseen(1942)}]{esseen1942liapunov}
Esseen CG (1942) On the liapunov limit error in the theory of probability.
  \emph{Ark. Mat. Astr. Fys.} 28:1--19.

\bibitem[{Feldman et~al.(2008)Feldman, Mandelbaum, Massey, \protect\BIBand{}
  Whitt}]{feldman2008staffing}
Feldman Z, Mandelbaum A, Massey WA, Whitt W (2008) Staffing of time-varying
  queues to achieve time-stable performance. \emph{Management Science}
  54(2):324--338.

\bibitem[{Fetzer \protect\BIBand{} Graeber(2021)}]{fetzer2021measuring}
Fetzer T, Graeber T (2021) Measuring the scientific effectiveness of contact
  tracing: Evidence from a natural experiment. \emph{Proceedings of the
  National Academy of Sciences} 118(33):e2100814118.

\bibitem[{Fox et~al.(2021)Fox, Bailey, Seamon, \protect\BIBand{}
  Miranda}]{fox2021response}
Fox MD, Bailey DC, Seamon MD, Miranda ML (2021) Response to a {COVID-19}
  outbreak on a {University Campus—Indiana, August 2020}. \emph{Morbidity and
  Mortality Weekly Report} 70(4):118.

\bibitem[{Gans et~al.(2003)Gans, Koole, \protect\BIBand{}
  Mandelbaum}]{gans2003telephone}
Gans N, Koole G, Mandelbaum A (2003) Telephone call centers: Tutorial, review,
  and research prospects. \emph{Manufacturing \& Service Operations Management}
  5(2):79--141.

\bibitem[{Gardner et~al.(2017{\natexlab{a}})Gardner, Harchol-Balter,
  Scheller-Wolf, \protect\BIBand{} Van~Houdt}]{gardner2017better}
Gardner K, Harchol-Balter M, Scheller-Wolf A, Van~Houdt B (2017{\natexlab{a}})
  A better model for job redundancy: Decoupling server slowdown and job size.
  \emph{IEEE/ACM transactions on networking} 25(6):3353--3367.

\bibitem[{Gardner et~al.(2017{\natexlab{b}})Gardner, Harchol-Balter,
  Scheller-Wolf, Velednitsky, \protect\BIBand{}
  Zbarsky}]{gardner2017redundancy}
Gardner K, Harchol-Balter M, Scheller-Wolf A, Velednitsky M, Zbarsky S
  (2017{\natexlab{b}}) Redundancy-d: The power of d choices for redundancy.
  \emph{Operations Research} 65(4):1078--1094.

\bibitem[{Garnett et~al.(2002)Garnett, Mandelbaum, \protect\BIBand{}
  Reiman}]{garnett2002designing}
Garnett O, Mandelbaum A, Reiman M (2002) Designing a call center with impatient
  customers. \emph{Manufacturing \& Service Operations Management}
  4(3):208--227.

\bibitem[{Gilbert \protect\BIBand{} Pollak(1960)}]{gilbert1960amplitude}
Gilbert E, Pollak H (1960) Amplitude distribution of shot noise. \emph{The Bell
  System Technical Journal} 39(2):333--350.

\bibitem[{Giufurta \protect\BIBand{} O'Connell(2021)}]{giufurta2021full}
Giufurta A, O'Connell S (2021) With full {Statler}, isolated students trickle
  into off-campus hotels. \emph{Cornell Daily Sun} .

\bibitem[{Gluckman(2021)}]{gluckman2021some}
Gluckman N (2021) Some universities have less space to isolate students this
  fall. {Is} that a problem? \emph{Chronicle of Higher Education} .

\bibitem[{Green et~al.(2007)Green, Kolesar, \protect\BIBand{}
  Whitt}]{green2007coping}
Green LV, Kolesar PJ, Whitt W (2007) Coping with time-varying demand when
  setting staffing requirements for a service system. \emph{Production and
  Operations Management} 16(1):13--39.

\bibitem[{Gupta et~al.(2022)Gupta, Starr, Farahani, \protect\BIBand{}
  Asgari}]{gupta2022om}
Gupta S, Starr MK, Farahani RZ, Asgari N (2022) Om forum—pandemics/epidemics:
  Challenges and opportunities for operations management research.
  \emph{Manufacturing \& Service Operations Management} 24(1):1--23.

\bibitem[{Gurvich et~al.(2010)Gurvich, Luedtke, \protect\BIBand{}
  Tezcan}]{gurvich2010staffing}
Gurvich I, Luedtke J, Tezcan T (2010) Staffing call centers with uncertain
  demand forecasts: A chance-constrained optimization approach.
  \emph{Management Science} 56(7):1093--1115.

\bibitem[{Halfin \protect\BIBand{} Whitt(1981)}]{halfin1981heavy}
Halfin S, Whitt W (1981) Heavy-traffic limits for queues with many exponential
  servers. \emph{Operations Research} 29(3):567--588.

\bibitem[{Harrison \protect\BIBand{} Resnick(1976)}]{harrison1976stationary}
Harrison JM, Resnick SI (1976) The stationary distribution and first exit
  probabilities of a storage process with general release rule.
  \emph{Mathematics of Operations Research} 1(4):347--358.

\bibitem[{Harrison \protect\BIBand{} Resnick(1978)}]{harrison1978recurrence}
Harrison JM, Resnick SI (1978) The recurrence classification of risk and
  storage processes. \emph{Mathematics of Operations Research} 3(1):57--66.

\bibitem[{Hong \protect\BIBand{} Wang(2021)}]{hong2021sharp}
Hong Y, Wang W (2021) Sharp waiting-time bounds for multiserver jobs.
  \emph{arXiv preprint arXiv:2109.05343} .

\bibitem[{Ibrahim et~al.(2016)Ibrahim, Ye, L’Ecuyer, \protect\BIBand{}
  Shen}]{ibrahim2016modeling}
Ibrahim R, Ye H, L’Ecuyer P, Shen H (2016) Modeling and forecasting call
  center arrivals: A literature survey and a case study. \emph{International
  Journal of Forecasting} 32(3):865--874.

\bibitem[{Jennings et~al.(1996)Jennings, Mandelbaum, Massey, \protect\BIBand{}
  Whitt}]{jennings1996server}
Jennings OB, Mandelbaum A, Massey WA, Whitt W (1996) Server staffing to meet
  time-varying demand. \emph{Management Science} 42(10):1383--1394.

\bibitem[{Jongbloed \protect\BIBand{} Koole(2001)}]{jongbloed2001managing}
Jongbloed G, Koole G (2001) Managing uncertainty in call centres using poisson
  mixtures. \emph{Applied Stochastic Models in Business and Industry}
  17(4):307--318.

\bibitem[{Kaspi(1984)}]{kaspi1984storage}
Kaspi H (1984) Storage processes with {Markov} additive input and output.
  \emph{Mathematics of Operations Research} 9(3):424--440.

\bibitem[{Kella \protect\BIBand{} Whitt(1999)}]{kella1999linear}
Kella O, Whitt W (1999) Linear stochastic fluid networks. \emph{Journal of
  Applied Probability} 36(1):244--260.

\bibitem[{Kim \protect\BIBand{} Whitt(2014)}]{kim2014call}
Kim SH, Whitt W (2014) Are call center and hospital arrivals well modeled by
  nonhomogeneous {Poisson} processes? \emph{Manufacturing \& Service Operations
  Management} 16(3):464--480.

\bibitem[{Kingman(2009)}]{kingman2009first}
Kingman J (2009) The first {Erlang} century—and the next. \emph{Queueing
  systems} 63(1):3--12.

\bibitem[{Lash et~al.(2021)Lash, Moonan, Byers, Bonacci, Bonner, Donahue,
  Donovan, Grome, Janssen, Magleby et~al.}]{lash2021covid}
Lash RR, Moonan PK, Byers BL, Bonacci RA, Bonner KE, Donahue M, Donovan CV,
  Grome HN, Janssen JM, Magleby R, et~al. (2021) Covid-19 case investigation
  and contact tracing in the us, 2020. \emph{JAMA network open}
  4(6):e2115850--e2115850.

\bibitem[{Levi et~al.(2020)Levi, Singhvi, \protect\BIBand{}
  Zheng}]{levi2020economically}
Levi R, Singhvi S, Zheng Y (2020) Economically motivated adulteration in
  farming supply chains. \emph{Management Science} 66(1):209--226.

\bibitem[{Lu \protect\BIBand{} Pang(2017)}]{lu2017heavy}
Lu H, Pang G (2017) Heavy-traffic limits for a fork-join network in the
  {Halfin}--{Whitt} regime. \emph{Stochastic Systems} 6(2):519--600.

\bibitem[{Massey \protect\BIBand{} Whitt(1996)}]{massey1996stationary}
Massey WA, Whitt W (1996) Stationary-process approximations for the
  nonstationary {Erlang} loss model. \emph{Operations Research} 44(6):976--983.

\bibitem[{Mathijsen et~al.(2018)Mathijsen, Janssen, van Leeuwaarden,
  \protect\BIBand{} Zwart}]{mathijsen2018robust}
Mathijsen BW, Janssen A, van Leeuwaarden JS, Zwart B (2018) Robust
  heavy-traffic approximations for service systems facing overdispersed demand.
  \emph{Queueing systems} 90(3):257--289.

\bibitem[{Mills et~al.(2013)Mills, Argon, \protect\BIBand{}
  Ziya}]{mills2013resource}
Mills AF, Argon NT, Ziya S (2013) Resource-based patient prioritization in
  mass-casualty incidents. \emph{Manufacturing \& Service Operations
  Management} 15(3):361--377.

\bibitem[{{National Association of County and City Health
  Officials}(2020)}]{naccho2020building}
{National Association of County and City Health Officials} (2020) Building
  covid-19 contact tracing in health departments to support reopening american
  society safely. \emph{NACCHO position statement}
  \urlprefix\url{https://www.naccho.org/uploads/full-width-images/Contact-Tracing-Statement-4-16-2020.pdf},
  {Accessed: 5/4/2023}.

\bibitem[{Neuts(1978)}]{neuts1978algorithmic}
Neuts MF (1978) An algorithmic solution to the {$GI/M/C$} queue with group
  arrivals. Technical report, Delaware Univ. Newark Dept. of Statistics and
  Computer Science.

\bibitem[{{NYC Department of Health and Mental Hygiene}(2023)}]{nyc2023nyc}
{NYC Department of Health and Mental Hygiene} (2023) Nyc coronavirus disease
  2019 (covid-19) data.
  \urlprefix\url{https://www.nyc.gov/site/doh/covid/covid-19-data.page},
  {Accessed: 5/4/2023}.

\bibitem[{{NYC Health + Hospitals}(2022)}]{nyc2022covid}
{NYC Health + Hospitals} (2022) {COVID}-19 contact tracing public report --
  reporting period: April 24, 2022 - april 29, 2022.
  \urlprefix\url{https://hhinternet.blob.core.windows.net/uploads/2022/05/public_weekly_report_04302022_updated.pdf},
  {Accessed: 5/4/2023}.

\bibitem[{{\"O}zkan \protect\BIBand{} Ward(2019)}]{ozkan2019control}
{\"O}zkan E, Ward AR (2019) On the control of fork-join networks.
  \emph{Mathematics of Operations Research} 44(2):532--564.

\bibitem[{Pang \protect\BIBand{} Whitt(2012)}]{pang2012impact}
Pang G, Whitt W (2012) The impact of dependent service times on large-scale
  service systems. \emph{Manufacturing \& Service Operations Management}
  14(2):262--278.

\bibitem[{Prabhu(2012)}]{prabhu2012stochastic}
Prabhu NU (2012) \emph{Stochastic storage processes: queues, insurance risk,
  dams, and data communication}, volume~15 (Springer Science \& Business
  Media).

\bibitem[{Rainisch et~al.(2022)Rainisch, Jeon, Pappas, Spencer, Fischer,
  Adhikari, Taylor, Greening, Moonan, Oeltmann et~al.}]{rainisch2022estimated}
Rainisch G, Jeon S, Pappas D, Spencer KD, Fischer LS, Adhikari BB, Taylor MM,
  Greening B, Moonan PK, Oeltmann JE, et~al. (2022) Estimated covid-19 cases
  and hospitalizations averted by case investigation and contact tracing in the
  us. \emph{JAMA network open} 5(3):e224042--e224042.

\bibitem[{Reed(2009)}]{reed2009g}
Reed J (2009) The {$G/GI/N$} queue in the {Halfin}--{Whitt} regime. \emph{The
  Annals of Applied Probability} 19(6):2211--2269.

\bibitem[{Rubinovitch \protect\BIBand{} Cohen(1980)}]{rubinovitch1980level}
Rubinovitch M, Cohen J (1980) Level crossings and stationary distributions for
  general dams. \emph{Journal of Applied Probability} 17(1):218--226.

\bibitem[{Ruebush et~al.(2021)Ruebush, Fraser, Poulin, Allen, Lane,
  \protect\BIBand{} Blumenstock}]{ruebush2021covid}
Ruebush E, Fraser MR, Poulin A, Allen M, Lane J, Blumenstock JS (2021) Covid-19
  case investigation and contact tracing: early lessons learned and future
  opportunities. \emph{Journal of Public Health Management and Practice}
  27(1):S87--S97.

\bibitem[{Ruzankin(2020)}]{ruzankin2020absolute}
Ruzankin PS (2020) On absolute central moments of poisson distribution.
  \emph{Journal of Statistical Theory and Practice} 14(4):1--6.

\bibitem[{Sullivan et~al.(1980)Sullivan, Crone, \protect\BIBand{}
  Jalickee}]{sullivan1980approximation}
Sullivan J, Crone L, Jalickee J (1980) Approximation of the unit step function
  by a linear combination of exponential functions. \emph{Journal of
  Approximation Theory} 28(4):299--308.

\bibitem[{Thomasian(2014)}]{thomasian2014analysis}
Thomasian A (2014) Analysis of fork/join and related queueing systems.
  \emph{ACM Computing Surveys (CSUR)} 47(2):1--71.

\bibitem[{Tirmazi et~al.(2020)Tirmazi, Barker, Deng, Haque, Qin, Hand,
  Harchol-Balter, \protect\BIBand{} Wilkes}]{tirmazi2020borg}
Tirmazi M, Barker A, Deng N, Haque ME, Qin ZG, Hand S, Harchol-Balter M, Wilkes
  J (2020) Borg: the next generation. \emph{Proceedings of the fifteenth
  European conference on computer systems}, 1--14.

\bibitem[{van Leeuwaarden et~al.(2019)van Leeuwaarden, Mathijsen,
  \protect\BIBand{} Zwart}]{van2019economies}
van Leeuwaarden JS, Mathijsen BW, Zwart B (2019) Economies-of-scale in
  many-server queueing systems: Tutorial and partial review of the {QED}
  {Halfin}--{Whitt} heavy-traffic regime. \emph{SIAM Review} 61(3):403--440.

\bibitem[{Vasan et~al.(2022)Vasan, Foote, \protect\BIBand{}
  Long}]{vasan2022ensuring}
Vasan A, Foote M, Long T (2022) Ensuring widespread and equitable access to
  treatments for covid-19. \emph{JAMA} 328(8):705--706.

\bibitem[{Wang et~al.(2019)Wang, Harchol-Balter, Jiang, Scheller-Wolf,
  \protect\BIBand{} Srikant}]{wang2019delay}
Wang W, Harchol-Balter M, Jiang H, Scheller-Wolf A, Srikant R (2019) Delay
  asymptotics and bounds for multitask parallel jobs. \emph{Queueing Systems}
  91(3):207--239.

\bibitem[{Wang et~al.(2022)Wang, Du, James, Fox, Lachmann, Meyers,
  \protect\BIBand{} Bhavnani}]{wang2022effectiveness}
Wang X, Du Z, James E, Fox SJ, Lachmann M, Meyers LA, Bhavnani D (2022) The
  effectiveness of covid-19 testing and contact tracing in a us city.
  \emph{Proceedings of the National Academy of Sciences} 119(34):e2200652119.

\bibitem[{Watson et~al.(2020)Watson, Cicero, Blumenstock, \protect\BIBand{}
  Fraser}]{watson2020national}
Watson C, Cicero A, Blumenstock JS, Fraser MR (2020) \emph{A national plan to
  enable comprehensive COVID-19 case finding and contact tracing in the US}
  (Johns Hopkins Bloomberg School of Public Health, Center for Health
  Security),
  \urlprefix\url{https://centerforhealthsecurity.org/sites/default/files/2023-02/200410-national-plan-to-contact-tracing.pdf},
  {Accessed: 5/4/2023}.

\bibitem[{Whitt(1992)}]{whitt1992understanding}
Whitt W (1992) Understanding the efficiency of multi-server service systems.
  \emph{Management Science} 38(5):708--723.

\bibitem[{Whitt(2007)}]{whitt2007you}
Whitt W (2007) What you should know about queueing models to set staffing
  requirements in service systems. \emph{Naval Research Logistics (NRL)}
  54(5):476--484.

\bibitem[{Wolff(1982)}]{wolff1982poisson}
Wolff RW (1982) Poisson arrivals see time averages. \emph{Operations Research}
  30(2):223--231.

\bibitem[{Yao(1985)}]{yao1985some}
Yao DD (1985) Some results for the queues {$M^X/M/c$ and $GI^X/G/c$}.
  \emph{Operations Research Letters} 4(2):79--83.

\bibitem[{Yao et~al.(1984)Yao, Chaudhry, \protect\BIBand{}
  Templeton}]{yao1984bounds}
Yao DD, Chaudhry M, Templeton J (1984) On bounds for bulk arrival queues.
  \emph{European Journal of Operational Research} 15(2):237--243.

\bibitem[{Yeo(1974)}]{yeo1974finite}
Yeo G (1974) A finite dam with exponential release. \emph{Journal of Applied
  Probability} 11(1):122--133.

\bibitem[{Yeo(1976)}]{yeo1976dam}
Yeo G (1976) A dam with general release rule. \emph{The ANZIAM Journal}
  19(4):469--477.

\bibitem[{Yom-Tov \protect\BIBand{} Mandelbaum(2014)}]{yomtov2014erlang}
Yom-Tov GB, Mandelbaum A (2014) Erlang-r: A time-varying queue with reentrant
  customers, in support of healthcare staffing. \emph{Manufacturing \& Service
  Operations Management} 16(2):283--299.

\bibitem[{Zeltyn \protect\BIBand{} Mandelbaum(2005)}]{zeltyn2005call}
Zeltyn S, Mandelbaum A (2005) Call centers with impatient customers:
  many-server asymptotics of the {M/M/n+ G} queue. \emph{Queueing Systems}
  51(3):361--402.

\bibitem[{Zhao(1994)}]{zhao1994analysis}
Zhao Y (1994) Analysis of the {$GI^X/M/c$} model. \emph{Queueing Systems}
  15(1-4):347--364.

\end{thebibliography}

\begin{APPENDIX}{}

%\tr{DON'T FORGET TO UPDATE!}

This appendix contains proofs of our main results, as well as supporting and auxiliary results that expand the story of this work. Let us outline those here. Appendix~\ref{batchProof} contains the proofs of the large batch limits contained in Section~\ref{batchSec}, and likewise Appendix~\ref{hybridProof} contains the proofs of the \edit{batch-and-rate} limits from Sections~\ref{hybridSec}. Appendix~\ref{support} then contains proofs of supporting and related results to these limits, as well as some additional results not yet shown in the paper. As mentioned in the main body of the text, \edit{Appendix~\ref{simApp} contains further details of the contact tracing simulation.} \edit{Furthermore,} Appendix~\ref{staffingSec} contains results and techniques for staffing the large batch arrival queue by way of staffing the storage process, and Appendix~\ref{techAppend} houses technical results supporting this\edit{, while Appendix~\ref{waitOptSec} develops an alternate arrival pattern control problem that uses the storage process framework}. Finally, Appendix~\ref{dependence} contains simulation experiments and discussion on the impacts of dependence between jobs within the same batch.

\section{Proofs of Large Batch Limits}\label{batchProof}

\subsection{Proof of Theorem~\ref{fBatchScale}}\label{fBatchScaleProof}

\proof{Proof.}
We will show the convergence of the batch scaling of the queue through analyzing its moment generating function. To begin, we note that the infinite server queue length can be expressed in terms of indicator functions as
\begin{align*}
Q_t(n)
&
=
\sum_{j=1}^{Q_0(n)} \mathbf{1}\{t < S_{0,j}\}
+
\sum_{i=1}^{N_t} \sum_{j=1}^{B_i(n)} \mathbf{1}\{t < A_i + S_{i,j}\}
,
\end{align*}
where $S_{i,j}$ is the service duration of the $j^\text{th}$ customer within the $i^\text{th}$ batch and $S_{0,j}$ is the remaining service time of the $j^\text{th}$ job that was in service at time 0. In this way, the first term on the right hand side represents the number of jobs in the system at time 0 that remain in the system at time $t$, whereas the double summation counts the number of jobs from each batch that remain in service at time $t$. Because there are infinitely many servers, we can note that the number jobs remaining since time 0 is independent from the number of jobs in the system that entered after time 0. Hence, we will consider these groups separately. Starting with those jobs initially present, we can note that since $\{S_{0,j} \mid 1 \leq j \leq Q_0(n)\}$ are the only stochastic terms, the law of large numbers yields that
$$
\frac{1}{n}\sum_{j=1}^{Q_0(n)} \mathbf{1}\{t < S_{0,j}\}
=
\frac{Q_0(n)}{n} \frac{1}{Q_0(n)}\sum_{j=1}^{Q_0(n)} \mathbf{1}\{t < S_{0,j}\}
\stackrel{a.s.}{\longrightarrow}
\psi_0 \bar{G}_0(t)
$$
Thus, without loss of generality, we will hereforward assume that the queue starts empty. We then write the moment generating function of $Q_t(n)$ at $\frac{\theta}n$ as
\begin{align*}
\E{e^{\frac{\theta Q_t(n)}n} }
&
=
\E{\mathrm{exp}\left( \frac{\theta}{n} \sum_{i=1}^{N_t} \sum_{j=1}^{B_i(n)} \mathbf{1}\{t < A_i + S_{i,j}\}\right)}
.
\end{align*}
By conditioning on the filtration of the counting process $\mathcal{F}_t^N$, total expectation yields that
\begin{align*}
\E{\mathrm{exp}\left(\frac{\theta}{n} \sum_{i=1}^{N_t} \sum_{j=1}^{B_i(n)} \mathbf{1}\{t < A_i + S_{i,j}\}\right)}
&
=
\E{\prod_{i=1}^{N_t}\E{\mathrm{exp}\left(\frac{\theta}{n} \sum_{j=1}^{B_i(n)} \mathbf{1}\{t < A_i + S_{i,j}\}\right) \Big| \mathcal{F}_t^N}}
.
\end{align*}
Focusing on the inner expectation, we again use the tower property. We now condition on the batch size $B_i(n)$, which leaves the service duration as the only uncertain quantity. The indicator is thus a Bernoulli random variable with success probability $\bar{G}(t-A_i)$, and since these are i.i.d. within the batch we have that
\begin{align*}
\E{\mathrm{exp}\left(\frac{\theta}{n} \sum_{j=1}^{B_i(n)} \mathbf{1}\{t < A_i + S_{i,j}\}\right) \Big| \mathcal{F}_t^N}
&
=
\E{\prod_{j=1}^{B_i(n)} \E{e^{\frac{\theta}{n}  \mathbf{1}\{t < A_i + S_{i,j}\}} \Big| \mathcal{F}_t^N, B_i(n)}
\Big| \mathcal{F}_t^N}
\\
&
=
\E{\left(G(t-A_i) + \bar{G}(t-A_i) e^{\frac{\theta}n}\right)^{B_i(n)}
\Big| \mathcal{F}_t^N}
\\
&
=
\E{\left(1 + \bar{G}(t-A_i) (e^{\frac{\theta}n}-1)\right)^{B_i(n)}
\Big| \mathcal{F}_t^N}
.
\end{align*}
By now using the identity $x = e^{\log(x)}$, we can transform this to
\begin{align*}
\E{\left(1 + \bar{G}(t-A_i) (e^{\frac{\theta}n}-1)\right)^{B_i(n)}
\Big| \mathcal{F}_t^N}
&
=
\E{
\mathrm{exp}\left(\log\left(\left(1 + \bar{G}(t-A_i) (e^{\frac{\theta}n}-1)\right)^{B_i(n)}\right)\right)
\Big| \mathcal{F}_t^N}
\\
&
=
\E{
e^{B_i(n) \log\left(1 + \bar{G}(t-A_i) (e^{\frac{\theta}n}-1)\right)}
\Big| \mathcal{F}_t^N}
,
\end{align*}
which we can now re-express further through two series expansions. Specifically, using a Taylor and a Mercator series expansion on $e^{\frac{\theta}n}-1$ and $\log\left(1+\bar{G}(t-A_i) (e^{\frac{\theta}n}-1)\right)$, respectively, we simplify to
\begin{align*}
\E{
e^{B_i(n) \log\left(1 + \bar{G}(t-A_i) (e^{\frac{\theta}n}-1)\right)}
\Big| \mathcal{F}_t^N}
&
=
\E{
e^{\frac{\theta B_i(n) \bar{G}(t-A_i)}n + O\left(\frac{B_i(n)}{n^2} \right)}
\Big| \mathcal{F}_t^N}
.
\end{align*}
Returning to the original expectation, we now have that
\begin{align*}
\E{\prod_{i=1}^{N_t}\E{\mathrm{exp}\left(\frac{\theta}{n} \sum_{j=1}^{B_i(n)} \mathbf{1}\{t < A_i + S_{i,j}\}\right) \Big| \mathcal{F}_t^N}}
&
=
\E{
e^{\sum_{i=1}^{N_t} \frac{\theta B_i(n) \bar{G}(t-A_i)}n + O\left(\frac{B_i(n)}{n^2} \right)}
}
,
\end{align*}
and as $n \to \infty$, this converges to
\begin{align*}
\E{
e^{\sum_{i=1}^{N_t} \frac{\theta B_i(n) \bar{G}(t-A_i)}n + O\left(\frac{B_i(n)}{n^2} \right)}
}
\longrightarrow
\E{
e^{\theta \sum_{i=1}^{N_t}  M_i \bar{G}(t-A_i) }
}
,
\end{align*}
which yields the stated result for the queue. 
%To now yield the specific form of the generating function when $N_t$ is a Poisson process, we note that when conditioned on the quantity $N_t$ we have
%\begin{align*}
%\E{
%e^{\sum_{i=1}^{N_t} \theta M_i \bar{G}(t-A_i) }
%}
%&
%=
%\E{
%\E{e^{\sum_{i=1}^{N_t} \theta M_i \bar{G}(t-A_i) } \mid N_t}
%}
%=
%\E{\E{e^{\theta M_1 \bar{G}(U_1(0,t))}}^{N_t}}
%,
%\end{align*}
%where $U_i(0,t) \sim \mathrm{Uni}(0,t)$ are i.i.d and independent of $M_i$. Then, conditioning on $M_1$, this inner expectation can be expressed
%$$
%\E{e^{\theta M_1 \bar{G}(U_1(0,t))}}
%=
%\E{\E{e^{\theta M_1 \bar{G}(U_1(0,t))}\mid M_1}}
%=
%\E{\frac{1}{t}\int_0^t e^{\theta M_1 \bar{G}(x)}\mathrm{d}x}
%.
%$$
%By exchanging the order of integration and expectation via Fubini's theorem and substituting into the moment generating function for the Poisson process, we achieve the corresponding stated form.
\hfill\Halmos\\
\endproof

\subsection{Proof of Theorem~\ref{fDelayConv}}\label{fDelayConvProof}

\proof{Proof.}
In a manner similar to the proof of the infinite server  to shot noise convergence in Theorem~\ref{fBatchScale}, we begin by decomposing the queue length process into a sum of indicators. By comparison to the infinite server decomposition however, these indicators depend not only on the batch arrival epochs and the individual service durations, but also on the lengths of time that jobs wait to begin service while the servers were occupied. Recalling that $W_{i,j}$ is the total time the $j^\text{th}$ job within the $i^\text{th}$ batch spends waiting, we can express the queue length in the delay model queue at time $t$ as
\begin{align}
Q_t^C(n)
&=
\sum_{i=1}^{N_t}\sum_{j=1}^{B_i(n)} \mathbf{1}\{t < A_i + S_{i,j}\}
+
\sum_{i=1}^{N_t} \sum_{j=1}^{B_i(n)} \mathbf{1}\{ A_i + S_{i,j} \leq t < A_i + S_{i,j} + W_{i,j}\}
\nonumber
\\
&
\quad
+
\sum_{j=1}^{(Q_0^C(n) \wedge cn)} \mathbf{1}\{t < S_{0,j}\}
+
\sum_{j=1}^{(Q_0^C(n) - cn)^+} \mathbf{1}\{ t < W_{\cdot, j} + S_{\cdot, j}\}
.
\label{delayDecomp}
\end{align}
One can interpret this decompositions as follows. The first double summation across arrival epochs and batch sizes gives an idealized infinite server representation that would be accurate if no jobs had to wait to begin service. The second double summation then corrects that under-counting for any jobs that had to wait and have not yet completed service at time $t$. The third and fourth terms then capture the initial state of the system, with the third term counting which jobs have remained in service from time 0 to time $t$ and with the fourth term counting the number of jobs that were waiting at time 0 and have not completed service by $t$. Here we use $S_{0,j}$ to represent the remaining service times of the jobs that are in service at time 0 and we use $W_{\cdot, j}$ and $S_{\cdot, j}$ to represent the waiting and service times for the jobs that are present in the system at time 0 but were not in service. In this notation, the residual service time $S_{0,j}$ need not be equivalent in distribution to $S_{i,j}$ for $i \in \mathbb{Z}^+$, whereas $S_{\cdot, j}$ is equivalent to $S_{i,j}$.

To begin moving towards the storage process limit, we first show a batch-arrival-queue analog of Proposition 2.1 from \citet{reed2009g}. That is, we seek to justify
\begin{align}
\int_0^t \left(Q_{t-s}^C - cn\right)^+ \mathrm{d} G(s)
\nonumber
&=
\sum_{i=1}^{N_t} \sum_{j=1}^{B_i(n)} \left(\bar{G}(t- A_i - W_{i,j}) - \bar{G}(t-A_i)\right)
\\
&
\quad
+
\sum_{j=1}^{(Q_0^C(n) - cn)^+} \left(\bar{G}(t - W_{\cdot,j}) - \bar{G}(t)\right)
,
\label{reedEq}
\end{align}
and this follows from a generalization of the arguments from \citet{reed2009g}. Starting with the summations over the tail CDF terms, one can re-express these in terms of integrals over the service distribution measure, and these integrals can then be adjusted to a standard interval of $[0,t]$ through the introduction of indicator functions:
\begin{align*}
&
\sum_{i=1}^{N_t} \sum_{j=1}^{B_i(n)} \left(\bar{G}(t- A_i - W_{i,j}) - \bar{G}(t-A_i)\right)
+
\sum_{j=1}^{(Q_0^C(n) - cn)^+} \left(\bar{G}(t - W_{\cdot,j}) - \bar{G}(t)\right)
\\
&
=
\sum_{i=1}^{N_t} \sum_{j=1}^{B_i(n)}  \int_{(t- A_i - W_{i,j})^+}^{t-A_i} \mathrm{d}G(s)
+
\sum_{j=1}^{(Q_0^C(n) - cn)^+} \int_{(t - W_{\cdot,j})^+}^t \mathrm{d}G(s)
\\
&
=
\sum_{i=1}^{N_t} \sum_{j=1}^{B_i(n)}  \int_{0}^{t} \mathbf{1}\{A_i \leq t - s < A_i + W_{i,j} \} \mathrm{d}G(s)
+
\sum_{j=1}^{(Q_0^C(n) - cn)^+} \int_{0}^{t} \mathbf{1}\{ t - s < W_{i,j} \} \mathrm{d}G(s)
.
\end{align*}
Then, one can recognize that the number of jobs waiting at an arbitrary time $u \geq 0$ can be written
$$
\left(Q_{u}^C(n) - cn\right)^+
=
\sum_{j=1}^{N_u} \sum_{j=1}^{B_i(n)} \mathbf{1}\{A_i \leq u < A_i + W_{i,j}\}
+
\sum_{j=1}^{(Q_0^C(n) - cn)^+} \mathbf{1}\{u < W_{\cdot,j}\}
,
$$
as the first term on the right-hand side captures the number jobs still waiting across each batch of arrivals and the second term captures the number of jobs that have been waiting since time 0. Thus, by exchanging the order of summation and integration, we can now observe that
\begin{align*}
&
\sum_{i=1}^{N_t} \sum_{j=1}^{B_i(n)}  \int_{0}^{t} \mathbf{1}\{A_i \leq t - s < A_i + W_{i,j} \} \mathrm{d}G(s)
+
\sum_{j=1}^{(Q_0^C(n) - cn)^+} \int_{0}^{t} \mathbf{1}\{ t - s < W_{i,j} \} \mathrm{d}G(s)
\\
&=
\int_{0}^{t} \left(\sum_{i=1}^{N_t} \sum_{j=1}^{B_i(n)}   \mathbf{1}\{A_i \leq t - s < A_i + W_{i,j} \}
+
\sum_{j=1}^{(Q_0^C(n) - cn)^+}  \mathbf{1}\{ t - s < W_{i,j} \} \right) \mathrm{d}G(s)
\\
&=
\int_{0}^{t} \left(Q_{t-s}^C(n) - cn\right)^+ \mathrm{d}G(s)
,
\end{align*}
and thus we achieve Equation~\eqref{reedEq}.

Returning now to the decomposition of the queue length in Equation~\eqref{delayDecomp}, we can use the equality from Equation~\eqref{reedEq} to re-express the queue length as
\begin{align*}
Q_t^C(n)
&=
\sum_{i=1}^{N_t}\sum_{j=1}^{B_i(n)} \mathbf{1}\{t < A_i + S_{i,j}\}
+
\sum_{i=1}^{N_t} \sum_{j=1}^{B_i(n)} \mathbf{1}\{ A_i + S_{i,j} \leq t < A_i + S_{i,j} + W_{i,j}\}
\nonumber
\\
&
\quad
-
\sum_{i=1}^{N_t} \sum_{j=1}^{B_i(n)} \left(\bar{G}(t- A_i - W_{i,j}) - \bar{G}(t-A_i)\right)
-
\sum_{j=1}^{(Q_0^C(n) - cn)^+} \left(\bar{G}(t - W_{\cdot,j}) - \bar{G}(t)\right)
\\
&
\quad
+
\int_{0}^{t} \left(Q_{t-s}^C(n) - cn\right)^+ \mathrm{d}G(s)
+
\sum_{j=1}^{(Q_0^C(n) \wedge cn)} \mathbf{1}\{t < S_{0,j}\}
+
\sum_{j=1}^{(Q_0^C(n) - cn)^+} \mathbf{1}\{ t < W_{\cdot, j} + S_{\cdot, j}\}
%%%%%%%%%%%%%%%%%%%%%%%%%%
\\
&=
\sum_{i=1}^{N_t}\sum_{j=1}^{B_i(n)} \mathbf{1}\{t < A_i + S_{i,j}\}
+
\sum_{j=1}^{(Q_0^C(n) \wedge cn)} \mathbf{1}\{t < S_{0,j}\}
+
\int_{0}^{t} \left(Q_{t-s}^C(n) - cn\right)^+ \mathrm{d}G(s)
\\
&
\quad
+
\left(Q_0^C(n) - cn \right)^+ \bar{G}(t)
+
\sum_{i=1}^{N_t} \sum_{j=1}^{B_i(n)} \left( \mathbf{1}\{ t - A_i - W_{i,j} <  S_{i,j}\} - \bar{G}(t- A_i - W_{i,j})\right)
\\
&
\quad
-
\sum_{i=1}^{N_t} \sum_{j=1}^{B_i(n)} \left( \mathbf{1}\{ t - A_i < S_{i,j} \} - \bar{G}(t-A_i)\right)
+
\sum_{j=1}^{(Q_0^C(n) - cn)^+} \left(\mathbf{1}\{ t - W_{\cdot, j} < S_{\cdot, j}\} - \bar{G}(t - W_{\cdot,j}) \right) .
\end{align*}
Through this decomposition, we will now prove that the batch scaling of the queue length converges to the generalized storage process. We proceed through induction on the arrival times, where one can suppose that we have conditioned on the filtration of the arrival process up to time $t$, like in the Proof of Theorem~\ref{fBatchScale}. For the base case, let $0 \leq t < A_1$. Then, the normalized queue length at time $t$ can be written
\begin{align*}
\frac{Q_t^C(n)}{n}
&=
\frac{1}{n}\sum_{j=1}^{(Q_0^C(n) \wedge cn)} \mathbf{1}\{t < S_{0,j}\}
+
\frac{1}{n}\int_{0}^{t} \left(Q_{t-s}^C(n) - cn\right)^+ \mathrm{d}G(s)
+
\frac{1}{n}\left(Q_0^C(n) - cn \right)^+ \bar{G}(t)
\\
&
\quad
+
\frac{1}{n}\sum_{j=1}^{(Q_0^C(n) - cn)^+} \left(\mathbf{1}\{ t - W_{\cdot, j} < S_{\cdot, j}\} - \bar{G}(t - W_{\cdot,j}) \right)
,
\end{align*}
which we now analyze piece by piece. By the law of large numbers, the assumption on the initial values, and the continuous mapping theorem, we have that as $n \to \infty$
$$
\frac{1}{n}\sum_{j=1}^{(Q_0^C(n) \wedge cn)} \mathbf{1}\{t < S_{0,j}\}
\stackrel{D}{\Longrightarrow}
\left(\psi_0^C \wedge c \right) \bar{G}_0 (t)
,
$$
where $\bar{G}_0(\cdot)$ is the complementary CDF of the residual service durations of the initial jobs in service at time 0. We can also similarly observe that
$$
\frac{1}{n}\left(Q_0^C(n) - cn \right)^+ \bar{G}(t)
\stackrel{D}{\Longrightarrow}
\left(\psi_0^C - c \right)^+ \bar{G}(t)
,
$$
as $n \to \infty$. Now, for the summation over jobs that were waiting to begin service at time $0$, we can employ a martingale argument such as that used in e.g.~\citet{andrews1988laws}. Let $\mathcal{S}_j$ for $0 \leq j \leq \left(Q_0^C(n) - cn \right)^+$ be the filtration generated by the collection of service times of the jobs initially in service at time 0 and of the first $j$ jobs to enter service after time 0, i.e.~$\mathcal{S}_j = \sigma\left(\{S_{0,1}, \dots, S_{0, cn}, S_{\cdot, 1}, \dots, S_{\cdot, j}\}\right)$. Then, one can note that for $j < \left(Q_0^C(n) - cn \right)^+$, $W_{\cdot, j+1}$ is $\mathcal{S}_j$ measurable, as the previous service durations dictate the time that this job waits. Thus, we can recognize that $\E{\mathbf{1}\{ t - W_{\cdot, j} < S_{\cdot, j}\} \mid \mathcal{S}_j} = \bar{G}(t - W_{\cdot,j})$. This implies that the summation is a martingale difference sequence, and thus we have that
$$
\frac{1}{n}\sum_{j=1}^{(Q_0^C(n) - cn)^+} \left(\mathbf{1}\{ t - W_{\cdot, j} < S_{\cdot, j}\} - \bar{G}(t - W_{\cdot,j}) \right)
\stackrel{D}{\Longrightarrow}
0
,
$$
as $n \to \infty$. Thus, as $n \to \infty$ the queue length on $0 \leq t < A_1$ converges to a process $z(\cdot)$ satisfying
$$
z(t)
=
\left(\psi_0^C \wedge c \right) \bar{G}_0 (t)
+
\left(\psi_0^C - c \right)^+ \bar{G}(t)
+
\int_{0}^{t} \left(z(t-s) - c\right)^+ \mathrm{d}G(s)
.
$$
We can observe that on this time interval each of these terms are deterministic, and thus Proposition 3.1 of \citet{reed2009g} yields that the function $z(\cdot)$ that solves this equation is unique. Since this matches the expression for $\psi_t^C$ on $0 \leq t < A_i$ as given by Equation~\eqref{fPsiCGen}, we have that $\frac{Q_t^C(n)}{n} \stackrel{D}{\Longrightarrow} \psi_t^C$ as $n \to \infty$ for $0 \leq t < A_1$. At the precise epoch of the first arrival, we can note that we furthermore have the convergence of the process immediately after the batch of jobs arrives, which is a direct consequence of the preceding arguments and assumption that $\frac{B_1(n)}{n} \stackrel{D}{\Longrightarrow} M_1$ as $n \to \infty$. Thus, $\frac{Q_t^C(n)}{n} \stackrel{D}{\Longrightarrow} \psi_t^C$ as $n \to \infty$ for $0 \leq t \leq A_1$, satisfying the base case of our inductive argument.

For the inductive step, we now assume that $\frac{Q_s^C(n)}{n} \stackrel{D}{\Longrightarrow} \psi_s^C$ as $n \to \infty$ for $s$ such that $0 \leq s \leq A_i$ and some $i \in \mathbb{Z}^+$. Let us now take $t$ such that $A_i \leq t < A_{i+1}$. We have established that we can decompose the normalized queue length as
\begin{align*}
\frac{Q_t^C(n)}{n}
&=
\frac{1}{n}\sum_{i=1}^{N_t}\sum_{j=1}^{B_i(n)} \mathbf{1}\{t < A_i + S_{i,j}\}
+
\frac{1}{n}\sum_{j=1}^{(Q_0^C(n) \wedge cn)} \mathbf{1}\{t < S_{0,j}\}
+
\frac{1}{n}\int_{0}^{t} \left(Q_{t-s}^C(n) - cn\right)^+ \mathrm{d}G(s)
\\
&
\quad
+
\frac{1}{n}\left(Q_0^C(n) - cn \right)^+ \bar{G}(t)
+
\frac{1}{n}\sum_{i=1}^{N_t} \sum_{j=1}^{B_i(n)} \left( \mathbf{1}\{ t - A_i - W_{i,j} <  S_{i,j}\} - \bar{G}(t- A_i - W_{i,j})\right)
\\
&
\quad
-
\frac{1}{n}\sum_{i=1}^{N_t} \sum_{j=1}^{B_i(n)} \left( \mathbf{1}\{ t - A_i < S_{i,j} \} - \bar{G}(t-A_i)\right)
+
\frac{1}{n}\sum_{j=1}^{(Q_0^C(n) - cn)^+} \left(\mathbf{1}\{ t - W_{\cdot, j} < S_{\cdot, j}\} - \bar{G}(t - W_{\cdot,j}) \right)
,
\end{align*}
and we can again analyze this piece-by-piece. By the batch scaling convergence of infinite server queues to shot noise processes in Theorem~\ref{fBatchScale}, we can observe that as $n\to\infty$
\begin{align*}
\frac{1}{n}\sum_{i=1}^{N_t}\sum_{j=1}^{B_i(n)} \mathbf{1}\{t < A_i + S_{i,j}\} \stackrel{D}{\Longrightarrow} \sum_{i=1}^{N_t} M_i \bar{G}(t-A_i)
,
\end{align*}
and
\begin{align*}
\frac{1}{n}\sum_{i=1}^{N_t} \sum_{j=1}^{B_i(n)} \left( \mathbf{1}\{ t - A_i < S_{i,j} \} - \bar{G}(t-A_i)\right) \stackrel{D}{\Longrightarrow} 0
.
\end{align*}
Similarly, analogous arguments to the base case show that the initial condition terms are such that
$$
\frac{1}{n}\sum_{j=1}^{(Q_0^C(n) \wedge cn)} \mathbf{1}\{t < S_{0,j}\}
\stackrel{D}{\Longrightarrow}
\left(\psi_0^C \wedge c \right) \bar{G}_0 (t)
,
$$
$$
\frac{1}{n}\left(Q_0^C(n) - cn \right)^+ \bar{G}(t)
\stackrel{D}{\Longrightarrow}
\left(\psi_0^C - c \right)^+ \bar{G}(t)
,
$$
and
$$
\frac{1}{n}\sum_{j=1}^{(Q_0^C(n) - cn)^+} \left(\mathbf{1}\{ t - W_{\cdot, j} < S_{\cdot, j}\} - \bar{G}(t - W_{\cdot,j}) \right)
\stackrel{D}{\Longrightarrow}
0
.
$$
For the remaining double summation over arrival epochs and batch sizes, we can again make use of a martingale structure. For $i \in \mathbb{Z}^+$ and $j \in \mathbb{Z}^+$, let us define the sigma algebra generated by the arrival times and service times of all jobs up to and including to the $j^{th}$ job within the $i^\text{th}$ batch, which is
\begin{align*}
\mathcal{S}_{i,j}
=
\sigma\Bigg(&\{S_{0,1}, \dots, S_{0, cn}, S_{\cdot, 1}, \dots, S_{\cdot, (cn - Q_0^C(n))^+}\} \cup \{A_1, \dots, A_i\}
\\
&\cup \left(\bigcup_{k=1}^{i-1} \{S_{k,1}, \dots, S_{k,B_k(n)} \} \right) \cup \{S_{i,1}, \cdots, S_{i,j}\} \Bigg)
.
\end{align*}
Then, we have that $W_{i,j+1}$ is $\mathcal{S}_{i,j}$ measurable since the queue is operating under first-come-first-serve, meaning that only the previous jobs determine how long the $j^\text{th}$ job in the $i^\text{th}$ batch waits. Thus, $\E{\mathbf{1}\{ t - A_i - W_{i,j + 1} <  S_{i,j + 1} \} \mid \mathcal{S}_{i,j}} = \bar{G}(t- A_i - W_{i,j})$. Therefore through martingale differences we have that
\begin{align*}
\frac{1}{n}\sum_{i=1}^{N_t} \sum_{j=1}^{B_i(n)} \left( \mathbf{1}\{ t - A_i - W_{i,j} <  S_{i,j}\} - \bar{G}(t- A_i - W_{i,j})\right)
\stackrel{D}{\Longrightarrow}
0
,
\end{align*}
as $n \to \infty$. Bringing these pieces together we now have that for $t \in [A_i, A_{i+1})$ the queue length process converges to a process $z_i(\cdot)$ satisfying
$$
z_i(t)
=
\left(\psi_0^C \wedge c\right) \bar{G}_0 (t)
+
\left(\psi_0^C - c\right)^+ \bar{G}(t)
+
\sum_{i=1}^{N_t} M_i \bar{G}(t-A_i)
+
\int_0^t \left( z_i(t-s) - c \right)^+ \mathrm{d} G(s)
.
$$
From the inductive hypothesis and the uniqueness given by Proposition 3.1 of \citet{reed2009g}, we have that $z_i(s) = \psi_s^C$ must hold for all $s \leq A_i$. One can then observe that $z_i(t)$ is deterministic for $A_i < t < A_{i+1}$, meaning that Proposition 3.1 of \citet{reed2009g} further implies that $z_i(t) = \psi_t^C$ on this interval as well. To complete the inductive step, we can note that by the given convergence of the batch sizes to the jump sizes, we also have that $Q_{A_{i+1}}^C(n) \slash n \stackrel{D}{\Longrightarrow} \psi_{A_{i+1}}^C$ as $n \to \infty$, and this completes the proof.
\hfill\Halmos\\
\endproof

\section{Proofs of Large \edit{Batch-and-Rate} Limits}\label{hybridProof}

\subsection{Proof of Theorem~\ref{bothLimitInf}}

\proof{Proof.}
As we did in the proof of Proposition~\ref{infMeanVar}, here we will invoke Proposition 2.4 of~\citet{daw2019distributions}. We know that the queue length is equivalent in distribution to a sum of scaled Poisson random variables, i.e.,
$$
Q^{\edit{\infty}}(m)
\stackrel{D}{=}
\sum_{j=1}^{ \edit{n_0}m^\nu }
j Y_j
,
$$
where $Y_j \sim \mathsf{Pois}\left({\edit{\lambda_0} m^{1-\nu}}\slash{j\mu}\right)$ are independent. Starting with the scaling with normalization by $m$, we consider two sequences of events indexed by $m$: $\{Q^{\edit{\infty}}(m)\slash m - \edit{\lambda_0 n_0}  \slash \mu > \epsilon\}$ and $\{\edit{\lambda_0 n_0}  \slash \mu - Q^{\edit{\infty}}(m)\slash m > \epsilon\}$, where $\epsilon > 0$. Beginning with the former, the sum of scaled Poissons decomposition allows us to invoke a Chernoff inequality to bound the probability of this event. That is,
\begin{align*}
\PP{
\frac{1}{m}Q^{\edit{\infty}}(m) - \frac{\edit{\lambda_0 n_0} }{\mu}
>
\epsilon
}
&
=
\PP{
\sum_{j=1}^{ \edit{n_0}m^\nu }
j Y_j
> 
\epsilon m
+
\frac{\edit{\lambda_0 n_0} m}{\mu}
}
\leq
\inf_{\theta > 0}
e^{-\theta(
\epsilon m
+
\frac{\edit{\lambda_0 n_0}  m}{\mu}
)
}
\prod_{j=1}^{ \edit{n_0}m^\nu }
\E{e^{\theta
j Y_j
}
}
.
\end{align*}
Through use of the Poisson moment generation function and Taylor expansions on $e^{j \theta}$, we can see that the product is equal to
\begin{align*}
\prod_{j=1}^{ \edit{n_0}m^\nu }
\E{e^{\theta
j Y_j
}
}
&
=
e^{\sum_{j=1}^{ \edit{n_0}m^\nu }\frac{\edit{\lambda_0} m^{1-\nu}}{j \mu}(e^{j \theta}-1)}
=
e^{\sum_{j=1}^{ \edit{n_0}m^\nu }\frac{\edit{\lambda_0} m^{1-\nu}}{\mu} \sum_{k=1}^\infty \frac{j^{k-1}\theta^k}{k!} }
.
\end{align*}
Now, because $\sum_{j=1}^b j^{k-1} \leq b^k$, we can bound the term in the exponent by 
$$
\sum_{j=1}^{ \edit{n_0}m^\nu }\frac{\edit{\lambda_0} m^{1-\nu}}{\mu} \sum_{k=1}^\infty \frac{j^{k-1}\theta^k}{k!} 
\leq
\frac{\edit{\lambda_0} m^{1-\nu}}{\mu} \sum_{k=1}^\infty \frac{(\theta  \edit{n_0}m^\nu )^k}{k!}
=
\frac{\edit{\lambda_0} m^{1-\nu}}{\mu} (e^{\theta  \edit{n_0}m^\nu } - 1)
.
$$
Hence, we can now see that this event's probability is bounded by
\begin{align*}
\PP{
\frac{1}{m}Q^{\edit{\infty}}(m) - \frac{\edit{\lambda_0 n_0} }{\mu}
>
\epsilon
}
&
\leq
\inf_{\theta > 0}
e^{-\theta(
\epsilon m
+
\frac{\edit{\lambda_0 n_0}  m}{\mu}
)
+
\frac{\edit{\lambda_0 }  m^{1-\nu}}{\mu} (e^{\theta  \edit{ n_0} m^\nu } - 1)}
,
\end{align*}
since at each $\theta > 0$ this initial Chernoff inequality expression is bounded above by the exponential function that we have just now reached. Conveniently, we can identify the value of this infimum since simple derivative checks show that the function in the exponent is minimized at
$$
\theta 
= 
\frac{1}{ \edit{ n_0} m^\nu }
\log\left(
\frac{
\epsilon m
+
\frac{\edit{\lambda_0 n_0}  m}{\mu}
}{
\frac{\edit{\lambda_0 n_0}  m}{\mu}   
}
\right)
=
\frac{1}{\edit{ n_0} m^\nu }
\log\left(
1
+
\frac{\mu \epsilon}{\edit{\lambda_0 n_0} }
\right)
.
$$
We now have
\begin{align*}
\PP{
\frac{1}{m}Q^{\edit{\infty}}(m) - \frac{\edit{\lambda_0 n_0} }{\mu}
>
\epsilon
}
&
\leq
\left(
1
+
\frac{\mu \epsilon}{\edit{\lambda_0 n_0} }
\right)^{
-m^{1-\nu}
\left(
\frac{\epsilon }{\edit{ n_0} }
+
\frac{\edit{\lambda_0 }   }{\mu}
\right)
}
e^{
\frac{ \epsilon m^{1-\nu}}{\edit{ n_0} }
}
=
\left(
\left(
1
+
\frac{\mu \epsilon}{\edit{\lambda_0 n_0} }
\right)^{
-
\left(
\frac{\epsilon }{\edit{ n_0} }
+
\frac{\edit{\lambda_0 }   }{\mu}
\right)
}
e^{
\frac{\epsilon}{\edit{ n_0} }
}
\right)^{m^{1-\nu}}
.
\end{align*}
Now, we can observe that the logarithm of the expression inside the widest parenthesis is negative, i.e.
\begin{align*}
\log
\left(
\left(
1
+
\frac{\mu \epsilon}{\edit{\lambda_0 n_0} }
\right)^{
-
\left(
\frac{\epsilon }{\edit{ n_0} }
+
\frac{\edit{\lambda_0 }   }{\mu}
\right)
}
e^{
\frac{\epsilon}{\edit{ n_0} }
}
\right)
&
=
\frac{\edit{\lambda_0 } }{\mu}
\left(
\frac{\mu\epsilon}{\edit{\lambda_0 n_0} }
-
\left(
1
+
\frac{\mu \epsilon}{\edit{\lambda_0 n_0} }
\right)
\log
\left(
1
+
\frac{\mu \epsilon}{\edit{\lambda_0 n_0} }
\right)
\right)
<
0
,
\end{align*}
since $z - (1+z)\log(1+z) < 0$ for all $z > 0$. Hence, because 
$
\sum_{k=1}^{\infty} x^{k^{p}} < \infty
$
for $x \in [0,1)$ and $p \in (0,1]$,\footnote{The lower end of this interval being open is akin to taking $\nu < 1$, matching the fact that there is a non-degenerate distribution yielded by the pure batch scaling limit.} we can see that
$$
\sum_{m=1}^\infty
\PP{
\frac{1}{m}Q^{\edit{\infty}}(m) - \frac{\edit{\lambda_0 n_0} }{\mu}
>
\epsilon
}
< 
\infty
,
$$
so we now turn our attention to the latter of the two event sequences indexed by $m$.

Since $Q^{\edit{\infty}}(m)$ is non-negative, here we restrict to $\epsilon < \frac{\edit{\lambda_0 n_0}  }{\mu}$. Again through a Chernoff inequality approach on the sum of scaled Poisson's decomposition, we see that
\begin{align*}
\PP{\frac{\edit{\lambda_0 n_0} }{\mu} - \frac{1}{m}Q(m) > \epsilon}
&=
\PP{\sum_{j=1}^{ \edit{ n_0} m^\nu } j Y_j < \frac{\edit{\lambda_0 n_0}  m}{\mu} - \epsilon m}
\leq
\inf_{\theta > 0}
e^{\theta\left(\frac{\edit{\lambda_0 n_0}  m}{\mu} - \epsilon m\right)}
e^{\sum_{j=1}^{ \edit{ n_0} m^\nu } \frac{\edit{\lambda_0 }  m^{1-\nu}}{j\mu} (e^{-\theta j} - 1)}
.
\end{align*}
Now, we can observe that because $e^{-x} - 1 \leq (e^{-kx}-1)\slash k$ for all $k \geq 1$ and all $x \geq 0$, we can bound the function in the rightmost exponent via
$$
\sum_{j=1}^{ \edit{ n_0} m^\nu } \frac{\edit{\lambda_0 }  m^{1-\nu}}{j\mu} (e^{-\theta j} - 1)
\leq
\sum_{j=1}^{ \edit{ n_0} m^\nu } \frac{\edit{\lambda_0 }  m^{1-\nu}}{\mu  \edit{ n_0}  m^\nu  } (e^{-\theta  \edit{ n_0}  m^\nu } - 1)
\leq
\frac{\edit{\lambda_0 }  m^{1-\nu}}{\mu } (e^{-\theta  \edit{ n_0}  m^\nu } - 1)
.
$$
Again, this yields an object that is more amenable to minimizing in close form. Here, direct calculus yields that 
$$
\theta 
= 
-
\frac{1}{ \edit{n_0}m^\nu }
\log
\left(
\frac{
\frac{\edit{\lambda_0 n_0} m}{\mu} - \epsilon m
}{
\frac{\edit{\lambda n_0} m}{\mu} 
}
\right)
=
-
\frac{1}{\edit{n_0}m^\nu}
\log
\left(
1 - \frac{\mu \epsilon }{\edit{\lambda_0 n_0}}
\right) 
.
$$
The upper bound on the probability of the event is now
\begin{align*}
\PP{\frac{\edit{\lambda_0 n_0}}{\mu} - \frac{1}{m}Q^{\edit{\infty}}(m) > \epsilon}
& 
\leq
\left(
1 - \frac{\mu \epsilon }{\edit{\lambda_0 n_0}}
\right) ^{
m^{1-\nu}
\left(\frac{\epsilon}{\edit{n_0}} - \frac{\edit{\lambda_0} }{\mu}  \right)}
e^{ - \frac{\epsilon m^{1-\nu}}{\edit{n_0} } 
}
\leq
\left(
\left(
1 - \frac{\mu \epsilon }{\edit{\lambda_0 n_0}}
\right) ^{
\frac{\epsilon}{\edit{n_0}}  - \frac{\edit{\lambda_0} }{\mu} 
}
e^{ 
- \frac{ \epsilon }{\edit{n_0} } 
}
\right)^{m^{1-\nu}}
.
\end{align*}
By once more taking the logarithm of the expression inside the widest parenthesis, we can see that the base of this exponent is less than 1:
\begin{align*}
\log
\left(
\left(
1 - \frac{\mu \epsilon }{\edit{\lambda_0 n_0}}
\right) ^{
\frac{\epsilon}{\edit{n_0}}  - \frac{\edit{\lambda_0} }{\mu} 
}
e^{ 
- \frac{ \epsilon }{\edit{n_0} } 
}
\right)
&
=
\frac{\epsilon}{\edit{n_0}}
\left(
\log\left(
1 - \frac{\mu \epsilon }{\edit{\lambda_0 n_0}}
\right)
-
1
\right)
-
\frac{\edit{\lambda_0}}{\mu}
\log\left(
1 - \frac{\mu \epsilon }{\edit{\lambda_0 n_0}}
\right)
<
0
,
\end{align*}
which follows immediately from the fact that $z(\log(1-z)-1) - \log(z) < 0$ for all $0 < z < 1$. Hence, again by the fact that $\sum_{k=1}^\infty x^{k^p}$ converges for $x \in [0,1)$ and $p \in (0,1]$, we have that
$$
\sum_{m=1}^\infty 
\PP{\frac{\edit{\lambda_0 n_0}}{\mu} - \frac{1}{m}Q^{\edit{\infty}}(m) > \epsilon}
< 
\infty
, 
$$
which further yields that 
\begin{align*}
\sum_{m=1}^\infty
\PP{\left| \frac{\edit{\lambda_0 n_0}}{\mu} - \frac{1}{m}Q^{\edit{\infty}}(m) \right| > \epsilon}
%=
%\PP{\left\{\frac{\lambda n}{\mu} - \frac{1}{m}Q(m) > \epsilon \right\} \bigcup \left\{\frac{1}{m}Q(m)  - \frac{\lambda n}{\mu} > \epsilon \right\}}
=
\sum_{m=1}^\infty
\PP{\frac{\edit{\lambda_0 n_0}}{\mu} - \frac{1}{m}Q^{\edit{\infty}}(m) > \epsilon } 
+ 
\sum_{m=1}^\infty
\PP{ \frac{1}{m}Q^{\edit{\infty}}(m)  - \frac{\edit{\lambda_0 n_0}}{\mu} > \epsilon }
<
\infty
,
\end{align*}
and thus by the Borel-Cantelli lemma, we have reached the stated result for scaling by $m$.

Turning now to the limit of the queue when centered by its mean and normalized by %$m^{(1+\nu)\slash2} $
$\sqrt{m^{1+\nu} + m\slash n}$, we will approach this through the moment generating function provided by the sum of scaled Poisson's representation. This leads us to the following closed-form expression for the MGF:
%\begin{align*}
%\E{e^{\theta m^{-\frac{1+\nu}{2}} \left(Q(m) - \frac{\lambda n m}{\mu}\right)}}
%&
%=
%e^{-\frac{\theta \lambda n}{\mu} m^{\frac{1-\nu}{2}}}
%\E{e^{\theta m^{-\frac{1+\nu}{2}} \sum_{j=1}^{ nm^\nu } j Y_j }}
%\\
%&
%=
%e^{-\frac{\theta \lambda n}{\mu} m^{\frac{1-\nu}{2}}}
%\prod_{j=1}^{ nm^\nu }
%\E{e^{\theta j m^{-\frac{1+\nu}{2}}  Y_j  }}
%\\
%&
%=
%e^{-\frac{\theta \lambda n}{\mu} m^{\frac{1-\nu}{2}}
%+
%\sum_{j=1}^{ nm^\nu } \frac{\lambda m^{1-\nu}}{j\mu} \left(e^{\theta  j m^{-\frac{1+\nu}{2}}} - 1 \right)}
%.
%\end{align*}
\begin{align*}
\E{e^{\theta \left(m^{1+\nu} + \frac{m}{\edit{n_0}} \right)^{-{1}\slash{2}} \left(Q^{\edit{\infty}}(m) - \frac{\edit{\lambda_0 n_0} m}{\mu}\right)}}
&
=
e^{-\theta \left(m^{1+\nu} + \frac{m}{\edit{n_0}} \right)^{-{1}\slash{2}} \frac{\edit{\lambda_0 n_0} m}{\mu}}
\E{e^{\theta \left(m^{1+\nu} + \frac{m}{\edit{n_0}} \right)^{-{1}\slash{2}}\sum_{j=1}^{\edit{n_0}m^\nu} j Y_j }}
\\
&
=
e^{-\theta \left(m^{1+\nu} + \frac{m}{\edit{n_0}} \right)^{-{1}\slash{2}} \frac{\edit{\lambda_0 n_0} m}{\mu}}
\prod_{j=1}^{\edit{n_0}m^\nu}
\E{e^{\theta \left(m^{1+\nu} + \frac{m}{\edit{n_0}} \right)^{-{1}\slash{2}} j Y_j }}
\\
&
=
e^{-\theta \left(m^{1+\nu} + \frac{m}{\edit{n_0}} \right)^{-{1}\slash{2}} \frac{\edit{\lambda_0 n_0} m}{\mu}
+
\sum_{j=1}^{ \edit{n_0}m^\nu } 
\frac{\edit{\lambda_0} m^{1-\nu}}{j\mu} \left(e^{\theta \left(m^{1+\nu} + \frac{m}{\edit{n_0}} \right)^{-{1}\slash{2}} j} - 1 \right)}
.
\end{align*}
To tackle the limit as $m \to \infty$, let us focus on the terms in the exponent. By expanding the nested exponential function of $\theta$ according to a Taylor series, we can push the sum over $j$ through and see that the first order terms will cancel with the mean centering term. Furthermore, the terms order three and above can all be seen to be no more than $m^{-3(1-\nu)\slash 2}$ up to constants. Thus, we have
%\begin{align*}
%&
%-
%\frac{\theta \lambda n}{\mu} m^{\frac{1-\nu}{2}}
%+
%\sum_{j=1}^{ nm^\nu } \frac{\lambda m^{1-\nu}}{j\mu} \left(e^{\theta  j m^{-\frac{1+\nu}{2}}} - 1 \right)
%=
%-
%\frac{\theta \lambda n}{\mu} m^{\frac{1-\nu}{2}}
%+
%\sum_{j=1}^{ nm^\nu } \frac{\lambda m^{1-\nu}}{j\mu} \sum_{k=1}^\infty \frac{1}{k!} \left(\theta  j m^{-\frac{1+\nu}{2}}\right)^k
%\\
%&
%\qquad
%=
%-
%\frac{\theta \lambda n}{\mu} m^{\frac{1-\nu}{2}}
%+
%\sum_{j=1}^{ nm^\nu } \frac{\lambda m^{1-\nu}}{\mu} 
%\left(
%\theta  m^{-\frac{1+\nu}{2}}
%+
%\frac{\theta^2 j}{2} m^{-(1+\nu)}
%+
%O\left(j^2 m^{-\frac{3}{2}(1+\nu)}\right)
%\right)
%.
%\end{align*}
\begin{align*}
&
\quad
-
\theta \left(m^{1+\nu} + \frac{m}{\edit{n_0}} \right)^{-{1}\slash{2}} \frac{\edit{\lambda_0 n_0} m}{\mu}
+
\sum_{j=1}^{ \edit{n_0}m^\nu } 
\frac{\edit{\lambda_0} m^{1-\nu}}{j\mu} 
\left(
\theta \left(m^{1+\nu} + \frac{m}{\edit{n_0}} \right)^{-{1}\slash{2}} j
+
\frac{\theta^2}{2} \left(m^{1+\nu} + \frac{m}{\edit{n_0}} \right)^{-1} j^2
\right.
\\
&
\left.
\qquad
+
\frac{\theta^3}{6} \left(m^{1+\nu} + \frac{m}{\edit{n_0}} \right)^{-{3}\slash{2}} j^3
+
\dots
\right)
\\
&
%\quad
=
-
\theta \left(m^{1+\nu} + \frac{m}{\edit{n_0}} \right)^{-{1}\slash{2}} \frac{\edit{\lambda_0 n_0} m}{\mu}
+
\frac{\lambda m^{1-\nu}}{\mu} 
\left(
\theta \left(m^{1+\nu} + \frac{m}{\edit{n_0}} \right)^{-{1}\slash{2}}  \edit{n_0}m^\nu
+
\frac{\theta^2}{2} \left(m^{1+\nu} + \frac{m}{\edit{n_0}} \right)^{-1} \frac{\edit{n_0}m^\nu (\edit{n_0}m^\nu + 1)}{2}
\right.
\\
&
\qquad
\left.
+
\frac{\theta^3}{6} \left(m^{1+\nu} + \frac{m}{\edit{n_0}} \right)^{-{3}\slash{2}} \frac{\edit{n_0}m^\nu (\edit{n_0}m^\nu +1)(2\edit{n_0}m^\nu + 1)}{6}
+
\dots
\right)
\\
&
%\quad
=
\frac{\edit{\lambda_0} m^{1-\nu}}{\mu} 
\left(
\frac{\edit{n_0^2}\theta^2}{4} \left(m^{1+\nu} + \frac{m}{\edit{n_0}} \right)^{-1} \left(m^{2\nu} + \frac{m^\nu}{\edit{n_0}}\right)
+
O\left(m^{-\frac{3(1-\nu)}{2}}\right) 
\right)
.
\end{align*}
%With these first and second order terms in focus, we can combine like terms and evaluate the sums over $j$ as follows
%\begin{align*}
%&
%-
%\frac{\theta \lambda n}{\mu} m^{\frac{1-\nu}{2}}
%+
%\sum_{j=1}^{ nm^\nu } \frac{\lambda m^{1-\nu}}{\mu} 
%\left(
%\theta  m^{-\frac{1+\nu}{2}}
%+
%\frac{\theta^2 j}{2} m^{-(1+\nu)}
%+
%O\left(j^2 m^{-\frac{3}{2}(1+\nu)}\right)
%\right)
%=
%\frac{\lambda \theta^2 }{4\mu} 
%\left(
%n^2
%+
%nm^{-\nu}
%\right)
%+
%O\left(  m^{-\frac{1-\nu}{2}}\right)
%.
%\end{align*}
Multiplying the leading $m^{1-\nu}$ through, we can further simplify to
\begin{align*}
\frac{\edit{\lambda_0} m^{1-\nu}}{\mu} 
\left(
\frac{\edit{n_0^2}\theta^2}{4} \left(m^{1+\nu} + \frac{m}{\edit{n_0}} \right)^{-1} \left(m^{2\nu} + \frac{m^\nu}{\edit{n_0}}\right)
+
O\left(m^{-\frac{3(1-\nu)}{2}}\right) 
\right)
=
\frac{\edit{\lambda_0 n_0^2}\theta^2}{4\mu} 
+
O\left(m^{-\frac{1-\nu}{2}}\right) 
.
\end{align*}
In this form, we can quickly see that if $\nu < 1$ as $m \to \infty$ the terms order three and above will vanish, leaving simply
\begin{align*}
\frac{\edit{\lambda_0 n_0^2}\theta^2}{4\mu} 
+
O\left(m^{-\frac{1-\nu}{2}}\right) 
\to
\frac{\edit{\lambda_0 n_0^2} \theta^2}{4\mu}
,
\end{align*}
thus implying that
$$
\E{e^{\theta \left(m^{1+\nu} + \frac{m}{\edit{n_0}} \right)^{-{1}\slash{2}} \left(Q(m) - \frac{\edit{\lambda_0 n_0} m}{\mu}\right)}}
\longrightarrow
e^{\frac{\edit{\lambda_0 n_0^2} \theta^2}{4\mu}}
$$
as $m \to \infty$ if $\nu \in [0,1)$.
\hfill\Halmos\\
\endproof

\subsection{Proof of Theorem~\ref{bothLimitMulti}}\label{bothLimitMultiProof}

\proof{Proof.}
As motivated by the proof of Proposition 1 in~\citet{halfin1981heavy}, let us first translate the exceedance probability into terms defined relative to the infinite server queue rather than the multi-server queue. For notational simplicity, let us temporarily repress the precise dependence on $m$ and briefly consider the arrival rate, batch size, and staffing as simply $\lambda$, $n$, and $c$, respectively. Looking at Lemma~\ref{matrixLemma} and comparing the transition dynamics for $Q^C$ and $Q^{\edit{\infty}}$, one can see that these two Markov chains should have the set of same balance equations up until state $c$. That is, for $\pi_k = \PP{Q^C = k}$ and $\tilde{\pi}_k = \PP{Q^{\edit{\infty}} = k}$ and all $k \leq c$, both systems
\begin{align*}
\pi_k 
&= 
\begin{cases}
\mathbf{v}_1^\mathsf{T} 
\prod_{j = n+1}^c
\mathbf{C}\left(\frac{\lambda}{j\mu}\right)
\begin{bmatrix}
\frac{\Gamma\left(\frac{\lambda}{\mu} + n\right)}{\Gamma\left(\frac{\lambda}{\mu}\right) \Gamma\left(n+1\right)} \\
\vdots \\
\frac{\lambda}{\mu}
\end{bmatrix}
\pi_0
,
& 
\text{for } n \leq k \leq c ,
\\
\\
\frac{\Gamma\left(\frac{\lambda}{\mu} + k\right)}{\Gamma\left(\frac{\lambda}{\mu}\right) \Gamma\left(k+1\right)} 
\pi_0
,
&
\text{for } 0 \leq k \leq n - 1
,
\end{cases}
\end{align*}
and
\begin{align*}
\tilde \pi_k 
&= 
\begin{cases}
\mathbf{v}_1^\mathsf{T} 
\prod_{j = n+1}^c
\mathbf{C}\left(\frac{\lambda}{j\mu}\right)
\begin{bmatrix}
\frac{\Gamma\left(\frac{\lambda}{\mu} + n\right)}{\Gamma\left(\frac{\lambda}{\mu}\right) \Gamma\left(n+1\right)} \\
\vdots \\
\frac{\lambda}{\mu}
\end{bmatrix}
\tilde \pi_0
,
& 
\text{for } n \leq k \leq c ,
\\
\\
\frac{\Gamma\left(\frac{\lambda}{\mu} + k\right)}{\Gamma\left(\frac{\lambda}{\mu}\right) \Gamma\left(k+1\right)} 
\tilde \pi_0
,
&
\text{for } 0 \leq k \leq n - 1
,
\end{cases}
\end{align*}
are true. So, by multiplying $\tilde \pi_0 \, {\Gamma\left({\lambda}\slash{\mu}+n+1\right)}\slash \left({\Gamma\left({\lambda}\slash{\mu}+1\right)\Gamma\left(n+1\right)}\right)$ in both the numerator and denominator of the exceedance probability expression from Lemma~\ref{matrixLemma}, we can see that
\begin{align*}
\PP{Q^C \geq c}
&=
\frac{
\tilde \pi_{c}
+
\frac{\lambda}{c\mu - \lambda n}
\sum_{i=1}^n (n+1-i) \tilde \pi_{c+1-i}
}{
\sum_{k=0}^{c} \tilde \pi_k
+
\frac{\lambda}{c\mu - \lambda n}
\sum_{i=1}^n(n+1-i) \tilde \pi_{c+1-i}
}
,
\end{align*}
since $[ n \,\, n-1 \,\, \dots \,\, 1] = \sum_{i=1}^n (n+1-i) \mathbf{v}_i^\mathsf{T}$. Re-expressing into events defined by $Q^{\edit{\infty}}$, this is
\begin{align*}
\frac{
\tilde \pi_{c}
+
\frac{\lambda}{c\mu - \lambda n}
\sum_{i=1}^n (n+1-i) \tilde \pi_{c+1-i}
}{
\sum_{k=0}^{c} \tilde \pi_k
+
\frac{\lambda}{c\mu - \lambda n}
\sum_{i=1}^n(n+1-i) \tilde \pi_{c+1-i}
}
&
%=
%\frac{
%\PP{Q^{\edit{\infty}} = c}
%+
%\frac{\lambda}{c\mu - \lambda n}
%\sum_{i=1}^n
%\left(\PP{Q^{\edit{\infty}} \geq c -i} - \PP{Q^{\edit{\infty}} \geq c}\right)
%}{
%\PP{Q^{\edit{\infty}} \leq c}
%+
%\frac{\lambda}{c\mu - \lambda n}
%\sum_{i=1}^n
%\left(\PP{Q^{\edit{\infty}} \geq c -i} - \PP{Q^{\edit{\infty}} \geq c}\right)
%}
%\\
%&
%\quad
=
\frac{
\PP{Q^{\edit{\infty}} = c}
+
\frac{\lambda}{c\mu - \lambda n}
\E{\left( Q^{\edit{\infty}} - c + n \right)^+ \mathbf{1}\{Q^{\edit{\infty}} \leq  c\} }
}{
\PP{Q^{\edit{\infty}} \leq c}
+
\frac{\lambda}{c\mu - \lambda n}
\E{\left( Q^{\edit{\infty}} - c + n\right)^+ \mathbf{1}\{Q^{\edit{\infty}} \leq  c\} }
}
,
\end{align*}
where $x^+ = \max\{x , 0\}$ for all $x \in \mathbb{R}$.
Returning to the scaling regime at hand with arrival rate $\edit{\lambda_0} m^{1-\nu}$, batch size $\edit{n_0}m^\nu$, and staffing level $\edit{\lambda_0 n_0} m \slash \mu + \delta m^{(1+\nu)\slash 2}$, let us first note that we have $\lim_{m \to \infty}\PP{Q^{\edit{\infty}}(m) = \edit{\lambda_0 n_0} m \slash \mu + \delta m^{(1+\nu)\slash 2}} = 0$ immediately from Theorem~\ref{bothLimitInf}. Hence, we can now attack the two remaining distinct objects in this expression separately. Knocking out the simpler piece first, we can quickly observe that Theorem~\ref{bothLimitInf} implies that the first term in the denominator converges to
\begin{align*}
\PP{Q^{\edit{\infty}}(m) \leq \frac{\edit{\lambda_0 n_0} m }{ \mu } + \delta m^{\frac{1+\nu} 2}}
&
=
\PP{\left(Q^{\edit{\infty}}(m) - \frac{\edit{\lambda_0 n_0} m }{ \mu }\right) m^{-\frac{1+\nu} 2} \leq  \delta }
\longrightarrow
\Phi\left(
\frac{\delta}{\sqrt{\frac{\edit{\lambda_0 n_0^2}}{2\mu}}}
\right)
,
\end{align*}
as $m \to \infty$. Now, turning to the remaining piece of the exceedance probability expression, 
\begin{align*}
&
\frac{\edit{\lambda_0} m^{1-\nu}}{\mu \delta m^{\frac{1+\nu}{2}}}
\E{\left( Q^{\edit{\infty}}(m) - \frac{\edit{\lambda_0 n_0} m}{\mu} - \delta m^{\frac{1+\nu}{2}} + \edit{n_0}m^\nu \right)^+ \mathbf{1}\left\{Q^{\edit{\infty}}(m) \leq  \frac{\edit{\lambda_0 n_0} m}{\mu} + \delta m^{\frac{1+\nu}{2}}\right\} }
\\
&
\qquad
=
\frac{\edit{\lambda_0} m^{1-\nu}}{\mu \delta }
\E{\left( 
\left(Q^{\edit{\infty}}(m) - \frac{\edit{\lambda_0 n_0} m}{\mu}\right)m^{-\frac{1+\nu}{2}} 
- 
\delta
+ 
\edit{n_0}m^{-\frac{1-\nu}{2}} 
%+ 
%m^{-\frac{1+\nu}{2}}
\right)^+ 
\mathbf{1}\left\{
\left(Q^{\edit{\infty}}(m) - \frac{\edit{\lambda_0 n_0} m}{\mu}\right)m^{-\frac{1+\nu}{2}} 
%<
\leq
\delta 
\right\} }
.
\end{align*}
%By multiplying and dividing by $ nm^\nu $, we can then this summation as a right-valued Riemman sum approximating a corresponding integral. That is, because the probability of the event is a monotone function in $x$, there exists some constant $0 < \mathcal{C}_1 \leq 1$ such that
%\begin{align*}
%&
%\frac{\lambda m^{\frac{1-3\nu}{2}} }{\mu \delta }
%\sum_{i=1}^{ nm^\nu }
%\PP{
%\delta  - i m^{-\frac{1+\nu} 2}
%\leq
%\left(
%Q(m)
%-
%\frac{\lambda n m }{ \mu } 
%\right)
%m^{-\frac{1+\nu} 2}
%< 
%\delta 
%}
%\\
%&
%\quad
%=
%\frac{\lambda n m^{\frac{1-\nu}{2}}  }{\mu \delta }
%\int_0^1
%\PP{
%\delta  - x  nm^{-\frac{1-\nu} 2}
%\leq
%\left(
%Q(m)
%-
%\frac{\lambda n m }{ \mu } 
%\right)
%m^{-\frac{1+\nu} 2}
%< 
%\delta 
%}
%\mathrm{d}x
%\\
%&
%\qquad
%+
%\mathcal{C}_1 
%m^{\frac{1-3\nu}{2}}
%\PP{
%\delta  - nm^{-\frac{1-\nu} 2}
%\leq
%\left(
%Q(m)
%-
%\frac{\lambda n m }{ \mu } 
%\right)
%m^{-\frac{1+\nu} 2}
%< 
%\delta 
%}
%\end{align*}
%As $m$ grows large, we can approximate this centered and scaled infinite server queue length by its Gaussian limit. Moreover, through the Berry-Esseen bound in Lemma~\ref{berryLemma}, we know that the error from this approximation no more than order $m^{-{(\nu \wedge 1 - \nu)}\slash 2}$ at any point of the CDF. Taking $\mathcal{C}_2 > 0$ as an appropriate constant to express this bound and $X \sim \mathsf{Norm}(0, \lambda n^2 \slash(2\mu))$, we then have
For simplicity, let $\bar Q_\delta(m) = \left(Q^{\edit{\infty}}(m) - \frac{\edit{\lambda_0 n_0} m}{\mu}\right)m^{-\frac{1+\nu}{2}} - \delta$. % and let $b_m = nm^{-\frac{1-\nu}{2}}$. 
Because $\PP{b \leq \bar{Q}_\delta (m) \leq a} \to \PP{b \leq X - \delta \leq a}$ as $m \to \infty$ for all $b \leq a$ where $X \sim \mathsf{Norm}(0, \edit{\lambda_0 n_0^2} \slash 2 \mu)$, we can observe that 
\begin{align*}
&
\Bigg|
\E{\left( 
\left(Q^{\edit{\infty}}(m) - \frac{\edit{\lambda_0 n_0} m}{\mu}\right)m^{-\frac{1+\nu}{2}} 
- 
\delta
+ 
\edit{n_0}m^{-\frac{1-\nu}{2}} 
%+ 
%m^{-\frac{1+\nu}{2}}
\right)^+ 
\mathbf{1}\left\{
\left(Q^{\edit{\infty}}(m) - \frac{\edit{\lambda_0 n_0} m}{\mu}\right)m^{-\frac{1+\nu}{2}} 
%<  
\leq
\delta 
\right\} }
\\
&
\qquad
-
\E{\left( 
X
- 
\delta
+ 
\edit{n_0}m^{-\frac{1-\nu}{2}} 
%+ 
%m^{-\frac{1+\nu}{2}}
\right)^+ 
\mathbf{1}\left\{
X
%<  
\leq
\delta 
\right\} }
\Bigg|
\\
=
&
\left|
\E{
\bar{Q}_\delta(m)
+
\edit{n_0}m^{-\frac{1-\nu}{2}}
%b_m
\,
\big|
\,
-
\edit{n_0}m^{-\frac{1-\nu}{2}}
%b_m
\leq
\bar{Q}_\delta(m)
\leq
%<
0
}
\PP{- 
\edit{n_0}m^{-\frac{1-\nu}{2}}
%b_m
\leq
\bar{Q}_\delta(m)
\leq
%<  
0
}
\right.
\\
&
\qquad
\left.
-
\E{
X
- 
\delta
+ 
\edit{n_0}m^{-\frac{1-\nu}{2}}
%b_m
\,
\big|
\,
-
\edit{n_0}m^{-\frac{1-\nu}{2}}
%b_m 
\leq
X
-
\delta
\leq
%<  
0
}
\PP{
-
\edit{n_0}m^{-\frac{1-\nu}{2}}
%b_m 
\leq
X
-
\delta
\leq
%<  
0
}
\right|
\\
\leq
\,\,
&
\bar{\mathcal{C}}
\edit{n_0}m^{-\frac{1-\nu}{2}}
%(nm^{-\frac{1-\nu}{2}} + m^{-\frac{1+\nu}{2}})
\left|
\PP{- 
\edit{n_0}m^{-\frac{1-\nu}{2}}
%b_m
\leq
\bar{Q}_\delta(m)
\leq
%<  
0
}
-
\PP{
-
\edit{n_0}m^{-\frac{1-\nu}{2}}
%b_m 
\leq
X
-
\delta
\leq
%<  
0
}
\right|
\\
%\leq
%\,\,
%&
%\mathcal{C}_1 (nm^{-\frac{1-\nu}{2}} + m^{-\frac{1+\nu}{2}})
%m^{-\frac{1-\nu}{2} - \epsilon}
%\\
\leq
\,\,
&
\bar{\mathcal{C}}
m^{-1+\nu - \epsilon}
\end{align*}
for some $\bar{\mathcal{C}} \geq 0$ and $\epsilon > 0$. Hence, we can instead consider the limit of
\begin{align*}
\frac{\lambda m^{1-\nu}}{\mu \delta }
\E{\left( 
X
- 
\delta
+ 
nm^{-\frac{1-\nu}{2}} 
%+ 
%m^{-\frac{1+\nu}{2}}
\right)^+ 
\mathbf{1}\left\{
X
\leq
%<  
\delta 
\right\} }
,
\end{align*}
as $m \to \infty$. 
Translating the expectation to an integral over the Gaussian density and converting this to a double integral, we see that
\begin{align*}
\frac{\edit{\lambda_0} m^{1-\nu}}{\mu \delta }
\E{\left( 
X
- 
\delta
+ 
\edit{n_0}m^{-\frac{1-\nu}{2}} 
%+ 
%m^{-\frac{1+\nu}{2}}
\right)^+ 
\mathbf{1}\left\{
X
\leq
%<  
\delta 
\right\} }
&
=
\frac{\edit{\lambda_0} m^{1-\nu}}{\mu \delta }
\int_{\delta - \edit{n_0}m^{-({1-\nu})/{2}} %b_m
}^\delta
\left( 
x
- 
\delta
+
\edit{n_0}m^{-\frac{1-\nu}{2}}
%b_m
\right)
\frac{
e^{-\frac{x^2}{\edit{\lambda_0 n_0^2} \slash \mu}}
}
{
\edit{n_0}\sqrt{\frac{\lambda \pi}{\mu}}
}
\mathrm{d}x
\\
&
=
\frac{\edit{\lambda_0} m^{1-\nu}}{\mu \delta }
\int_{\delta -\edit{n_0}m^{-({1-\nu})/{2}} %b_m
}^\delta
\int_0^{
x
- 
\delta
+
\edit{n_0}m^{-({1-\nu})/{2}}
%b_m
}
\frac{
e^{-\frac{x^2}{\edit{\lambda_0 n_0^2} \slash \mu}}
}
{
\edit{n_0}\sqrt{\frac{\edit{\lambda_0} \pi}{\mu}}
}
\mathrm{d}y \,
\mathrm{d}x
\\
&
=
\frac{\edit{\lambda_0} m^{1-\nu}}{\mu \delta }
\int_0^{
\edit{n_0}m^{-({1-\nu})/{2}}
%b_m
}
\int_{y+\delta - \edit{n_0}m^{-({1-\nu})/{2}} %b_m
}^\delta
\frac{
e^{-\frac{x^2}{\edit{\lambda n_0^2} \slash \mu}}
}
{
\edit{n_0}\sqrt{\frac{\edit{\lambda_0} \pi}{\mu}}
}
\mathrm{d}x \, \mathrm{d}y
.
\end{align*}
Now, by introducing a change of variables so that $y = (1-z)\edit{n_0}m^{-({1-\nu})/{2}}$, we can re-write this integral as
\begin{align*}
\frac{\edit{\lambda_0} m^{1-\nu}}{\mu \delta }
\int_0^{
\edit{n_0}m^{-({1-\nu})/{2}}
%b_m
}
\int_{y+\delta - \edit{n_0}m^{-({1-\nu})/{2}} %b_m
}^\delta
\frac{
e^{-\frac{x^2}{\edit{\lambda_0 n_0^2} \slash \mu}}
}
{
\edit{n_0}\sqrt{\frac{\edit{\lambda_0} \pi}{\mu}}
}
\mathrm{d}x \, \mathrm{d}y
&
=
\frac{\edit{\lambda_0} \edit{n_0}m^{\frac{1-\nu}{2}} }{\mu \delta }
\int_0^{
1
}
\int_{\delta - z \edit{n_0}m^{-({1-\nu})/{2}} %b_m
}^\delta
\frac{
e^{-\frac{x^2}{\edit{\lambda_0 n_0^2} \slash \mu}}
}
{
\edit{n_0}\sqrt{\frac{\edit{\lambda_0} \pi}{\mu}}
}
\mathrm{d}x \, \mathrm{d}z
%\\
%&
%=
%\frac{\lambda (nm^{\frac{1-\nu}{2}} + m^{\frac{1-3\nu}{2}} )}{\mu \delta }
%\int_0^{
%1
%}
%\int_{\delta - z b_m
%}^\delta
%\frac{
%e^{-\frac{x^2}{\lambda n^2 \slash \mu}}
%}
%{
%n\sqrt{\frac{\lambda \pi}{\mu}}
%}
%\mathrm{d}x \, \mathrm{d}z
.
\end{align*}
To work towards identifying the limit, let us take a Taylor expansion of the exponential inside the double integral and evaluate the inner integral. This yields
\begin{align*}
&
\frac{\edit{\lambda_0 n_0}m^{\frac{1-\nu}{2}} }{\mu \delta }
\int_0^{
1
}
\int_{\delta - z \edit{n_0}m^{-({1-\nu})/{2}}
}^\delta
\frac{
e^{-\frac{x^2}{\edit{\lambda_0} \edit{n_0^2} \slash \mu}}
}
{
\edit{n_0}\sqrt{\frac{\edit{\lambda_0} \pi}{\mu}}
}
\mathrm{d}x \, \mathrm{d}z
\\
&
\quad
=
\frac{ m^{\frac{1-\nu}{2}}  }{ \delta }
\sqrt{\frac{\edit{\lambda_0}}{\mu \pi}}
\int_0^{
1
}
\int_{\delta - z \edit{n_0}m^{-({1-\nu})/{2}}
}^\delta
\left(
1
+
\sum_{k=1}^{\infty}
\frac{x^{2k}}{k!}
\left(\frac{-1}{\edit{\lambda_0} \edit{n_0^2} \slash \mu}\right)^k
\right)
\mathrm{d}x \, \mathrm{d}z
\\
&
\quad
=
\frac{ m^{\frac{1-\nu}{2}}  }{ \delta }
\sqrt{\frac{\edit{\lambda_0}}{\mu \pi}}
\int_0^1
\left(
z \edit{n_0}m^{-({1-\nu})/{2}}
+
\sum_{k=1}^{\infty}
\frac{1}{k!}
\left(\frac{-1}{\edit{\lambda_0} \edit{n_0^2} \slash \mu}\right)^k
\frac{1}{2k+1}
\left(
\delta^{2k+1}
-
(\delta - z \edit{n_0}m^{-({1-\nu})/{2}})^{2k+1}
\right)
\right)
\mathrm{d}z
.
\end{align*}
Let us now apply a binomial theorem expansion to $(\delta - z \edit{n_0}m^{-({1-\nu})/{2}})^{2k+1}$. Because the zero-order $z\edit{n_0}m^{-({1-\nu})/{2}}$ term will cancel with the $\delta^{2k+1}$, we can see that we are left with the first order $z\edit{n_0}m^{-({1-\nu})/{2}}$ term and orders two and above. The first order terms can then be collected back into a Taylor expansion of an exponential function, leaving a remainder that features $\edit{n_0}m^{-({1-\nu})/{2}}$ of at least quadratic order.
\begin{align*}
&
\frac{ m^{\frac{1-\nu}{2}}  }{ \delta }
\sqrt{\frac{\edit{\lambda_0}}{\mu \pi}}
\int_0^1
\left(
z \edit{n_0}m^{-({1-\nu})/{2}}
+
\sum_{k=1}^{\infty}
\frac{1}{k!}
\left(\frac{-1}{\edit{\lambda_0} \edit{n_0^2} \slash \mu}\right)^k
\frac{1}{2k+1}
\left(
\delta^{2k+1}
-
(\delta - z \edit{n_0}m^{-({1-\nu})/{2}})^{2k+1}
\right)
\right)
\mathrm{d}z
\\
&
\quad
=
%\frac{ nm^{\frac{1-\nu}{2}} + m^{\frac{1-3\nu}{2}} }{n \delta }
\frac{ m^{\frac{1-\nu}{2}}  }{ \delta }
\sqrt{\frac{\edit{\lambda_0}}{\mu \pi}}
\int_0^1
\left(
z \edit{n_0}m^{-({1-\nu})/{2}}
-
\sum_{k=1}^{\infty}
\frac{1}{k!}
\left(\frac{-1}{\edit{\lambda_0} \edit{n_0^2} \slash \mu}\right)^k
\frac{1}{2k+1}
\left(
\sum_{i=1}^{2k+1}
{2k+1 \choose i}
(-z \edit{n_0}m^{-({1-\nu})/{2}})^{i} \delta^{2k+1 - i}
\right)
\right)
\mathrm{d}z
\\
&
\quad
=
%\frac{ nm^{\frac{1-\nu}{2}} + m^{\frac{1-3\nu}{2}} }{n \delta }
\frac{ m^{\frac{1-\nu}{2}}  }{ \delta }
\sqrt{\frac{\edit{\lambda_0}}{\mu \pi}}
\int_0^1
\left(
z \edit{n_0}m^{-({1-\nu})/{2}}
-
\sum_{k=1}^{\infty}
\frac{1}{k!}
\left(\frac{-1}{\edit{\lambda_0} \edit{n_0^2} \slash \mu}\right)^k
\frac{1}{2k+1}
\left(
-
(2k+1) z \edit{n_0}m^{-({1-\nu})/{2}}\delta^{2k}
+
O\left(m^{-({1-\nu})}\right)
\right)
\right)
\mathrm{d}z
\\
&
\quad
=
%\frac{ nm^{\frac{1-\nu}{2}} + m^{\frac{1-3\nu}{2}} }{n \delta }
\frac{ m^{\frac{1-\nu}{2}}  }{ \delta }
\sqrt{\frac{\edit{\lambda_0}}{\mu \pi}}
\left(
\frac{\edit{n_0}m^{-({1-\nu})/{2}}}{2} e^{\frac{-\delta^2}{\edit{\lambda_0} \edit{n_0^2} \slash \mu}}
+
O\left(m^{-({1-\nu})}\right)
\right)
.
\end{align*}
%Recalling that $b_m = nm^{-\frac{1-\nu}{2}} + m^{-\frac{1+\nu}{2}}$, we can see that 
As $m \to  \infty$, we find
$$
%\frac{ nm^{\frac{1-\nu}{2}} + m^{\frac{1-3\nu}{2}} }{n \delta }
\frac{ m^{\frac{1-\nu}{2}}  }{ \delta }
\sqrt{\frac{\edit{\lambda_0}}{\mu \pi}}
\left(
\frac{\edit{n_0}m^{-({1-\nu})/{2}}}{2} e^{\frac{-\delta^2}{\edit{\lambda_0} \edit{n_0^2} \slash \mu}}
+
O\left(m^{-({1-\nu})}\right)
\right)
\longrightarrow
\frac{ \edit{n_0} }{2 \delta }
\sqrt{\frac{\edit{\lambda_0}}{\mu \pi}}
 e^{\frac{-\delta^2}{\edit{\lambda_0} \edit{n_0^2} \slash \mu}}
 ,
$$
implying that
\begin{align*}
\frac{\edit{\lambda_0} m^{1-\nu}}{\mu \delta m^{\frac{1+\nu}{2}}}
\E{\left( Q^{\edit{\infty}}(m) - \frac{\edit{\lambda_0} \edit{n_0} m}{\mu} - \delta m^{\frac{1+\nu}{2}} + \edit{n_0}m^\nu \right)^+ \mathbf{1}\left\{Q^{\edit{\infty}}(m) \leq  \frac{\edit{\lambda_0} \edit{n_0} m}{\mu} + \delta m^{\frac{1+\nu}{2}}\right\} }
&
\longrightarrow
\frac{ \edit{n_0} }{2 \delta }
\sqrt{\frac{\edit{\lambda_0}}{\mu \pi}}
 e^{\frac{-\delta^2}{\edit{\lambda_0} \edit{n_0^2} \slash \mu}}
 ,
\end{align*}
and, moreover, yielding the stated result for the limit of the exceedance probability.
\hfill\Halmos\\
\endproof

\section{Proofs of Preliminary, Auxiliary, and Supporting Results}\label{support}

\subsection{\edit{Simple Demonstration of the Inherent Staffing Differences under Batches}}\label{batchDemo}

%\tr{update}

 What is a good place to start \edit{for a first pass at staffing a batch arrival queue}? One tempting approach could be to recognize the {effective} customer arrival rate as the product of the batch arrival rate and the batch size, and then plug this effective rate into classical individual arrival staffing formulas. On the other hand, another tempting approach could be to recognize that a batch of size $n$ could be split to $n$ separate queueing systems. Hence, we could find the staffing level for a queue with individual arrival rate given by the true batch arrival rate and then calculate the batch staffing level by multiplying this individual staffing level by the batch size. In Figure~\ref{linFig}, we can see that both of these approaches miss the mark, with the former leading to severe under-staffing, and the latter, over-staffing.

\begin{figure}[htb]
\centering
\includegraphics[width=\textwidth]{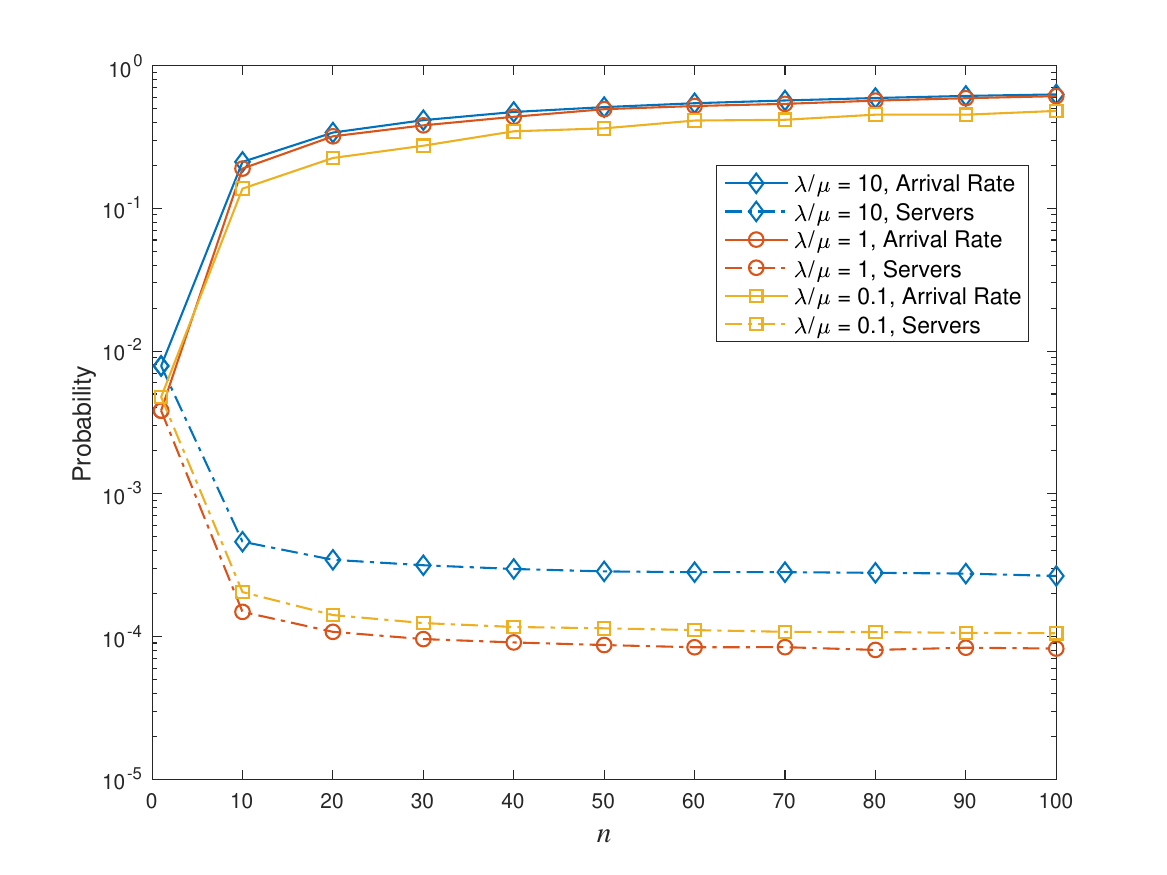}
\caption{Steady-state probability of the queue length exceeding the number of servers in the $\boldsymbol{M^n/M/c}$ system using $\boldsymbol{M/M/c}$ staffing levels. For a target probability of $\boldsymbol{\epsilon = 0.01}$, the ``Arrival Rate'' staffing levels are calculated with arrival rate $\boldsymbol{\lambda n}$ and service rate $\boldsymbol{\mu}$, while the ``Servers'' staffing levels are calculated by multiplying the $\boldsymbol{\lambda}$ and $\boldsymbol{\mu}$ level by $\boldsymbol{n}$.} \label{linFig}
\end{figure}

In hindsight, it's not hard to find counter-arguments for each of these approaches. For the first method, while it is true that the arrival rate is the average rate that customers enter the service system, that does not address the fact that many of these customers enter simultaneously, leaving many of the effective inter-arrival times equal to 0. This will lead to under-staffing because it under-estimates the variability of the customer arrival process. Then, in the second method, it is true that $n$ customers arrive at once, but this neglects the potential pooling benefits from having a centralized queue that receives all customers rather than each server having their own separate, dedicated queues.\footnote{This single-centralized-queue versus multiple-separated-queues comparison hints at some of the important differences between the batch arrival queueing model and the fork-join queueing model. We discuss the differences between these models in more depth in Section~\ref{model}.} Hence, Figure~\ref{linFig} illustrates the importance and intrigue of this problem. As the batch size increases, the performance of each individual staffing heuristic weakens. However, in the second method, in which the staffing level is directly proportional to the batch size or, equivalently, to the effective arrival rate, the staffing exceedance probability does not appear to be converging to 0, which would be the case in individual arrival systems according to the QD regime. Intuiting that this means that large batch arrivals create distinct challenges not reproduced by large arrival rates, we will approach the staffing problem by studying the limit of the queueing model as the batch size increases. Through our analysis, we will be able to explain precisely why each of the methods in Figure~\ref{linFig} won't work and, more importantly, rigorously identify how to properly staff a batch arrival queue.

\subsection{What's a Burst to a Batch?}\label{burst}

At various points we have claimed that our staffing analysis will also apply to rapid bursts of arrivals, rather than just to batches of arrivals. This arrival structure may occur quite naturally in many applications.
%For example, \citet{mirzaeian2018queueing} studies a model of \emph{platoons} of autonomous vehicles, and the tight proximity of these vehicles may imply that when one vehicle needs assistance, the others will need assistance soon afterwards. 
Intuitively, if a service system receives a burst or cluster of arrivals  that is very nearly simultaneous, then this should be quite similar to a truly simultaneous batch of arrivals. Here, we will introduce a pair of stylized models and make brief arguments in favor of this. 

Consider two parsimonious Markov models of bursty arrivals; one exogenously driven and one endogenously driven, each ephemeral. For the exogenous case, suppose that at time 0 an external event occurs which spurs arrivals downstream according to a Poisson process with rate $\alpha_\mathsf{Exo} > 0$. However, at some independently and exponentially distributed expiration time, the external event concludes and the Poisson arrival stream ceases, ending the burst. Suppose that this expiration time has mean ${1}\slash{\beta_\mathsf{Exo}} > 0$. Let $\tau_\mathsf{Exo} \geq 0$ be the duration, or the time of the last arrival, of the exogenously driven burst, and let $Z_\mathsf{Exo} \geq 0$ be the size, or the number of arrivals that occur, of this burst.

Now, for the endogenously driven model, we will mirror the exogenous model in a self-exciting fashion.\footnote{This is modeled after the structure of the ephemerally self-exciting process (ESEP) studied in \citet{daw2021ephemerally}.} Let an initial arrival occur at time 0. Suppose this arrival generates downstream arrivals according to a Poisson process at rate $\alpha_\mathsf{Endo} > 0$, and suppose further that all future arrivals do as well, with all Poisson streams being mutually independent. Each of these streams cease after an independently and exponentially distributed time has past since the given stream's initial arrival, and we will suppose that the rate of these exponential random variables is $\beta_\mathsf{Endo} > \alpha_\mathsf{Endo}$. Like in the exogenous model setting, let $\tau_\mathsf{Endo}$ be the duration of the burst and $Z_\mathsf{Endo}$ its size.

\begin{proposition}\label{burstToBatch}
Suppose that $\alpha_\mathsf{Exo}, \beta_\mathsf{Exo} \to \infty$ and $\alpha_\mathsf{Endo}, \beta_\mathsf{Endo} \to \infty$ with ${\alpha_\mathsf{Exo}}\slash{\beta_\mathsf{Exo}}$ and  ${\alpha_\mathsf{Endo}}\slash{\beta_\mathsf{Endo}}$ fixed. Then, the bursts become instantaneous, i.e.
\begin{align}
\tau_\mathsf{Exo} \stackrel{\mathsf{p}}{\longrightarrow} 0
\qquad \text{ and } \qquad
\tau_\mathsf{Endo} \stackrel{\mathsf{p}}{\longrightarrow} 0
, 
\end{align}
while the distributions of $Z_\mathsf{Exo}$ and $Z_\mathsf{Endo}$ are unchanged.
\end{proposition}
\proof{Proof.}
To formalize the limit, let us introduce a scaling parameter $\eta$ such that all four parameters (the two $\alpha$'s and the two $\beta$'s) are multiplied by $\eta$.
We will start with the exogenous burst model. Here the duration $\tau_\mathsf{Exo}$ is upper bounded by the exponentially distributed expiration time by definition, and in this scaling regime this exponential distribution has rate parameter $\eta \beta_\mathsf{Exo}$. Letting $T_{\mathsf{Exo},\beta} \sim \mathsf{Exp}(\eta \beta_\mathsf{Exo})$, we can quickly see that
\begin{align*}
%\sum_{\eta = 1}^\infty
%\PP{\tau_{\mathsf{Exo},\eta} > \epsilon}
%\leq
%\sum_{\eta = 1}^\infty
%\PP{T_{\mathsf{Exo},\eta} > \epsilon}
%=
%\sum_{\eta = 1}^\infty
%e^{-\eta \beta_\mathsf{Exo} \epsilon}
%=
%\frac{e^{-\beta_\mathsf{Exo} \epsilon}}{1 - e^{-\beta_\mathsf{Exo} \epsilon}}
%<
%\infty
%,
\E{\tau_{\mathsf{Exo},\eta}} \leq \E{T_{\mathsf{Exo},\eta}} = \frac{1}{\eta\beta_\mathsf{Exo}} 
\longrightarrow
0
,
\end{align*}
as $\eta \to \infty$ for all $\beta_\mathsf{Exo} > 0$. Hence by Markov's inequality we can see that for any $\epsilon > 0$, 
\begin{align*}
\lim_{\eta\to\infty}\PP{\tau_{\mathsf{Exo},\eta} > \epsilon}
\leq
\lim_{\eta\to\infty} \frac{1}{\epsilon}\E{\tau_{\mathsf{Exo},\eta}}
=
0
,
\end{align*}
which shows the convergence in probability.\footnote{It is straightforward to make a stronger statement for the exogenous burst duration and achieve almost sure convergence to 0 through the Borel-Cantelli lemma, since $\PP{T_{\mathsf{Exo},\eta} > \epsilon} = e^{-\eta \beta_\mathsf{Exo} \epsilon}$.} For the distribution of the size of the burst, we can see that given the expiration time $T_{\mathsf{Exo},\eta}$, the total number of arrivals follows a Poisson distribution with mean $\eta \alpha_\mathsf{Exo} T_{\mathsf{Exo},\eta}$. This Poisson-exponential mixture is known to yield a geometric distribution, which can be easily observed through manipulating the MGF:
\begin{align*}
\E{e^{\theta Z_{\mathsf{Exo}}}}
=
\E{\E{e^{\theta Z_{\mathsf{Exo}}}\mid T_{\mathsf{Exo},\eta}}}
=
\E{ e^{\eta \alpha_\mathsf{Exo} T_{\mathsf{Exo},\eta} (e^{\theta}-1) }}
=
\frac{
\eta \beta_\mathsf{Exo}
}{
\eta \beta_\mathsf{Exo}
-
\eta \alpha_\mathsf{Exo} (e^{\theta}-1) 
}
=
\frac{
 \frac{\beta_\mathsf{Exo}}{\alpha_\mathsf{Exo} + \beta_\mathsf{Exo}}
}{
 \frac{\beta_\mathsf{Exo}}{\alpha_\mathsf{Exo} + \beta_\mathsf{Exo}}
-
 \frac{\alpha_\mathsf{Exo}}{\alpha_\mathsf{Exo} + \beta_\mathsf{Exo}} e^{\theta}
}
.
\end{align*}
Here we can observe that the probability parameter in the geometric distribution is $ {\beta_\mathsf{Exo}}\slash({\alpha_\mathsf{Exo} + \beta_\mathsf{Exo}})$, which does not depend on $\eta$. This completes the proof for the exogenous case, so we turn to the endogenous model.

Here we can invoke Propositions 3.3 and 3.5 of \citet{daw2021ephemerally} to note that the mean duration of the endogenously driven burst is 
\begin{align*}
\E{\tau_{\mathsf{Endo},\eta}}
=
\frac{1}{\eta \alpha_\mathsf{Endo}}
\log\left(
\frac{\beta_\mathsf{Endo}}{\beta_\mathsf{Endo} - \alpha_\mathsf{Endo}}
\right)
,
\end{align*}
while the distribution of the size of the burst is given by
\begin{align*}
\PP{Z_\mathsf{Endo} = z}
=
\frac{1}{z}{2z-2 \choose z - 1}\left(\frac{\beta_\mathsf{Endo}}{\alpha_\mathsf{Endo} + \beta_\mathsf{Endo}}\right)^{z}\left(\frac{\alpha_\mathsf{Endo}}{\alpha_\mathsf{Endo} + \beta_\mathsf{Endo}}\right)^{z-1}
,
\end{align*}
for each $z \in \mathbb{Z}^+$. Each of these can be obtained by viewing the endogenous burst model as a random walk with absorption. The probability mass function of $Z_\mathsf{Endo}$ already has no dependence on $\eta$, so we are only left to show the convergence of the duration. Again through a simple Markov inequality for any $\epsilon > 0$, we can see that
\begin{align*}
\lim_{\eta \to \infty} \PP{\tau_{\mathsf{Endo},\eta} > \epsilon}
\leq
\lim_{\eta \to \infty} \frac{1}{\epsilon} \E{\tau_{\mathsf{Endo},\eta}}
=
\frac{1}{\eta \alpha_\mathsf{Endo} \epsilon}
\log\left(
\frac{\beta_\mathsf{Endo}}{\beta_\mathsf{Endo} - \alpha_\mathsf{Endo}}
\right)
=
0
,
\end{align*}
hence the duration converges to 0 in probability.
\hfill\Halmos\\
\endproof

These models are both simple, but the  underlying principle should hold merit in  greater generality.
In each of these continuous Markov chain models, the $\alpha$ and $\beta$ rates can be thought of as the underlying burst events occurring on smaller and smaller timescales.  Relative to the timescale of staffing decisions, we think this is quite natural. That is, bursts may take place on the order of seconds or minutes, whereas staffing decisions typically last for several hours. Matching that reasoning, Proposition~\ref{burstToBatch} suggests that we can treat such a burst as a batch, as the size distributions do not change throughout the scaling. Of course, the arrival stream need not be exclusively bursts nor exclusively batches, and may actually be bursts of batches. Again, in such cases, Proposition~\ref{burstToBatch} suggests we can think of this as batches of batches, which are simply batches with a different size distribution. Hence, we will stick to strictly studying batch arrivals. 

%\subsection{Proof of Proposition~\ref{burstToBatch}}

\subsection{Proof of Proposition~\ref{infMeanVar}}\label{infMeanVarProof}

\proof{Proof.}
These can both be seen as near immediate consequences of Proposition 2.4 of~\citet{daw2019distributions}, which provides that $Q^\infty(m) \stackrel{D}{=} \sum_{j=1}^{nm^\nu} j Y_j$ where $Y_j \sim \mathsf{Pois}(\lambda m^{1-\nu} \slash (j \mu))$ are independent. For the mean, this implies
$$
\E{Q^\infty(m)}
=
\sum_{j=1}^{n m^\nu}
j \E{Y_j}
= 
n m^\nu \frac{\lambda m^{1-\nu}}{\mu}
=
\frac{\lambda n m}{\mu}
.
$$
Then, for the variance, we can leverage the independence of the Poisson random variables to similarly see
$$
\Var{Q^\infty(m)}
=
\sum_{j=1}^{n m^\nu}
j^2 \Var{Y_j}
=
\sum_{j=1}^{n m^\nu}
\frac{\lambda j m^{1-\nu}}{\mu}
=
\frac{\lambda n m (n m^{\nu} + 1)}{2\mu}
,
$$
and so we complete the proof.
\hfill\Halmos\\
\endproof

\subsection{Matrix Calculation of Exceedance Probability for the Markovian Queueing Model}

By convention, we will take a matrix product as evaluated with the largest index at the left-most position, followed by the second largest to its immediate right, and so on. Furthermore, we will take a matrix product over an empty set of indices to be the identity matrix.

\begin{lemma}\label{matrixLemma}
In the $M^n/M/c$ queue with arrival rate $\lambda$, batch size $n$, and staffing level $c \geq n$, the steady-state exceedance probability is equal to
\begin{align}
\PP{Q^C \geq c}
&=
\frac{
\frac{\lambda}{c\mu - \lambda n}
[
\frac{c\mu}{\lambda}
\,\,
n-1
\,\,
\dots
\,\,
2
\,\,
1
]
\prod_{j=n+1}^{c} \mathbf{C}\left(\frac{\lambda}{j \mu}\right) \mathbf{s}
}{
1
+
\sum_{k=n+1}^{c-1} \mathbf{v}_1^\mathsf{T} \prod_{j=n+1}^{k} \mathbf{C}\left(\frac{\lambda}{j \mu}\right) \mathbf{s}
+
\frac{\lambda}{c\mu - \lambda n}
[
\frac{c\mu}{\lambda}
\,\,
n-1
\,\,
\dots
\,\,
2
\,\,
1
]
%\frac{\lambda}{c\mu - \lambda n}
%[ n \,\, n-1 \,\, \dots \,\, 1]
\prod_{j=n+1}^{c} \mathbf{C}\left(\frac{\lambda}{j \mu}\right) \mathbf{s}
}
,
\end{align}
where $\mathbf{v}_i$ for $1 \leq i \leq n$ is a $n$-dimensional unit column vector in the $i^\mathrm{th}$ coordinate, $\mathbf{C}(x)$ in $\mathbb{R}^{n \times n}$ for $x \in \mathbb{R}$ is the companion matrix given by
\begin{align}
\mathbf{C}(x)
&
=
\begin{bmatrix}
x && x && x & \dots & x && x \\
1 && 0 && 0 & \dots & 0 && 0\\
0 && 1 && 0 & \dots & 0 && 0\\
0 && 0 && 1 & \dots & 0 && 0\\
\vdots && \vdots && \vdots & \ddots & \vdots && \vdots\\
0 && 0 && 0 & \dots & 1 && 0
\end{bmatrix}
,
\end{align}
and $\mathbf{s}$ is a $n$-dimensional column vector such that 
\begin{align}
\mathbf{s}_i
=
\frac{\frac{\lambda}{\mu}\Gamma\left(\frac{\lambda}{\mu} + i\right)\Gamma\left(n+1\right)}{\Gamma\left(\frac{\lambda}{\mu} + n + 1\right)\Gamma\left(i+1\right)} 
%=
%\begin{cases}
%\frac{\Gamma\left(\frac{\lambda}{\mu} + i\right)}{\Gamma\left(\frac{\lambda}{\mu}\right)\Gamma\left(i+1\right)} 
%\left(
%\frac{\Gamma\left(\frac{\lambda}{\mu} + (n \wedge c) + 1\right)}{\Gamma\left(\frac{\lambda}{\mu}+1\right)\Gamma\left((n \wedge c) +1\right)}
%+
%\frac{c\mu}{\lambda}\left(\left(1+\frac{\lambda}{c\mu}\right)^{(c-n)^+ +1} - 1 - \frac{\lambda}{c\mu}\right) 
%\frac{\Gamma\left(\frac{\lambda}{\mu} + c\right)}{\Gamma\left(\frac{\lambda}{\mu}\right)\Gamma\left(c+1\right)}  
%\right)^{-1}
%&
%i \leq c
%\\
%\left(1 + \frac{\lambda}{c\mu}\right)^{i-c}
%\frac{\Gamma\left(\frac{\lambda}{\mu} + c\right)}{\Gamma\left(\frac{\lambda}{\mu}\right)\Gamma\left(c+1\right)} 
%\left(
%\frac{\Gamma\left(\frac{\lambda}{\mu} + (n \wedge c) + 1\right)}{\Gamma\left(\frac{\lambda}{\mu}+1\right)\Gamma\left((n \wedge c) +1\right)}
%+
%\frac{c\mu}{\lambda}\left(\left(1+\frac{\lambda}{c\mu}\right)^{(c-n)^+ +1} - 1 - \frac{\lambda}{c\mu}\right) 
%\frac{\Gamma\left(\frac{\lambda}{\mu} + c\right)}{\Gamma\left(\frac{\lambda}{\mu}\right)\Gamma\left(c+1\right)}  
%\right)^{-1}
%&
%i \geq c + 1
%\end{cases}
\end{align}
for each coordinate $1 \leq i \leq n$.
\end{lemma}
\proof{Proof.}
Let us point to \citet{neuts1978algorithmic} for origins of matrix analytic calculations for the distributions of batch arrival exponential service queueing systems; we provide this lemma here for simplicity and completeness of the paper, as we are not aware of this precise expression being available previously. Standard CTMC techniques yield that the multi-server steady-state probabilities $\pi_i = \PP{Q^C = i}$ will satisfy the balance equations 
\begin{align*}
\pi_{k+1} = \frac{\lambda + \mu (k \wedge c)}{\mu (k+1 \wedge c)} \pi_k - \frac{\lambda}{\mu(k+1 \wedge c)} \pi_{k-n}
,
\end{align*}
where $\pi_k = 0$ for all $k < 0$. Equivalently, $\pi_k = \frac{\lambda}{\mu(k\wedge c)} \sum_{i=1}^n \pi_{k-i}$. Therefore, any $n$-dimensional vector of consecutive steady-state probabilities will satisfy
\begin{align*}
\begin{bmatrix}
\pi_{k+1} \\
\vdots\\
\pi_{k-n+2}
\end{bmatrix}
=
\mathbf{C}\left(\frac{\lambda}{\mu(k+1\wedge c)}\right)
\begin{bmatrix}
\pi_{k} \\
\vdots\\
\pi_{k-n+1}
\end{bmatrix}
.
\end{align*}
Furthermore, for $k \leq n$, this also implies that
\begin{align*}
\pi_k
=
\frac{\Gamma\left(\frac{\lambda}{\mu} + k\right)}{\Gamma\left(\frac{\lambda}{\mu}\right)\Gamma\left(k+1\right)} \pi_0
.
\end{align*}
%and moreover for $k \leq n$ in general we have
%\begin{align*}
%\pi_k
%=
%\left(1 + \frac{\lambda}{c\mu}\right)^{(k-c)^+}
%\frac{\Gamma\left(\frac{\lambda}{\mu} + (k \wedge c)\right)}{\Gamma\left(\frac{\lambda}{\mu}\right)\Gamma\left((k \wedge c) +1\right)} \pi_0
%.
%\end{align*}
Now, we can combine these facts and see that for any $k \geq n+1$,
\begin{align*}
\pi_k
&=
\mathbf{v}_1^\mathsf{T}
\begin{bmatrix}
\pi_{k} \\
\vdots\\
\pi_{k-n+1}
\end{bmatrix}
%=
%\mathbf{v}_1^\mathsf{T}
%\prod_{j=1}^{k-n}
%\mathbf{C}\left(\frac{\lambda}{\mu (j\wedge c)}\right)
%\begin{bmatrix}
%\pi_{n} \\
%\vdots\\
%\pi_{1}
%\end{bmatrix}
=
\mathbf{v}_1^\mathsf{T}
\prod_{j=n+1}^{k}
\mathbf{C}\left(\frac{\lambda}{\mu (j\wedge c)}\right)
\begin{bmatrix}
%\left(1 + \frac{\lambda}{c\mu}\right)^{(n-c)^+}
\frac{\Gamma\left(\frac{\lambda}{\mu} + n\right)}{\Gamma\left(\frac{\lambda}{\mu}\right)\Gamma\left(n+1\right)} 
 \\
\vdots\\
\frac{\lambda}{\mu}
\end{bmatrix}
\pi_0
.
\end{align*}
So, we have that for any $n+1 \leq k \leq c$,
\begin{align*}
\pi_k
=
\mathbf{v}_1^\mathsf{T}
\prod_{j=n+1}^{k}
\mathbf{C}\left(\frac{\lambda}{j \mu}\right)
\begin{bmatrix}
\frac{\Gamma\left(\frac{\lambda}{\mu} + n\right)}{\Gamma\left(\frac{\lambda}{\mu}\right)\Gamma\left(n+1\right)} 
 \\
\vdots\\
\frac{\lambda}{\mu}
\end{bmatrix}
\pi_0
,
\end{align*}
while for $k \geq c$, the arguments of the companion matrices cease to change, leaving
\begin{align*}
\pi_k
=
\mathbf{v}_1^\mathsf{T}
\mathbf{C}\left(\frac{\lambda}{c \mu}\right)^{k-c}
\prod_{j=n+1}^{c}
\mathbf{C}\left(\frac{\lambda}{j \mu}\right)
\begin{bmatrix}
\frac{\Gamma\left(\frac{\lambda}{\mu} + n\right)}{\Gamma\left(\frac{\lambda}{\mu}\right)\Gamma\left(n+1\right)} 
 \\
\vdots\\
\frac{\lambda}{\mu}
\end{bmatrix}
\pi_0
.
\end{align*}
This gives us all we need to simplify to the stated expression. In particular, we can note that
\begin{align*}
\PP{Q^C \geq c} 
&
=
\sum_{k=c}^\infty
\pi_k
\\
&
=
\mathbf{v}_1^\mathsf{T}
\sum_{k=c}^\infty
\mathbf{C}\left(\frac{\lambda}{c \mu}\right)^{k-c}
\prod_{j=n+1}^{c}
\mathbf{C}\left(\frac{\lambda}{j \mu}\right)
\begin{bmatrix}
\frac{\Gamma\left(\frac{\lambda}{\mu} + n\right)}{\Gamma\left(\frac{\lambda}{\mu}\right)\Gamma\left(n+1\right)} 
 \\
\vdots\\
\frac{\lambda}{\mu}
\end{bmatrix}
\pi_0
\\
&
=
\mathbf{v}_1^\mathsf{T}
\left(
\mathbf{I}
-
\mathbf{C}\left(\frac{\lambda}{c \mu}\right)
\right)^{-1}
\prod_{j=n+1}^{c}
\mathbf{C}\left(\frac{\lambda}{j \mu}\right)
\begin{bmatrix}
\frac{\Gamma\left(\frac{\lambda}{\mu} + n\right)}{\Gamma\left(\frac{\lambda}{\mu}\right)\Gamma\left(n+1\right)} 
 \\
\vdots\\
\frac{\lambda}{\mu}
\end{bmatrix}
\pi_0
\\
&
=
\begin{bmatrix}
\frac{1}{1-\frac{\lambda n}{c\mu}}
&
\frac{\frac{\lambda (n-1)}{c\mu}}{1-\frac{\lambda n}{c\mu}}
&
\dots
&
\frac{\frac{2 \lambda }{c\mu}}{1-\frac{\lambda n}{c\mu}}
&
\frac{\frac{\lambda }{c\mu}}{1-\frac{\lambda n}{c\mu}}
\end{bmatrix}
\prod_{j=n+1}^{c}
\mathbf{C}\left(\frac{\lambda}{j \mu}\right)
\begin{bmatrix}
\frac{\Gamma\left(\frac{\lambda}{\mu} + n\right)}{\Gamma\left(\frac{\lambda}{\mu}\right)\Gamma\left(n+1\right)} 
 \\
\vdots\\
\frac{\lambda}{\mu}
\end{bmatrix}
\pi_0
\\
%&
%=
%\begin{bmatrix}
%\frac{c\mu}{c\mu-{\lambda n}}
%&
%\frac{{\lambda (n-1)}}{c\mu-{\lambda n}}
%&
%\dots
%&
%\frac{{2 \lambda }}{c\mu-{\lambda n}}
%&
%\frac{{\lambda }}{c\mu-{\lambda n}}
%\end{bmatrix}
%\prod_{j=n+1}^{c}
%\mathbf{C}\left(\frac{\lambda}{j \mu}\right)
%\begin{bmatrix}
%\frac{\Gamma\left(\frac{\lambda}{\mu} + n\right)}{\Gamma\left(\frac{\lambda}{\mu}\right)\Gamma\left(n+1\right)} 
% \\
%\vdots\\
%\frac{\lambda}{\mu}
%\end{bmatrix}
%\pi_0
%\\
&
=
\frac{\lambda}{c\mu - \lambda n}
\begin{bmatrix}
\frac{c\mu}{\lambda}
&
n-1
&
\dots
&
2
&
1
\end{bmatrix}
\prod_{j=n+1}^{c}
\mathbf{C}\left(\frac{\lambda}{j \mu}\right)
\begin{bmatrix}
\frac{\Gamma\left(\frac{\lambda}{\mu} + n\right)}{\Gamma\left(\frac{\lambda}{\mu}\right)\Gamma\left(n+1\right)} 
 \\
\vdots\\
\frac{\lambda}{\mu}
\end{bmatrix}
\pi_0
,
\end{align*}
and $\pi_0$ can be found through the fact that the full distribution must sum to 1. That is,
\begin{align*}
1
&
= 
\sum_{k=0}^\infty 
\pi_k
\\
&
=
\pi_0 
+ 
\sum_{k=1}^n \frac{\Gamma\left(\frac{\lambda}{\mu} + k\right)}{\Gamma\left(\frac{\lambda}{\mu}\right)\Gamma\left(k+1\right)} 
\pi_0
+
\sum_{k=n+1}^{c-1}
\mathbf{v}_1^\mathsf{T}
\prod_{j=n+1}^{k}
\mathbf{C}\left(\frac{\lambda}{j \mu}\right)
\begin{bmatrix}
\frac{\Gamma\left(\frac{\lambda}{\mu} + n\right)}{\Gamma\left(\frac{\lambda}{\mu}\right)\Gamma\left(n+1\right)} 
 \\
\vdots\\
\frac{\lambda}{\mu}
\end{bmatrix}
\pi_0
\\
&
\qquad
+
\frac{\lambda}{c\mu - \lambda n}
\begin{bmatrix}
\frac{c\mu}{\lambda}
&
n-1
&
\dots
&
2
&
1
\end{bmatrix}
\prod_{j=n+1}^{c}
\mathbf{C}\left(\frac{\lambda}{j \mu}\right)
\begin{bmatrix}
\frac{\Gamma\left(\frac{\lambda}{\mu} + n\right)}{\Gamma\left(\frac{\lambda}{\mu}\right)\Gamma\left(n+1\right)} 
 \\
\vdots\\
\frac{\lambda}{\mu}
\end{bmatrix}
\pi_0
,
\end{align*}
hence, 
\begin{align*}
\pi_0
&
=
\left(
\sum_{k=0}^n \frac{\Gamma\left(\frac{\lambda}{\mu} + k\right)}{\Gamma\left(\frac{\lambda}{\mu}\right)\Gamma\left(k+1\right)} 
+
\sum_{k=n+1}^{c-1}
\mathbf{v}_1^\mathsf{T}
\prod_{j=n+1}^{k}
\mathbf{C}\left(\frac{\lambda}{j \mu}\right)
\begin{bmatrix}
\frac{\Gamma\left(\frac{\lambda}{\mu} + n\right)}{\Gamma\left(\frac{\lambda}{\mu}\right)\Gamma\left(n+1\right)} 
 \\
\vdots\\
\frac{\lambda}{\mu}
\end{bmatrix}
\right.
\\
&
\qquad
\left.
+
\frac{\lambda}{c\mu - \lambda n}
\begin{bmatrix}
\frac{c\mu}{\lambda}
&
n-1
&
\dots
&
2
&
1
\end{bmatrix}
\prod_{j=n+1}^{c}
\mathbf{C}\left(\frac{\lambda}{j \mu}\right)
\begin{bmatrix}
\frac{\Gamma\left(\frac{\lambda}{\mu} + n\right)}{\Gamma\left(\frac{\lambda}{\mu}\right)\Gamma\left(n+1\right)} 
 \\
\vdots\\
\frac{\lambda}{\mu}
\end{bmatrix}
\right)^{-1}
.
\end{align*}
Finally, by recognizing that 
$$
\sum_{k=0}^n 
%\sum_{k=1}^n 
\frac{\Gamma\left(\frac{\lambda}{\mu} + k\right)}{\Gamma\left(\frac{\lambda}{\mu}\right)\Gamma\left(k+1\right)} 
= 
 \frac{\Gamma\left(\frac{\lambda}{\mu} + n + 1\right)}{\Gamma\left(\frac{\lambda}{\mu}+1\right)\Gamma\left(n+1\right)} 
 ,
$$
we simplify to the stated expression.
\hfill\Halmos\\
\endproof

\subsection{\edit{Proof of Proposition~\ref{coincideProp} and a Supporting Berry-Esseen Type Bound}}

\edit{Before proceeding with proving the asymptotic coincidence of the some-wait and all-wait exceedances in the hybrid limit, we first will prove a Berry-Esseen type bound that we will employ in the proof of Proposition~\ref{coincideProp}.}

\begin{lemma}\label{berryLemma}
\edit{
For $m \in \mathbb{Z}^+$, let $Q^\infty(m)$ be the queue length of a $M^n/M/\infty$ system with arrival rate $\lambda(m) = \lambda_0 m^{1-v}$ and batch size $n(m) = n_0 m^\nu$ for $\nu \in (0,1)$. Additionally, let $X \sim \mathsf{Norm}(0, \frac{\lambda_0 n_0^2}{2\mu})$. Then, for $F_m(\cdot)$ as the CDF of $Q^\infty(m)$ and $\Phi(\cdot)$ as the CDF of a standard Normal random variable, 
\begin{align}
\sup_{x \in \mathbb{R}}
\left| 
F_m\left(\frac{x - {\lambda_0 n_0 m}/{\mu}}{m^{\frac{1+\nu}{2}}}\right)
-
\Phi\left(\frac{x}{n_0 \sqrt{{\lambda_0}\slash({2\mu})}}\right)
\right|
\leq
\frac{\hat{\mathcal{C}}}{m^{\frac{(1-\nu \wedge \nu)}{2}}}
,
\end{align}
for some constant $\hat{\mathcal{C}} > 0$.}
\end{lemma}
\proof{Proof.}
\edit{Here we simply invoke that the classical Berry-Esseen bound for the $m^{(1+\nu)\slash 2}$ scaling, specifically through the bound for sums of nonidentical random variables from \citet{esseen1942liapunov}. Together with the sum of scaled Poisson's decomposition used in the proof of Theorem~\ref{bothLimitInf}, this implies that the bound is proportional to the ratio
$$
\frac{
\sum_{j=1}^{ n_0 m^\nu } j^3 \E{ \left| Y_j - \frac{\lambda_0 m^{1-\nu}}{j \mu} \right|^3}
}{
\left(\sum_{j=1}^{ n_0 m^\nu } j^2 \Var{Y_j} \right)^{{3}\slash{2}}
}
.
$$
Starting with the denominator, immediately from the equidispersion of Poisson random variables, we find
\begin{align*}
\left(\sum_{j=1}^{ n_0 m^\nu } j^2 \Var{Y_j} \right)^{{3}\slash{2}}
&
=
\left(\sum_{j=1}^{ n_0 m^\nu }  \frac{\lambda_0 j m^{1-\nu}}{\mu}\right)^{{3}\slash{2}}
=
\left(\frac{\lambda_0   n_0 m ( n_0 m^\nu +1)}{2\mu} \right)^{{3}\slash{2}}
.
\end{align*}
For the numerator, from Corollary 1 of~\citet{ruzankin2020absolute}, we have that 
\begin{align*}
j^3
\E{ \left| Y_j - \frac{\lambda_0 m^{1-\nu}}{j \mu} \right|^3}
&
=
\frac{2\lambda_0 j^2 m^{1-\nu}}{\mu}
\left(
\left(
\frac{\lambda_0 m^{1-\nu}}{j \mu}
-
\left\lfloor
\frac{\lambda_0 m^{1-\nu}}{j \mu}
\right\rfloor
\right)^2
+
2
\left\lfloor
\frac{\lambda_0 m^{1-\nu}}{j \mu}
\right\rfloor
+
1
\right)
\PP{Y_j = \left\lfloor
\frac{\lambda_0 m^{1-\nu}}{j \mu}
\right\rfloor}
\\
&
\qquad
+
\frac{\lambda_0 j^2 m^{1-\nu}}{\mu} 
\left(
1
-
2\PP{Y_j \leq \frac{\lambda_0 m^{1-\nu}}{j \mu} }
\right)
.
\end{align*}
We bound this as follows. Starting with two relatively crude upper bounds, we can note that by definition $x - \lfloor x \rfloor \leq 1$ for all $x$, and similarly $1 - 2 \PP{X \leq x} \leq 1$ for any $X$ and $x$. Then, by Stirling's approximation, we can see that
$$
\frac{\lambda_0 m^{1-\nu}}{j \mu}
\PP{Y_j = \left\lfloor
\frac{\lambda_0 m^{1-\nu}}{j \mu}
\right\rfloor}
\leq
\sqrt{
\frac{1}{2\pi}\left\lceil\frac{\lambda_0 m^{1-\nu}}{j \mu}\right\rceil
}
.
$$
Together, this now provides
\begin{align*}
j^3
\E{ \left| Y_j - \frac{\lambda_0 m^{1-\nu}}{j \mu} \right|^3}
&
\leq
\frac{\lambda_0 j^2 m^{1-\nu}}{\mu} 
\left(
5
+
\sqrt{
\frac{2}{\pi}\left\lceil\frac{\lambda_0 m^{1-\nu}}{j \mu}\right\rceil
}
\right)
=
\frac{5\lambda_0 j^2 m^{1-\nu}}{\mu} 
+
\frac{\lambda_0 j^2 m^{1-\nu}}{\mu} 
\sqrt{
\frac{2}{\pi}\left\lceil\frac{\lambda_0 m^{1-\nu}}{j \mu}\right\rceil
}
,
\end{align*}
which after summing across $j$ yields that
\begin{align*}
\sum_{j=1}^{ n_0m^\nu } j^3 \E{ \left| Y_j - \frac{\lambda_0 m^{1-\nu}}{j \mu} \right|^3}
&
\leq
\sum_{j=1}^{ n_0m^\nu }
\left(
\frac{5\lambda_0 j^2 m^{1-\nu}}{\mu} 
+
\frac{\lambda_0 j^2 m^{1-\nu}}{\mu} 
\sqrt{
\frac{2}{\pi}\left\lceil\frac{\lambda_0 m^{1-\nu}}{j \mu}\right\rceil
}
\right)
\\
&
\leq
\frac{5\lambda_0  n_0m^{1+2\nu}  }{\mu} 
+
  n_0m^{1+3\nu/2}  \lceil m^{1-\nu}\rceil^{\frac{1}{2}} 
\sqrt{
\frac{2}{\pi}\left\lceil\frac{\lambda_0 }{\mu}\right\rceil
}
.
\end{align*}
Returning now to the ratio of the third and second moment terms, we see
\begin{align*}
\frac{
\sum_{j=1}^{ n_0m^\nu } j^3 \E{ \left| Y_j - \frac{\lambda_0 m^{1-\nu}}{j \mu} \right|^3}
}{
\left(\sum_{j=1}^{ n_0m^\nu } j^2 \Var{Y_j} \right)^{{3}\slash{2}}
}
\leq
\frac{
\frac{5\lambda_0  n_0m^{1+2\nu}  }{\mu} 
+
n_0m^{1+3\nu/2} \lceil m^{1-\nu}\rceil^{\frac{1}{2}} 
\sqrt{
\frac{2}{\pi}\left\lceil\frac{\lambda_0 }{\mu}\right\rceil
}
}{
\left(\frac{\lambda_0   n_0 m ( n_0 m^\nu +1)}{2\mu} \right)^{{3}\slash{2}}
}
\leq
\frac{ \hat{\mathcal{C}}^0 }{ m^{\frac{1-\nu}{2}} }
+
\frac{ \hat{\mathcal{C}}^1 }{  n_0m^{\nu/2}  }
,
\end{align*}
for some $\hat{\mathcal{C}}^0  > 0$ and $\hat{\mathcal{C}}^1 > 0$. Hence, asymptotically the dependence on $m$ relies on the smaller of these two powers, yielding the stated result.
%\tr{
%\begin{align*}
%&
%\frac{2\lambda j^2 m^{1-\nu}}{\mu}
%\left(
%\left(
%\frac{\lambda m^{1-\nu}}{j \mu}
%-
%\left\lfloor
%\frac{\lambda m^{1-\nu}}{j \mu}
%\right\rfloor
%\right)^2
%+
%2
%\left\lfloor
%\frac{\lambda m^{1-\nu}}{j \mu}
%\right\rfloor
%+
%1
%\right)
%\PP{Y_j = \left\lfloor
%\frac{\lambda m^{1-\nu}}{j \mu}
%\right\rfloor}
%\\
%&
%\qquad
%+
%\frac{\lambda j^2 m^{1-\nu}}{\mu} 
%\left(
%1
%-
%2\PP{Y_j \leq \frac{\lambda m^{1-\nu}}{j \mu} }
%\right)
%\\
%\leq
%\,
%&
%\frac{2\lambda m^{1-\nu}}{\mu}
%\left(
%\left(
%\frac{\lambda m^{1-\nu}}{\mu}
%-
%\left\lfloor
%\frac{\lambda m^{1-\nu}}{ \mu}
%\right\rfloor
%\right)^2
%+
%\frac{2 \lambda m^{1-\nu}}{\mu}
%+
%\lceil nm^\nu \rceil
%\right)
%\PP{Y_j = \left\lfloor
%\frac{\lambda m^{1-\nu}}{j \mu}
%\right\rfloor}
%\\
%&
%\qquad
%+
%\frac{\lambda \lceil nm^\nu \rceil^2 m^{1-\nu}}{\mu} 
%\left(
%\PP{Y_j \geq \left\lceil\frac{\lambda m^{1-\nu}}{j \mu} \right\rceil }
%-
%\PP{Y_j \leq \left\lfloor \frac{\lambda m^{1-\nu}}{j \mu} \right\rfloor }
%\right)
%\end{align*}
%}
\hfill\Halmos\\
\endproof
}

\edit{While we do not expect the bound in Lemma~\ref{berryLemma} to be the tightest possible, it is sufficient for our goal of proving Proposition~\ref{coincideProp}. We do so now.\\}

\proof{Proof of Proposition~\ref{coincideProp}.}
\edit{%bound by $Q^\infty$ by bounding $\pi_0$ and $\tilde \pi_0$ and then use berry esseen\\
Let us begin by bounding the exceedance probability differences by that of the corresponding event for the infinite server system. For simplicity, let us think of the difference of some-wait and all-wait probabilities as the probability of single event, i.e. 
$
\PP{Q^C(m) + n(m) > c(m)}
-
\PP{Q^C(m) \geq c(m)}
=
\PP{c(m) - n(m) < Q^C(m) < c(m)}
$. Now, letting $\pi_k = \PP{Q^C(m) = k}$ as in the proofs of Theorem~\ref{bothLimitMulti} and Lemma~\ref{matrixLemma}, we can write this probability as
\begin{align*}
\PP{c(m) - n(m) < Q^C(m) < c(m)}
&=
\sum_{k=c(m)-n(m)+1}^{c(m)-1} \pi_k
=
\sum_{k=c(m)-n(m)+1}^{c(m)-1} f_k \pi_0
,
\end{align*}
where the coefficients $f_k$ are given by
\begin{align*}
f_k
&=
\begin{cases}
\mathbf{v}_1^\mathsf{T} 
\prod_{j = n(m)+1}^{c(m)}
\mathbf{C}\left(\frac{\lambda(m)}{j\mu}\right)
\begin{bmatrix}
\frac{\Gamma\left(\frac{\lambda(m)}{\mu} + n(m)\right)}{\Gamma\left(\frac{\lambda(m)}{\mu}\right) \Gamma\left(n(m)+1\right)} \\
\vdots \\
\frac{\lambda(m)}{\mu}
\end{bmatrix}
,
& 
\text{for } n(m) \leq k \leq c(m) ,
\\
\\
\frac{\Gamma\left(\frac{\lambda(m)}{\mu} + k\right)}{\Gamma\left(\frac{\lambda(m)}{\mu}\right) \Gamma\left(k+1\right)} 
,
&
\text{for } 0 \leq k \leq n(m) - 1
,
\end{cases}
\end{align*}
and $\pi_0 = \PP{Q^C(m) = 0}$. (Note that here, by comparison to the proofs of Theorem~\ref{bothLimitMulti} and Lemma~\ref{matrixLemma}, we have not suppressed the dependence on $m$.) Now, as also used in the proof of Theorem~\ref{bothLimitMulti}, we can see that the same coefficients appear in the same places within the balance equations for $Q^\infty$ at states $c(m)$ and below, and, moreover, the analogous probability satisfies a nearly identical equation,
\begin{align*}
\PP{c(m) - n(m) < Q^\infty(m) < c(m)}
&=
\sum_{k=c(m)-n(m)+1}^{c(m)-1} f_k \tilde \pi_0
,
\end{align*}
with the only change being that we are now using $\tilde \pi_0 = \PP{Q^\infty(m) = 0}$ instead of $\pi_0$.}

\edit{Now, invoking the approach and notation of Lemma~\ref{matrixLemma}, we can see that the difference between $\pi_0$ and $\tilde \pi_0$ lies in the terms associated with states above $c$. That is, subtracting the inverse of each probability yields
\begin{align*}
&
\frac{1}{\pi_0} - \frac{1}{\tilde \pi_0}
\nonumber
\\
&
\quad
=
\mathbf{v}_1^\mathsf{T}
\sum_{k=c(m)+1}^\infty
\mathbf{C}\left(\frac{\lambda(m)}{c(m) \mu}\right)^{k-c(m)}
\prod_{j=n(m)+1}^{c(m)}
\mathbf{C}\left(\frac{\lambda(m)}{j \mu}\right)
\mathbf{s}(m)
-
\mathbf{v}_1^\mathsf{T}
\sum_{k=c(m)+1}^\infty
\prod_{j=n(m)+1}^{k}
\mathbf{C}\left(\frac{\lambda(m)}{j \mu}\right)
\mathbf{s}(m)
\nonumber
\\
&
\quad
=
\mathbf{v}_1^\mathsf{T}
\sum_{k=c(m)+1}^\infty
\left(
\mathbf{C}\left(\frac{\lambda(m)}{c(m) \mu}\right)^{k-c(m)}
-
\prod_{j=c(m)+1}^{k}
\mathbf{C}\left(\frac{\lambda(m)}{j \mu}\right)
\right)
\prod_{j=n(m)+1}^{c(m)}
\mathbf{C}\left(\frac{\lambda(m)}{j \mu}\right)
\mathbf{s}(m)
,
\end{align*}
where $\mathbf{s}(m) \in \mathbb{R}_+^{n(m)}$ is a column vector defined with entries
\begin{align*}
\mathbf{s}(m)
&=
\begin{bmatrix}
\frac{\Gamma\left(\frac{\lambda(m)}{\mu} + n(m)\right)}{\Gamma\left(\frac{\lambda(m)}{\mu}\right)\Gamma\left(n(m)+1\right)} 
 \\
\vdots\\
\frac{\lambda(m)}{\mu}
\end{bmatrix}
.
\end{align*}
Let us observe that $\mathbf{s}(m)$, each companion matrix, and $\mathbf{v}_1$ all have non-negative values at every coordinate, leaving us to consider the difference of companion matrix products that appear inside the summation. For any given $k \geq c(m)$, we can notice that within
both the left-hand and right-hand products contain $k - c(m)$ multiplied terms, and, moreover, because $j > c(m)$, every element in a given $\mathbf{C}(\lambda(m) / j\mu)$ is non-negative and no more than the element at the same position in $\mathbf{C}(\lambda(m)/c(m)\mu)$. Hence, we have
\begin{align*}
\mathbf{C}\left(\frac{\lambda(m)}{c(m) \mu}\right)^{k-c(m)}
-
\prod_{j=c(m)+1}^{k}
\mathbf{C}\left(\frac{\lambda(m)}{j \mu}\right)
\geq 
0
,
\end{align*}
yielding that $\pi_0 \leq \tilde \pi_0$, and, more broadly, 
\begin{align*}
\PP{c(m) - n(m) < Q^C(m) < c(m)}
\leq
\PP{c(m) - n(m) < Q^\infty(m) < c(m)}
.
\end{align*}
Hence, we now switch to analyzing the infinite server queue only.
}

\edit{Plugging in the values of $c(m)$, $n(m)$, and $\lambda(m)$ and rearranging the expression of the event, we have
\begin{align*}
\PP{c(m) - n(m) < Q^\infty(m) < c(m)}
&=
\PP{
 \delta  - n_0 m^{-\frac{1-\nu}{2}}
< 
\left(Q^\infty(m) - \frac{\lambda_0 n_0 m }{ \mu} \right) m^{-\frac{1+\nu}{2}}
< 
\delta 
}
.
\end{align*}
By Lemma~\ref{berryLemma}, we have that 
\begin{align*}
\PP{
 \delta  - n_0 m^{-\frac{1-\nu}{2}}
< 
\left(Q^\infty(m) - \frac{\lambda_0 n_0 m }{ \mu} \right) m^{-\frac{1+\nu}{2}}
< 
\delta 
}
\leq
\PP{
 \delta  - n_0 m^{-\frac{1-\nu}{2}}
< 
X
< 
\delta 
}
+
\hat{\mathcal{C}} m^{-\frac{(1-\nu\wedge\nu)}{2}}
,
\end{align*}
where $X \sim \mathsf{Norm}(0, \lambda_0 n_0^2 / 2\mu)$ and $\hat{\mathcal{C}}$ is some positive constant. Then, as $m \to \infty$, we can see that both $\PP{
 \delta  - n_0 m^{-\frac{1-\nu}{2}}
< 
X
< 
\delta 
} 
\to 
0$
and
$\hat{\mathcal{C}} m^{-{(1-\nu\wedge\nu)}/{2}} \to 0$; thus we complete the proof.
\Halmos\\
\endproof
}

\subsection{\edit{Proof of Proposition~\ref{serveCostProp}}}

\proof{Proof.}
\edit{
From Equation~\eqref{serveCostObj}, we have that we are aiming to minimize $\mathcal{C}_0 ( {m}/{\mu} + \delta n \sqrt{\lambda} ) + \mathcal{C}_1 \lambda$. Because $m = \lambda n$, we can recognize that $n \sqrt{\lambda}$ can be equivalent written $m / \sqrt{\lambda}$. Hence, the service cost objective becomes
\begin{align*}
\mathcal{C}_0 \left( \frac{m}{\mu} + \delta n \sqrt{\lambda} \right) + \mathcal{C}_1 \lambda
&=
\mathcal{C}_0 \left( \frac{m}{\mu} + \frac{\delta m}{\sqrt{\lambda}} \right) + \mathcal{C}_1 \lambda
.
\end{align*}
The first and second derivatives of this expression with respect to $\lambda$ are
\begin{align*}
\frac{\partial}{\partial \lambda}\left(
\mathcal{C}_0 \left( \frac{m}{\mu} + \frac{\delta m}{\sqrt{\lambda}} \right) + \mathcal{C}_1 \lambda
\right)
&=
-
\frac{\mathcal{C}_0\delta m}{2{\lambda}^{3/2}} 
+
\mathcal{C}_1 ,
\\
\frac{\partial^2}{\partial \lambda^2}\left(
\mathcal{C}_0 \left( \frac{m}{\mu} + \frac{\delta m}{\sqrt{\lambda}} \right) + \mathcal{C}_1 \lambda
\right)
&=
\frac{3\mathcal{C}_0\delta m}{4{\lambda}^{5/2}} 
.
\end{align*}
We can then immediately see that $\lambda = \left(m \delta \mathcal{C}_0 \slash 2\mathcal{C}_1\right)^{{2}/{3}}$ is the unique first order solution, and moreover, by the second derivative, this point must be a minimum.
\Halmos\\
\endproof
}

\subsection{\edit{Asymptotic Waiting Times in the Large Batch Limit with Exponential Services}}\label{waitApp}

%\tr{be sure to define/remind all parameters within this subsection}

%\tr{next batch waiting time for $G_t^{B(n)}/M/cn$ conditioned on present state}

\edit{In this subsection of the appendix, we establish results for the limit of the waiting time per customer in the large batch regime analyzed in Section~\ref{batchSec}. Throughout this subsection, we will assume exponentially distributed service durations, i.e.~$S_{i,j} \stackrel{\mathsf{iid}}{\sim} \mathsf{Exp}(\mu)$ for some $\mu > 0$. To begin, we prove that the distribution of the normalized total waiting time for all customers in a batch in the $G_t^{B(n)}/M/cn$ converges to a difference of squared exceedances of the storage process over the staffing level.}

\begin{proposition}\label{waitLimit}
\edit{
Suppose that a batch arrives at time $t$ in the $G_t^{B(n)}/M/cn$ system, and let $W_{1,j}$ be the waiting time of customer $j$ within the batch. Then, the waiting within the batch converges to
\begin{align}
\frac{1}{n}\sum_{j=1}^{B_1(n)} W_{1,j}
\stackrel{D}{\Longrightarrow}
\frac{1}{2c\mu}
\left(
\left(\psi^C_{t^-} + M_1 - c\right)_+^2 - \left(\psi^C_{t^-} - c\right)_+^2
\right)
,
\end{align}
as $n \to \infty$, where $X_{t^-}$ is the limit from the left, i.e. $X_{t^-} = \lim_{s \nearrow t}X_s$, and $\left(x\right)^2_+ = \left(\max\{x,0\}\right)^2$. 
}
\end{proposition}
\proof{Proof.}
\edit{
Let us decompose each customer's wait into two potential sources: the wait for the batch-start, which is experienced by all the customers in the batch, and the wait from the batch start until the given customer's start of service. We will denote these by $W_0$ (the same for all customers in the batch, so no $j$ index) and $W_{1,j}'$, respectively. Let us start with the former.}

\edit{Because the service discipline is first-come-first-serve, $W_0$ is independent of the arriving batch size, $B_1(n)$, and instead depends only on the present queue length. Letting $Q_{t}^C(n)$ be the number of customers in system just before the batch arrives, we can express $W_0$ as
\begin{align*}
W_0
=
\sum_{i=1}^{(Q_{t}^C(n)-cn)^+}
S_i
,
\end{align*}
where $S_i$ is the service duration (excluding any time before $t$) for the $i$th customer to complete service after the batch arrives but before its first customer starts service. By the memoryless property of the service distribution, $S_i \stackrel{\mathsf{iid}}{\sim} \mathsf{Exp}(cn\mu)$. Conditioning on the present value of the queue length, we can see
\begin{align*}
\E{e^{\theta W_0} \mid Q_t^C(n)}
&=
\prod_{i=1}^{(Q_{t}^C(n)-cn)^+}
\E{e^{\theta S_i}}
=
\left(
\frac{cn\mu}{cn\mu - \theta}
\right)^{(Q_{t}^C(n)-cn)^+}
=
e^{(Q_{t}^C(n)-cn)^+ \log \left(1 + \frac{\theta}{cn\mu - \theta}\right)}
.
\end{align*}
Now, because $\log(1+x) \leq x$ for $x > - 1$, we can observe that 
\begin{align*}
\E{e^{\theta W_0} }
=
\E{\E{e^{\theta W_0} \mid Q_t^C(n)}}
=
\E{
e^{(Q_t^C(n) - cn)^+ \log \left(1 + \frac{\theta}{cn\mu - \theta}\right)}
}
\leq
\E{
e^{\frac{\theta (Q_t^C(n) - cn)^+}{cn\mu - \theta}}
}
\longrightarrow
\E{e^{{\theta (\psi_{t^-}^C} - c)^+/{c\mu}}}
,
\end{align*}
as $n \to \infty$ by Theorem~\ref{fDelayConv}, and similarly since $\log(1+x) \geq x - x^2/2$ for $x > -1$, we further have 
\begin{align*}
\E{e^{\theta W_0} }
%=
%\E{\E{e^{\theta W_0} \mid Q_t^C(n)}}
=
\E{
e^{(Q_t^C(n) - cn)^+ \log \left(1 + \frac{\theta}{cn\mu - \theta}\right)}
}
\geq
\E{
e^{\frac{\theta (Q_t^C(n) - cn)^+}{cn\mu - \theta} - \frac{\theta^2 (Q_t^C(n) - cn)^+}{(cn\mu - \theta)^2}}
}
\longrightarrow
\E{e^{\theta (\psi_{t^-}^C - c)^+/{c\mu}}}
,
\end{align*}
as $n \to \infty$, again by Theorem~\ref{fDelayConv}, and thus we have $W_0 \stackrel{D}{\Longrightarrow}  (\psi_{t^-}^C - c)^+/{c\mu}$.}

\edit{Turning to $W_{1,j}'$, let us observe that it may not be the case that all $B_1(n)$ customers in the batch will wait. For brevity of notation, let us define $B_W(n)$ as the number of customer in the batch that wait, which can be found via $B_W(n) = (Q_t^C(n) + B_1(n) - cn)^+ - (Q_t^C(n) - cn)^+$. Then, we can again leverage the first-come-first-serve discipline to observe that
\begin{align*}
\frac{1}{n}\sum_{j=1}^{B_1(n)} W_{1,j}'
&=
\frac{1}{n}\sum_{j=1}^{B_W(n)} W_{1,j}'
=
\frac{1}{n}\sum_{j=1}^{B_W(n)} \left(B_W(n) +1 - j\right) S_j
,
\end{align*}
where $S_j \stackrel{\mathsf{iid}}{\sim} \mathsf{Exp}(cn\mu)$. Conditioning on $B_W(n)$,  the moment generating function for the exponential distribution once again allows us to simply, finding
\begin{align*}
\E{e^{\frac{\theta}{n}\sum_{j=1}^{B_W(n)} \left(B_W(n) +1 - j\right) S_j} \mid B_W(n)}
&=
\prod_{j=1}^{B_W(n)} \frac{cn\mu}{cn\mu - \frac{\theta}{n}\left(B_W(n) +1 - j\right)}
=
e^{-\sum_{j=1}^{B_W(n)} \log\left(1 - \frac{\theta}{cn^2\mu}\left(B_W(n) +1 - j\right) \right) }
. 
\end{align*}
Then, employing the tower property and the Mercator series expansion of $\log(1-x) = \sum_{\ell = 1}^\infty = (-x)^\ell / \ell$, we have
\begin{align*}
\E{e^{\frac{\theta}{n}\sum_{j=1}^{B_1(n)} W_{1,j}'}}
&=
\E{e^{-\sum_{j=1}^{B_W(n)} \log\left(1 - \frac{\theta}{cn^2\mu}\left(B_W(n) +1 - j\right) \right) }}
\\
&=
\E{e^{-\sum_{j=1}^{B_W(n)} \sum_{\ell=1}^\infty \frac{1}{\ell}\left( - \frac{\theta}{cn^2\mu}\left(B_W(n) +1 - j\right) \right)^\ell }}
\\
&=
\E{e^{
\frac{1}{n}\sum_{j=1}^{B_W(n)} \frac{\theta}{c\mu}\left(\frac{B_W(n)}{n} + \frac{1}{n} - \frac{j}{n}\right) 
-
\sum_{j=1}^{B_W(n)} \sum_{\ell=2}^\infty \frac{1}{\ell}\left( - \frac{\theta}{cn^2\mu}\left(B_W(n) +1 - j\right) \right)^\ell 
}}
.
\end{align*}
Because $B_W(n) / n \stackrel{D}{\Longrightarrow} (\psi_{t^-}^C - M_1 + c)^+ - (\psi_{t^-}^C - c)^+$ and $B_W(n) / n^\ell \stackrel{p}{\longrightarrow} 0$ for $\ell \geq 2$ as $n \to \infty$ by Theorem~\ref{fDelayConv}, we can see by continuous mapping that
\begin{align*}
\E{e^{
\frac{1}{n}\sum_{j=1}^{B_W(n)} \frac{\theta}{c\mu}\left(\frac{B_W(n)}{n} + \frac{1}{n} - \frac{j}{n}\right) 
-
\sum_{j=1}^{B_W(n)} \sum_{\ell=2}^\infty \frac{1}{\ell}\left( - \frac{\theta}{cn^2\mu}\left(B_W(n) +1 - j\right) \right)^\ell 
}}
\longrightarrow
\E{
e^{\frac{\theta}{2c\mu} \left((\psi_{t^-}^C + M_1 - c)^+ - (\psi_{t^-}^C - c)^+\right)^2 }
}
,
\end{align*}
as $n \to \infty$.}

\edit{Together, this yields that
\begin{align*}
\frac{1}{n}\sum_{j=1}^{B_1(n)} W_{1,j}
&=
\frac{B_W(n)}{n} W_0 
+ 
\frac{1}{n}\sum_{j=1}^{B_W(n)} W_{1,j}'
\\
&
\stackrel{D}{\Longrightarrow}
\left((\psi_{t^-}^C + M_1 - c)^+ - (\psi_{t^-}^C - c)^+\right)
 \frac{(\psi_{t^-}^C - c)^+}{c\mu}
 +
 \frac{1}{2c\mu} \left((\psi_{t^-}^C + M_1 - c)^+ - (\psi_{t^-}^C - c)^+\right)^2
 ,
\end{align*}
which by expanding the quadratic and distributing we can find this to be
\begin{align*}
&
\left((\psi_{t^-}^C + M_1 - c)^+ - (\psi_{t^-}^C - c)^+\right)
 \frac{(\psi_{t^-}^C - c)^+}{c\mu}
 +
 \frac{1}{2c\mu} \left((\psi_{t^-}^C + M_1 - c)^+ - (\psi_{t^-}^C - c)^+\right)^2
 \\&\quad=
 \frac{1}{c\mu}
 \left((\psi_{t^-}^C + M_1 - c)^+(\psi_{t^-}^C - c)^+ - (\psi_{t^-}^C - c)_+^2\right)
 \\&\qquad
 +
 \frac{1}{2c\mu} \left((\psi_{t^-}^C + M_1 - c)_+^2 - 2(\psi_{t^-}^C + M_1 - c)^+(\psi_{t^-}^C - c)^+ + (\psi_{t^-}^C - c)_+^2 \right)
 \\&\quad=
 \frac{1}{2c\mu} \left((\psi_{t^-}^C + M_1 - c)_+^2 -  (\psi_{t^-}^C - c)_+^2 \right)
 ,
\end{align*}
which is the stated result.
\Halmos\\
\endproof
}

%\tr{define $\lambda$}

%\tr{waiting time identity for $M^{B(n)}/M/cn$ system in steady-state}

\edit{Following the distributional result in Proposition~\ref{waitLimit}, we prove that if the arrival process is stationary Poisson at rate $\lambda > 0$, then the steady-state mean waiting time in the $M^{B(n)}/M/cn$ system converges to a simple expression in terms of the expected steady-state storage exceedance.}

\begin{proposition}\label{waitID}
\edit{Suppose that a batch arrives to the $M^{B(n)}/M/cn$ system in steady-state, and let $W_{1,j}$ be the waiting time of customer $j$ within the batch. Then, the mean wait within the batch converges to
\begin{align}
\E{\frac{1}{n}\sum_{j=1}^{B_1(n)} W_{1,j}}
\longrightarrow
\frac{1}{\lambda}
\E{(\psi^C - c)^+}
,
\end{align}
as $n \to \infty$.}
\end{proposition}
\proof{Proof.}
\edit{
By Theorem~\ref{inter} we are justified in taking the interchange of limits and moving straight to analyzing the storage process  Leveraging the Markovian nature of this model form, we can recognize that the infinitesimal generator of the storage process yields that
\begin{align*}
\frac{\mathrm{d}}{\mathrm{d}t}\E{(\psi_t - c)_+^2}
&=
\lambda \E{(\psi_t^C + M_1 - c)_+^2 - (\psi_t^C - c)_+^2}
-
2c\mu \E{(\psi_t^C - c)^+}
,
\end{align*}
since the model is a piecewise deterministic Markov process \citep[see, e.g.,][]{davis1984piecewise}. In steady-state, this ODE yields an equilibrium equation of 
\begin{align*}
0
&=
\lambda \E{(\psi^C + M_1 - c)_+^2 - (\psi^C - c)_+^2}
-
2c\mu \E{(\psi^C - c)^+}
,
\end{align*}
and thus by Proposition~\ref{waitLimit} we simplify to the stated expression.
\Halmos\\
\endproof
}

\section{\edit{Further Details of the Section~\ref{caseStudy2} Contact Tracing Simulations}}\label{simApp}

\edit{In this section of the appendix, we provide further details of the contact tracing simulations. First, in Algorithm~\ref{batchBreak}, we give the stick-breaking-type procedure for splitting the true batch size data according to the synthetic arrival rate in the $M^B/M/c$ contact tracing simulation model in the Section~\ref{caseStudy2} experiment.}

\begin{algorithm}[htb]
\SetAlgoLined
\edit{
\textbf{Input:} Daily arrival rate $\lambda > 0$, true daily batch size sequence $B_1, \dots, B_7$ for the week.
}

\edit{
\textbf{Output:} Batch sizes $\tilde B_1, \dots, \tilde B_N$, where $N \sim (\mathsf{Pois}(7\lambda) \wedge 1)$.
}

\begin{enumerate}
\item \edit{Generate the total number of batches in the week, $N \sim (\mathsf{Pois}(7\lambda) \wedge 1)$.}\\ \edit{If $N = 1$, \textbf{return} $\tilde B_1 = \sum_{i=1}^7 B_i$.}
\item \edit{Generate $U_{(1)} < \dots < U_{(N-1)}$ as ordered i.i.d.~$\mathsf{Uni}(0,7)$ random variables.}
\item \edit{Find the the new cumulative batch sizes according to the ordered uniforms:}
\begin{enumerate}
\item \edit{Set $\bar B_1 = \sum_{i=1}^{\lfloor U_{(1)}\rfloor} B_i + (U_{(1)} - \lfloor U_{(1)}\rfloor) B_{\lceil U_{(1)}\rceil}$.}
\item \edit{For $2 \leq \ell \leq N-1$, set $\bar B_\ell = \sum_{i=1}^{\lfloor U_{(\ell)}\rfloor} B_i + (U_{(\ell)} - \lfloor U_{(\ell)}\rfloor) B_{\lceil U_{(\ell)}\rceil}$.}
\item \edit{Set $\tilde B_N = \sum_{i=1}^7 B_i $.}
\end{enumerate}
\item \edit{\textbf{return} the split batch sizes $\tilde B_\ell = \bar B_\ell - \bar B_{\ell-1}$ for each $\ell \in \{1, \dots, N\}$, with $\bar B_0 = 0$.}
\end{enumerate}
 \caption{\edit{Randomized  Splitting of a Week's Batch Sizes in Contact Tracing Simulation}}
 \label{batchBreak}
\end{algorithm}

\edit{This pseudocode is not overly complex; hence the aim of its inclusion is for completeness and clarity. In particular, let us note that this sub-routine both preserves the total number of cases in each of the 74 weeks and maintains any day-to-day heterogeneity in the \citet{blaney2022covid} and \citet{nyc2023nyc} data. That is, say for example that the batch size for the first day of a given week was much larger than the other six days, i.e.~$B_1 \gg B_i$ for $2 \leq i \leq 7$, and say that $U_{(\ell)} = \ell/3$ for $1 \leq \ell  < N = 21$. Then, $\tilde B_1$, $\tilde B_2$, and $\tilde B_3$ will also each be larger than $\tilde B_\ell$ for $\ell \geq 4$. 
}

\begin{figure}[htb]
\centering
\includegraphics[width=\textwidth]{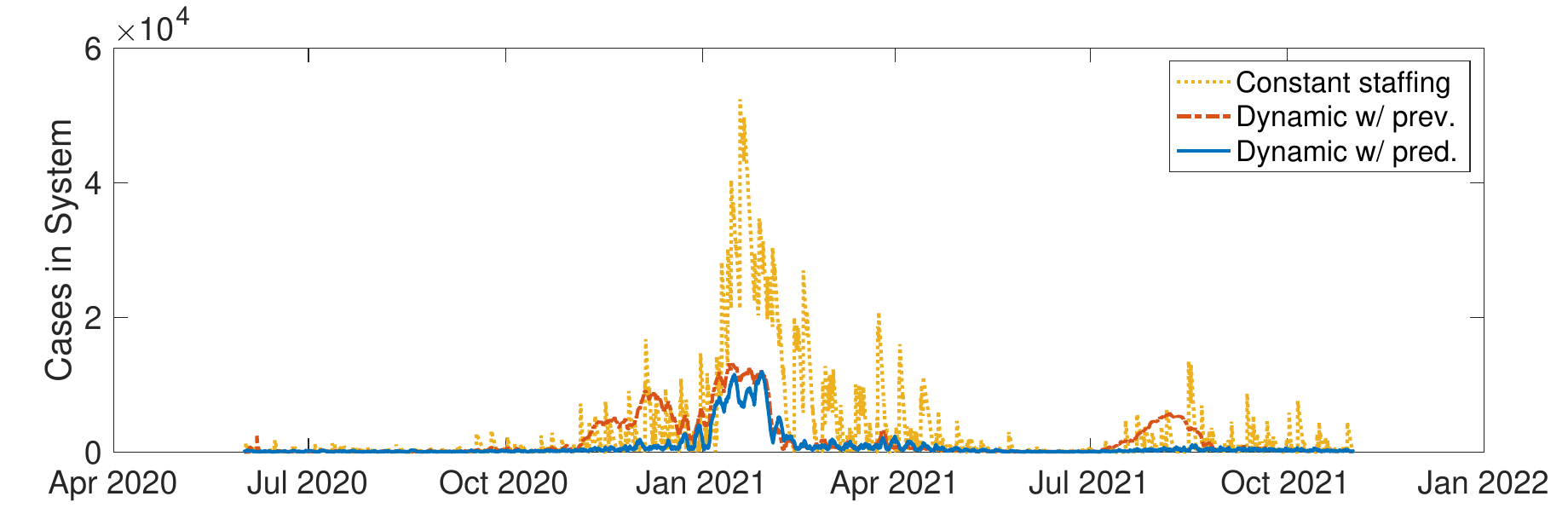}
\caption{\edit{Sample paths of the number of cases in system under the three rate-staffing policies given in the Section~\ref{caseStudy2} experiment.}}\label{CTsamplePath1}
\end{figure}

\edit{For further intuition on the Section~\ref{caseStudy2} experiment, in Figure~\ref{CTsamplePath1} we show one $Q_t^C$ sample path under each of the three rate-staffing policies. As Figure~\ref{CTdynFig} summarizes and formalizes, we can see in each of these individual replications that the caseload, and thus, the waiting, is higher under the $(1, 937)$ policy than under either of the dynamic policies. The only notable exceptions to this come in a late 2020 stretch and Fall 2021 stretch, during which times the constant policy has a lower number in system relative to the dynamic policy that sets the arrival rate and staffing according to last week's arrival volume. By observing the left-most plot in Figure~\ref{CTdynFig}, we can see that this is precisely when case counts are rising, and because the prediction lags in this particular dynamic policy. }

\begin{figure}[htb]
\centering
\includegraphics[width=.8\textwidth]{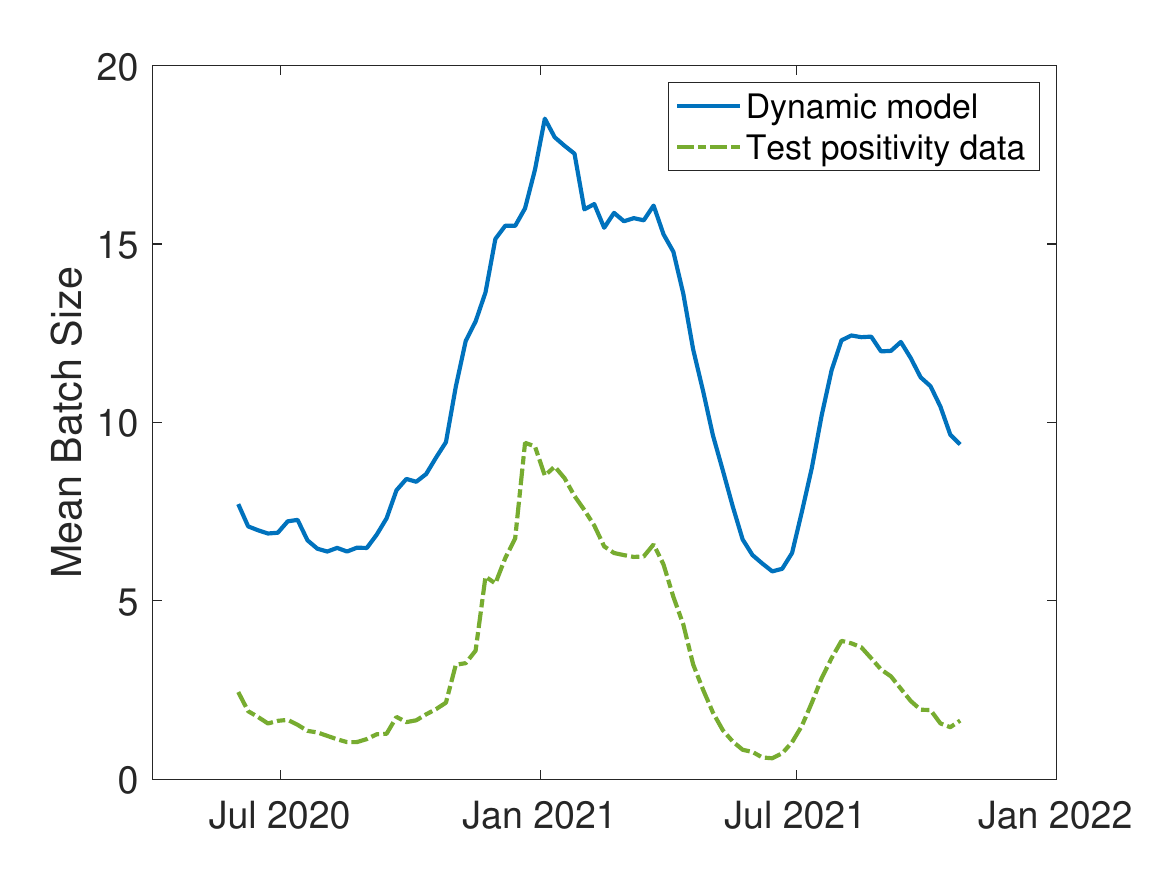}
\caption{\edit{Comparison of the implied mean batch sizes under the predictive policies in the Section~\ref{caseStudy2} experiment and the most realistic smallest mean batch size according to test positivity data.}}\label{batchSizeFig}
\end{figure}

\edit{Finally, in Figure~\ref{batchSizeFig}, we plot the batch size that results according to the $(m^{2/3}, (m/\mu + \delta m^{2/3} \wedge 937)$ policy and compare it to what may be the smallest feasible batch size in reality. That is, using the daily test positivity data \citep{nyc2023nyc} and the 96-well standard of PCR tests \citep{emery2004real}, the green dash-dot curve in Figure~\ref{batchSizeFig} gives what would be essentially the mean batch size if the results of all PCR tests were reported directly to Test \& Trace in real time. As we can see, the dynamic rate-staffing policy is close to this level, but there is still some implied aggregation.}

%\tr{other things to add: stick-breaking pseudocode? offered load?}
\edit{To that end, let us note that the waiting time metrics used throughout Section~\ref{contactSec} exclusively measure the wait from when Test \& Trace receives the cases to when the case investigation begins. However, for the actual patients that Test \& Trace serves, there is also waiting from when the test is taken to when Test \& Trace receives the results. While we do not model or quantify this explicitly, it is clear that more frequent batches would also reduce this wait, as aggregation also incurs waiting upstream of Test \& Trace.}

\edit{Finally, let us also refer here to the extended contact tracing simulation in Appendix~\ref{caseStudy3}. This experiment extends to data beyond the \citet{blaney2022covid} window to assess the substantial challenges of the Omicron wave. Furthermore, it is also an example of an arrival pattern control problem with a different objective, as this follows the ``all-hands-on-deck'' philosophy and leaves staffing at $c = 937$. Simply by controlling the arrival-rate-batch-size tradeoff, managers can optimize costs that arrive from processing batches and from cases waiting.}

\section{Calculating Staffing Levels through Storage Processes}\label{staffingSec}

%A key tool in our calculations will be the convergence of a sum of exponential functions to the indicator function $\mathbf{1}\{x \leq c\}$, as shown in \citet{sullivan1980approximation}. This is of particular use to us in calculating the probability of a storage process exceeding a threshold, through which we drive our staffing analysis. 
%One can note that an alternative approach to this would be to leverage asymptotic normality results, such as those in \citet{lane1984central,rice1977generalized} for shot noise processes. While this may work well in some settings, we can note that for systems that require a very rapid service rate,  such as in the look-ahead service, the rate of arrivals may not be fast enough to justify a Gaussian approximation. This is of particular concern for approximating the tails of the distribution, which is at the heart of this problem. However, such an approximation could be promising in areas large enough to have a significant number of miles driven, and thus we discuss a normal-based approximation in our numerical studies in Section~\ref{numerSec}. Moreover, this simple approximation helps us build intuition for our general results on how batches impact service systems.

Now that we have developed an understanding of queues with large batch arrivals through connections to storage processes, in this section we will leverage this insight and use the storage process models to staff the batch arrival queue. In Section~\ref{model}, we defined the staffing problem as finding a staffing threshold $c$ such that the desired exceedance probability,
$
\PP{Q_t^C(n) \geq cn}
$,
is smaller than some target $\epsilon > 0$. We have also discussed how one could consider the probability of other, stricter events, like $\PP{Q_t^C(n) + B_i(n) \geq cn}$. By normalizing these events by $n$, we can see that the batch scaling limit in Theorem~\ref{fDelayConv} yields that
$$
\PP{Q_t^C(n) \geq cn} \to \PP{\psi_t^C > c}
\quad
\text{and}
\quad
\PP{Q_t^C(n) + B_i(n) \geq cn} \to \PP{\psi_t^C + M_i > c}
,
$$
as $n \to \infty$, allowing us to ``staff'' the storage process instead. One general approach to this problem would be to take a simulation-based approach, such as the well-known iterative staffing algorithm introduced in \citet{feldman2008staffing}. It is thus worth noting that the results of Theorems~\ref{fBatchScale} and~\ref{fDelayConv} have an immediate consequence of greatly simplifying the simulation of batch arrival queueing systems. For large batch sizes, one can simply simulate a storage process instead. This only requires generating random variables for the arrival epochs and jump sizes; one need not simulate service durations. In the large batch setting, this can deliver substantial savings in computation complexity, as large batches mean that a large number of service durations must be generated.

To draw upon results from the storage process literature and calculate explicit staffing levels, we will now assume that we are in the Markovian setting with $N_t$ as a Poisson process with rate $\lambda > 0$ and with exponential service at rate $\mu > 0$. In this case, we are able to make use of a closed form expression for the moment generating function of the shot noise process, which is
\begin{align}
\E{e^{\theta \psi_t}}
=
e^{\theta\psi_0 \bar{G}_0(t) + \lambda \int_0^t \left(\E{e^{\theta M_1 \bar{G}(x)}}-1\right)\mathrm{d}x}
\label{psiMGF}
.
\end{align}
Following standard stability assumptions for multi-server queueing models we will also suppose
$
\lambda \E{B_1(n)} < c n \mu
$
for all $n \in \mathbb{Z}^+$ and we suppose that in the limit we have $\lambda \E{M_1} < c \mu$ as well. Thus, the objects we use to determine the staffing levels will be the storage and shot noise processes in steady-state. We denote these as $\psi^C$ and $\psi^\infty$, respectively. We now cite a result from the storage process literature providing integral equations for the steady-state densities of  $\psi^\infty$ and $\psi^C$ in Lemma~\ref{intEqSNandC}.

\begin{lemma}\label{intEqSNandC}
The steady-state density of the shot noise process $f_\infty(\cdot)$ exists and is given by the unique solution to the integral equation
\begin{align}
f_\infty(x) = \frac{\lambda}{\mu x}\int_0^x \PP{M_1 > x- y} f_\infty(y) \mathrm{d} y,
\end{align}
for all $x > 0$. Furthermore, the steady-state density of the storage process $f_C(\cdot)$ exists and is given by the unique solution to the integral equation
\begin{align}
f_C(x) = \frac{\lambda}{\mu(x \wedge c)}\int_0^x \PP{M_1 > x- y} f_C(y) \mathrm{d} y,
\end{align}
for all $x > 0$.
\end{lemma}
\proof{Proof.}
This follows directly from Theorem 5 of \citet{brockwell1982storage}.
\Halmos\\
\endproof

As an alternate representation of the integrals in Lemma~\ref{intEqSNandC}, we can observe that in the case of the threshold storage process, for example, we have
\begin{align}
\int_0^x \PP{M_1 > x- y} f_C(y) \mathrm{d} y
=
\PP{M_1 + \psi^C > x}
-
\PP{\psi^C > x}
,
\label{threshAlt}
\end{align}
since $\int_0^\infty \PP{M_1 > x- y} f_C(y) \mathrm{d} y = \PP{M_1 + \psi^C > x}$ and $\PP{M_1 > x- y} = 1$ for all $y > x$. This expression will be of use to us in relating the two processes, further enabling us to use the shot noise process to understand the threshold storage process, just as we have used the infinite server queue to understand the multi-server queue. To begin, in Theorem~\ref{inter} we will now use this alternate expression to justify our study of the stationary setting through a validation of the interchange of the limits of time and of the batch scaling.

\begin{theorem}\label{inter}
In the stationary Markovian infinite server and delay queueing models, the interchange of limits of time and batch scaling is justified. That is,
\begin{align}
\lim_{n \to \infty} \lim_{t \to \infty} \PP{\frac{Q_t^\infty(n)}{n} \leq x} = \lim_{t \to \infty} \lim_{n \to \infty} \PP{\frac{Q_t^\infty(n)}{n} \leq x},
\label{interIS}
\end{align}
and
\begin{align}
\lim_{n \to \infty} \lim_{t \to \infty} \PP{\frac{Q_t^C(n)}{n} \leq x} = \lim_{t \to \infty} \lim_{n \to \infty} \PP{\frac{Q_t^C(n)}{n} \leq x},
\label{interC}
\end{align}
for all $x > 0$.
\end{theorem}
\proof{Proof.}
For the infinite server queueing model, this interchange can be quickly observed through differential equations for the moment generating functions of $Q_t^\infty(n)$ and $\psi_t^\infty$. Let $\mathcal{M}^n(\theta, t)$ be the moment generating function of the scaled Markovian infinite server queue, i.e. $\mathcal{M}^n(\theta, t) = \E{e^{\frac{\theta}{n} Q_t^\infty(n)}}$. Then,  $\mathcal{M}^n(\theta, t)$ satisfies
$$
\frac{\partial  \mathcal{M}^n(\theta, t)}{\partial t}
=
\lambda \left( \E{e^{\frac{\theta}{n} B_1(n)}} - 1 \right) \mathcal{M}^n(\theta, t)
+
n \mu \left( e^{-\frac{\theta}{n}} - 1 \right) \frac{\partial  \mathcal{M}^n(\theta, t)}{\partial \theta}
,
$$
since $\frac{\partial  \mathcal{M}^n(\theta, t)}{\partial \theta} = \E{\frac{Q_t(n)}{n} e^{\frac{\theta}{n} Q_t^\infty(n)}}$. Then, we have that for any $n \in \mathbb{Z}^+$ the moment generating function of the steady-state queue, say $\mathcal{M}^n(\theta, \infty)$, will be given by the solution to the time-equilibrium ordinary differential equation
$$
0
=
\lambda \left( \E{e^{\frac{\theta}{n} B_1(n)}} - 1 \right) \mathcal{M}^n(\theta, \infty)
+
n \mu \left( e^{-\frac{\theta}{n}} - 1 \right) \frac{\mathrm{d}  \mathcal{M}^n(\theta, \infty)}{\mathrm{d} \theta}
.
$$
As $n \to \infty$, the limiting steady-state object will then satisfy
$$
0
=
\lambda \left( \E{e^{\theta M_1}} - 1 \right) \mathcal{M}^\infty(\theta, \infty)
-
\mu \theta \frac{\mathrm{d}  \mathcal{M}^\infty(\theta, \infty)}{\mathrm{d} \theta}
.
$$
By comparison, the moment generating function of the shot noise process that yielded in the Markovian case of the batch scaling in Theorem~\ref{fDelayConv}, say $\mathcal{M}^\psi(\theta, t)$, will satisfy
$$
\frac{\partial  \mathcal{M}^\psi(\theta, t)}{\partial t}
=
\lambda \left( \E{e^{\theta M_1}} - 1 \right) \mathcal{M}^\psi(\theta, t)
-
\mu \theta \frac{\partial  \mathcal{M}^\psi(\theta, t)}{\partial \theta}
,
$$
which implies that in steady-state the shot noise process moment generating function, say $\mathcal{M}^\psi(\theta, \infty)$, is given by the solution to
$$
0
=
\lambda \left( \E{e^{\theta M_1}} - 1 \right) \mathcal{M}^\psi(\theta, \infty)
-
\mu \theta \frac{\partial  \mathcal{M}^\psi(\theta, \infty)}{\partial \theta}
.
$$
Hence, $\mathcal{M}^\psi(\theta, \infty) = \mathcal{M}^\infty(\theta, \infty)$, justifying Equation~\eqref{interIS}. To now prove Equation~\eqref{interC}, we start with describing the balance equations for the queue. Letting $\pi_i^n = \lim_{t \to \infty} \PP{Q_t^\infty(n) = i}$ for every $i \in \mathbb{N}$, we have that these steady-state probabilities satisfy
$$
\left(\lambda + \mu(i \wedge cn) \right) \pi_i^n = \lambda \sum_{j=1}^i \PP{B_1(n) = j} \pi_{i-j}^n + \mu (i + 1 \wedge cn) \pi_{i+1}^n
,
$$
for any $n \in \mathbb{Z}^+$. By induction, we can observe that this implies that the probabilities satisfy the recurrence relation
$$
\pi_i^n
=
\frac{\lambda}{\mu (i \wedge cn)} \sum_{j = 1}^i \PP{B_1(n) \geq j} \pi_{i-j}^n
,
$$
for all $i \in \mathbb{Z}^+$. At $i = 1$ this follows immediately from the global balance equation for $\pi_0^n$, so we proceed to the inductive step and assume that the hypothesis holds on $i \in \{1, \dots, k\}$ for some $k \in \mathbb{Z}^+$. Then, through this assumption and the balance equation for $\pi_k^n$, we can observe that
$$
\lambda \pi_k^n
+
\lambda \sum_{j = 1}^k \PP{B_1(n) \geq j} \pi_{k-j}^n
=
\lambda \sum_{j=1}^k \PP{B_1(n) = j} \pi_{k-j}^n + \mu (k + 1 \wedge cn) \pi_{k+1}^n
,
$$
and since $\PP{B_1(n)  \geq 1}$  this simplifies to
$$
\mu (k + 1 \wedge cn) \pi_{k+1}^n
=
\lambda \pi_k^n
+
\lambda \sum_{j = 1}^k \PP{B_1(n) \geq j + 1} \pi_{k-j}^n
=
\lambda \sum_{j = 1}^{k+1} \PP{B_1(n) \geq j} \pi_{k+1-j}^n
,
$$
which completes the induction. With this confirmation of the recursion, let us now observe an alternate representation of the summation within it. That is, for $Q^C(n)$ as the delay model in steady-state, one can note through the law of total probability that
\begin{align*}
\PP{B_1(n) + Q^C(n) \geq i}
&=
\sum_{j=0}^\infty \PP{B_1(n) \geq i - j}\pi_j^n
%\\
%&=
%\sum_{j=0}^{i-1} \PP{B_1(n) \geq i - j}\pi_j^n
%+
%\sum_{j=i}^\infty \PP{B_1(n) \geq i - j}\pi_j^n
\\
&=
\sum_{j=0}^{i-1} \PP{B_1(n) \geq i - j}\pi_j^n
+
\PP{Q^C(n)  \geq i}
,
\end{align*}
since $\PP{B_1(n) \geq i - j} = 1$ for all $j \geq i$. This then implies that one can re-express the recurrence relation as
$$
\pi_i^n
=
\frac{\lambda}{\mu (i \wedge cn)} \left(\PP{B_1(n) + Q^C(n) \geq i} - \PP{Q^C(n)  \geq i}\right)
,
$$
and we can now use this to give a representation for $F^n(x) \equiv \PP{Q^C(n) \leq xn}$. Since $F^n(x) = \sum_{i=0}^{\lfloor xn \rfloor} \pi_i^n$, we have that
\begin{align*}
F^n(x)
=
\pi_0^n
+
\sum_{i=1}^{\lfloor xn \rfloor} \frac{\lambda}{\mu (i \wedge cn)} \left(\PP{B_1(n) + Q^C(n) \geq i} - \PP{Q^C(n)  \geq i}\right)
.
\end{align*}
By changing the step size of the summation to being in increments of $\frac{1}{n}$, this sum becomes
\begin{align*}
F^n(x)
=
\pi_0^n
+
\sum_{ \stackrel{ i=\frac{1}{n},}{ \Delta = \frac{1}{n}}}^{\lfloor xn \rfloor \slash n} \frac{\lambda}{n \mu (i \wedge c)} \left(\PP{\frac{B_1(n)}{n} + \frac{Q^C(n)}{n} \geq i} - \PP{\frac{Q^C(n)}{n}  \geq i}\right)
.
\end{align*}
Letting $Y$ be equivalent in distribution to the limiting object of $\frac{Q^C(n)}{n}$ as $n \to \infty$, we have that  $F^\infty(x) = \PP{Y \leq x}$ is given by
\begin{align}
F^\infty(x)
&=
\int_0^x
\frac{\lambda}{\mu (z \wedge c)} \left(\PP{M_1 + Y \geq z} - \PP{Y \geq z}\right)
\mathrm{d}z
,
\label{delayAlt}
\end{align}
for all $x > 0$, since $\pi_0^n \to 0$ and $\frac{B_1(n)}{n} \stackrel{D}{\Longrightarrow} M_1$ as $n \to \infty$. Using Lemma~\ref{intEqSNandC} and the alternate representation of the integral in Equation~\eqref{threshAlt}, one can note that $F_C(x) = \PP{\psi^C \leq x}$ will be given by
$$
F_C(x)
=
\int_0^x
\frac{\lambda}{\mu (z \wedge c)} \left(\PP{M_1 + \psi^C \geq z} - \PP{\psi^C  \geq z}\right)
\mathrm{d}z.
$$
From  Lemma~\ref{intEqSNandC} we have that $F_C(x)$ is the unique distribution satisfying this equation and thus $Y \stackrel{D}{=} \psi^C$, completing the proof.
\Halmos\\
\endproof

Having now justified the interchange of limits, it is worth noting that in specific settings the integral equations in Lemma~\ref{intEqSNandC} can yield results directly. An example of this is in the case of exponential distributed marks, which arise as the limit of geometrically distributed batches. \edit{We solve the integral equation explicitly and use it to study another optimal arrival pattern problem in Appendix~\ref{waitOptSec}. Deterministic jump sizes also hold a good deal of practical tractability. In this case where $\PP{M_1=1}=1$, Equation~\eqref{threshAlt} implies that the integral equation for $f_C(\cdot)$ becomes
\begin{align}
f_C(x)
&=
\frac{\lambda}{\mu(x\wedge c)}
\left(
F_C(x)
-
F_C(x-1)
\right)
,
\label{deterIntEq}
\end{align}
where $F_C(\cdot)$ is the CDF of the storage process in steady-state. Because $f_C(x) = \frac{\mathrm{d}}{\mathrm{d}x}F_C(x)$, Equation~\eqref{deterIntEq} can be viewed as a delay differential equation for $F_C(\cdot)$. Since $F_C(x-1) = 0$ for $x < 1$, we can find the initial condition that $F_C(x) = k_0 x^{\lambda / \mu}$ for some constant $k_0 > 0$. It is then straightforward to obtain proportional solutions of $F_C(\cdot)$ in each integer interval, and the normalization constant can be numerically approximated by considering a sufficiently large range of $x$ values. This technique is how the storage process values are computed in Figure~\ref{storageVsGaussianFig}. In Section~\ref{genStaff},} we will now develop an asymptotic approach to calculate the exceedance probabilities for general batch sizes. 

%\tr{add that this creates a delay differential equation for constant jump sizes and an explicit solution for exponential jump sizes -- add!}

\subsection{Asymptotic Analysis for General Batch Sizes}\label{genStaff}

To calculate the exceedance probabilities for $\psi^C$, we will again draw upon its relationship with the tractable shot noise process, $\psi^\infty$. Furthermore, we will also make use of a transform method for computing the cumulative distribution function and truncated expectation of a random variable through use of orthogonal Legendre polynomials. This approach is based on an generalization of \citet{sullivan1980approximation}, in which the authors provide a representation for the indicator function through a sum of exponential functions. In Section~\ref{techAppend} of the Appendix, we extend this result for use in studying continuous random variables. Through use of the resulting Lemma~\ref{legendreLemma}, we derive the following expressions for the exceedance probabilities in Theorem~\ref{delayGen}.

\begin{theorem}\label{delayGen}
In the Markovian case, the threshold exceedance probabilities for $\psi^C$ are given by
\begin{align}
%\lim_{m \to \infty}
%\frac{\frac{\lambda}{\mu} \E{M_1}\left(1 - \frac{\lambda}{c\mu} \E{M_1}\right) - \sigma_{m,c} }{c\left(1 - \frac{\lambda}{c\mu} \E{M_1}\right) - \sigma_{m,c} }
% \lim_{m \to \infty}
%1 - \sigma_{m,c}^{(1)}
%\leq
\PP{\psi^C > c}
=
\lim_{\ell \to \infty}
\frac{\frac{\lambda}{\mu} \E{M_1}   - \sigma_{\ell,c}^{(C1)} }{c  - \sigma_{\ell,c}^{(C1)} }
,
\label{delayGenEq1}
\end{align}
and
\begin{align}
\PP{\psi^C + M_1 > c}
=
\lim_{\ell \to \infty}
\frac{
\frac{\lambda}{\mu}\E{M_1}
-
\sigma_{\ell,c}^{(C2)}
}{
c
-
\sigma_{\ell,c}^{(C2)}
}
+
\frac{
\sigma_{\ell,c}^{(C1)}\left( \frac{c\mu}{\lambda}   - \E{M_1}  \right)
}{
\left(c  - \sigma_{\ell,c}^{(C1)} \right) \left(c-\sigma_{\ell,c}^{(C2)} \right)
}
,
\label{delayGenEq2}
\end{align}
where for $\ell \in \mathbb{Z}^+$ and $c$ as the capacity threshold, $\sigma_{\ell,c}^{(C1)}$ is given by
\begin{align}
\sigma_{\ell,c}^{(C1)}
&=
\sum_{k=1}^\ell \frac{ c \lambda   }{\mu k} \left(1 - \E{e^{-\frac{k}{c} M_1}} \right) \frac{ a_k^\ell e^{-\lambda \int_0^\infty \left(1 - \E{e^{-\frac{k}{c} M_1 e^{-\mu x}}}\right) \mathrm{d}x}}
{
\sum_{i=1}^\ell  a_i^\ell e^{-\lambda \int_0^\infty \left(1 - \E{e^{-\frac{i}{c} M_1 e^{-\mu x}}}\right) \mathrm{d}x}
}
,
\end{align}
and $\sigma_{\ell,c}^{(C2)}$ is given by
\begin{align}
\sigma_{\ell,c}^{(C2)}
&=
\sum_{k=1}^\ell
\frac{
\E{M_1 e^{-\frac{k}{c} M_1}}
+
\E{e^{-\frac{k}{c} M_1}}\frac{ c \lambda  }{\mu k} \left(1 - \E{e^{-\frac{k}{c} M_1}} \right)
}{
\sum_{i=1}^\ell a_i^\ell
\E{e^{-\frac{i}{c} M_1}} e^{-\lambda \int_0^\infty \left(1 - \E{e^{-\frac{i}{c} M_1 e^{-\mu x}}}\right) \mathrm{d}x }
}
a_k^\ell
e^{-\lambda \int_0^\infty \left(1 - \E{e^{-\frac{k}{c} M_1 e^{-\mu x}}}\right) \mathrm{d}x }
,
\end{align}
with $a_k^\ell$ as defined in Equation~\eqref{akDef}.
\end{theorem}
\proof{Proof.}
To begin, we first recall that Equation~\eqref{threshAlt} gives us that
$$
(x \wedge c) f_C(x)
=
\frac{\lambda}{\mu}
\left(
\PP{M_1 + \psi^C > x}
-
\PP{\psi^C > x}
\right)
,
$$
and by integrating each side across all $x$ this further implies that
$$
\E{\psi^C \wedge c}
=
\frac{\lambda}{\mu}
\left(
\E{M_1 + \psi^C }
-
\E{\psi^C }
\right)
 = 
 \frac{\lambda}{\mu} \E{M_1}
.
$$
This same expectation can also be expressed through conditioning as
\begin{align*}
\E{\psi^C \wedge c} = c \PP{\psi^C > c} + \E{\psi^C \mid \psi^C \leq c}\left(1 - \PP{\psi^C > c}\right)
,
\end{align*}
and thus by setting these two expressions equal to one another we find that
\begin{align}
\PP{\psi^C > c} = \frac{\frac{\lambda}{\mu} \E{M_1} - \E{\psi^C \mid \psi^C \leq c}}{c - \E{\psi^C \mid \psi^C \leq c}}
.
\label{truncEq}
\end{align}
Although we do not know this truncated mean of $\psi^C$ in closed form,  we can observe that
$$
\E{\psi^C \mid \psi^C \leq c} = \E{\psi^\infty \mid \psi^\infty \leq c} ,
$$
because the integral equations of the these truncated densities are equivalent for all $x \in (0, c]$, as can be observed through Lemma~\ref{intEqSNandC}. Now, by total probability we can recognize that
$$
\E{\psi^\infty \mid \psi^\infty \leq c}
=
\frac{\E{\psi^\infty \mathbf{1}\{\psi^\infty \leq c\}} }{ \PP{\psi^\infty \leq c}}
.
$$
For $\ell \in \mathbb{Z}^+$, we now define the quantities $\sigma_{\ell,c}^{(1)}$ and $\sigma_{\ell,c}^{(2)}$ as
\begin{align*}
\sigma_{\ell,c}^{(1)}
&=
\sum_{k=1}^\ell  a_k^\ell e^{-\lambda \int_0^\infty \left(1 - \E{e^{-\frac{k}{c} M_1 e^{-\mu x}}}\right) \mathrm{d}x}
,
\end{align*}
and
\begin{align*}
\sigma_{\ell,c}^{(2)}
&=
\sum_{k=1}^\ell \frac{ c \lambda  a_k^\ell }{\mu k} \left(1 - \E{e^{-\frac{k}{c} M_1}} \right) e^{-\lambda \int_0^\infty \left(1 - \E{e^{-\frac{k}{c} M_1 e^{-\mu x}}}\right) \mathrm{d}x}
.
\end{align*}
Using Theorem~\ref{fBatchScale} and Lemma~\ref{legendreLemma}, we have that $\sigma_{\ell,c}^{(1)} \to \PP{ \psi^\infty \leq c }$ and $\sigma_{\ell,c}^{(2)} \to \E{\psi^\infty \mathbf{1}\{ \psi^\infty \leq c \}}$ as $\ell \to \infty$. Thus, by substituting $\sigma_{\ell,c}^{(C1)} = \sigma_{\ell,c}^{(2)} \slash \sigma_{\ell,c}^{(1)}$ into Equation~\eqref{truncEq} and simplifying, we achieve the stated form in Equation~\eqref{delayGenEq1}.

To now prove Equation~\eqref{delayGenEq2}, we start by finding an identity for $\E{\psi^C + M_1 \wedge c}$. Because Lemma~\ref{intEqSNandC} implies that the threshold storage process density $f_C(x)$ satisfies
$$
(x \wedge c)f_C(x)
=
\frac{\lambda}{\mu}\left(\PP{\psi^C + M_1 > x} - \PP{\psi^C > x}\right)
,
$$
we are able to observe that
\begin{align*}
\E{\psi^C \mathbf{1}\{\psi^C < c\}}
&=
\int_0^c (x \wedge c) f_C(x) \mathrm{d}x
\\
&=
\frac{\lambda}{\mu} \int_0^c \PP{\psi^C + M_1 > x} \mathrm{d}x
-
\frac{\lambda}{\mu} \int_0^c \PP{\psi^C > x} \mathrm{d}x
\\
&=
\frac{\lambda}{\mu} \E{\psi^C + M_1 \wedge c}
-
\frac{\lambda}{\mu}  \E{\psi^C \wedge c}
.
\end{align*}
Because we know that $\E{\psi^C \wedge c} = \frac{\lambda}{\mu}\E{M_1}$ and $\E{\psi^C \mid \psi^C \leq c} = \E{\psi^\infty \mid \psi^\infty \leq c}$, we can note that this now implies that expectation of the minimum of the threshold and the storage process plus a jump is equal to
$$
\E{\psi^C + M_1 \wedge c}
=
\frac{\mu}{\lambda}\E{\psi^\infty \mid \psi^\infty \leq c} \PP{\psi^C \leq c}
+
\frac{\lambda}{\mu}\E{M_1}
,
$$
all of which on the right-hand side we now know how to calculate. Then, by mimicking the conditioning decomposition we used previously on $\E{\psi^C \wedge c}$, we can note that $\E{\psi^C + M_1 \wedge c}$ is also equal to
$$
\E{\psi^C + M_1 \wedge c}
=
c \PP{\psi^C + M_1 > c}
+
\E{\psi^C + M_1 \mid \psi^C + M_1 \leq c}\left(1 - \PP{\psi^C + M_1 > c}\right)
.
$$
By setting these two expressions for $\E{\psi^C + M_1 \wedge c}$ equal to one another and solving for $\PP{\psi^C + M_1 > c}$, we have that
\begin{align}
\PP{\psi^C + M_1 > c}
=
\frac{
\frac{\mu}{\lambda}\E{\psi^\infty \mid \psi^\infty \leq c} \PP{\psi^C \leq c}
+
\frac{\lambda}{\mu}\E{M_1}
-
\E{\psi^C + M_1 \mid \psi^C + M_1 \leq c}
}{
c
-
\E{\psi^C + M_1 \mid \psi^C + M_1 \leq c}
}
\end{align}
Again through the integral equations, we can recognize that $\E{\psi^C + M_1 \mid \psi^C + M_1 \leq c} = \E{\psi^\infty + M_1 \mid \psi^\infty + M_1 \leq c}$. Because $M_1$ is independent from the state of the shot noise process $\psi^\infty$, we have that
$$
\E{e^{\theta(\psi^\infty + M_1)}}
=
\E{e^{\theta \psi^\infty}}\E{e^{\theta M_1}}
=
\E{e^{\theta M_1}} e^{-\lambda \int_0^\infty \left(1 - \E{e^{\theta M_1 e^{-\mu x}}}\right) \mathrm{d}x}
,
$$
by use of Theorem~\ref{fBatchScale} and Equation~\eqref{psiMGF}. Then, for $\ell \in \mathbb{Z}^+$ let us additionally define $\sigma_{\ell,c}^{(3)}$ and $\sigma_{\ell,c}^{(4)}$ such that
$$
\sigma_{\ell,c}^{(3)}
=
\sum_{k=1}^\ell a_k^\ell
\E{e^{-\frac{k}{c} M_1}} e^{-\lambda \int_0^\infty \left(1 - \E{e^{-\frac{k}{c} M_1 e^{-\mu x}}}\right) \mathrm{d}x }
,
$$
and
$$
\sigma_{\ell,c}^{(4)}
=
\sum_{k=1}^\ell a_k^\ell
\left(
\E{M_1 e^{-\frac{k}{c} M_1}}
+
\E{e^{-\frac{k}{c} M_1}}\frac{ c \lambda  }{\mu k} \left(1 - \E{e^{-\frac{k}{c} M_1}} \right)
\right)
e^{-\lambda \int_0^\infty \left(1 - \E{e^{-\frac{k}{c} M_1 e^{-\mu x}}}\right) \mathrm{d}x }
.
$$
Through these definitions, Lemma~\ref{legendreLemma} yields that $\sigma_{\ell,c}^{(3)} \to \PP{\psi^\infty + M_1 \leq c}$ and $\sigma_{\ell,c}^{(4)} \to \E{(\psi^\infty + M_1) \mathbf{1}\{ \psi^\infty + M_1 \leq c\}}$ as $\ell \to \infty$. Thus we have that $\sigma_{\ell,c}^{(C2)} = \sigma_{\ell,c}^{(4)}\slash\sigma_{\ell,c}^{(3)} \to \E{\psi^\infty + M_1 \mid \psi^\infty + M_1 \leq c}$, and this completes the proof.
\hfill\Halmos\\
\endproof

%The derivations behind this computational methodology also enable us to provide a closed form expression for the  utilization of the teleoperators, meaning the ratio between the mean number of busy servers and the total number employed. Because the number of busy servers is the minimum of the number of jobs in the system and the staffing level, we can make use of the expected value of the minimum of the storage process and the threshold $c$. In Corollary~\ref{utilCor}, we observe that with this expectation in hand, the utilization is easy to compute.
%
%
%\begin{corollary}\label{utilCor}
%In the Markovian case, the steady-state utilization of the teleoperators in the large batch setting is given by $\frac{1}{c}\E{\psi^C \wedge c} = \frac{\lambda}{c\mu}\E{M_1}$.
%\begin{proof}
%In the proof of Theorem~\ref{delayGen}, we have seen that $\E{\psi^C \wedge c} = \frac{\lambda}{\mu}\E{M_1}$. Through the batch scaling in Theorem~\ref{delayConv}, the teleoperation utilization converges to
%$$
%\E{\frac{1}{cn}\left(Q_\infty^C(n) \wedge cn\right)}
%\longrightarrow
%\frac{1}{c} \E{\psi^C \wedge c}
%=
%\frac{\lambda}{c\mu}\E{M_1}
%,
%$$
%as $n \to \infty$.
%\end{proof}
%\end{corollary}

%It is worth noting that in some experiments in Section~\ref{numerSec} we make use of log-normally distributed marks in the storage process. Although the moment generating function of the log normal is not known in closed form, we are able to use the approximation provided in \citet{asmussen2016laplace}, which performs quite well in simulation comparisons.

As a side consequence of the proof of Theorem~\ref{delayGen}, we can also identify a practical, closed-form upper bound on $\PP{\psi^C > c}$. To do so we bound first find a lower bound for the truncated mean $\E{\psi^C \mid \psi^C \leq c} = \E{\psi^\infty \mid \psi^\infty \leq c}$. Letting $\bar{f}(x)$ be the truncated density on $(0,c]$, through Lemma~\ref{intEqSNandC} we then have that
$$
\E{\psi^\infty \mid \psi^\infty \leq c}
=
\int_0^c
x \bar{f}(x) \mathrm{d}x
=
\int_0^c
\frac{\lambda}{\mu}
\left(
\PP{M_1 + {\psi}^\infty > x \mid \psi^\infty \leq c}
-
\PP{{\psi}^\infty > x \mid \psi^\infty \leq c}
\right)
\mathrm{d}x
.
$$
Because $\int_0^c \PP{\psi^\infty > x \mid \psi^\infty \leq c} \mathrm{d}x = \E{\psi^\infty \mid \psi^\infty \leq c}$, we have
$$
\E{\psi^\infty \mid \psi^\infty \leq c}
=
\frac{\lambda}{\lambda + \mu}
\int_0^c
\PP{M_1 + {\psi}^\infty > x \mid \psi^\infty \leq c}
\mathrm{d}x
.
$$
Then, by observing that $\PP{M_1 + \psi^\infty > x \mid \psi^\infty \leq c} \geq \PP{M_1 > x}$ through the independence of the two quantities and the fact that each is positive, we furthermore have
$$
\E{\psi^\infty \mid \psi^\infty \leq c}
\geq
\frac{\lambda}{\lambda + \mu}
\int_0^c
\PP{M_1 > x}
\mathrm{d}x
=
\frac{\lambda}{\lambda + \mu}
\E{M_1 \wedge c}
.
$$
Using the decomposition in Equation~\eqref{truncEq}, this now yields the upper bound
$$
\PP{\psi^C > c}
\leq
\frac{\frac{\lambda}{\mu}\E{M_1} - \frac{\lambda}{\lambda + \mu}\E{M_1 \wedge c}}
{c - \frac{\lambda}{\lambda + \mu}\E{M_1 \wedge c}}
.
$$
This bound is most helpful in cases of small $\lambda$, as in that case $\psi^\infty$ is likely to be small.

\section{\edit{Optimal Arrival Patterns for Wait-Based Objectives with Constant Staffing}}\label{waitOptSec}

%\tr{$M^M/M/cn$ system -- geometric batches -- exponential marks -- can get distribution in closed form }

\edit{In this section of the appendix, we will establish a second notion of \emph{optimal arrival patterns}. Here, we will now be optimizing for wait-based costs rather than the staffing-based costs that we studied in Section~\ref{caseStudy2}. The aim of this analysis and case study is once again illustrative, and so to enable straightforward computations we will develop this concept for the $M^M/M/cn$ queueing system. That is, let us assume that we have batches of geometric size: $B_i(n) \sim \mathsf{Geo}(\alpha/n)$ for some $\alpha > 0$. For stability of the system, we will assume $\lambda < \alpha c \mu$. }

%\edit{Assuming $\lambda < \alpha c \mu$}

\subsection{\edit{Large Batch Limit of the Markovian System with Geometric Batches}}

\edit{To begin, we specify the large batch limit of this particular system. To leverage Theorem~\ref{fDelayConv}, let us first interpret the jump size distribution for the limit of the batch sizes themselves. For any $\theta < -n\log(1- \alpha/n)$, we find that
\begin{align}
\E{e^{\theta B_1(n) / n}}
&=
\frac{
\frac{\alpha}{n}e^{\theta / n}
}{
1 - \left(1 - \frac{\alpha}{n}\right)e^{\theta / n}
}
=
\frac{
\alpha
}{
\alpha - n(1 - e^{-\theta / n})
}
\longrightarrow
\frac{\alpha}{\alpha - \theta}
,
\end{align}
and thus the limiting marks are exponentially distributed: $M_i \sim \mathsf{Exp}(\alpha)$. As one might expect, this distribution grants considerable tractability. Indeed, in Proposition~\ref{expJumpProp} we obtain the density of the steady-state storage process in closed form.
}

\begin{proposition}\label{expJumpProp}
\edit{
As $n \to \infty$, the batch scaling of the $M^{M(n)}/M/cn$ steady-state queue $Q^C(n)$ yields $Q^C(n) / n \Longrightarrow \psi^C$, where $\psi^C$ has density $f(x)$ given by
\begin{align}
f(x)
&=
\frac{
1
 }{
\alpha^{-\frac{\lambda}{\mu}} \gamma \left(\frac{\lambda}{\mu}, \alpha c \right) 
+ 
\frac{c^{\frac{\lambda}{\mu} - 1}e^{-\alpha c} }  {  \alpha - \frac{\lambda}{c\mu} }
}
(x \wedge c)^{\frac{\lambda}{\mu}-1} e^{-\left(\alpha - \frac{\lambda}{c\mu}\right) x - \frac{\lambda}{c\mu}(x \wedge c)}
,
\label{expJumpPDF}
\end{align}
for all $x > 0$, with $\gamma(\cdot)$ as the lower incomplete gamma function.
}
\end{proposition}
\proof{Proof.}
\edit{
From Theorem~\ref{inter}, we are justified in taking the interchange of limits, and we can manipulate Lemma~\ref{intEqSNandC} to yield that $f(x)$ satisfies the simplified integral equation
\begin{align*}
f(x)
&=
\frac{\lambda}{\mu (x \wedge c)} \int_0^x e^{-\alpha(x-y)} f(y) \mathrm{d}y
=
\frac{\lambda e^{-\alpha x}}{\mu (x \wedge c)} \int_0^x e^{\alpha y} f(y) \mathrm{d}y
.
\end{align*}
Letting $h(x) = e^{\alpha x} f(x)$, this means that we have
\begin{align*}
(x\wedge c)h(x) 
&=
\frac{\lambda}{\mu}\int_0^x h(y) \mathrm{d}y
,
\end{align*}
and this integral equation becomes tractable two solve case-wise as an ordinary differential equation. For $x \in (0, c)$, we have $h(x) + x h'(x) = \frac{\lambda}{\mu}h(x)$, or, equivalently,
\begin{align*}
h'(x)
&=
\frac{1}{x}\left(\frac{\lambda}{\mu} - 1 \right) h(x)
,
\end{align*}
and this yields that $h(x) = k_1 x^{\frac{\lambda}{\mu}-1}$ for some constant $k_1$. Then, for $x \geq c$, we simply have
\begin{align*}
h'(x)
&=
\frac{\lambda}{c\mu}h(x)
,
\end{align*}
which provides $h(x) = k_2 e^{\frac{\lambda}{c\mu}x}$ for another constant $k_2$. Between the two solutions and the fact that $h$ must be continuous, we now have
\begin{align*}
k_1 c^{\frac{\lambda}{\mu}-1}
&=
h(c)
=
k_2 e^{\frac{\lambda}{\mu}}
,
\end{align*}
and thus we see that $k_2 = k_1 c^{\frac{\lambda}{\mu}-1}e^{-\frac{\lambda}{\mu}}$. Hence, from the definition of $h(x)$, we have that $f(x) = k_1 (x \wedge c)^{\frac{\lambda}{\mu}-1} e^{-\left(\alpha - \frac{\lambda}{c\mu}\right) x - \frac{\lambda}{c\mu}(x \wedge c)}$.
}

\edit{
We are then left to find $k_1$. Integrating the density sans constant, we see that
\begin{align*}
\int_0^c
(x \wedge c)^{\frac{\lambda}{\mu}-1} e^{-\left(\alpha - \frac{\lambda}{c\mu}\right) x - \frac{\lambda}{c\mu}(x \wedge c)}
\mathrm{d}x
&=
\int_0^c
x^{\frac{\lambda}{\mu}-1} e^{-\alpha x}
\mathrm{d}x
=
\frac{1}{\alpha^{\frac{\lambda}{\mu}}}
\gamma\left(
\frac{\lambda}{\mu}
,
\alpha c
\right)
,
\end{align*}
and 
\begin{align*}
\int_c^\infty
(x \wedge c)^{\frac{\lambda}{\mu}-1} e^{-\left(\alpha - \frac{\lambda}{c\mu}\right) x - \frac{\lambda}{c\mu}(x \wedge c)}
\mathrm{d}x
&=
c^{\frac{\lambda}{\mu}-1} e^{-\frac{\lambda}{\mu}}
\int_c^\infty
e^{-\left(\alpha - \frac{\lambda}{c\mu}\right) x}
\mathrm{d}
x
=
\frac{c^{\frac{\lambda}{\mu}-1} e^{-\frac{\lambda}{\mu}}}{\alpha - \frac{\lambda}{c\mu}} e^{-\left(\alpha - \frac{\lambda}{c\mu}\right)c}
,
\end{align*}
and thus we simplify to the stated expression for $f$.
}
\Halmos\\
\endproof

%\tr{remark on conditionally gamma below, conditionally exponential above}

\edit{Equation~\eqref{expJumpPDF} may be even more simple and interpretable than it appears. As can be gleamed from the proof of Proposition~\ref{expJumpProp}, this distribution is conditionally exponential above $c$ and conditionally gamma below it. We can exploit this for two quick additional results. First, in Corollary~\ref{ExpExceedCor}, we obtain the exceedance probability in closed form.}

\begin{corollary}\label{ExpExceedCor}
\edit{As $n \to \infty$ in the batch scaling of the $M^{M(n)}/M/cn$, the steady-state exceedance probability converges to
\begin{align}
\PP{Q^C(n) \geq cn}
\longrightarrow
\frac{
(\alpha c)^{\frac{\lambda}{\mu} }e^{-\alpha c} 
}{
c \left(\alpha - \frac{\lambda}{c\mu}\right)
 \gamma \left(\frac{\lambda}{\mu}, \alpha c \right) 
+ 
(\alpha c)^{\frac{\lambda}{\mu} }e^{-\alpha c} 
}
,
\end{align}
where $\gamma(\cdot)$ is the lower incomplete gamma function.
}
\end{corollary}

\edit{Then, leaning in particular on the conditionally exponential distribution of $\psi^C$ above $c$, we can further refine the waiting time results in Appendix~\ref{waitApp} and obtain the conditional distribution of the asymptotic notion of the full-batch waiting time, meaning the time until an arriving batch first enters service.}

\begin{proposition}\label{expJumpCondWait}
\edit{Let $\tau_W$ be the time until the first start of service for a newly arriving jump within the storage process $\psi^C$ in steady-state. Then, $\tau_W \mid \{\tau_W > 0\} \sim \mathsf{Exp}(\alpha c \mu - \lambda)$ and
\begin{align}
\E{\tau_W \mid \tau_W > 0}
&=
\frac{1}{\alpha c \mu - {\lambda}}
,
\end{align}
where $\PP{\tau_W > 0} = \PP{\psi^C > c}$ is as  given in Corollary~\ref{ExpExceedCor}.
}
\end{proposition}
\proof{Proof.}
\edit{From Proposition~\ref{expJumpProp} and Corollary~\ref{ExpExceedCor}, we can recognize that $\psi^C - c \mid (\psi^C \geq c) \sim \mathsf{Exp}(\alpha  - \frac{\lambda}{c\mu})$. Because arrivals are Poisson, the event $\{\tau_W > 0\}$ is equivalent to the event $\{\psi^C > c\}$, and, moreover, the amount by which $\psi^C$ exceeds $c$ will exactly specify the wait until the start of the batch. (This is formalized by Proposition~\ref{waitLimit}.)  That is, given $\psi^C >c$, the wait will be the excess divided by the service rate, $\tau_W = (\psi^C - c) / (c\mu)$. Taking the expected value using the conditional exponential distribution and simplifying, we achieve the stated expression.}
\Halmos\\
\endproof

\edit{Naturally, this closely resembles the distribution of the conditional waiting time in the classic Erlang-C model. Using the clean tractability offered by the exponentially distributed jumps (and inter-arrival times), we will now develop a corresponding objective for the arrival pattern control problem and extend the contact tracing case study.}

\subsection{\edit{Contact Tracing Experiment Extended: Adjusting Operations for the Omicron Wave}}\label{caseStudy3}

\edit{
Like in Section~\ref{caseStudy2}, here we will solve an arrival pattern control problem, in which a central controller can decide the arrival rate and the mean batch size, so long as the overall arrival volume is maintained. By comparison to the service-cost objective in Equation~\eqref{serveCostObj} and the resulting optimal arrival pattern in Proposition~\ref{serveCostProp}, here we will treat the staffing level as fixed. Because this implies that the staffing depends only on the effective arrival volume and not the underlying arrival rate and batch sizes, this places us in the large batch staffing regime. Hence, we will pose the arrival pattern control problem in the limiting form, meaning in terms of the storage process rather than the queueing system. 
}

\edit{
Using Proposition~\ref{expJumpCondWait}, we will define the \emph{waiting cost objective} for the exponential jump Markovian storage process as 
\begin{align}
\mathcal{C}_2 \E{\tau_W \mid \tau_W > 0} + \mathcal{C}_3 \lambda
&=
\mathcal{C}_2 \frac{1}{\alpha c \mu - \lambda} + \mathcal{C}_3 \lambda
,
\label{waitCostObj}
\end{align}
where $\mathcal{C}_2, \mathcal{C}_3 > 0$. Like in Equation~\eqref{serveCostObj}, here there is a fixed cost for processing batches or jumps, and this results in a linear cost in the arrival rate. Unlike Equation~\eqref{serveCostObj}, the tradeoff in this case is on the waiting time, where there is a penalty for the mean time until service begins for jumps that arrive to a system in excess of the storage capacity. This yields the following optimal arrival pattern.
}

\begin{proposition}\label{waitCostProp}
\edit{In the arrival pattern control problem for the Markovian storage process with $\mathsf{Exp}(\alpha)$ jumps, fixed effective arrival rate $m \in \mathbb{R}_+$ where $m = \lambda / \alpha$, and constant staffing $c \in \mathbb{R}_+$ where $c \mu > m$, the mean steady-state waiting-related costs are minimized if and only if 
\begin{align}
\lambda = \mathcal{C}_{**} \sqrt{ \frac{m}{c\mu - m}}
\qquad
\text{and}
\qquad
\alpha = \mathcal{C}_{**}\sqrt{ \frac{1}{m(c\mu - m)}}
,
\label{waitCostSoln}
\end{align}
where $\mathcal{C}_{**} = \sqrt{{\mathcal{C}_2}/{\mathcal{C}_3}}$.}
\end{proposition}
\proof{Proof.}\edit{Using Equation~\eqref{waitCostObj} and the fact that $m = \lambda / \alpha$, we can re-express the waiting-cost objective as 
\begin{align*}
\mathcal{C}_2 \frac{1}{\alpha c \mu - \lambda} + \mathcal{C}_3 \lambda
&=
\frac{\mathcal{C}_2}{\lambda} \frac{1}{\frac{c \mu}{m} - 1} + \mathcal{C}_3 \lambda
=
\frac{\mathcal{C}_2}{\lambda} \frac{m}{{c \mu} - m} + \mathcal{C}_3 \lambda
,
\end{align*}
and this can now be viewed as a function of the arrival rate. Taking the derivative with respect to $\lambda$, we find
\begin{align*}
\frac{\partial}{\partial \lambda}\left(
\frac{\mathcal{C}_2}{\lambda} \frac{m}{{c \mu} - m} + \mathcal{C}_3 \lambda
\right)
&=
-\frac{\mathcal{C}_2}{\lambda^2} \frac{m}{{c \mu} - m} + \mathcal{C}_3
.
\end{align*}
By setting this derivative equal to 0 and solving for $\lambda$, we can see that Equation~\eqref{waitCostSoln} provides the unique solution, and furthermore it follows immediately from the second derivative that this critical point is in fact the minimum.
\Halmos\\
\endproof
}

%\tr{stylized for simple closed form solutions, but the idea is clear}

\edit{Certainly, this problem has been stylized to enable simple closed form solutions, but we believe the idea is clear. If the service capacity $c\mu$ is substantially larger than the effective arrival volume $m$ (which we would expect for the low utilization large batch regime), then the optimal arrival pattern in Proposition~\ref{waitCostProp} is fairly intuitive: the arrival rate and mean jump size are both roughly the square root of the total arrival volume. However, if $m$ is near $c\mu$, Equation~\eqref{waitCostSoln} sets the arrival rate much faster, and thus the batches will be smaller. Using the mean arrival volume across all days, this is what prescribes the light orange, dash-dot line in Figure~\ref{lamIncFig}.}

\edit{While that curve was included in Figure~\ref{lamIncFig} for the sake of comparison to to other fixed arrival rate levels, we acknowledge that the overall arrival volume $\bar{m}$ is of course not known a priori in a pandemic. So, in the following extended contact tracing experiment, we will take the spirit of the low utilization form of Proposition~\ref{waitCostProp} and dynamically set the arrival rate at square root order of the daily arrival volume for the present week. Like in the Section~\ref{caseStudy2} experiments, this means that the arrival volume is time-varying, but we will make control decisions within each week as though it is in steady-state. Thus, this is effectively a pointwise stationary approximation approach, and this likely could be improved in practice. Again, the aim here is simply to illustrate the impacts of batches and illuminate the tradeoff between the batch size and arrival rate. We leave the time-varying staffing problem to future work.}

\edit{In this final case study component, we expand to consider a data set that spans from June 1, 2020 to May 1, 2022, rather than just the June 1, 2020 to October 31, 2021 timeframe we used in Section~\ref{contactSec}. In doing so, we extend to a timeline that includes not only the start of the Text \& Trace operations, the pre-vaccine surges in Winter 2020-21, and the Delta variant outbreak that were all within the \citet{blaney2022covid} period, but also the highly contagious Omicron variant that came after.\footnote{\edit{A comprehensive timeline of Covid-19 in the United States is available through the CDC Museum Covid-19 Timeline: \url{https://www.cdc.gov/museum/timeline/covid19.html}.}} From publicly available Test \& Trace reports, we know that this era was particularly challenging for the contact tracing effort \citep{nyc2022covid}. As indicated on slides 2 and 3 therein, in the date range corresponding to Section~\ref{contactSec}, approximately 90\% of cases responded to contact and successfully completed investigation. (\citet{blaney2022covid} reports 89.4\% of case investigations were reached.) This is quite consistent across the first year and a half of Test \& Trace data, suggesting something of a fundamental limit for public receptiveness towards contact tracing efforts. However, as the Omicron surge begins at the end of 2021 and start of 2022, the percentage of cases reached craters from approximately 90\% to under 25\%. Naturally, Test \& Trace had to adapt to the significantly more contagious form of the virus. As stated on slide 8 of \citet{nyc2022covid}, ``[s]tarting the week of December 19, 2021, operational changes were made to Trace in response to case surges from the Omicron variant.'' In service operations language, these changes appear to have constituted a reduction in the mean service duration. Through Test \& Trace's adjustments, the number of cases with text-messaging as the method of contact rises from none before Omicron to as many as hundreds of thousands at the Omicron-induced peak. Previously, contact had been exclusively made over the phone or in person.}

%\tr{performance data: public report slides \citet{nyc2022covid}: ``Starting the week of December 19, 2021, operational changes were made to Trace in response to case surges from the Omicron variant.''}

%\tr{slide 3: percentage of cases reached craters from approximately 90\% to under 25\%. to adjust, the number of cases with text message as the method of contact rises from none before Omicron to as much as hundreds of thousands at the Omicron-induced peak. hence, we model the mean service duration at the previous presumed 80 minutes, but also at 50 minutes and 20 minutes.}

%\tr{time-varying, but we will staff within each week as though it is steady-state -- effectively a pointwise stationary approximation, could be improved in practice -- again, aim here is simply to illustrate the impacts of batches}

\begin{figure}[htb]
\centering
\includegraphics[width=\textwidth]{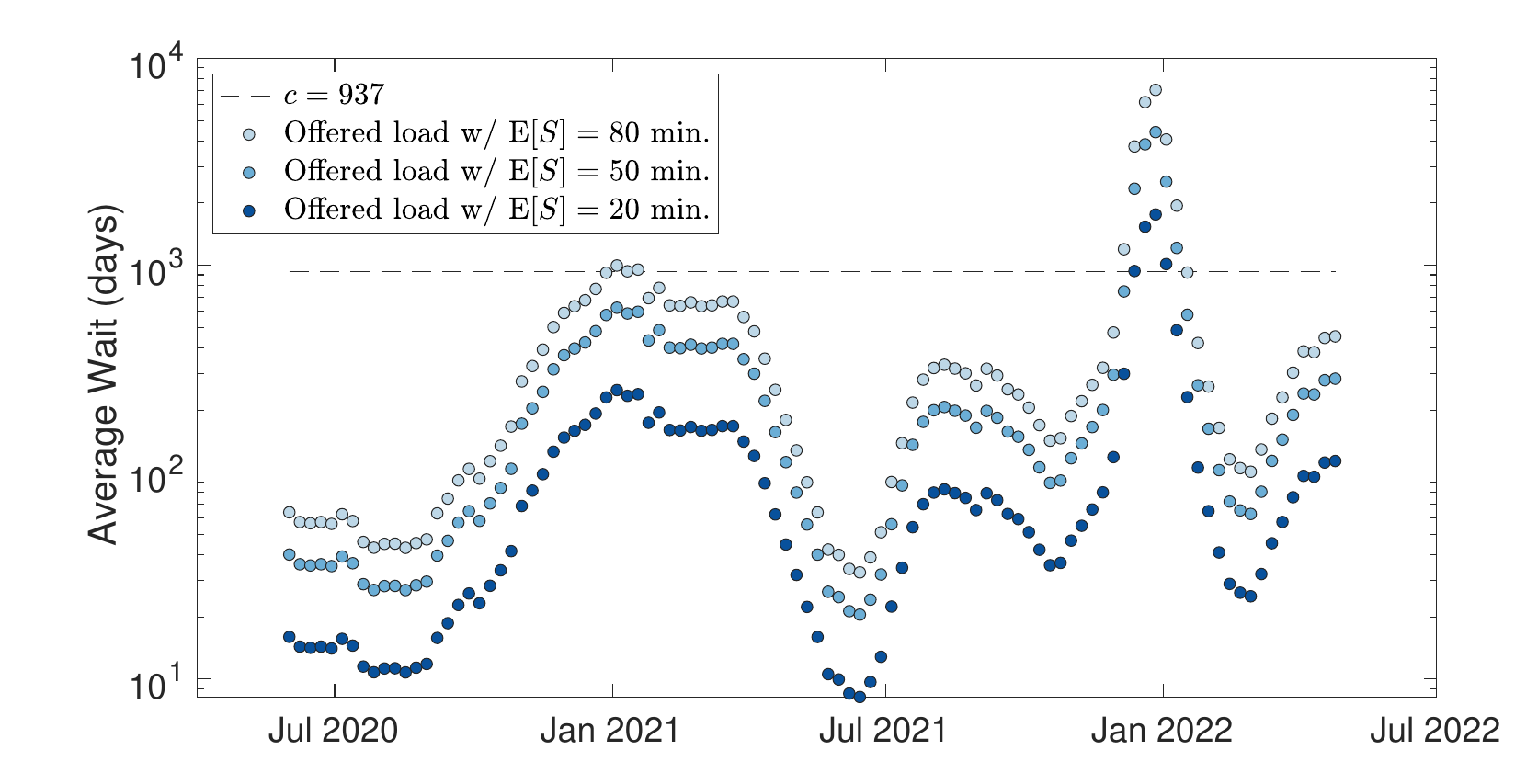}
\caption{\edit{Mean daily case investigation offered load in each week from June 1, 2020 to May 1, 2022 under each of the hypothetical mean service duration scenarios.}}\label{fullOLfig}
\end{figure}

\edit{To model the change to this more efficient communication method, we will simulate the case investigation queueing system's mean service duration at not just the previous presumed 80 minutes, but also at the two shorter options considered in the Section~\ref{caseStudy1} sensitivity analysis: 50 minutes and 20 minutes. Rather than modeling a switch in duration, we simply study each of the three across the full horizon. In Figure~\ref{fullOLfig}, we plot the mean offered load for each of these three mean service durations. By comparison to the queue's constant staffing level, $c = 937$, each duration still leads to the system being overloaded for multiple weeks at the onset of the Omicron surge. Hence, this variant presents an unavoidably large challenge that cannot be entirely mitigated with further reduction of the service duration or an increase of staff. It is possible that one or both of those measures were pursued. However, given the focus of this paper, what we will now study in this simulation experiment is how much can be smoothed by exploiting the arrival-rate-batch-size tradeoff.}

\begin{figure}[htb]
\centering
\includegraphics[width=\textwidth]{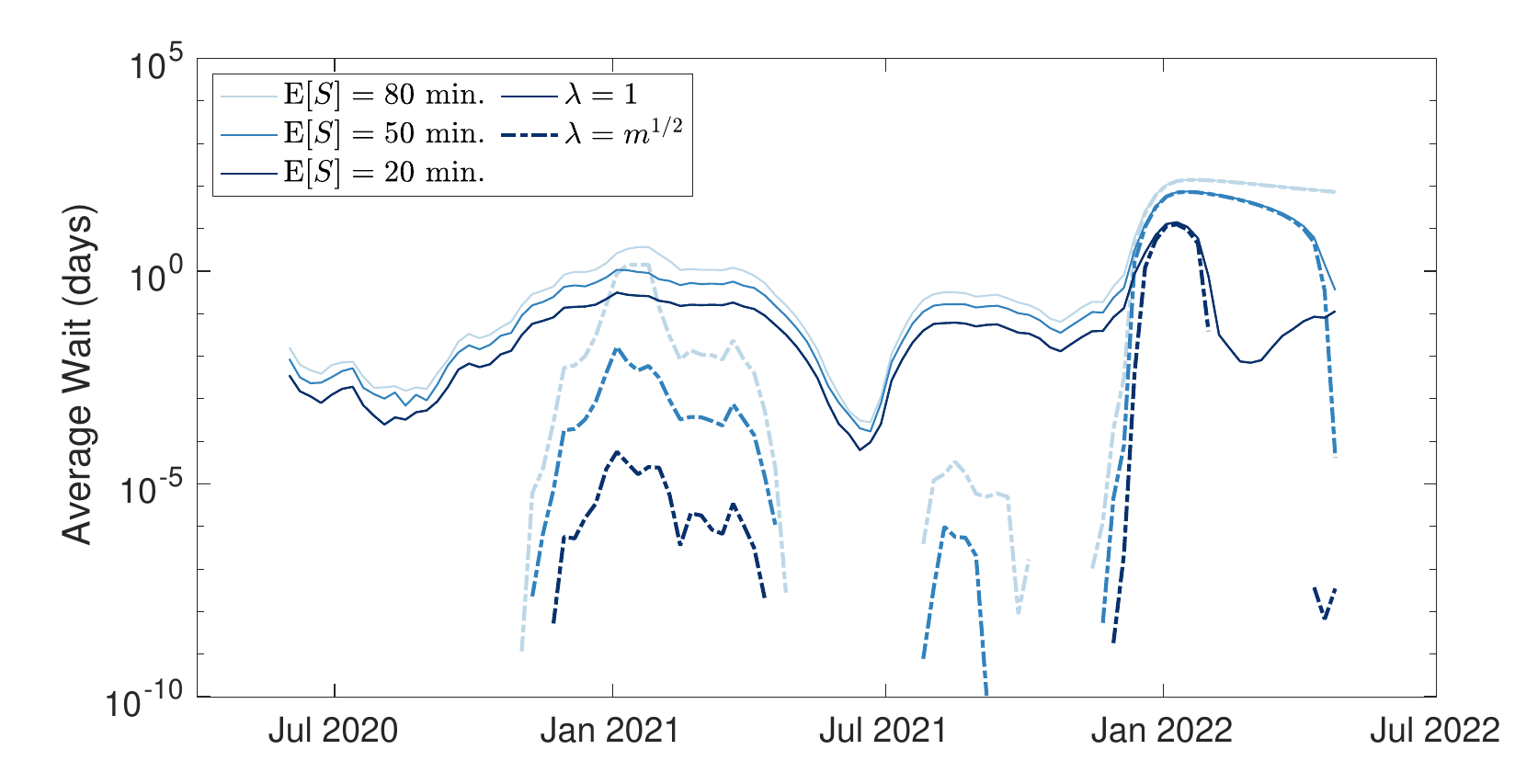}
\caption{\edit{Average wait per case within each day's arrivals under each of the hypothetical mean service durations and the two arrival pattern policies.} }\label{fullWaitFig}
\end{figure}

\edit{Here we again simulate the case investigation as an $M^B/M/c$ queue, as we did in Section~\ref{caseStudy1}. By comparison to that section of the case study, however, we do not vary staffing from week-to-week; only the arrival rate (and, by consequence, the batch size) are controlled across time. Batches are once again sampled in order from the data and split according to the arrival rate via Algorithm~\ref{batchBreak}. We consider two different styles of arrival rate, $\lambda = 1$ and $\lambda = m^{1/2}$. }

\edit{In Figure~\ref{fullWaitFig}, we plot the average wait for each day's cases under the three service durations and two arrival rate scenarios. As Figure~\ref{fullOLfig} suggested, all size duration-rate combinations are overwhelmed by the Omicron surge. In this, immediately we see a contrast between the two arrival rate policies. Across the three mean duration options, waiting persists for the full horizon if $\lambda = 1$, but there are many stretches where $\lambda = m^{1/2}$ eliminates waiting entirely. This becomes particularly notable in the recovery from surges. By comparison, in the pre-vaccine surge in Winter 2020-21, across all three durations, the dynamic arrival rate delivers a mean wait under 0.1 days several weeks ahead of when the daily policy does. Furthermore, at the Delta surge in Summer 2021, the dynamic policy achieves virtually no wait, while the constant policy has wait at least as high as the service duration itself.}

\edit{This becomes more extreme at the Omicron surge in Winter 2021-22. For $\E{S} = 80$ minutes, the system essentially never recovers within the simulation window, regardless of the arrival policy. At $\E{S} = 50$ minutes, the dynamic policy meaningfully separates from the constant arrival rate only at the end of the simulation window. However, at $\E{S} = 20$ minutes, the dynamic policy is able to completely eliminate waiting early into 2022, while the constant policy remains mired in backlog.}

%\tr{obviously departs from reality -- a case that waits more than a few days has no public health value to trace}

\edit{Obviously, all of these simulations depart from reality, but this is especially true for $\E{S} = 50$ and $\E{S} = 80$ minutes. In those settings, the average wait exceeds 10 and 100 days, respectively, for several weeks. In reality, there is no public health value to conduct contact tracing for a case that waits more than just a few days. Hence, the true operation could likely recover and stabilize the wait earlier than Figure~\ref{fullWaitFig} indicates through the expiration of cases or blocking of arrivals. To capture this in the simulation, we could instead model the queue with a structure like a deterministic abandonment time for each customer. There could then be another form of an arrival pattern control problem that aims to minimize the rate of abandonment. That is an interesting model in its own right, but even the one studied here holds practical insight for the design of contact tracing. What the $\E{S} = 20$ minutes values reveal is, even though Figure~\ref{fullOLfig} shows that the overall case volume is unchanged, the system performance can be salvaged or recovered through managing the arrival pattern. By speeding up the arrivals and reducing the batch size, Figure~\ref{fullWaitFig} shows that the waiting drops considerably relative to the fixed daily arrival pattern. As in the previous two sections of this case study, this emphasizes what we believe is the primary takeaway from this paper: batch arrivals place a dangerous and deceptive stress on service systems, and thus necessitate careful management. }

\section{A Technical Lemma Based on Legendre Polynomials}\label{techAppend}

To support our general batch analysis we will now introduce a  technical lemma that extends \citet{sullivan1980approximation} to a probabilistic context. In \citet{sullivan1980approximation}, the authors use shifted, asymmetric Legendre polynomials to produce a sum of exponential functions of $x \geq 0$ that converges to the indicator function $\mathbf{1}\{x \leq c\}$ for any constant $c > 0$. These approximations make use of the generalized hypergeometric function $\,_3F_2(\cdot)$, which is defined
$$
\,_3F_2(a_1, a_2, a_3, b_1, b_2, x)
=
\sum_{i=0}^\infty
\frac{(a_1)_i (a_2)_i (a_3)_i}{ (b_1)_i (b_2)_i } \frac{x^i}{i!}
,
$$
where $(c)_i = \prod_{j=0}^{i-1} (c+j)$ is a rising factorial. By use of the dominated convergence theorem, in Lemma~\ref{legendreLemma} we generalize this result using a sum of moment generating functions of a continuous non-negative random variable. We find convergence to the cumulative distribution function of the random variable, as well as to the expectation of the product between the random variable and an indicator function. Therefore, this lemma provides a method to find this cumulative probability and expectation when one only has access to the moment generating function of the random variable. This is paramount to our  staffing analysis, and because of its generality we believe it may also be of use in other applications. For clarity's sake, we note that the moment generating functions used in this technique are for strictly negative space parameters and thus will exist for all distributions. These functions can thus be viewed as Laplace transforms of the density with real, negative arguments. It is worth noting that the batch scalings enable us to use this lemma, as the storage processes satisfy the required condition of continuous support but the queueing models do not.

\begin{lemma}\label{legendreLemma}
Let $X$ be a non-negative continuous random variable and let $\mathcal{M}(\cdot)$ be its moment generating function and let $\mathcal{M}'(\cdot)$ be its first derivative, i.e.
$
\mathcal{M}(z) = \E{e^{z X}}
$
and
$
\mathcal{M}'(z) = \frac{\mathrm{d}}{\mathrm{d}\theta} \E{e^{\theta X}} |_{\theta = z}
$.
Then, for the sequence $\{a_k^\ell \mid \ell, k \in \mathbb{Z}^+\}$ given by
\begin{align}
a_{k}^m
=
\left(-1\right)^{k+1}
{\ell \choose k}{\ell + k \choose k}
\,_3 F_2 \left(k, -\ell, \ell+1; 1, k+1; \frac{1}{e}\right)
\label{akDef}
,
\end{align}
the summation over the products between $a_k^\ell$ and $\mathcal{M}\left(-\frac{k}c\right)$ is such that
\begin{align}
\lim_{\ell \to \infty} \, \sum_{k=1}^\ell a_k^\ell \, \mathcal{M}\left(-\frac{k}c\right) = \PP{ X \leq c}
,
\end{align}
whereas the summation over the products between $a_k^\ell$ and $\mathcal{M}'\left(-\frac{k}c\right)$ is such that
\begin{align}
\lim_{\ell \to \infty} \, \sum_{k=1}^\ell a_k^\ell \, \mathcal{M}'\left(-\frac{k}c\right) = \E{X \mathbf{1}\{ X \leq c\}}
,
\end{align}
for all $c > 0$.
\end{lemma}
\proof{Proof.}
For $x \geq 0$ and $\ell \in \mathbb{Z}^+$, let the function $\mathcal{L}_\ell(x)$ be defined as
\begin{align}
\mathcal{L}_\ell(x)
=
\sum_{k=1}^\ell a_k^\ell e^{-\frac{k x} {c}}
,
\label{LmDef1}
\end{align}
where each $a_k^\ell$ is as given in Equation~\eqref{akDef}. By \citet{sullivan1980approximation}, we have that
$$
\int_0^\infty \left(\mathcal{L}_\ell(x) - \mathbf{1}\{x \leq c\}\right)^2 \mathrm{d}x \longrightarrow 0
,
$$
as $\ell \to \infty$, which implies that
$
\mathcal{L}_\ell(x) \longrightarrow \mathbf{1}\{x \leq c\}
$
pointwise for $x \in [0, c)$ and $x \in (c, \infty)$ as $m \to \infty$. Furthermore, from \citet{sullivan1980approximation} we also have that $\mathcal{L}_{\ell}(x)$ can be equivalently expressed
\begin{align}
\mathcal{L}_\ell(x)
=
- \int_0^1 \tilde{P}_\ell\left(\frac{w}{e}\right) \frac{\mathrm{d}}{\mathrm{d}w}\tilde{P}_\ell\left(w e^{-\frac{x}c}\right) \mathrm{d}w
,
\label{LmDef2}
\end{align}
where $\tilde{P}_\ell(\cdot)$  is a shifted, asymmetric Legendre polynomial defined by
$$
\tilde{P}_\ell(w)
=
\sum_{k=0}^\ell {\ell \choose k}{\ell+k \choose k} (-w)^k
,
$$
for $w \in [0,1]$. For reference, this can be connected to a standard Legendre polynomial $P_\ell(\cdot)$ via the transformation
$
\tilde{P}_\ell(w)
=
P_\ell(1 - 2w)
.
$
To employ the dominated convergence theorem, we now bound $|\mathcal{L}_\ell(x)|$ as follows. Via the integral definition in Equation~\eqref{LmDef2}, we can observe that the values of this function at the origin are $\mathcal{L}_\ell(0) = 1 + (-1)^{\ell+1} \tilde{P}_\ell(1/ e)$, meaning that $\mathcal{L}_\ell(0) \in (0,2)$ for all $\ell$. Hence, we now focus on the quantity when $x$ is positive. In this case, we can see that
$$
\sup_{x > 0} \,
\left|
\int_0^1 \tilde{P}_\ell\left(\frac{w}{e}\right) \frac{\mathrm{d}}{\mathrm{d}w}\tilde{P}_\ell\left(w e^{-\frac{x}c}\right) \mathrm{d}w
\right|
\,
\leq
\,
\sup_{x > 0} \,
\left|
\int_0^1  \frac{\mathrm{d}}{\mathrm{d}w}\tilde{P}_\ell\left(w e^{-\frac{x}c}\right) \mathrm{d}w
\right|
,
$$
which can be explained as follows. Note $x$ dictates how much or how little to integrate along $\frac{\mathrm{d}}{\mathrm{d}w}\tilde{P}_\ell(w e^{-\frac{x}c})$. That is, at $x = 0$, the integral evaluates $\frac{\mathrm{d}}{\mathrm{d}w}\tilde{P}_\ell(w)$ at every point in its domain $[0,1]$ but for positive $x$ the derivative is only evaluated from 0 to $e^{-\frac{x}{c}}$.  Because we know that the shifted Legendre polynomial is bounded on $-1 \leq \tilde{P}_\ell(\cdot) \leq 1$, the integral on the left hand side is subject to negative values in both $\tilde{P}_\ell(w/e)$ and $\frac{\mathrm{d}}{\mathrm{d}w}\tilde{P}_\ell(w e^{-\frac{x}c})$, whereas the right hand side only has $\frac{\mathrm{d}}{\mathrm{d}w}\tilde{P}_\ell(w e^{-\frac{x}c})$. Note furthermore that $\tilde{P}_\ell(w/e)$ and $\frac{\mathrm{d}}{\mathrm{d}w}\tilde{P}_\ell(w e^{-\frac{x}c})$ cannot match in sign at every $w \in [0,1]$, as $\tilde{P}_\ell(w/e)$ is a polynomial of degree $\ell$ while $\frac{\mathrm{d}}{\mathrm{d}w}\tilde{P}_\ell(w e^{-\frac{x}c})$ is a polynomial of degree $\ell-1$. Thus, any interval that the integral on the left hand side evaluates over can be improved upon in the right hand side by evaluating only on a subinterval in which the derivative is positive, and it does so with a larger value as $\tilde{P}_\ell(w/e) \leq 1$. Integrating on the right hand side now leads us to the simpler form
$$
\sup_{x > 0} \,
\left|
\int_0^1  \frac{\mathrm{d}}{\mathrm{d}w}\tilde{P}_\ell\left(w e^{-\frac{x}c}\right) \mathrm{d}w
\right|
=
\sup_{x > 0} \,
\left|
1 - \tilde{P}_\ell\left(e^{-\frac{x}c}\right)
\right|
\leq
2,
$$
where the final bound again follows through the observation that $-1 \leq \tilde{P}_\ell(\cdot) \leq 1$.
With this bound in hand, to use the dominated convergence theorem we now review the specific convergence from \citet{sullivan1980approximation}. From \citet{sullivan1980approximation}, we have that $\mathcal{L}_\ell(x) \to \mathbf{1}\{x \leq c\}$ pointwise for $x \in [0, c)$ and $x \in (c, \infty)$. At the point of discontinuity in the indicator function at $x = c$, it can be observed that $\mathcal{L}_\ell(c) \to \frac{1}{2}$ as $\ell \to \infty$. Because the random variable $X$ is assumed to be continuous, the singleton $\{c\}$ is of measure 0 and thus $\mathcal{L}_\ell(x) \to \mathbf{1}\{x \leq c\}$ almost everywhere, justifying use of the dominated convergence theorem. Using this, we now have that
$$
\E{\mathcal{L}_\ell\left(X\right)}
\longrightarrow
\E{\mathbf{1} \{ X \leq c \} }
=
\PP{X \leq c}
\quad
\text{and}
\quad
\E{X \mathcal{L}_\ell\left(X\right)}
\longrightarrow
\E{X \mathbf{1} \{ X \leq c \} }
,
$$
as $\ell \to \infty$. Using the definition of $\mathcal{L}_\ell(x)$ in Equation~\eqref{LmDef1} and linearity of expectation, one can write
$$
\E{\mathcal{L}_\ell\left(X\right)}
=
\sum_{k=1}^\ell a_k^\ell \E{ e^{-\frac{k X} {c}} }
\quad
\text{and}
\quad
\E{X \mathcal{L}_\ell\left(X\right)}
=
\sum_{k=1}^\ell a_k^\ell \E{X e^{-\frac{k X} {c}} }
,
$$
and by observing that $\mathcal{M}'\left(-\frac{k}c\right) = \E{X e^{-\frac{k X} {c}} }$, we complete the proof.
\Halmos\\
\endproof

\begin{figure}[htb]
\vspace{-1.25in}
\centering
\includegraphics[width=0.9\columnwidth]{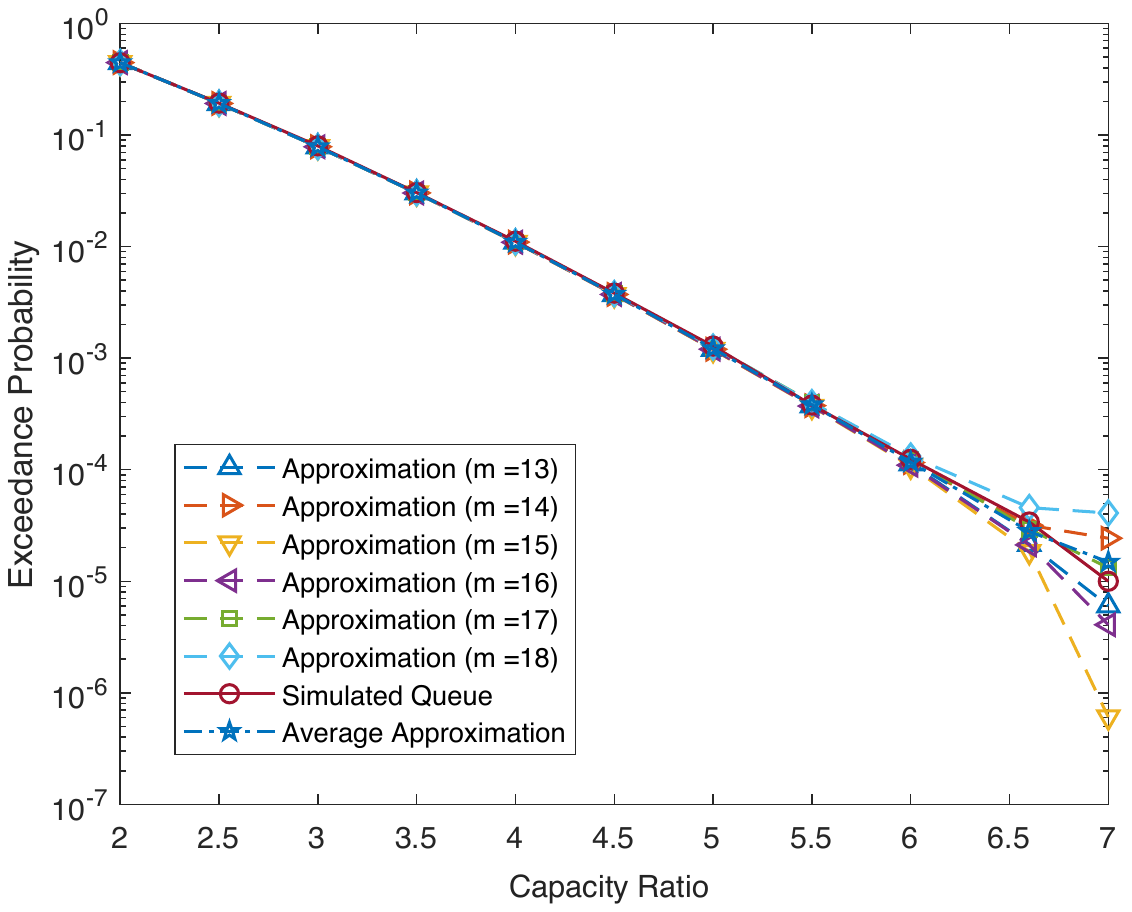}
\vspace{-1.45in}
%\vspace{-1.3pc}
\caption{Comparison of Legendre approximations and the empirical exceedance probability in a simulated queue with fixed size batches of size $n = 100$, $\lambda = 3$, and $\mu = 2$ (where $m$ is a stand-in for $\ell$ for typographic ease).}\label{fig:SimComp2}
\end{figure}

As a related numerical discussion, let us demonstrate how we perform approximate implementations of the expressions in Theorem~\ref{delayGen} as based on the Legendre exponential forms given in Lemma~\ref{legendreLemma}. As an initial observation, we can note that as $\ell$ grows large, calculations of the coefficients given in Equation~\eqref{akDef} become subject to numerical inaccuracies, such as overflow, due to the large binomial coefficients. While this could potentially be assuaged by use of Stirling's approximation or something similar, in our numerical experiments we have seen that such techniques may not be necessary for strong performance. However, we can note that the convergences in these results need not be monotone, hence we will not simply take the expression for the largest $\ell$ before numerical instability is observed. To explain through example, we will calculate the empirical  exceedance probability in the delay queueing model via simulation and compare it to various approximate Legendre sums. Based on Theorem~\ref{delayGen}, we have that
$$
\PP{Q^C(n) > c n}
\approx
\PP{\psi^C > c}
\approx
\frac{\frac{\lambda}{\mu} \E{M_1}   - \sigma_{\ell,c}^{(C)} }{c  - \sigma_{\ell,c}^{(C)} }
,
$$
and so we will consider candidate $\ell$ values, which we plot in Figure~\ref{fig:SimComp2}.

As one can see, for relatively small values of $\ell$ the approximation performs quite well, as the simulated values and the approximation are virtually indistinguishable before the true probability is approximately of order $10^{-5}$. However, if desired we can improve this further by taking the average among the candidate approximations. We can see that this does well in this example, and  we can quickly show it will do no worse than the worst individual approximation. For $p$ as the true probability and $p_{\sigma_\ell}$ as the approximation at $\ell$, by the triangle inequality we have that
$$
\left| \sum_{k=\ell_0}^{\ell_1} \frac{p_{\sigma_k}}{\ell_1 - \ell_0+1} - p \right|
=
\left| \sum_{k=\ell_0}^{\ell_1} \frac{p_{\sigma_k} - p }{\ell_1 - \ell_0+1}  \right|
\leq
 \sum_{k=\ell_0}^{\ell_1}  \frac{ \left| p_{\sigma_k} - p \right| }{\ell_1 - \ell_0+1}
 \leq
\max_{\ell_0 \leq k \leq \ell_1} \left| p_{\sigma_k} - p \right|
.
$$
Thus, a loose description of an approximation heuristic based on these Legendre limits is as follows: compute multiple candidate approximations, remove clear errors caused by numerical instabilities and pre-convergence gaps, and take the average of the remaining candidates. While our  experiments suggest that this simple approach does well, we can note that it could be possible to develop more sophisticated numerical approximations based on these limits and we find this to be an interesting direction of future research.

\section{Exploration of Service Time Dependence within Batches}\label{dependence}

\begin{figure}[ht!]
\centering
\includegraphics[width=\textwidth]{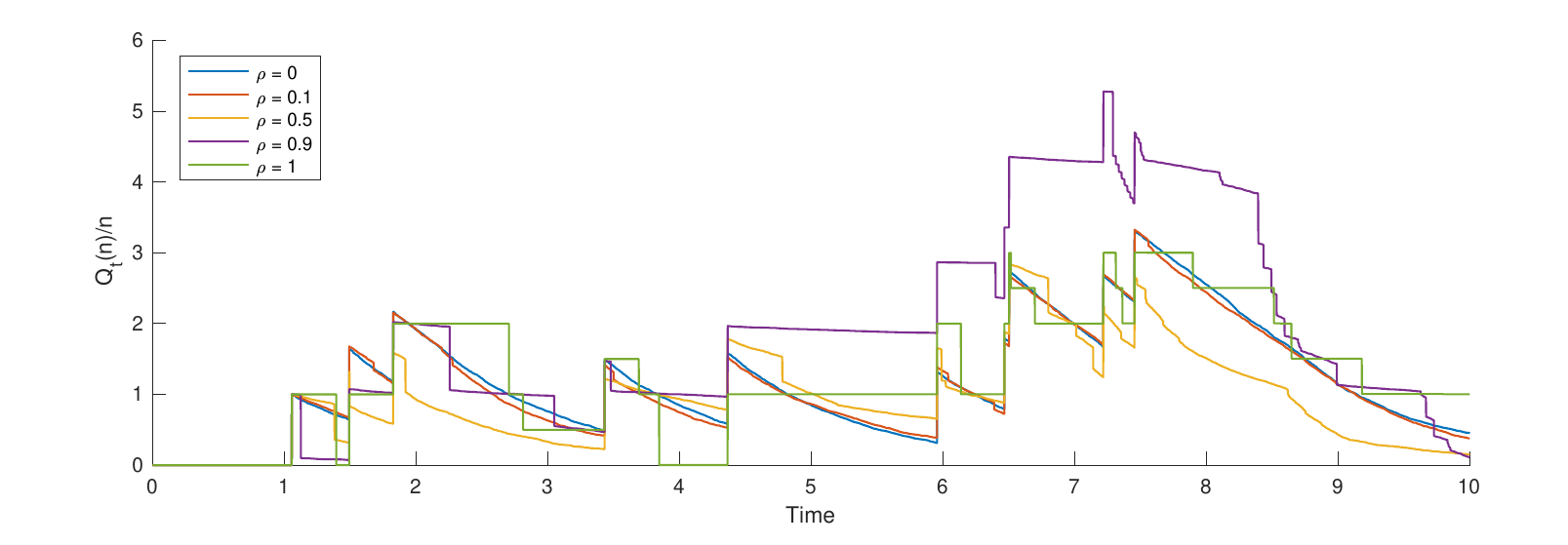}
\\
\footnotesize{(a)}
\\
\includegraphics[width=\textwidth]{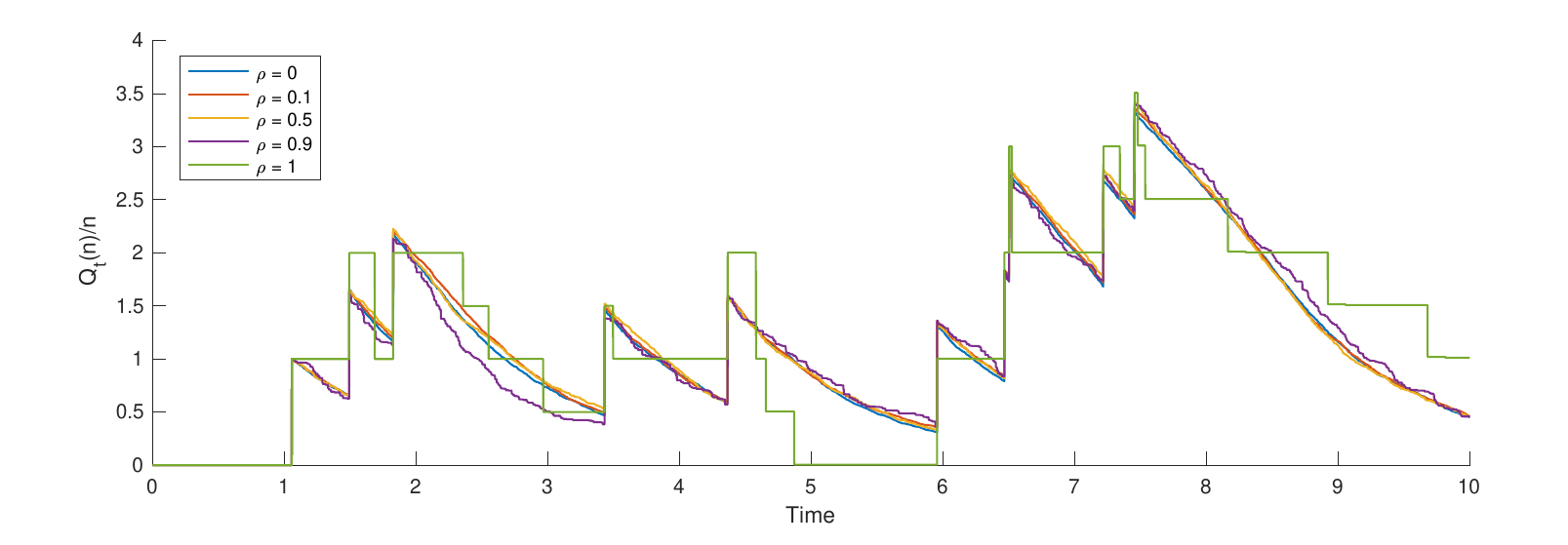}
\\
\footnotesize{(b)}
\\
\includegraphics[width=\textwidth]{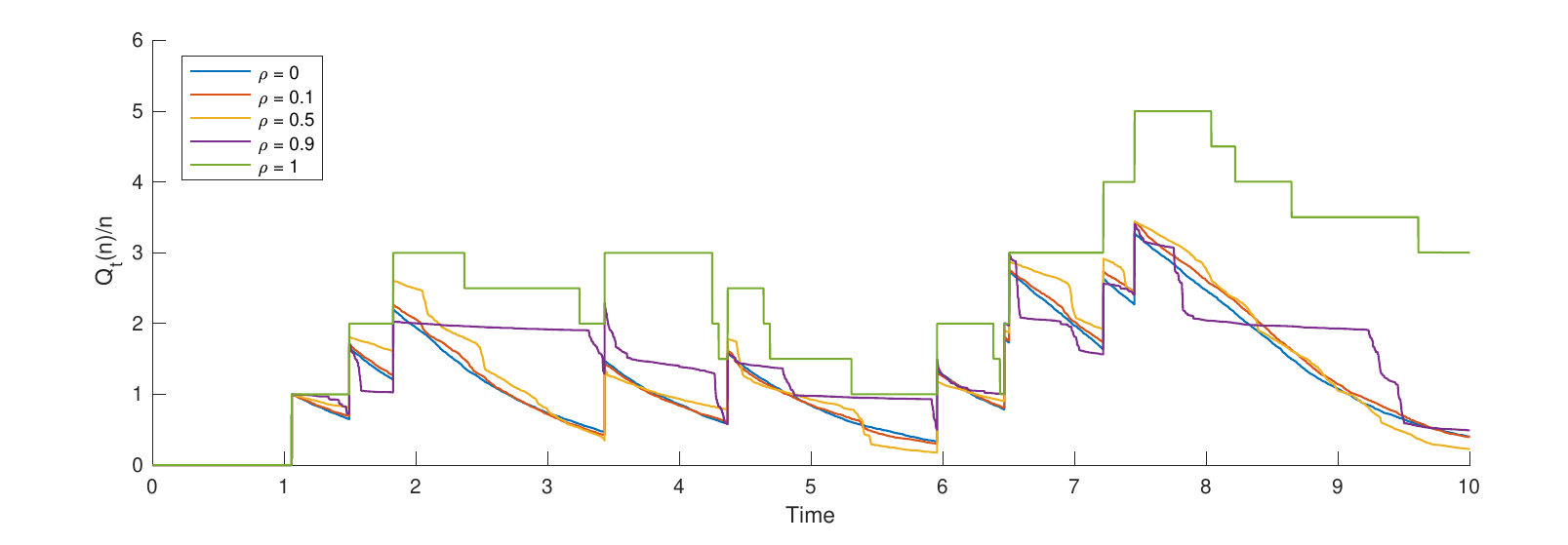}
\\
\footnotesize{(c)}
\caption{A comparison of queue length sample paths with dependent service durations within each arriving batch for various dependence structures. In all experiments, the service distribution is unit rate exponential service and the batch sizes are deterministic, with $n = 1000$ and $c = 1.5$.}\label{dependFig}
\end{figure}

For a final numerical experiment, let us also explore dependence within  batches of jobs.  
%Because the simulated future scenarios are generated from a common initial state in the look-ahead teleoperation, it is possible that the operators' response times could be correlated. 
As an empirical exploration of this, in Figure~\ref{dependFig} we plot normalized queue length processes under three different dependency structures. In each setting, there is a probability $\varrho \in [0,1]$ that each successive service time will be dependent.  In the first case, (a), each service time in a batch has probability $\varrho$ of being equal to the first duration within that batch and otherwise will be drawn independently, i.e. for an arbitrary batch $i$ and $j \geq 2$,
$$
S_{i,j}
=
\begin{cases}
S_{i,1} & \text{with probability $\varrho$,}\\
\tilde{S}_{i,j} & \text{otherwise,}
\end{cases}
$$
where $\tilde{S}_{i,j}$ is an independent draw from the service distribution. In case (b) this is instead equal to the previous time with probability $\varrho$ and independent otherwise, meaning
$$
S_{i,j}
=
\begin{cases}
S_{i,j-1} & \text{with probability $\varrho$,}\\
\tilde{S}_{i,j} & \text{otherwise,}
\end{cases}
$$
and in (c) each time is an average over all previous service times within the batch with probability $\varrho$ and again otherwise independently drawn:
$$
S_{i,j}
=
\begin{cases}
\frac{1}{j-1}\sum_{k=1}^{j-1}S_{i,k} & \text{with probability $\varrho$,}\\
\tilde{S}_{i,j} & \text{otherwise.}
\end{cases}
$$

In each of these settings, we plot simulated sample paths for $\varrho \in \{0,0.1,0.5,0.9,1\}$ and we hold the arrival epochs fixed across all the experiments. When $\varrho = 0$ all durations are independent regardless of the dependency setting, and thus these processes are effectively identical on this sample path due to the results in Theorem~\ref{fDelayConv}. Similarly, if $\varrho = 1$ the service times within each batch are perfectly correlated. Moreover, these processes are equivalently distributed across the dependency settings. In the case of infinitely many servers, the normalized queue length process can be trivially identified as a piecewise constant jump process (and in the case of deterministic batch sizes, this is an infinite server queue). However in the multi-server case the batch arrival queue is not as easily understood, and even more insight is lost in the intermediate settings of $\varrho \in \{0.1, 0.5, 0.9\}$. The previous time dependency in case (b) shows a subdued system level dependency, as the difference in sample paths between $\varrho = 0$ and $\varrho \in \{0.1, 0.5, 0.9\}$ is not as pronounced as in cases (a) and (c). This illustrates that batch scaling limits subject to dependency within batches may merit its own future study. It is worth noting though that in the infinite server setting there are immediately available extensions of Theorem~\ref{fBatchScale}. For example, for each arriving batch in case (a), there is a binomially distributed number of jobs that are identical in duration, with the remaining jobs independently drawn. Under the batch scaling limit, this means that a $\varrho$ fraction of each jump will contribute to a piecewise constant jump process while the remaining $1-\varrho$ will function as part of a shot noice process. Moreover, because the limits in Theorem~\ref{fBatchScale} make use of the law of large numbers, one could recover the infinite server batch scaling if the service durations are weakly dependent.

%It is also worth noting that even though the jobs within look-ahead assistance are simulated from the same initial states, evidence from the pioneering implementations of this methodology suggests that job durations can vary significantly even on the same simulated future scenario. In fact, \citet{lundgard2018bolt} finds that  the best of the instantaneously crowdsourced decisions often come from the operators who take the most time to respond. With this variance within batch in mind, perhaps it is more appropriate to consider a generalized model that addresses the effect of the underlying initial state through service duration distributions that change across batches. In the infinite server model this can still be understood in the large batch setting, as this would correspond to a service distribution that changes with each arrival epoch, i.e. $\bar{G}_i(\cdot)$ for the $i^\text{th}$ arrival. In this way, Theorem~\ref{fBatchScale} can be naturally and immediately extended. This generalization is more challenging to address in the multi-server setting, however, and thus this non-identically distributed model makes for an interesting and important direction of future study.

\printendnotes

\end{APPENDIX}

\end{document}